%% file: main.tex
\documentclass[apjl,ip,twocolumn]{aastex61}
\usepackage{amsmath}
\usepackage{comment}
\usepackage{ifthen}
\usepackage[switch]{lineno}

\input{newcommand_gen.tex}

\newcommand{\loopand}{\ifnum\value{planetcounter}=2 and \else\fi}
\newcommand{\loopcomma}{\ifnum\value{planetcounter}<2 ,\else. \fi}
\newcommand{\loopcommanoperiod}{\ifnum\value{planetcounter}<2 ,\else \space\fi}
\newcommand{\loopcommanospace}{\ifnum\value{planetcounter}<2 ,\else \fi}

\input{newcommand_spec_multi.tex}

\newcounter{planetcounter}

\newboolean{emulateapj}
\setboolean{emulateapj}{true}

\newboolean{rvtablelong}
\setboolean{rvtablelong}{true}

\newboolean{astroph}
\setboolean{astroph}{true}

\shortauthors{Hartman et al.}
\shorttitle{Transiting Giant Planets Around M Dwarfs}
\ifthenelse{\boolean{emulateapj}}{
    
}{
    
}

\begin{document}

\title{
TOI~4201\,b and TOI~5344\,b: Discovery of Two Transiting Giant Planets Around M Dwarf Stars and Revised Parameters for Three Others
}

\correspondingauthor{Joel Hartman}
\email{jhartman@astro.princeton.edu}

\author[0000-0001-8732-6166]{J. D. Hartman}
\affiliation{Department of Astrophysical Sciences, Princeton University, 4 Ivy Lane, Princeton, NJ 08544, USA}

\author[0000-0001-7204-6727]{G. \'A. Bakos}
\affiliation{Department of Astrophysical Sciences, Princeton University, 4 Ivy Lane, Princeton, NJ 08544, USA}

\author{Z. Csubry}
\affiliation{Department of Astrophysical Sciences, Princeton University, 4 Ivy Lane, Princeton, NJ 08544, USA}

\author[0000-0001-8638-0320]{A. W. Howard}
\affil{Department of Astronomy, California Institute of Technology, Pasadena, CA 91125, USA}

\author[0000-0002-0531-1073]{H. Isaacson}
\affil{Department of Astronomy, University of California at Berkeley, Berkeley, CA 94720, USA}

\author[0000-0002-8965-3969]{S. Giacalone}
\affil{Department of Astronomy, University of California at Berkeley, Berkeley, CA 94720, USA}

\author[0000-0003-1125-2564]{A. Chontos}
\affiliation{Department of Astrophysical Sciences, Princeton University, 4 Ivy Lane, Princeton, NJ 08544, USA}

\author[0000-0001-8511-2981]{N. Narita}
\affil{Komaba Institute for Science, The University of Tokyo, 3-8-1 Komaba, Meguro, Tokyo 153-8902, Japan}
\affil{Astrobiology Center, 2-21-1 Osawa, Mitaka, Tokyo 181-8588, Japan}
\affil{Instituto de Astrof\'isica de Canarias (IAC), E-38200 La Laguna, Tenerife, Spain}

\author[0000-0002-4909-5763]{A. Fukui}
\affil{Komaba Institute for Science, The University of Tokyo, 3-8-1 Komaba, Meguro, Tokyo 153-8902, Japan}
\affil{Instituto de Astrof\'isica de Canarias (IAC), E-38200 La Laguna, Tenerife, Spain}

\author[0000-0002-6424-3410]{J. P. de Leon}
\affil{Department of Multi-Disciplinary Sciences, Graduate School of Arts and Sciences, The University of Tokyo, 3-8-1 Komaba, Meguro, Tokyo 153-8902, Japan}

\author[0000-0002-7522-8195]{N. Watanabe}
\affil{Department of Multi-Disciplinary Sciences, Graduate School of Arts and Sciences, The University of Tokyo, 3-8-1 Komaba, Meguro, Tokyo 153-8902, Japan}

\author[0000-0003-1368-6593]{M. Mori}
\affil{Department of Multi-Disciplinary Sciences, Graduate School of Arts and Sciences, The University of Tokyo, 3-8-1 Komaba, Meguro, Tokyo 153-8902, Japan}

\author[0000-0002-5331-6637]{T. Kagetani}
\affil{Department of Multi-Disciplinary Sciences, Graduate School of Arts and Sciences, The University of Tokyo, 3-8-1 Komaba, Meguro, Tokyo 153-8902, Japan}

\author[0000-0002-9436-2891]{I. Fukuda}
\affil{Department of Multi-Disciplinary Sciences, Graduate School of Arts and Sciences, The University of Tokyo, 3-8-1 Komaba, Meguro, Tokyo 153-8902, Japan}

\author[0000-0002-0488-6297]{Y. Kawai}
\affil{Department of Multi-Disciplinary Sciences, Graduate School of Arts and Sciences, The University of Tokyo, 3-8-1 Komaba, Meguro, Tokyo 153-8902, Japan}

\author[0000-0002-5658-5971]{M. Ikoma}
\affil{Division of Science, National Astronomical Observatory of Japan, 2-21-1 Osawa, Mitaka, Tokyo 181-8588, Japan}

\author[0000-0003-0987-1593]{E. Palle}
\affil{Instituto de Astrof\'isica de Canarias (IAC), E-38200 La Laguna, Tenerife, Spain}
\affil{Departamento de Astrof\'isica, Universidad de La Laguna (ULL), E-38206 La Laguna, Tenerife, Spain}

\author{F. Murgas}
\affil{Instituto de Astrof\'isica de Canarias (IAC), E-38200 La Laguna, Tenerife, Spain}
\affil{Departamento de Astrof\'isica, Universidad de La Laguna (ULL), E-38206 La Laguna, Tenerife, Spain}

\author[0000-0002-2341-3233]{E. Esparza-Borges}
\affil{Instituto de Astrof\'isica de Canarias (IAC), E-38200 La Laguna, Tenerife, Spain}
\affil{Departamento de Astrof\'isica, Universidad de La Laguna (ULL), E-38206 La Laguna, Tenerife, Spain}

\author[0000-0001-5519-1391]{H. Parviainen}
\affil{Instituto de Astrof\'isica de Canarias (IAC), E-38200 La Laguna, Tenerife, Spain}
\affil{Departamento de Astrof\'isica, Universidad de La Laguna (ULL), E-38206 La Laguna, Tenerife, Spain}

\author[0000-0002-0514-5538]{L. G. Bouma}
\affil{Department of Astronomy, California Institute of Technology, Pasadena, CA 91125, USA}

\author{M. Cointepas}
\affil{Univ. Grenoble Alpes, CNRS, IPAG, F-38000 Grenoble, France}
\affil{Observatoire de Gen\`eve, Département d’Astronomie, Universit\'e de Gen\`eve, Chemin Pegasi 51b, 1290 Versoix, Switzerland}
\author[0000-0001-9003-8894]{X. Bonfils}
\affil{Univ. Grenoble Alpes, CNRS, IPAG, F-38000 Grenoble, France}

\author[0000-0003-3208-9815]{J.M. Almenara}
\affil{Observatoire de Gen\`eve, Département d’Astronomie, Universit\'e de Gen\`eve, Chemin Pegasi 51b, 1290 Versoix, Switzerland}

\author[0000-0001-6588-9574]{Karen A.\ Collins}
\affiliation{Center for Astrophysics \textbar \ Harvard \& Smithsonian, 60 Garden Street, Cambridge, MA 02138, USA}

\author[0000-0003-2781-3207]{Kevin I.\ Collins}
\affiliation{George Mason University, 4400 University Drive, Fairfax, VA, 22030 USA}

\author[0009-0009-5132-9520]{Howard M. Relles}
\affiliation{Center for Astrophysics \textbar \ Harvard \& Smithsonian, 60 Garden Street, Cambridge, MA 02138, USA}

\author[0000-0003-1464-9276]{Khalid Barkaoui}
\affiliation{Astrobiology Research Unit, Universit\'e de Li\`ege, 19C All\'ee du 6 Ao\^ut, 4000 Li\`ege, Belgium}
\affiliation{Department of Earth, Atmospheric and Planetary Science, Massachusetts Institute of Technology, 77 Massachusetts Avenue, Cambridge, MA 02139, USA}
\affil{Instituto de Astrof\'isica de Canarias (IAC), E-38200 La Laguna, Tenerife, Spain}

\author[0000-0001-8227-1020]{Richard P. Schwarz}
\affiliation{Center for Astrophysics \textbar \ Harvard \& Smithsonian, 60 Garden Street, Cambridge, MA 02138, USA}

\author{Ghachoui Mourad}
\affiliation{Oukaimeden Observatory, High Energy Physics and Astrophysics Laboratory, Cadi Ayyad University, Marrakech, Morocco}

\author{Mathilde Timmermans}
\affiliation{Astrobiology Research Unit, University of Li\`ege, All\'ee du 6 ao\^ut, 19, 4000 Li\`ege (Sart-Tilman), Belgium}

\author{Georgina Dransfield}
\affiliation{School of Physics \& Astronomy, University of Birmingham, Edgbaston, Birmingham B15 2TT, United Kingdom}

\author[0000-0001-9892-2406]{Artem Burdanov}
\affiliation{Department of Earth, Atmospheric and Planetary Sciences, MIT, 77 Massachusetts Avenue, Cambridge, MA 02139, USA}

\author[0000-0003-2415-2191]{Julien de Wit}
\affiliation{Department of Earth, Atmospheric and Planetary Sciences, MIT, 77 Massachusetts Avenue, Cambridge, MA 02139, USA}

\author{Emmanu\"el Jehin}
\affiliation{Space Sciences, Technologies and Astrophysics Research (STAR) Institute, Universit\'e de Li\`ege, All\'ee du 6 Ao\^ut 19C, B-4000 Li\`ege, Belgium}

\author[0000-0002-5510-8751]{Amaury H.M.J. Triaud}
\affiliation{School of Physics \& Astronomy, University of Birmingham, Edgbaston, Birmingham B15 2TT, United Kingdom}

\author[0000-0003-1462-7739]{Micha\"el Gillon}
\affiliation{Astrobiology Research Unit, University of Li\`ege, All\'ee du 6 ao\^ut, 19, 4000 Li\`ege (Sart-Tilman), Belgium}

\author[0000-0001-6285-9847]{Zouhair Benkhaldoun}
\affiliation{Oukaimeden Observatory, High Energy Physics and Astrophysics Laboratory, Faculty of sciences Semlalia, Cadi Ayyad University, Marrakech, Morocco
}

\author[0000-0003-1728-0304]{Keith Horne}
\affiliation{SUPA School of Physics and Astronomy, University of St~Andrews, North Haugh, St~Andrews KY16~9SS, Scotland, UK}

\author[0000-0003-3904-6754]{Ramotholo Sefako} 
\affiliation{South African Astronomical Observatory, P.O. Box 9, Observatory, Cape Town 7935, South Africa}

\author[0000-0002-5389-3944]{A. Jord\'an}
\affil{Facultad de Ingenier\'ia y Ciencias, Universidad Adolfo Ib\'{a}\~{n}ez, Av. Diagonal las Torres 2640, Pe\~{n}alol\'{e}n, Santiago, Chile}
\affil{Millennium Institute for Astrophysics, Chile}
\affil{Data Observatory Foundation, Chile}
\affil{El Sauce Observatory---Obstech, Chile}

\author[0000-0002-9158-7315]{R. Brahm}
\affil{Facultad de Ingenier\'ia y Ciencias, Universidad Adolfo Ib\'{a}\~{n}ez, Av. Diagonal las Torres 2640, Pe\~{n}alol\'{e}n, Santiago, Chile}
\affil{Millennium Institute for Astrophysics, Chile}
\affil{Data Observatory Foundation, Chile}

\author[0000-0001-7070-3842]{V. Suc}
\affil{Facultad de Ingenier\'ia y Ciencias, Universidad Adolfo Ib\'{a}\~{n}ez, Av. Diagonal las Torres 2640, Pe\~{n}alol\'{e}n, Santiago, Chile}
\affil{El Sauce Observatory---Obstech, Chile}

\author[0000-0002-2532-2853]{Steve~B.~Howell}
\affil{NASA Ames Research Center, Moffett Field, CA 94035, USA}

 \author[0000-0001-9800-6248]{E. Furlan}
\affiliation{NASA Exoplanet Science Institute, Caltech/IPAC, Mail Code 100-22, 1200 E. California Blvd., Pasadena, CA 91125, USA}

\author[0000-0001-5347-7062]{J.E. Schlieder}
\affiliation{NASA Goddard Space Flight Center, 8800 Greenbelt Road, Greenbelt, MD 20771, USA}

\author[0000-0002-5741-3047]{D. Ciardi}
\affiliation{NASA Exoplanet Science Institute, Caltech/IPAC, Mail Code 100-22, 1200 E. California Blvd., Pasadena, CA 91125, USA}

\author[0000-0001-7139-2724]{T. Barclay}
\affiliation{NASA Goddard Space Flight Center, 8800 Greenbelt Road, Greenbelt, MD 20771, USA}

\author[0000-0002-9329-2190]{E.J. Gonzales}
\affiliation{Department of Astronomy and Astrophysics, University of California Santa Cruz, Santa Cruz, CA 95064, USA}

\author[0000-0002-1835-1891]{I. Crossfield}
\affiliation{Department of Physics \& Astronomy, University of Kansas, KS 66045, USA }

\author[0000-0001-8189-0233]{C. D. Dressing}
\affil{Department of Astronomy, University of California at Berkeley, Berkeley, CA 94720, USA}

\author[0000-0003-2228-7914]{M. Goliguzova}
\affil{Sternberg Astronomical Institute, Lomonosov Moscow State University, Universitetskii, 13, Moscow, Russia}

\author[0000-0002-4398-6258]{A. Tatarnikov}
\affil{Sternberg Astronomical Institute, Lomonosov Moscow State University, Universitetskii, 13, Moscow, Russia}

\author[0000-0003-2058-6662]{George~R.~Ricker}
\affiliation{Department of Physics and Kavli Institute for Astrophysics and Space Research, Massachusetts Institute of Technology, Cambridge, MA 02139, USA}

\author[0000-0001-6763-6562]{Roland~Vanderspek}
\affiliation{Department of Physics and Kavli Institute for Astrophysics and Space Research, Massachusetts Institute of Technology, Cambridge, MA 02139, USA}

\author[0000-0001-9911-7388]{David~W.~Latham}
\affiliation{Center for Astrophysics \textbar \ Harvard \& Smithsonian, 60 Garden Street, Cambridge, MA 02138, USA}

\author[0000-0002-6892-6948]{S.~Seager}
\affiliation{Department of Physics and Kavli Institute for Astrophysics and Space Research, Massachusetts Institute of Technology, Cambridge, MA 02139, USA}
\affiliation{Department of Earth, Atmospheric and Planetary Sciences, Massachusetts Institute of Technology, Cambridge, MA 02139, USA}
\affiliation{Department of Aeronautics and Astronautics, MIT, 77 Massachusetts Avenue, Cambridge, MA 02139, USA}

\author[0000-0002-4265-047X]{Joshua~N.~Winn}
\affiliation{Department of Astrophysical Sciences, Princeton University, 4 Ivy Lane, Princeton, NJ 08544, USA}

\author[0000-0002-4715-9460]{Jon~M.~Jenkins}
\affiliation{NASA Ames Research Center, Moffett Field, CA 94035, USA}

\author[0009-0008-5145-0446]{Stephanie Striegel}
\affiliation{SETI Institute, 189 Bernardo Ave, Suite 200, Mountain View, CA 94043, USA}
\affiliation{NASA Ames Research Center, Moffett Field, CA 94035, USA}

\author[0000-0002-1836-3120]{Avi~Shporer}
\affiliation{Department of Physics and Kavli Institute for Astrophysics and Space Research, Massachusetts Institute of Technology, Cambridge, MA 02139, USA}

\author[0000-0001-7246-5438]{Andrew~Vanderburg}
\affiliation{Department of Physics and Kavli Institute for Astrophysics and Space Research, Massachusetts Institute of Technology, Cambridge, MA 02139, USA}

\author[0000-0001-8172-0453]{Alan~M.~Levine}
\affiliation{Department of Physics and Kavli Institute for Astrophysics and Space Research, Massachusetts Institute of Technology, Cambridge, MA 02139, USA}

\author[0000-0001-9786-1031]{Veselin B. Kostov}
\affiliation{NASA Goddard Space Flight Center, 8800 Greenbelt Road, Greenbelt, MD 20771, USA}
\affiliation{SETI Institute, 189 Bernardo Ave, Suite 200, Mountain View, CA 94043, USA}

\author[0000-0002-3555-8464]{David Watanabe}
\affiliation{Planetary Discoveries in Valencia, CA 91354, USA}

\begin{abstract}

\setcounter{footnote}{10}
We present the discovery from the {\em TESS} mission of two giant
planets transiting M dwarf stars: \hatcurb{4201} and
\hatcurb{5344}. We also provide precise radial velocity
measurements and updated system parameters for three other
M dwarfs with transiting giant planets: \hatcur{519}, \hatcur{3629} and
\hatcur{3714}. We measure planetary masses of
\hatcurPPm{519}\,\mjup, \hatcurPPm{3629}\,\mjup,
\hatcurPPm{3714}\,\mjup, \hatcurPPm{4201}\,\mjup, and
\hatcurPPm{5344}\,\mjup\ for \hatcurb{519}, \hatcurb{3629},
\hatcurb{3714}, \hatcurb{4201}, and \hatcurb{5344}, respectively. The corresponding stellar masses are \hatcurISOm{519}\,\msun,
\hatcurISOm{3629}\,\msun, \hatcurISOm{3714}\,\msun,
\hatcurISOm{4201}\,\msun\ and \hatcurISOm{5344}\,\msun. All five hosts have super-solar
metallicities, providing further support for recent findings that,
like for solar-type stars, close-in giant planets are preferentially
found around metal-rich M dwarf host stars.
Finally, we
describe a procedure for accounting for systematic errors in stellar
evolution models when those models are included directly in fitting a
transiting planet system. 
\setcounter{footnote}{0}
\end{abstract}

\keywords{
    planetary systems ---
    stars: individual (
\setcounter{planetcounter}{1}
\hatcur{519},
\setcounter{planetcounter}{2}
\hatcur{3629},
\setcounter{planetcounter}{3}
\hatcur{3714},
\setcounter{planetcounter}{4}
\hatcur{4201},
\setcounter{planetcounter}{5}
\hatcur{5344},
\setcounter{planetcounter}{6}
) 
    techniques: spectroscopic, photometric
}


\section{Introduction}
\label{sec:introduction}

M dwarf stars are the most common type of star in the Galaxy, making
up approximately 70\% of the local stellar population
\citep{henry:2018}. These stars often host short-period
Earth-size planets \citep{dressing:2015}, but they are not expected to
be common hosts of giant planets. Theoretical predictions based on the
core-accretion model of planet formation suggest that the occurrence rate
of giant planets around $0.5$\,\msun\ stars is only 10\% that of giant
planets around G dwarfs, and that it drops to essentially zero for
stars with masses less than $\sim 0.4$\,\msun\ \citep{burn:2021}.
However, other work by \citet{mercer:2020} has indicated that the
formation of giant planets via gravitational instabilities may actually be
enhanced around M dwarfs.

The vast majority of stars currently known to host giant planets have
F, G or K spectral types. Based on the NASA Exoplanet Archive \citep{ps}\footnote{Accessed on 2023-06-22 at 14:16}, only 18 of the 733 giant
planets with $M_{p}\sin i > 0.1$\,M$_{J}$ discovered to
  date by the radial velocity (RV) method orbit M dwarf stars\footnote{For the purposes of this discussion we take M dwarfs to be stars with
  $\teffstar < 4000$\,K, which may include late K dwarfs as well.}, and only 16 of the 621 confirmed giant
  planets discovered to date by the transit method orbit M
  dwarfs. The low representation of M dwarf hosts
  within the current sample of giant planets suggests
  that giant planets are not common around M dwarfs. However,
  such observations can be misleading because the current
  sample derives from surveys that are biased toward
  solar-type host stars. 

Statistical analyses of early RV surveys found hints that the
occurrence rate of giant planets is indeed smaller for M dwarfs
than for solar-type stars \citep{johnson:2010}, but the small sample
size left large uncertainties on the occurrence estimate. More
recently, \citet{sabotta:2021} and \citet{pass:2023} have reported,
based on RV surveys, occurrence rates for giant planets around M
dwarfs that are below that for solar mass host stars. Occurrence rates
for giant planets around M dwarfs have also been determined from
observations by the NASA {\em TESS} mission \citep{ricker:2015} by
\citet{bryant:2023} and \citet{gan:2023}, with these limits being focused on much shorter
period planets ($P < 10$\,d) than the RV surveys which are often sensitive out to $P \sim
1000$\,d. \citet{bryant:2023} find an occurrence rate of $0.194 \pm 0.072$\% for close-in giant planets around M dwarfs, while \citet{gan:2023} find a rate of $0.27 \pm 0.09$\%. The rate of hot Jupiters around M dwarfs thus appears to be lower than the rate for solar-type stars by a
factor of $\sim 2.5$--$5$. For mid-to-late M dwarfs, the rate is still $\gtrsim 0.1$\%, which is perhaps larger than theoretical predictions.

The very large population of M dwarfs in the Galaxy, coupled with
a giant planet occurrence rate that is small, but not vanishingly small, means
that there should be a substantial number of giant planets
hosted by such stars. Assuming the occurrence rate of giant planets
around M dwarfs is 1/5th that of giant planets around G stars, we
might expect $\sim$1/3rd of giant planets in the Galaxy to orbit M
dwarfs, whereas only 2--3\% of the existing sample of giant planets
have M dwarf hosts.

Finding this missing population of giant planets around M dwarfs would
not only allow for a better determination of the occurrence rate of
these planets, and thus better tests of planet formation theories, it
would also create better opportunities to characterize the physical
properties of a poorly studied class of exoplanets. This could include
leveraging the very deep transits of giant planets that transit M
dwarfs to study the atmospheres of these planets via transmission
spectroscopy \citep{charbonneau:2002}, or leveraging starspot crossing events in transiting
systems to measure the obliquities of the stellar hosts in these
systems \citep[e.g.,][]{sanchisojeda:2011}. To date no such studies have been published for the currently
small sample of giant planets known to transit M dwarfs. 

In recent years there has been an increase in the rate of finding giant planets around M dwarfs. This increase is due in part to RV surveys focused on M
dwarfs \citep{reiners:2018}, as well as transiting systems discovered by {\em
  Kepler} \citep{johnson:2012}, HATSouth \citep{hartman:2015:hats6,bakos:2018:hats71,jordan:2022:hats74hats77}, NGTS \citep{bayliss:2018}, and
{\em TESS} \citep{canas:2020,artigau:2021,kanodia:2021,parviainen:2021,gan:2022,canas:2022,hobson:2023,kanodia:2022,kanodia:2023}. Key factors contributing to the recent yield
of such systems have been the development of IR sensitive
high-precision RV instruments \citep[e.g.,][]{tamura:2012,artigau:2014,quirrenbach:2014}, transit surveys
that cover a larger number of faint M dwarf stars compared to earlier
surveys, efforts to use RV instruments on large telescopes, such as
Keck-I/HIRES, VLT/ESPRESSO, Subaru/IRD, and Magellan/PFS, to provide
RV confirmation of candidate transiting giant planets around these
faint M dwarfs, and the development of instruments capable of
simultaneous, multi-band, time-series photometry.

In this paper we present the discovery of two new transiting giant
planets around M dwarf stars by the {\em TESS} mission: \hatcurb{4201}
and \hatcurb{5344}. These planets were identified by {\em TESS},
followed up with a variety of ground-based photometric time-series
facilities, and confirmed via precise RV observations with
Keck-I/HIRES.  We also present new Keck-I/HIRES RV observations for
three other previously confirmed transiting giant planets around M
dwarf stars designated as \hatcur{519} \citep{parviainen:2021}, and \hatcur{3629} and
\hatcur{3714} \citep{canas:2022}. We combine the new RVs with new photometric
time-series observations presented here, and with previously
published observations, to update the physical parameters for these
three systems.

In analyzing these systems we follow a typical method of utilizing
theoretical stellar evolution models to constrain the properties of
the host stars. This method can yield higher precision parameter
values than purely empirical techniques, however it can also lead to
parameter constraints that are much tighter than systematic errors in
the evolution models themselves
\citep[e.g.,][]{tayar:2022,hobson:2023}. Systematic uncertainties
are often reported along with the statistical uncertainties \citep{hobson:2023},
but sometimes the systematic uncertainties are overestimates for parameters with empirical constraints,
such as the mean
stellar density \citep{eastman:2022}. In
this paper we also describe a method for incorporating systematic
errors in stellar evolution models into the analysis of transiting
planet systems in a manner that self-consistently allows for tighter
empirical constraints on parameters when available.

In the following section we discuss the observations that are used to
confirm and characterize each planetary system. In
section~\ref{sec:analysis} we describe the analysis methods. In
section~\ref{sec:discussion} we discuss the results.

\section{Observations}
\label{sec:obs}

All five of the systems presented here were first detected as
transiting planet candidates based on observations by the NASA {\em
  TESS} mission. Additional ground-based light curves and
high-precision RV measurements were gathered for each system and
combined with catalog astrometric and photometric data to determine
the parameters of the system (Section~\ref{sec:analysis}). Three of the
systems discussed here (\hatcur{519}, \hatcur{3629} and \hatcur{3714})
have already been studied in the literature. For these systems we
briefly describe the published data that we employed and then
describe our new RV observations and light curves.  For the two systems that are new discoveries (\hatcur{4201} and
\hatcur{5344}) we describe in greater detail the space-based and
ground-based light curves as well as the RV observations used to
confirm the objects as transiting planet systems. The light curves
used in the analysis of each system are summarized in
Table~\ref{tab:phfusummary}, while the RV data are summarized in
Table~\ref{tab:rvsummary}.

\subsection{TOI~519}

\citet{parviainen:2021} announced the discovery of
\hatcurb{519}, and \citet{kagetani:2023} published a mass
measurement for the planet based on Subaru/IRD RV
measurements. Transits of this system were first detected in the
Sector~7 observations from {\em TESS}. The {\em TESS} observations were
obtained at 120\,s cadence, and reduced to trend-filtered light
curves by the NASA Science Processing Operations Center (SPOC)
Pipeline at NASA Ames Research Center
\citep{jenkinsSPOC2016,jenkins:2010}. Multiple threshold crossing
events were detected and the target passed all of the data validation
tests conducted by the pipeline. The Sector~7 light curve was combined
with a variety of multi-band ground-based light curves to rule out a
stellar mass companion to \hatcur{519}. Here we combine the light
curves that were previously used by \citet{parviainen:2021} with new
RV observations from Keck-I/HIRES to determine the parameters of the
system.

\subsubsection{High-Contrast Imaging}

High-contrast imaging observations to search for resolved stellar
companions of \hatcur{519} have been reported on ExoFOP-TESS\footnote{\url{https://exofop.ipac.caltech.edu/tess/index.php}}, but were
not included in \citet{parviainen:2021}. These observations include
Adaptive Optics (AO) imaging in the $K$-band with the NIRC2 instrument on
the Keck-II~10\,m telescope on the night of 2019 Apr 7, speckle
imaging at 692\,nm and 880\,nm with the DSSI instrument on the
LDT~4.3\,m telescope on 2020 Feb 10, and speckle imaging at 562\,nm
and 832\,nm with the \'Alopeke instrument on the Gemini~8\,m
telescope on the night of 2020 Feb 18. No companions were detected in
any of these observations. The LDT/DSSI observations were reported in \citet{clark:2022}. The NIRC2 observations and reductions followed the approach described by \citet{schlieder:2021}. The estimated
PSF width of the resulting image is 0\farcs103, and a contrast of $\Delta K = 7.742$\,mag at
0\farcs5 separation is achieved. For the \'Alopeke observations, data were reduced as described in \citet{howell:2011} and the
$5\sigma$ contrast achieved is $\Delta m = 4.86$\,mag at 0\farcs5 for
the 832\,nm observation, and $\Delta m = 4.01$\,mag at 0\farcs5 for
the 562\,nm observation.

\subsubsection{Light Curves}
\label{sec:519lc}

In addition to the 120\,s cadence {\em TESS} light curve from Sector~7
that was produced by SPOC and included in the analysis of
\citet{parviainen:2021}, \hatcur{519} was also observed by {\em TESS}
at 120\,s cadence during Sectors~34 and~61, as well as at 30\,min cadence via the full-frame-images (FFIs) gathered during
Sector~8. We obtained the 120\,s cadence SPOC light curves from MAST
for the three relevant sectors, while for Sector~8 an FFI light curve for this source was not publicly available, so we do not include those data in the analysis.
We used the Pre-search Data
Conditioning (PDC) light curves from the SPOC project
\citep{2014PASP..126..100S,2012PASP..124.1000S,Stumpe2012}.

Ground-based follow-up light curves of \hatcur{519} obtained with the
Las Cumbres Observatory Global Telescope (LCOGT) 1\,m network
\citep{brown:2013:lcogt} and with the MuSCAT2 multicolor imager
\citep{narita:2019} on the 1.52\,m Telescopio {\em Carlos S\'{a}nchez}
(TCS) at Teide Observatory have all been presented and discussed by
\citet{parviainen:2021}, who made these data available to us. 

A light curve of \hatcur{519} was obtained with the Chilean-Hungarian
Automated Telescope (CHAT) 0.7\,m telescope at Las Campanas
Observatory in Chile. These observations were carried out
through an $i$-band filter on the night of 24 March 2019, and were
reduced to trend-filtered light curves following \citet{jordan:2019}.

A total of six distinct transit events of \hatcurb{519} were observed
with the Exoplanets in Transits and their Atmospheres (ExTrA) facility
\citep{bonfils:2015}. Five of the events were observed by two of the
three 0.6\,m ExTrA telescopes, leading to a total of eleven independent
transit lights curves from ExTrA that we include in the analysis of
this system. Spectro-photometric observations were carried out over a
wavelength range of 0.85\,$\mu$m to 1.55\,$\mu$m, and were reduced to
light curves integrated over the full bandpass following
\citet{cointepas:2021}.

An additional transit of \hatcurb{519} was observed in $r$ and
$z_{s}$-bands on 2022 Mar 19 with MuSCAT2. These data were reduced
to light curves following \citet{parviainen:2020}.

\subsubsection{Radial Velocities}
\label{sec:519rv}

We obtained Keck-I/HIRES \citep{vogt:1994} observations of
\hatcur{519} between 2021 Oct 26 and 2022 Jan 08. A total of seven
observations were obtained through the I$_{2}$ cell,
together with a single I$_{2}$-free template exposure. Observations
were carried out through the California Planet Search
\citep[CPS;][]{howard:2010,howard:2016} queue, and were made using the
C2 decker with an exposure time of 900\,s. The seeing ranged between
0\farcs9 and 1\farcs7, and the total counts recorded by the HIRES
exposure meter for each observation were between 600 and 3200.

The HIRES spectra were reduced to relative radial velocity
measurements and corrected for barycentric motion following
standard CPS procedures. Spectral-line bisector span (BS) measurements were
determined following \citet{torres:2007:hat3}. The data are included in
Table~\ref{tab:rvs} and plotted in Figure~\ref{fig:toi519}.

\citet{kagetani:2023} published 18 RVs of \hatcur{519} derived from
mid-IR spectra obtained with Subaru/IRD
\citep{tamura:2012,kotani:2018}. We incorporated these RV measurements
into our joint analysis of this system.

\subsection{TOI~3629}

\hatcur{3629} was identified as as a transiting planet candidate by
\citet{canas:2022} who carried out a custom reduction of the
30\,min cadence {\em TESS} observations performed during Sector~17 of
the mission. This object was also independently selected as a
transiting planet candidate by the {\em TESS} Quick Look Pipeline
(QLP) \citep{huang:2020a}. \hatcur{3629} was also confirmed as a
transiting planet by \citet{canas:2022} based on RV observations with
the Habitable-zone Planet Finder
\citep[HPF;][]{mahadevan:2012,mahadevan:2014} and NEID
\citep{halverson:2016,schwab:2016} spectrographs.

We obtained Keck-I/HIRES observations \hatcur{3629}. Here we combine
the light curves and RVs for this system from \citet{canas:2022} with
the new Keck-I/HIRES RVs, and an additional Sector of {\em TESS}
observations, and a set of ground-based light curves not analyzed by
\citet{canas:2022} to update the parameters for the system.

\subsubsection{High-Contrast Imaging}

High-contrast imaging of \hatcur{3629} was reported by
\citet{canas:2022}, who ruled out bright companions with a magnitude difference $\Delta
      {\rm m} < 4$ at separations between 0\farcs2 and 1\farcs2 from
      \hatcur{3629}.

\subsubsection{Light Curves}

We make use of the 30\,min cadence {\em TESS} Sector~17 light curve of
\hatcur{3629} produced by the QLP, and made accessible on MAST. For
our analysis we use the detrended time series denoted with the keyword KSPSAP (i.e., it was produced through the {\em Kepler} Spline Simple Aperture Photometry method). In addition to
the Sector~17 observations, \hatcur{3629} was also observed at 120\,s
cadence during Sector~57 of the {\em TESS} mission. We make use of the
PDC light curve produced by SPOC from these data, which we obtained
from MAST.

Ground-based follow-up light curves of \hatcur{3629} have
been made available on the ExoFOP-TESS archive maintained by the NASA
Exoplanet Archive at IPAC. None of these light curves were previously
included in the analysis of \citet{canas:2022}, so we describe these
here. We do not include light curves that did not cover the transit event in this discussion.

An ingress was observed in $i^{\prime}$ with the SINISTRO imager on one of the LCOGT 1\,m telescopes at Teide Observatory, in Spain, on the night of 17 Oct 2021. Light curves were derived from these observations using {\sc AstroImageJ} \citep{collins:2017}.

A full transit of \hatcur{3629} was observed using the MuSCAT2 imager on the 1.52\,m TCS on the night of 24 Oct 2021. Observations were performed simultaneously in $g$, $r$, $i$, and $z_{s}$. Light curves were produced from the observations in a similar fashion to \hatcur{519} (Section~\ref{sec:519lc}).

While we restrict our analysis to the light curves discussed above, we
note that \citet{canas:2022} presented additional follow-up transit light curves
of \hatcur{3629} from the RBO~0.6\,m telescope, and the 1.55\,m Kuiper
Telescope. 

\subsubsection{Radial Velocities}

\citet{canas:2022} published RV observations of \hatcur{3629} from HPF
and NEID. These include a total of 23 HPF RVs obtained between 2021
Jan 18 and 22 Jan 14, and five NEID RVs obtained between 2021 Sep 21
and 2021 Nov 28. We included these published RVs in the re-analysis of
the system presented in Section~\ref{sec:analysis}.

We carried out spectroscopic observations of \hatcur{3629} with
Keck-I/HIRES between 2022 Jun 07 and 2022 Sep 18. A total of nine
spectra were gathered through the I$_{2}$ cell, and a single
I$_{2}$-free template spectrum was also obtained. Observations were
made through the C2 decker with seeing between 0\farcs9 and
1\farcs3. The template spectrum had an exposure time of 1200\,s, while
most of the I$_{2}$-in observations had exposure times of 900\,s. Two
of the I$_{2}$-in observations had shorter exposure times of 748\,s
and 796\,s. The exposure meter recorded between 4600 and 10,000 counts
during each observation. The observations were reduced to
high-precision RVs and BS measurements following the same methods as
discussed for TOI~519.

\subsection{TOI~3714}

\hatcur{3714} was selected as a candidate transiting planet system by the QLP project based on 30\,min cadence {\em TESS} FFI observations carried out in Sector~19. Like \hatcur{3629}, \hatcur{3714} was also previously confirmed by
\citet{canas:2022} based on HPF and NEID RVs. We independently
observed \hatcur{3714} with Keck-I/HIRES. Here we revise the parameters
for this system by combining the data from \citet{canas:2022} with the
new RVs.

\subsubsection{High-Contrast Imaging}

High-contrast imaging of \hatcur{3714} was reported by
\citet{canas:2022}, who ruled out bright companions with $\Delta {\rm
  m} < 4$ at separations between 0\farcs2 and 1\farcs2 from
\hatcur{3714}. \hatcur{3714}, however, does have a white dwarf
companion that is reported in Gaia DR3 \citep{gaiadr3} at a separation of
2\farcs67, and with $\Delta G = 4.56$ relative to \hatcur{3714}. \citet{canas:2022} measure a mass of $\sim
1.07$\,\msun\ and a cooling age of $\sim 2.4$\,Gyr for this white
dwarf. They adopted this as the age of the \hatcur{3714} system.

\subsubsection{Light Curves}
\label{sec:3714lc}

At the time of our analysis the only {\em TESS} observations available for \hatcur{3714} were the Sector~19 FFI data. We made use of the QLP light curve derived from these data, and available on MAST, and used the KSPSAP detrended time series in our analysis. 

An ingress of \hatcur{3714} was observed using the MUSCAT3 instrument \citep{Narita2020}
on the LCOGT 2.0\,m telescope at Haleakala on the night of 2021 Sep
03. Observations were gathered simultaneously through $g^{\prime}$,
$r^{\prime}$, $i^{\prime}$ and $z_{s}$ filters. 
The data reduction and differential photometry was performed using the pipeline described in \citet{Fukui2011}. 

An egress was observed
using MuSCAT2 on 2021 Aug 28 through $g$, $i$ and $z_{s}$ bands, while
a full transit was observed with this instrument on 2021 Sep 25
in the $g$, $r$, $i$ and $z_{s}$ bands. 
These data were reduced to light curves in a similar fashion to the MuSCAT2 observations of \hatcur{519}.

Two full transits of \hatcur{3714} were observed on the nights of 2022 Sep 21 and 2022 Oct 30 using the 0.6\,m TRAPPIST-North (TRAnsiting Planets and PlanetesImals Small Telescope, \citealp{Jehin2011,Gillon2011,Barkaoui2019_TN}) telescope. The first night of observations were carried out in the $I+z$-band, while the observations on the second night were performed using a $z$-band filter. Observations were scheduled using the tools of \citet{jensen:2013}, and the data were reduced to light curves following \citet{garcia:2022}. 

Two full transits of \hatcur{3714} were observed using the
SPECULOOS-North 1.0\,m telescope \citep{delrez:2018,sebastian:2021,Burdanov2022} and an Andor ikon-L imager on the nights of 2022 Oct 19 and 2022 Oct
30. The first transit was observed in $g^{\prime}$, while the second
was observed in $r$. The observations were scheduled following
\citet{sebastian:2021}, and reduced to light curves following
\citet{murray:2020} and \citet{garcia:2021,garcia:2022}.

In addition to the above transit observations, \citet{canas:2022} also
reported RBO 0.6\,m observations of \hatcur{3714} on the nights of
2021 Aug 16 and 2021 Nov 19, and observations with the ARCTIC imager
on the ARC 3.5\,m telescope at Apache Point Observatory on the night
of 2021 Nov 21. These data have not been published in an
electronically accessible form, and we do not include them in our
analysis of the system.

\subsubsection{Radial Velocities}

\citet{canas:2022} published RV observations of \hatcur{3714} from HPF
and NEID. These include a total of 12 HPF RVs obtained between 2021
Aug 24 and 2021 Dec 23, and eight NEID RVs obtained between 2021 Sep 22
and 2022 Jan 01. We included these published RVs in the re-analysis of
the system presented in Section~\ref{sec:analysis}.

We carried out spectroscopic observations of \hatcur{3714} with
Keck-I/HIRES between 2021 Sep 21 and 2021 Nov 28. A total of seven
spectra were gathered through the I$_{2}$ cell, and a single
I$_{2}$-free template spectrum was also obtained. Observations were
made through the C2 decker with seeing between 1\farcs0 and
1\farcs6. The template spectrum had an exposure time of 619\,s, while
most of the I$_{2}$-in observations had exposure times of 900\,s. One
of the I$_{2}$-in observations had an exposure time of 945\,s. The
exposure meter recorded between 2200 and 5000 counts during each
observation. The observations were reduced to high-precision RVs and
BS measurements following the same methods as discussed for TOI~519 (Section~\ref{sec:519rv}).

\subsection{TOI~4201}

\hatcur{4201} was identified as a transiting planet candidate by QLP based on a search of the {\em TESS} FFI observations gathered during Sector~6 of the mission. The {\em TESS} Science Office (TSO) reviewed the vetting information and issued an alert on 2021 July 12 following the process described by \citet{guerrero:2021}. Since confirmation of this object as a transiting planet has not yet been published, we describe all of the RV and photometric observations that we use to confirm the existence of this planet.

\subsubsection{High-Contrast Imaging}

High-contrast speckle imaging was performed with the Zorro instrument on the Gemini~8\,m telescope \citep{scott:2021}. Observations were obtained at 832\,nm and 562\,nm on 2023 Apr 24. The data were processed following the methods of \citep{howell:2011}. No companions were detected, and contrast limits of 5.28\,mag and 3.52\,mag are achieved at separations greater than 0\farcs5 in the 832\,nm and 562\,nm filters respectively. Furthermore, no neighbors are listed within 10\arcsec\ of \hatcur{4201} in the Gaia~DR3 catalog either.

\subsubsection{Light Curves}

We extracted an image-subtraction-based light curve from the {\em
  TESS} FFI Sector~6 observations of \hatcur{4201} following the
methods of the CDIPS project \citep{bouma:2019:cdips}. We decorrelated against trends in the time-series using a B-spline, and then applied the TFA algorithm to filter additional systematic variations from the data.

\hatcur{4201} was observed by the LCOGT~1\,m telescopes at Siding
Spring Observatory (SSO), the South Africa Astronomical Observatory
(SAAO) and Cerro Tololo Inter-American Observatory (CTIO) on the
nights of 2021 Sep 1, 2021 Sep 26, and 2021 Oct 3, respectively. The
first night was observed using an $i^{\prime}$ filter, while
observations on the second and third nights were obtained using both
$i^{\prime}$ and $g^{\prime}$ filters. Observations on the first night
were out of transit, so we do not include these data in the
analysis. The second night covered an ingress, while the third night
covered a full transit event. These data were reduced to light curves
using {\sc AstroImageJ} \citep{collins:2017}.

An full transit of \hatcur{4201} was observed in $z^{\prime}$ on 2023
Feb 24 using the SPECULOOS-North 1\,m telescope. These observations
and reductions were carried out in a similar fashion to the SPECULOOS
observations of \hatcur{3714} (Section~\ref{sec:3714lc}).

A full transit was observed on 2022 Jan 30 in $g$, $r$, and $z$-bands
with MuSCAT \citep{Narita2015}. These were reduced to light curves in a similar fashion
to the reduction of the MuSCAT3 observation of \hatcur{3714} (Section~\ref{sec:3714lc}).

Transits were observed with the ExTrA facility on 2022 Nov 2, 2022 Dec
15, 2023 Jan 20, and 2023 Mar 4. A total of eight independent
$0.85$\,$\mu$m--1.55\,$\mu$m light curves were obtained with the
various telescopes from this facility. The data were reduced in a similar manner as \hatcur{519} (Section~\ref{sec:519lc}).

\subsubsection{Radial Velocities}

We carried out spectroscopic observations of \hatcur{4201} with
Keck-I/HIRES between 2022 Sep 7 and 2023 Jan 10. A total of twelve
exposures were taken through the I$_{2}$ cell, and a single
I$_{2}$-free template spectrum was also obtained. Observations were
made through the C2 decker with seeing between 1\farcs0 and
1\farcs8. The template spectrum had an exposure time of 1200\,s, while
the I$_{2}$-in observations had exposure times between 900\,s and
1300\,s. The exposure meter recorded between 2500 and 5,000 counts
during each observation. The observations were reduced to
high-precision RVs and BS measurements following the same methods as
discussed for TOI~519 (Section~\ref{sec:519rv}).

\subsection{TOI~5344}

\hatcur{5344} was identified as a transiting planet candidate by the QLP based on a search of the {\em TESS} FFI observations gathered for this target during Sectors~43 and~44 of the mission. The TSO reviewed the vetting information and issued an alert on 2022 Feb 28. Since confirmation of this object as a transiting planet has not yet been published, we describe all of the RV and photometric observations that we use to confirm the existence of the planet.

\subsubsection{High-Contrast Imaging}

High-contrast speckle interferometric $I$-band imaging of \hatcur{5344} was obtained with the 2.5\,m telescope at the Caucasian Observatory of Sternberg Astronomical Institute (SAI) of Lomonosov Moscow State University \citep{strakhov:2023} on 2023 Jan 27. No companion is detected to a contrast limit of $\Delta I < 4$\,mag at separations of 0\farcs2 or more and $\Delta I < 7$\,mag at separations of 1\arcsec or more.

\subsubsection{Light Curves}

We use the QLP light curves of \hatcur{5344} derived from the {\em
  TESS} FFI observations from Sectors~43 and~44 of the mission. We
accessed these data from MAST, and make use of the KSPSAP detrended
time series for each sector.

A full transit of \hatcur{5344} was observed using the SBIG imager on
the LCOGT 0.4\,m telescope at Teide Observatory on the night of 2022
Mar 1. The observations were performed using an $i^{\prime}$ filter,
and were reduced to light curves using {\sc AstroImageJ} \citep{collins:2017}.

A full transit was also observed using the TRAPPIST-North 0.6\,m
telescope and an $I+z$ filter on the night of 2022 Aug
26. Observations were scheduled and reduced in a similar manner to
TRAPPIST-North observations of \hatcur{3714} (Section~\ref{sec:3714lc}).

Two transits were observed with the ExTrA facility on 2022 Dec 15 and
2023 Jan 22. A total of five independent $0.85$\,$\mu$m--1.55\,$\mu$m
light curves were obtained with the various telescopes from this
facility. The data were reduced in a similar fashion as \hatcur{519} (Section~\ref{sec:519lc}).

A full transit was observed with MuSCAT2 simultaneously in the $g$,
$r$, $i$, and $z_{s}$ bands on 2022 Dec 18. The observations were
carried out and reduced in a similar manner to \hatcur{519} (Section~\ref{sec:519lc}).

Finally, a full transit was observed with the SPECULOOS-North 1.0\,m
telescope through a $g^{\prime}$ filter on 2023 Jun 6. The
observations and reductions were carried out in a similar fashion to
the SPECULOOS observations and reductions of \hatcur{3714} (Section~\ref{sec:3714lc}).

\subsubsection{Radial Velocities}

We carried out spectroscopic observations of \hatcur{5344} with
Keck-I/HIRES between 2022 Sep 1 and 2023 Jan 10. A total of thirteen
exposures were taken through the I$_{2}$ cell, and a single
I$_{2}$-free template spectrum was also obtained. Observations were
made through the C2 decker with seeing between 0\farcs9 and
1\farcs6. The template spectrum had an exposure time of 1200\,s, while
the I$_{2}$-in observations had exposure times between 900\,s and
1200\,s. The exposure meter recorded between 3,000 and 5,000 counts
during each observation. The observations were reduced to
high-precision RVs and BS measurements following the same methods as
discussed for TOI~519 (Section~\ref{sec:519rv}).

\startlongtable
\ifthenelse{\boolean{emulateapj}}{
    \begin{deluxetable*}{llrrrr}
}{
    \begin{deluxetable}{llrrrr}
}
\tablewidth{0pc}
\tabletypesize{\scriptsize}
\tablecaption{
    Summary of photometric observations
    \label{tab:phfusummary}
}
\tablehead{
    \multicolumn{1}{c}{Instrument/Field\tablenotemark{a}} &
    \multicolumn{1}{c}{Date(s)} &
    \multicolumn{1}{c}{\# Images\tablenotemark{b}} &
    \multicolumn{1}{c}{Cadence\tablenotemark{c}} &
    \multicolumn{1}{c}{Filter} &
    \multicolumn{1}{c}{Precision\tablenotemark{d}} \\
    \multicolumn{1}{c}{} &
    \multicolumn{1}{c}{} &
    \multicolumn{1}{c}{} &
    \multicolumn{1}{c}{(sec)} &
    \multicolumn{1}{c}{} &
    \multicolumn{1}{c}{(mmag)}
}
\startdata
\sidehead{\textbf{\hatcur{519}}}
~~~~TESS/Sector 7 & 2019 Jan--2019 Feb & 16,240 & 120 & $T$ & 32.1 \\
~~~~TESS/Sector 34 & 2021 Jan--2021 Feb & 16,835 & 120 & $T$ & 21.5 \\
~~~~TESS/Sector 61 & 2023 Jan--2023 Feb & 15,453 & 120 & $T$ & 21.4 \\
~~~~CHAT~0.7\,m & 2019 Mar 24 & 31 & 258 & $i$ & 7.3 \\
~~~~LCOGT~1.0\,m & 2019 Mar 29 & 108 & 86 & $g^{\prime}$ & 20.4 \\
~~~~LCOGT~1.0\,m & 2019 Apr 1 & 53 & 176 & $i^{\prime}$ & 7.0 \\
~~~~LCOGT~1.0\,m & 2019 Apr 16 & 51 & 176 & $B$ & 26.9 \\
~~~~LCOGT~1.0\,m & 2019 Apr 16 & 77 & 176 & $z_{s}$ & 4.3 \\
~~~~MuSCAT2 & 2019 Nov 23 & 166 & 63 & $g$ & 19.4 \\
~~~~MuSCAT2 & 2019 Nov 23 & 166 & 63 & $r$ & 9.4 \\
~~~~MuSCAT2 & 2019 Nov 23 & 166 & 63 & $i$ & 5.9 \\
~~~~MuSCAT2 & 2019 Nov 23 & 166 & 63 & $z_{s}$ & 5.2 \\
~~~~MuSCAT2 & 2020 Jan 9 & 118 & 63 & $g$ & 56.5 \\
~~~~MuSCAT2 & 2020 Jan 9 & 116 & 63 & $r$ & 16.8 \\
~~~~MuSCAT2 & 2020 Jan 9 & 119 & 63 & $i$ & 8.3 \\
~~~~MuSCAT2 & 2020 Jan 9 & 118 & 63 & $z_{s}$ & 6.2 \\
~~~~MuSCAT2 & 2020 Jan 14 & 130 & 63 & $g$ & 53.9 \\
~~~~MuSCAT2 & 2020 Jan 14 & 131 & 63 & $r$ & 16.3 \\
~~~~MuSCAT2 & 2020 Jan 14 & 131 & 63 & $i$ & 8.0 \\
~~~~MuSCAT2 & 2020 Jan 14 & 131 & 63 & $z_{s}$ & 4.7 \\
~~~~MuSCAT2 & 2020 Feb 29 & 525 & 16 & $g$ & 60.2 \\
~~~~MuSCAT2 & 2020 Feb 29 & 137 & 60 & $r$ & 11.8 \\
~~~~MuSCAT2 & 2020 Feb 29 & 138 & 60 & $i$ & 7.5 \\
~~~~MuSCAT2 & 2020 Feb 29 & 137 & 60 & $z_{s}$ & 6.2 \\
~~~~ExTrA - tel2 & 2021 Jan 30 & 164 & 62 & 0.85--1.55\,$\mu$m & 15.5 \\
~~~~ExTrA - tel3 & 2021 Jan 30 & 163 & 62 & 0.85--1.55\,$\mu$m & 11.8 \\
~~~~ExTrA - tel2 & 2021 Feb  9 & 178 & 62 & 0.85--1.55\,$\mu$m & 18.6 \\
~~~~ExTrA - tel3 & 2021 Feb  9 & 179 & 62 & 0.85--1.55\,$\mu$m & 13.4 \\
~~~~ExTrA - tel2 & 2021 Mar  4 & 188 & 62 & 0.85--1.55\,$\mu$m & 12.1 \\
~~~~ExTrA - tel3 & 2021 Mar  4 & 189 & 62 & 0.85--1.55\,$\mu$m & 11.5 \\
~~~~ExTrA - tel2 & 2021 Mar  9 & 149 & 62 & 0.85--1.55\,$\mu$m & 11.5 \\
~~~~ExTrA - tel3 & 2021 Mar  9 & 147 & 62 & 0.85--1.55\,$\mu$m & 11.6 \\
~~~~ExTrA - tel2 & 2021 Mar 23 & 190 & 62 & 0.85--1.55\,$\mu$m & 14.9 \\
~~~~ExTrA - tel3 & 2021 Mar 23 & 191 & 62 & 0.85--1.55\,$\mu$m & 12.2 \\
~~~~ExTrA - tel2 & 2021 Apr 16 & 186 & 62 & 0.85--1.55\,$\mu$m & 16.9 \\
~~~~MuSCAT2 & 2022 Mar 19 & 150 & 91 & $r$ & 25.5 \\
~~~~MuSCAT2 & 2022 Mar 19 & 152 & 91 & $z_{s}$ & 7.2 \\
\sidehead{\textbf{\hatcur{3629}}}
~~~~TESS/Sector 17 & 2019 Oct 9--31 & 693 & 1800 & $T$ & 2.0 \\
~~~~TESS/Sector 57 & 2022 Sep--2022 Oct & 17,966 & 120 & $T$ & 5.9 \\
~~~~LCOGT~1.0\,m & 2021 Oct 17 & 244 & 90 & $i^{\prime}$ & 1.7 \\
~~~~MuSCAT2 & 2021 Oct 25 & 176 & 60 & $g$ & 4.4 \\
~~~~MuSCAT2 & 2021 Oct 25 & 215 & 51 & $r$ & 2.9 \\
~~~~MuSCAT2 & 2021 Oct 25 & 677 & 16 & $i$ & 3.9 \\
~~~~MuSCAT2 & 2021 Oct 25 & 420 & 26 & $z_{s}$ & 3.2 \\
\sidehead{\textbf{\hatcur{3714}}}
~~~~TESS/Sector 19 & 2019 Nov--2019 Dec & 812 & 1800 & $T$ & 3.9 \\
~~~~MuSCAT2 & 2021 Aug 28 & 96 & 91 & $g$ & 46.1 \\
~~~~MuSCAT2 & 2021 Aug 28 & 100 & 91 & $i$ & 19.5 \\
~~~~MuSCAT2 & 2021 Aug 28 & 72 & 91 & $z_{s}$ & 20.6 \\
~~~~MuSCAT3/LCOGT~2.0\,m & 2021 Sep 3 & 35 & 303 & $g^{\prime}$ & 1.6 \\
~~~~MuSCAT3/LCOGT~2.0\,m & 2021 Sep 3 & 98 & 111 & $r^{\prime}$ & 2.0 \\
~~~~MuSCAT3/LCOGT~2.0\,m & 2021 Sep 3 & 186 & 59 & $i^{\prime}$ & 1.9 \\
~~~~MuSCAT3/LCOGT~2.0\,m & 2021 Sep 3 & 184 & 60 & $z_{s}$ & 1.6 \\
~~~~MuSCAT2 & 2021 Sep 25 & 422 & 35 & $g$ & 8.6 \\
~~~~MuSCAT2 & 2021 Sep 25 & 214 & 71 & $r$ & 4.6 \\
~~~~MuSCAT2 & 2021 Sep 25 & 577 & 26 & $i$ & 4.0 \\
~~~~MuSCAT2 & 2021 Sep 25 & 581 & 26 & $z_{s}$ & 3.7 \\
~~~~TRAPPIST-North & 2022 Sep 21 & 256 & 70 & $I+z$ & 3.1 \\
~~~~SPECULOOS-North & 2022 Oct 19 & 149 & 130 & $g^{\prime}$ & 4.0 \\
~~~~SPECULOOS-North & 2022 Oct 30 & 222 & 67 & $r^{\prime}$ & 3.3 \\
~~~~TRAPPIST-North & 2022 Oct 30 & 160 & 100 & $z$ & 3.5 \\
\sidehead{\textbf{\hatcur{4201}}}
~~~~TESS/Sector 6 & 2018 Dec--2019 Jan & 977 & 1800 & $T$ & 8.7 \\
~~~~LCOGT~1.0\,m & 2021 Sep 26 & 23 & 545 & $g^{\prime}$ & 9.1 \\
~~~~LCOGT~1.0\,m & 2021 Sep 26 & 24 & 545 & $i^{\prime}$ & 6.5 \\
~~~~LCOGT~1.0\,m & 2021 Oct 3 & 26 & 546 & $g^{\prime}$ & 3.7 \\
~~~~LCOGT~1.0\,m & 2021 Oct 3 & 25 & 546 & $i^{\prime}$ & 4.2 \\
~~~~MuSCAT & 2022 Jan 30 & 124 & 121 & $g$ & 3.7 \\
~~~~MuSCAT & 2022 Jan 30 & 300 & 51 & $r$ & 2.4 \\
~~~~MuSCAT & 2022 Jan 30 & 298 & 51 & $z$ & 2.0 \\
~~~~ExTrA - tel2 & 2022 Nov 2 & 162 & 62 & 0.85--1.55\,$\mu$m & 6.5 \\
~~~~ExTrA - tel1 & 2022 Dec 15 & 152 & 62 & 0.85--1.55\,$\mu$m & 5.5 \\
~~~~ExTrA - tel2 & 2022 Dec 15 & 152 & 62 & 0.85--1.55\,$\mu$m & 6.5 \\
~~~~ExTrA - tel3 & 2022 Dec 15 & 152 & 62 & 0.85--1.55\,$\mu$m & 6.5 \\
~~~~ExTrA - tel3 & 2023 Jan 20 & 210 & 62 & 0.85--1.55\,$\mu$m & 7.1 \\
~~~~SPECULOOS-North & 2023 Feb 24 & 513 & 30 & $z^{\prime}$ & 3.9 \\
~~~~ExTrA - tel1 & 2023 Mar 4 & 184 & 62 & 0.85--1.55\,$\mu$m & 6.6 \\
~~~~ExTrA - tel2 & 2023 Mar 4 & 182 & 62 & 0.85--1.55\,$\mu$m & 7.8 \\
~~~~ExTrA - tel3 & 2023 Mar 4 & 184 & 62 & 0.85--1.55\,$\mu$m & 10.1 \\
\sidehead{\textbf{\hatcur{5344}}}
~~~~TESS/Sector~43 & 2021 Sep--2021 Oct & 2815 & 600 & $T$ & 7.5 \\
~~~~TESS/Sector~44 & 2021 Oct--2021 Nov & 2756 & 600 & $T$ & 6.0 \\
~~~~LCOGT~0.4\,m & 2022 Mar 1 & 26 & 417 & $i^{\prime}$ & 4.6 \\
~~~~TRAPPIST-North & 2022 Aug 26 & 150 & 80 & $I+z$ & 3.1 \\
~~~~ExTrA - tel1 & 2022 Dec 15 & 161 & 62 & 0.85--1.55\,$\mu$m & 5.1 \\
~~~~ExTrA - tel2 & 2022 Dec 15 & 160 & 62 & 0.85--1.55\,$\mu$m & 4.8 \\
~~~~ExTrA - tel3 & 2022 Dec 15 & 161 & 62 & 0.85--1.55\,$\mu$m & 4.7 \\
~~~~MuSCAT2 & 2022 Dec 18 & 341 & 41 & $g$ & 7.7 \\
~~~~MuSCAT2 & 2022 Dec 18 & 340 & 41 & $r$ & 4.5 \\
~~~~MuSCAT2 & 2022 Dec 18 & 649 & 21 & $i$ & 4.8 \\
~~~~MuSCAT2 & 2022 Dec 18 & 652 & 21 & $z_{s}$ & 5.3 \\
~~~~SPECULOOS-North & 2023 Jan 6 & 108 & 170 & $g^{\prime}$ & 4.0 \\
~~~~ExTrA - tel2 & 2023 Jan 22 & 103 & 62 & 0.85--1.55\,$\mu$m & 5.8 \\
~~~~ExTrA - tel3 & 2023 Jan 22 & 102 & 62 & 0.85--1.55\,$\mu$m & 5.1 \\
\enddata 
\tablenotetext{a}{ For {\em TESS} data we list the Sector from which the observations are taken.
}
\tablenotetext{b}{ Excluding any outliers or other data not included in the modelling. }
\tablenotetext{c}{ The median time between consecutive images rounded
  to the nearest second. Due to factors such as weather, the
  day--night cycle, guiding and focus corrections the cadence is only
  approximately uniform over short timescales.  } 
\tablenotetext{d}{
  The RMS of the residuals from the best-fit model. Note that in the case of {\em TESS} observations the transit may appear artificially shallower due to over-filtering and/or blending from unresolved neighbors. As a result the S/N of the transit may be less than what would be calculated from $\rpl/\rstar$ and the RMS estimates given here. }
\ifthenelse{\boolean{emulateapj}}{
    \end{deluxetable*}
}{
    \end{deluxetable}
}

%
%
\ifthenelse{\boolean{emulateapj}}{
    \begin{figure*}[!ht]
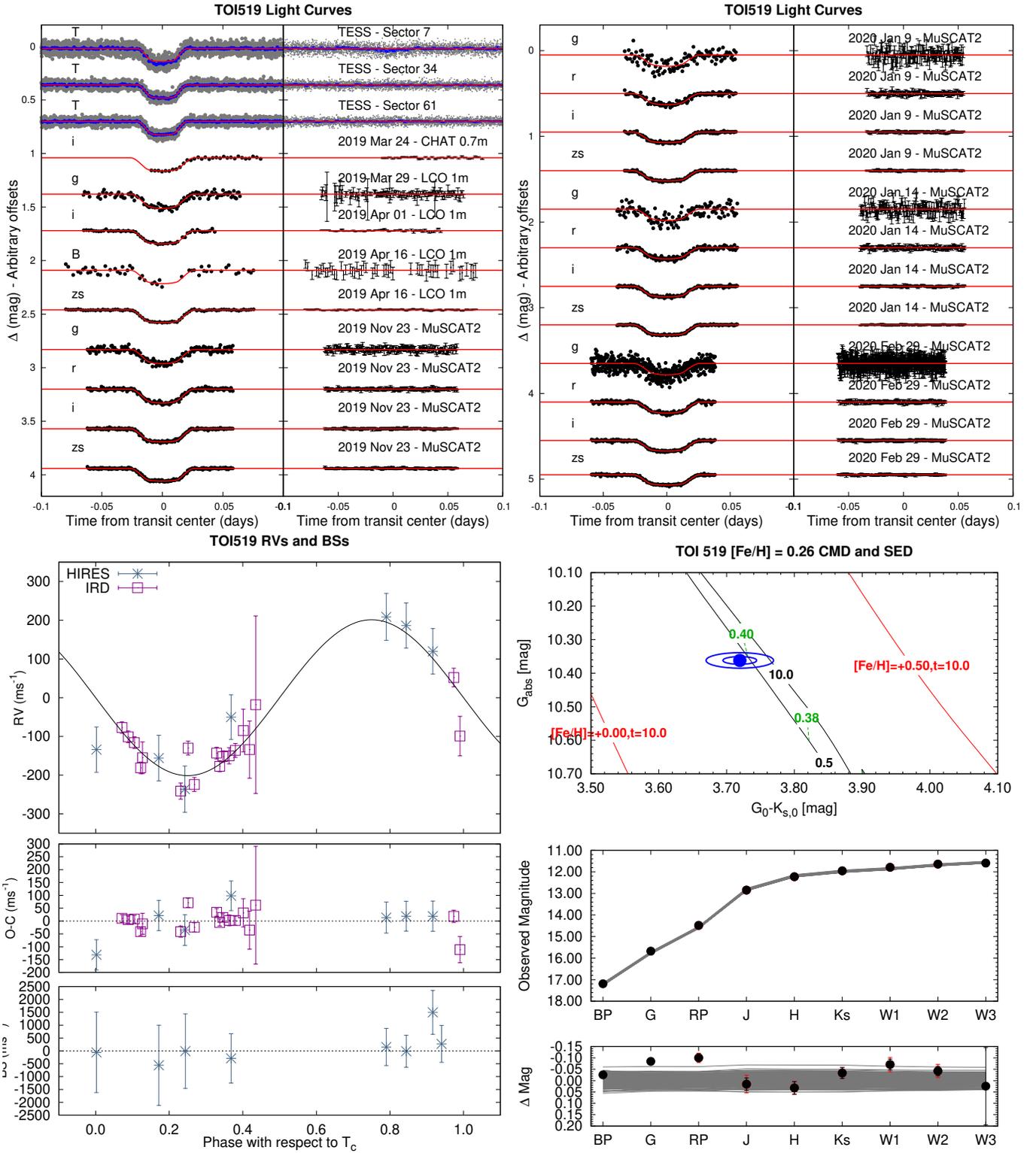

}{
    \begin{figure}[!ht]
}
 {
 \centering
 \leavevmode
 \includegraphics[width={1.0\linewidth}]{\hatcurhtr{519}-banner}
}
 {
 \centering
 \leavevmode
 \includegraphics[width={0.5\linewidth}]{\hatcurhtr{519}-lc}%
 \hfil
 \includegraphics[width={0.5\linewidth}]{\hatcurhtr{519}-lc2}%
 }
 {
 \centering
 \leavevmode
 \includegraphics[width={0.5\linewidth}]{\hatcurhtr{519}-rv}%
 \hfil
 \includegraphics[width={0.5\linewidth}]{\hatcurhtr{519}-iso-gk-gabs-isofeh-SED}%
 }                        
\caption{ Observations incorporated into the analysis of the
  transiting planet system \hatcur{519}. {\em Top:} Transit light
  curves with fitted model overplotted.
    The dates, filters and instruments used are indicated.  Additional light curves for this system are shown in Figure~\ref{fig:toi519extralcs}. (Caption continued on next page.) 
\label{fig:toi519}
}
\ifthenelse{\boolean{emulateapj}}{
    \end{figure*}
}{
    \end{figure}
}

%
%
\addtocounter{figure}{-1}
\ifthenelse{\boolean{emulateapj}}{
    \begin{figure*}[!ht]
}{
    \begin{figure}[!ht]
}
\caption{
    (Caption continued from previous page.)
For {\em TESS} we phase-fold the data, and plot the un-binned observations in grey, with the phase-binned values overplotted in blue. The residuals for each light curve are shown on the right-hand-side in the same order as the original light curves.  The error bars represent the photon
    and background shot noise, plus the readout noise. Note that these
    uncertainties are increased by a common factor in the fitting procedure to achieve a
    reduced $\chi^2$ of unity, but the uncertainties shown in the plot
    have not been scaled.
{\em Bottom Left:}
High-precision RVs phased with respect to the mid-transit time. 
The top panel shows the phased measurements together with the best-fit model.
The center-of-mass velocity has been subtracted. The middle panel shows the velocity $O\!-\!C$ residuals.
The error bars include the estimated jitter, which is varied as a free parameter in the fitting. The bottom panel shows the spectral line bisector spans {\em Bottom Right:} Color-magnitude diagram (CMD) and spectral energy distribution (SED). The top panel shows the absolute $G$ magnitude vs.\ the de-reddened $G - K_{S}$ color compared to
  theoretical isochrones (black lines) and stellar evolution tracks
  (green lines) from the MIST models interpolated at
  the best-estimate value for the metallicity of the host. The age
  of each isochrone is listed in black in Gyr, while the mass of each
  evolution track is listed in green in solar masses. The solid red lines show isochrones at higher and lower metallicities than the best-estimate value, with the metallicity and age in Gyr of each isochrone labelled on the plot. The filled
  blue circle shows the measured reddening- and distance-corrected
  values from Gaia DR2 and 2MASS, while the blue lines indicate
  the $1\sigma$ and $2\sigma$ confidence regions, including the
  estimated systematic errors in the photometry. The middle panel shows the SED as measured via broadband photometry through the listed filters. Here we plot the observed magnitudes without correcting for distance or extinction. Overplotted are 200 model SEDs randomly selected from the MCMC posterior distribution produced through the global analysis (gray lines). 
The model makes use of the predicted absolute magnitudes in each bandpass from the MIST isochrones, the distance to the system (constrained largely via Gaia DR2) and extinction (constrained from the SED with a prior coming from the {\sc mwdust} 3D Galactic extinction model).  
The bottom panel shows the $O\!-\!C$ residuals from the best-fit model SED. The errors listed in the catalogs for the broad-band photometry measurements are shown with black lines, while the errors including an assumed 0.02\,mag systematic uncertainty, which is added in quadrature to the listed uncertainties, are shown with red lines. These latter uncertainties are what we use in the fit.
\label{fig:toi519:labcontinue}}
\ifthenelse{\boolean{emulateapj}}{
    \end{figure*}
}{
    \end{figure}
}

\begin{figure}[!ht]
 {
 \centering
 \leavevmode
 \includegraphics[width={1.0\linewidth}]{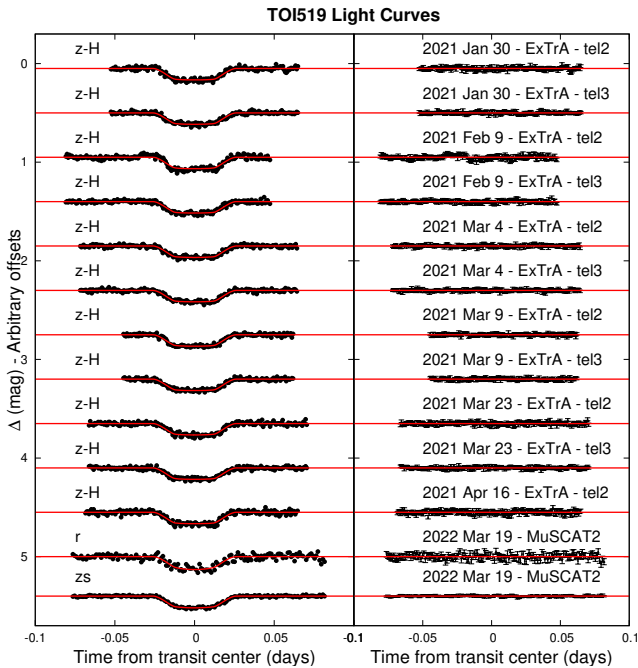}%
 }
\caption{Additional light curves used in the analysis of \hatcur{519} displayed in the same manner as the other light curves shown in Fig.~\ref{fig:toi519}.
}
\label{fig:toi519extralcs}
\end{figure}

%
%
\ifthenelse{\boolean{emulateapj}}{
    \begin{figure*}[!ht]
}{
    \begin{figure}[!ht]
}
 {
 \centering
 \leavevmode
 \includegraphics[width={1.0\linewidth}]{\hatcurhtr{3629}-banner}
}
 {
 \centering
 \leavevmode
 \includegraphics[width={0.5\linewidth}]{\hatcurhtr{3629}-TESSlc}%
 \hfil
 \includegraphics[width={0.5\linewidth}]{\hatcurhtr{3629}-lc}%
 }
 {
 \centering
 \leavevmode
 \includegraphics[width={0.5\linewidth}]{\hatcurhtr{3629}-rv}%
 \hfil
 \includegraphics[width={0.5\linewidth}]{\hatcurhtr{3629}-iso-gk-gabs-isofeh-SED}%
 }                        
\caption{
    Same as Figure~\ref{fig:toi519}, here we show the observations of \hatcur{3629} together with our best-fit model.
\label{fig:toi3629}
}
\ifthenelse{\boolean{emulateapj}}{
    \end{figure*}
}{
    \end{figure}
}

%
%
\ifthenelse{\boolean{emulateapj}}{
    \begin{figure*}[!ht]
}{
    \begin{figure}[!ht]
}
 {
 \centering
 \leavevmode
 \includegraphics[width={1.0\linewidth}]{\hatcurhtr{3714}-banner}
}
 {
 \centering
 \leavevmode
 \includegraphics[width={0.5\linewidth}]{\hatcurhtr{3714}-lc}%
 \hfil
 \includegraphics[width={0.5\linewidth}]{\hatcurhtr{3714}-lc2}%
 }
 {
 \centering
 \leavevmode
 \includegraphics[width={0.5\linewidth}]{\hatcurhtr{3714}-rv}%
 \hfil
 \includegraphics[width={0.5\linewidth}]{\hatcurhtr{3714}-iso-gk-gabs-isofeh-SED}%
 }                        
\caption{
    Same as Figure~\ref{fig:toi519}, here we show the observations of \hatcur{3714} together with our best-fit model.
\label{fig:toi3714}
}
\ifthenelse{\boolean{emulateapj}}{
    \end{figure*}
}{
    \end{figure}
}

%
%
\ifthenelse{\boolean{emulateapj}}{
    \begin{figure*}[!ht]
}{
    \begin{figure}[!ht]
}
 {
 \centering
 \leavevmode
 \includegraphics[width={1.0\linewidth}]{\hatcurhtr{4201}-banner}
}
 {
 \centering
 \leavevmode
 \includegraphics[width={0.5\linewidth}]{\hatcurhtr{4201}-lc}%
 \hfil
 \includegraphics[width={0.5\linewidth}]{\hatcurhtr{4201}-lc2}%
 }
 {
 \centering
 \leavevmode
 \includegraphics[width={0.5\linewidth}]{\hatcurhtr{4201}-rv}%
 \hfil
 \includegraphics[width={0.5\linewidth}]{\hatcurhtr{4201}-iso-gk-gabs-isofeh-SED}%
 }                        
\caption{
    Same as Figure~\ref{fig:toi519}, here we show the observations of \hatcur{4201} together with our best-fit model.
\label{fig:toi4201}
}
\ifthenelse{\boolean{emulateapj}}{
    \end{figure*}
}{
    \end{figure}
}

%
%
\ifthenelse{\boolean{emulateapj}}{
    \begin{figure*}[!ht]
}{
    \begin{figure}[!ht]
}
 {
 \centering
 \leavevmode
 \includegraphics[width={1.0\linewidth}]{\hatcurhtr{5344}-banner}
}
 {
 \centering
 \leavevmode
 \includegraphics[width={0.5\linewidth}]{\hatcurhtr{5344}-TESSlc}%
 \hfil
 \includegraphics[width={0.5\linewidth}]{\hatcurhtr{5344}-lc}%
 }
 {
 \centering
 \leavevmode
 \includegraphics[width={0.5\linewidth}]{\hatcurhtr{5344}-rv}%
 \hfil
 \includegraphics[width={0.5\linewidth}]{\hatcurhtr{5344}-iso-gk-gabs-isofeh-SED}%
 }                        
\caption{
    Same as Figure~\ref{fig:toi519}, here we show the observations of \hatcur{5344} together with our best-fit model.
\label{fig:toi5344}
}
\ifthenelse{\boolean{emulateapj}}{
    \end{figure*}
}{
    \end{figure}
}

\ifthenelse{\boolean{emulateapj}}{
    \begin{deluxetable*}{llrrrl}
}{
    \begin{deluxetable}{llrrrl}
}
\tablewidth{0pc}
\tabletypesize{\scriptsize}
\tablecaption{
    Summary of radial velocity observations.
    \label{tab:rvsummary}
}
\tablehead{
    \multicolumn{1}{c}{Instrument}          &
    \multicolumn{1}{c}{UT Date(s)}             &
    \multicolumn{1}{c}{\# Spec.}   &
    \multicolumn{1}{c}{S/N Range\tablenotemark{a}}           &
    \multicolumn{1}{c}{RV Precision\tablenotemark{b}} & 
    \multicolumn{1}{c}{Source} \\
    &
    &
    &
    &
    \multicolumn{1}{c}{(\ms)} &
}
\startdata
\noalign{\vskip -3pt}
\sidehead{\textbf{\hatcur{519}}}\\
\noalign{\vskip -13pt}
Subaru/IRD & 2021 May 4 -- 2022 Jan 24 & 18 & $\cdots$ & $42$ & \citet{kagetani:2023} \\
Keck-I/HIRES & 2021 Oct 26 & 1 & $44$ & $\cdots$ & this paper \\
Keck-I/HIRES+I$_{2}$ & 2021 Nov 19 -- 2022 Jan 8 & 7 & 24--57 & $70$ & this paper \\
\noalign{\vskip -3pt}
\sidehead{\textbf{\hatcur{3629}}}\\
\noalign{\vskip -13pt}
Hobby-Eberly/HPF & 2021 Jan 18 -- 2022 Jan 14 & 23 & $\cdots$ & $18$ & \citet{canas:2022} \\
WIYN 3.5m/NEID & 2021 Sep 21 -- 2021 Nov 28 & 5 & $\cdots$ & $13$ & \citet{canas:2022} \\
Keck-I/HIRES & 2022 Jun 7 & 1 & $100$ & $\cdots$ & this paper \\
Keck-I/HIRES+I$_{2}$ & 2022 Aug 11 -- 2022 Sep 18 & 9 & 68--77 & $16$ & this paper \\
\noalign{\vskip -3pt}
\sidehead{\textbf{\hatcur{3714}}}\\
\noalign{\vskip -13pt}
Hobby-Eberly/HPF & 2021 Aug 24 -- 2021 Dec 23 & 12 & $\cdots$ & $33$ & \citet{canas:2022} \\
Keck-I/HIRES+I$_{2}$ & 2021 Sep 21 & 1 & $71$ & $\cdots$ & this paper \\
WIYN 3.5m/NEID & 2021 Sep 22 -- 2022 Jan 8 & 8 & $\cdots$ & $11$ & \citet{canas:2022} \\
Keck-I/HIRES+I$_{2}$ & 2021 Sep 24 -- 2021 Nov 28 & 7 & 47--70 & $18$ & this paper \\
\noalign{\vskip -3pt}
\sidehead{\textbf{\hatcur{4201}}}\\
\noalign{\vskip -13pt}
Keck-I/HIRES & 2022 Sep 7 & 1 & $66$ & $\cdots$ & this paper \\
Keck-I/HIRES+I$_{2}$ & 2022 Sep 9 -- 2023 Jan 10 & 12 & 50--71 & $41$ & this paper \\
\noalign{\vskip -3pt}
\sidehead{\textbf{\hatcur{5344}}}\\
\noalign{\vskip -13pt}
Keck-I/HIRES+I$_{2}$ & 2022 Sep 1 -- 2023 Jan 10 & 13 & 55--71 & $15$ & this paper \\
Keck-I/HIRES & 2022 Sep 7 & 1 & $68$ & $1$ & this paper \\
\enddata 
\tablenotetext{a}{
    S/N from the exposure meter for the Keck-I/HIRES observation.
}
\tablenotetext{b}{
    The scatter in the RV residuals from the
    best-fit orbit which may include astrophysical jitter. We do not an RV precision estimate for the Keck-I/HIRES template I$_{2}$-free template observations.
}
\ifthenelse{\boolean{emulateapj}}{
    \end{deluxetable*}
}{
    \end{deluxetable}
}



\section{Analysis}
\label{sec:analysis}

\subsection{Derivation of Stellar Atmospheric Parameters}
\label{sec:atmosphericparams}

The stellar atmospheric parameters \teffstar\ and \feh\ were adopted or derived for each system as follows. We did not determine $\vsini$ values for any of the stars studied here, as the processes used to determine the atmospheric parameters do not yield reliable $\vsini$ values for M dwarf stars. We visually confirm that none of the spectra show notable rotational broadening, from which we estimate an upper limit of $\vsini \lesssim 10$\,\kms\ for each system.

\paragraph{TOI~519:} We applied the SpecMatch-Empirical procedure \citep{yee:2017} to the HIRES spectra of \hatcur{519} to measure $\teffstar = \hatcurSMEiteff{519}$\,K and $\feh = \hatcurSMEizfeh{519}$. For comparison, \citet{parviainen:2021} list $\teffstar = 3350 \pm 200$\,K. They obtained this value by determining a spectral type of M3.0--M4.5 based on a low-resolution spectrum from the Alhambra Faint Object Spectrograph and Camera (ALFOSC) on the Nordic Optical Telescope (NOT). This was then translated to an estimate of the effective temperature using the spectral type--\teffstar\ relation from \citet{houdebine:2019}.

\paragraph{TOI~3629:} We adopted $\teffstar = \hatcurSMEiteff{3629}$\,K and $\feh = \hatcurSMEizfeh{3629}$ from \citet{canas:2022} who applied the HPF-SpecMatch package \citep{stefansson:2020} to the HPF spectra of \hatcur{3629}. For comparison, applying the SpecMatch-Empirical procedure to the HIRES spectra gives $\teffstar = 3789 \pm 70$\,K and $\feh = 0.48 \pm 0.09$, consistent with the HPF values.

\paragraph{TOI~3714:} We adopted $\teffstar = \hatcurSMEiteff{3714}$\,K and $\feh = \hatcurSMEizfeh{3714}$ from \citet{canas:2022} who measured these parameters for \hatcur{3714} in the same fashion as they did for \hatcur{3629}. For comparison, applying the SpecMatch-Empirical procedure to the HIRES spectra gives $\teffstar = 3594 \pm 70$\,K and $\feh = 0.12 \pm 0.09$, consistent with the HPF values.

\paragraph{TOI~4201:} We applied the SpecMatch-Empirical procedure to the HIRES spectra of \hatcur{4201} to measure $\teffstar = \hatcurSMEiteff{4201}$\,K and $\feh = \hatcurSMEizfeh{4201}$.

\paragraph{TOI~5344:} We applied the SpecMatch-Empirical procedure to the HIRES spectra of \hatcur{5344} to measure $\teffstar = \hatcurSMEiteff{5344}$\,K and $\feh = \hatcurSMEizfeh{5344}$.

\subsection{Transiting Planet Modelling}
\label{sec:transitmodel}

We performed a joint analysis for each system of the light curves, RV
observations, astrometric parallaxes, catalog broad-band photometric
measurements, and spectroscopic atmospheric parameters
(Tables~\ref{tab:stellarobserved1} and~\ref{tab:stellarobserved2}). To do this we followed the methods of
\citet{hartman:2019:hats6069} and \citet{bakos:2020:hats71}, but with a significant
modification to account for systematic errors in the theoretical
stellar evolution models used in the fit. We discuss this modification
in more detail below.

We model the light curves using the model of
\citet{mandel:2002}, assuming a quadratic limb darkening law for the
host star.  We vary the limb darkening coefficients in the fit, with
Gaussian priors based on the theoretical tabulations of
\citet{claret:2012,claret:2013,claret:2018}. The RV observations are
modeled assuming a Keplerian orbit. We use version 1.2 of the MIST
theoretical stellar evolution models
\citep{paxton:2011,paxton:2013,paxton:2015,choi:2016,dotter:2016} to
model the broad-band photometry and atmospheric parameters, and to
provide a constraint on the allowed combinations of the stellar bulk
density, effective temperature and metallicity. We also use the MWDUST
3D Galactic extinction model \citep{bovy:2016} to place a Gaussian
prior on the line-of-sight extinction $A_{V}$ as a function of
distance, and to set a maximum allowed extinction. The
extinction in each band-pass is calculated from $A_{V}$ assuming a
$R_{V} = 3.1$ law. For each system we perform an analysis assuming the
orbit is circular and a separate analysis allowing
$\sqrt{e}\sin\omega$ and $\sqrt{e}\cos\omega$ to vary in the fit.  A
differential evolution Markov Chain Monte Carlo procedure is used to
fit the observations and determine the uncertainties on the varied
parameters (see \citet{hartman:2019:hats6069} for a list of
parameters, and priors).

For M dwarf stars, which have main sequence lifetimes that exceed the
current age of the Universe, the spread in the theoretical main
sequence is typically much smaller than the observational
uncertainties. As a result, using theoretical stellar evolution models
in the joint analysis of a transiting planet system can lead to very
small uncertainties on the stellar and planetary masses and radii,
which are often well below the estimated systematic uncertainties in
the evolution models themselves (\citealp{tayar:2022}; see also
\citealp{hobson:2023} for an example). As argued by
\citet{eastman:2022}, the uncertainties on the stellar masses and
radii may be smaller than the estimates of \citet{tayar:2022} when
using $\rhostar$, as determined from the transit light curves, to
self-consistently determine the other stellar parameters. The argument
of \citet{eastman:2022} holds for a purely empirical determination of
the system parameters, however when theoretical stellar evolution
models are used in constraining the fit to the observations, the
systematic errors on the models must still be accounted for.

The key parameters that we vary in the fit which directly impact the
stellar physical parameters are: (1) \teffstar, (2) \feh, (3) the
square of the impact parameter $b^{2}$, (4) $\zrstar$, (5)
$\rpl/\rstar$, (6) $\sqrt{e}\cos\omega$, (7) $\sqrt{e}\sin\omega$, (8)
$T_{c,0}$ and (9) $T_{c,N}$. Here $T_{c,0}$ is the time of transit
center for some initial epoch, and $T_{c,N}$ is the time of transit
center for a final epoch. Together $T_{c,0}$ and $T_{c,N}$ determine
the orbital period. The parameter $\zrstar$ is explained in a footnote to Table~\ref{tab:planetparam1}. The parameters (3) through (9) together determine
$\rhostar$ which is the parameter that is used, together with
\teffstar\ and \feh, to determine the stellar physical parameters via
interpolation within a pre-computed grid of MIST stellar evolutionary
models. In addition to these parameters, the distance modulus $\mu$
and extinction $A_{V}$ are also varied, and are used to determine the
predicted magnitude in each bandpass to be compared to the catalog magnitudes.

Here we augment the above parameters with four new systematic error
parameters: $\Delta$[Fe/H]$_{\rm sys}$, $\Delta T_{\rm eff,sys}$, $\Delta M_{\rm
  \star,sys}$, and $\Delta M_{\rm bol,sys}$. Here $\Delta$[Fe/H]$_{\rm sys}$
represents the systematic error on \feh, in dex, $\Delta T_{\rm eff,sys}$ is
the fractional systematic error in \teffstar, $\Delta M_{\rm \star,sys}$ is
the fractional systematic error in \mstar, and $\Delta M_{\rm bol,sys}$ is
the systematic error in the bolometric magnitude, in units of
magnitude. These act as hidden parameters that allow broader
distributions in \feh, \teffstar, \mstar, and $M_{\rm bol}$ beyond
what would be allowed by the stellar evolution model. The systematic
error parameters are allowed to vary in the fit assuming Gaussian
priors with mean values of $0$, and standard deviations of
$0.08$\,dex, $0.024$, $0.05$, and 0.021\,mag, respectively. These values for the systematic uncertainties are adopted from \citet{tayar:2022}, and are based on current uncertainties on the measured interferometric angular diameters and bolometric fluxes of stars, and comparisons between different grids of stellar evolution models.

Within each link in the Markov Chain we take the combination of
(\teffstar, \feh, \rhostar, $\Delta$[Fe/H]$_{\rm sys}$, $\Delta T_{\rm eff,sys}$, and
$\Delta M_{\rm \star,sys}$) and perform the look-up in the isochrone at the
values ($\teffstar(1+\Delta T_{\rm eff,sys})$, \feh$+\Delta$[Fe/H]$_{\rm sys}$,
\rhostar). This yields the following set of predicted,
isochrone-based, stellar parameters: ($M_{\rm \star,iso}$, $R_{\rm
  \star,iso}$, $L_{\rm \star,iso}$, $\log g_{\rm \star,iso}$, $V_{\rm
  mag,iso}$, $T_{\rm eff,iso}$).

We then correct the predicted, isochrone-based, stellar parameters for
$\Delta M_{\rm \star,sys}$, and $\Delta T_{\rm eff,sys}$ to obtain the adopted parameters as follows:
\begin{align}
M_{\rm \star,adopted} & = (1.0 + \Delta M_{\rm \star,sys})M_{\rm star,iso} \\
R_{\rm \star,adopted} & = (1.0 + \Delta M_{\rm \star,sys})^{1/3}R_{\rm star,iso} \\
\log L_{\rm \star,adopted} & = \log L_{\rm \star,iso} + \frac{2}{3}\log(1.0 + \Delta M_{\rm \star,sys}) \nonumber\\
 & + 4.0(\log(\teffstar) - \log(T_{\rm eff,iso}))\\
\log g_{\rm \star,adopted} & = \log g_{\rm \star,iso} + \frac{1}{3}\log(1.0 + \Delta M_{\rm \star,sys}) \\
V_{\rm mag,adopted} & = V_{\rm mag,iso} - \frac{5}{3}\log(1.0 + \Delta M_{\rm \star,sys}) \nonumber\\
 & - 10(\log(\teffstar) - \log(T_{\rm eff,iso})) \\
T_{\rm eff,adopted} & = \teffstar
\end{align}
The adopted values are used to calculate the posterior
distributions for the various stellar parameters. To account for
$\Delta M_{\rm bol,sys}$ we add this parameter to the predicted magnitudes in
each filter before determining $\chi^{2}$ for the model. Here $\Delta M_{\rm
  bol,sys}$ accounts for the systematic uncertainty in calculating the
bolometric magnitude from the theoretical luminosity of the star.

This procedure forces the model to have $\rhostar$ equal to the value
determined from the transit parameters, but allows $\mstar$ and
$\rstar$ to differ from the values predicted from the theoretical
stellar evolution model. Because we adjust the values for $L_{\rm
  \star,adopted}$, and the associated magnitudes, accordingly, it is
possible that the measured catalog photometry and parallax values will
constrain $\rstar$, and thus $\mstar$ (through $\rhostar$), to better
than what one would calculate simply from the a priori systematic
uncertainty on $\mstar$. The method that we employ here is able to
self-consistently handle both systematic uncertainties present in
stellar models, and the full set of observations that might provide
empirical constraints that are tighter than those systematic uncertainties. One limitation to our approach is that this procedure assumes that $\Delta T_{\rm eff,sys}$, $\Delta M_{\rm \star,sys}$, $\Delta$[Fe/H]$_{\rm sys}$ and $\Delta M_{\rm bol,sys}$ are uncorrelated, whereas the systematic errors in the stellar evolution models may very well be strongly correlated.

Figures~\ref{fig:toi519}--\ref{fig:toi5344} compares the best-fit
models to the observational data. The adopted stellar parameters are
listed in Tables~\ref{tab:stellarderived1} and~\ref{tab:stellarderived2}, while the adopted planetary
parameters are listed in Tables~\ref{tab:planetparam1} and~\ref{tab:planetparam2}.



\section{Discussion}
\label{sec:discussion}

In this paper we have presented the discovery of two new transiting
giant planets that orbit M dwarf stars, and have updated the parameters
for three other systems with new RVs and light curves. The system
parameters are compared to the parameters of other transiting planet systems from the
NASA Exoplanet Archive in Figures~\ref{fig:plmassradteqrad}
and~\ref{fig:stmassplmassstteffstfeh}. Here the planet equilibrium temperature $T_{\rm eq}$ is calculated assuming zero albedo, and full redistribution of heat, making it effectively a proxy for the incoming flux from the star at the orbital distance of the planet. We find that, like other giant
planets found around M dwarfs, these planets' radii are consistent with
theoretical mass-radius relationships and show no evidence of
inflation. Neither would we expect radius inflation for these planets;
radius inflation has only been observed for
planets with $T_{\rm eq} \gtrsim 1000$\,K \citep[e.g.,][]{thorngren:2016}, and
due to the low luminosities of M dwarf host stars all giant planets
found so far around such stars have $T_{\rm eq} \lesssim 1000$\,K.

The two newly discovered planets, \hatcurb{4201} and
\hatcurb{5344}, are both around high-metallicity host
stars. \hatcur{4201} has a spectroscopically measured \feh$ =
\hatcurSMEizfeh{4201}$, and an a posteriori value of \feh$ =
\hatcurISOzfeh{4201}$ based on the joint analysis, while \hatcur{5344}
has a spectroscopic metallicity of \feh$ = \hatcurSMEizfeh{5344}$ and
an a posteriori value of \feh$ = \hatcurISOzfeh{5344}$. We also
confirm high metallicities for the re-analyzed systems, with
spectroscopic metallicities from HIRES of $\hatcurSMEizfeh{519}$, $0.48 \pm
0.09$, and $0.12 \pm 0.09$ for \hatcur{519}, \hatcur{3629} and
\hatcur{3714}, respectively, and a posteriori values of
$\hatcurISOzfeh{519}$, $\hatcurISOzfeh{3629}$, and
$\hatcurISOzfeh{3714}$. As shown in
Fig.~\ref{fig:stmassplmassstteffstfeh} other giant-planet-hosting M
dwarf stars also tend to have super-solar metallicities, and in fact
there may be a trend toward higher metallicities for cooler
short-period giant-planet hosting stars. A strong correlation between
host star metallicity and short-period giant planet occurrence has previously been
established for FGK host stars \citep{fischer:2005}. The emerging set of giant-planet-hosting M
dwarfs appears to show this trend as well \citep[e.g.,][]{hirano:2018,gan:2022,kagetani:2023}.  The newly discovered systems
bolster this conclusion. A caveat is that the metallicities that are shown in Fig.~\ref{fig:stmassplmassstteffstfeh} are taken directly from the NASA Exoplanet Archive, and have been measured in an inhomogeneous fashion. Moreover, the determination of M dwarf metallicities from optical spectra may be subject to systematic errors. Indeed, the systematic differences between the metallicities measured from the optical spectra and the a posteriori metallicities for the five targets studied in this paper are typical for M dwarfs, and indicative of the difficulty of accurately measuring metallicities for M dwarfs based on either of these methods. An effort to homogeneously measure the metallicities of giant-planet-hosting M dwarfs using NIR spectra might allow for a more robust conclusion.

Table~\ref{tab:paramcompare} compares the stellar and planetary masses and radii for \hatcur{519}, \hatcur{3629} and \hatcur{3714} to the previously published values. We find that for \hatcur{3629} and \hatcur{3714} our values are within the error bars of the previous measurements. For both of these systems the planetary radius uncertainties are reduced slightly, but the stellar mass uncertainties are larger due to our method of treating systematic uncertainties. For \hatcur{519}, our stellar mass estimate is approximately $2\sigma$ higher than that of \citet{kagetani:2023} (when using our uncertainty; it is $4.6\sigma$ higher when using their uncertainty), but consistent within the uncertainty of the value from \citet{parviainen:2021}. The other values are consistent within the uncertainties, which are now somewhat smaller.  We note that the stellar mass estimate from \citet{kagetani:2023} is close to the value of $0.3398 \pm 0.0086$\,\msun\ that comes from using the empirical relation between stellar mass, absolute $K_{S}$ magnitude, and metallicity determined by \citet{mann:2019}.

The uncertainties on the parameters that we measure for all of the
systems studied in this paper account for systematic uncertainties in
the stellar evolution models, in a manner that self-consistently
allows for empirical constraints on the parameters that may reduce
the uncertainties below the systematic errors. We measure stellar
masses of \hatcurISOmlong{519}\,\msun, \hatcurISOmlong{3629}\,\msun,
\hatcurISOmlong{3714}\,\msun, \hatcurISOmlong{4201}\,\msun, and
\hatcurISOmlong{5344}\,\msun, for \hatcur{519}, \hatcur{3629},
\hatcur{3714}, \hatcur{4201}, and \hatcur{5344}, respectively.  The
relative uncertainties on the masses are thus 4.8\%, 5.0\%, 5.4\%,
5.3\%, and 5.6\%, respectively, and comparable to the assumed 5\%
systematic uncertainty on the stellar masses. For comparison, we find
typical uncertainties of $\sim 1$--$2$\% when we do not account for
systematic errors in the analysis.

We can also compare the stellar masses to the values predicted from the empirical relation between stellar mass, absolute $K_{S}$ magnitude, and metallicity determined by \citet{mann:2019}. Using this relation we find masses of $0.3398 \pm 0.0086$\,\msun, $0.604 \pm 0.016$, $0.503 \pm 0.012$\,\msun, $0.603\pm0.016$\,\msun, and $0.574 \pm 0.015$\,\msun\ for \hatcur{519}, \hatcur{3629}, \hatcur{3714}, \hatcur{4201}, and \hatcur{5344}, respectively. The masses from the empirical relations are systematically lower than the masses that we adopt based on the MIST isochrones. They are consistent with our values to within $1\sigma$ to $2\sigma$. Our adopted uncertainties are also about twice as large as the uncertainty estimates based on the empirical relations.

The stellar effective temperatures that come from our modelling are more tightly constrained than the assumed systematic
uncertainty. We measure respective a posteriori temperatures of
\hatcurISOteff{519}\,K, \hatcurISOteff{3629}\,K,
\hatcurISOteff{3714}\,K, \hatcurISOteff{4201}\,K and
\hatcurISOteff{5344}\,K, with corresponding fractional uncertainties
of 1.9\%, 2.0\%, 2.2\%, 1.7\%, and 1.7\%, whereas the assumed
systematic uncertainty is 2.4\%.  This uncertainty is still much higher
than the typical $\sim 0.2$\% uncertainty that results when systematic
errors are not accounted for in the modelling. The constraints on
$\rhostar$ from the transit observations, together with constraints on
the luminosity from the observed magnitudes and parallaxes, allows the
effective temperature to be more tightly constrained than the assumed
systematic error on this parameter in the stellar evolutionary models.

The systematic errors feed in to most of the other parameters as well. The
median stellar radius and planetary radius fractional uncertainties
are 1.9\%, and 2.0\%, respectively, whereas we typically find $\sim
0.5$\% and 1.0\% respective uncertainties when not including
systematic errors. For comparison, \citet{tayar:2022} suggest a
systematic stellar radius uncertainty of 4.2\%, which would set a
comparable uncertainty on the planetary radius, however this
uncertainty does not account for other empirical constraints that can
reduce the uncertainty below this limit, as we do here. We find that
the planetary mass uncertainties are largely set by the uncertainty on
the RV semiamplitude $K$, which exceeds the stellar mass uncertainty
for most of the systems. The only exception is \hatcurb{3714} for
which the fractional uncertainty on $K$ is 2.5\%, so that the
planetary mass fractional uncertainty of 4.4\% is largely set by the
stellar mass uncertainty.

Due to the small radii of the host stars, giant planets transiting M dwarfs are among the most favorable targets for performing transmission spectroscopy on low equilibrium temperature giant planets. Fig~\ref{fig:tsm} compares the Transmission Spectroscopy Metric \citep[TSM;][]{kempton:2018} of the five systems discussed in this paper to other transiting giant planets from the NASA Exoplanet Archive. When compared against other planets of comparable mass, giant planets around M dwarfs do not have particularly high values of TSM. This is because the relatively small scale heights of these cool, uninflated planets lead to lower TSM values compared to highly inflated hot Jupiters. However, when comparing against other planets with T$_{\rm eq} \lesssim 1000$\,K, giant planets around M dwarfs tend to have relatively high values of TSM. Among the targets discussed in this paper, \hatcurb{519} is the most favorable target for transmission spectroscopy, with TSM$ = 186 \pm 28$, while the more massive planet \hatcurb{4201} is relatively unfavorable with TSM$ = 20 \pm 2$.

%
%
\ifthenelse{\boolean{emulateapj}}{
    \begin{figure*}[!ht]
}{
    \begin{figure}[!ht]
}
 {
 \centering
 \leavevmode
 \includegraphics[width={0.5\linewidth}]{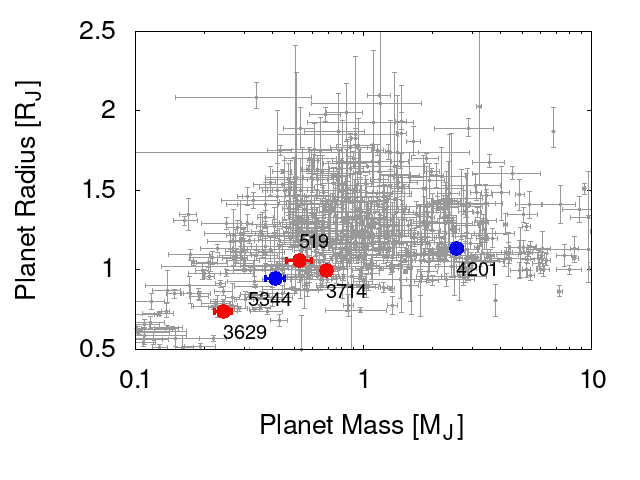}%
 \hfil
 \includegraphics[width={0.5\linewidth}]{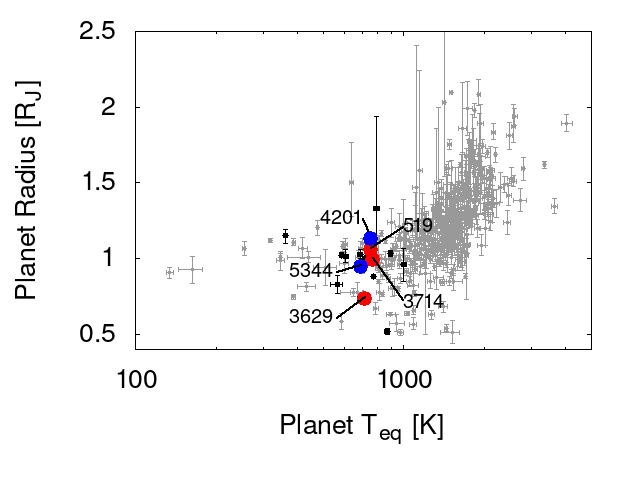}
 }
\caption{
    {\em Left:} Planet radius vs.\ mass. The two new planet discoveries and the three other systems that we re-analyze are indicated. Small grey points show all transiting giant planets with $M_{p} > 0.1$\,\mjup\ and $R_{p} > 0.5$\,\rjup\ from the NASA Exoplanet Archive. {\em Right:} Planet radius vs.\ equilibrium temperature for transiting giant planets. The equilibrium temperature is estimated assuming zero albedo and full redistribution of heat, so that it is effectively a proxy for the flux received from the star. The small black points show planets that orbit M dwarfs ($\teffstar < 4000$\,K), while the gray points show all other planets. None of the transiting giant planets discovered around M dwarfs to date appear to be inflated, but all have low equilibrium temperatures where planet inflation has not been observed for hotter host stars either.
\label{fig:plmassradteqrad}
}
\ifthenelse{\boolean{emulateapj}}{
    \end{figure*}
}{
    \end{figure}
}

%
%
\ifthenelse{\boolean{emulateapj}}{
    \begin{figure*}[!ht]
}{
    \begin{figure}[!ht]
}
 {
 \centering
 \leavevmode
 \includegraphics[width={0.5\linewidth}]{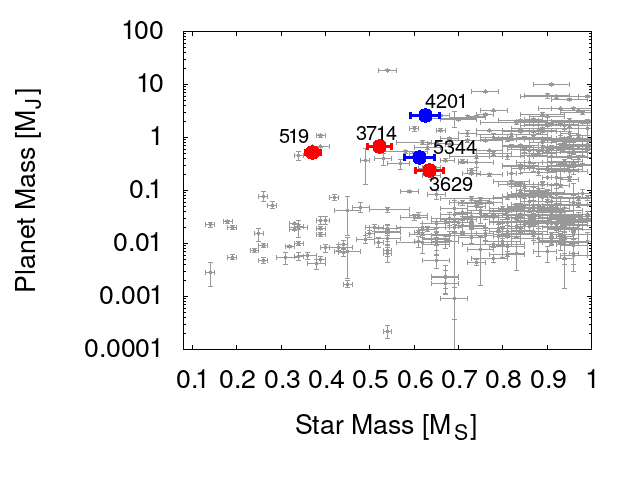}%
 \hfil
 \includegraphics[width={0.5\linewidth}]{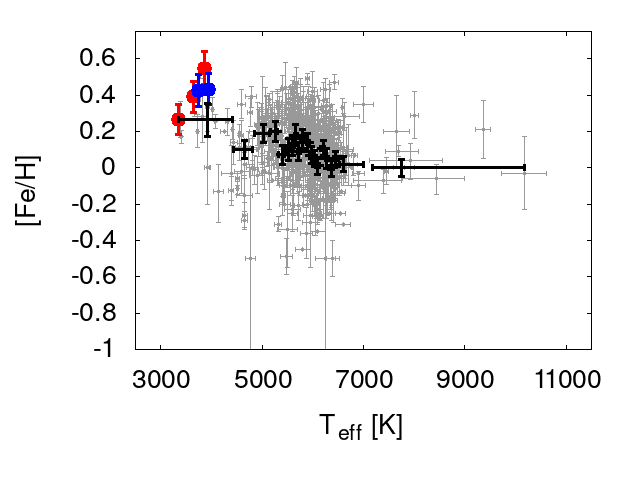}
 }
\caption{
    {\em Left:} Planet mass vs.\ host star mass. The two new planet
    discoveries and the three other systems that we re-analyze are
    indicated. Small gray points show all transiting planets with host
    star masses less than $1$\,\msun\ from the NASA Exoplanet
    Archive. {\em Right:} Metallicity vs.\ effective temperature for
    the host stars of transiting short period giant planets with $P <
    10$\,d, $R_{P} > 0.5$\,\rjup\ and $M_{p} > 0.1$\,\mjup\ (small
    gray points). The host stars of the five systems analyzed in this
    paper are indicated with the same colors as in the left panel. The
    black points show the median \feh\ vs. \teff\ when binning the
    sample into bins of 25 stars. The vertical error bars show the
    standard deviation within the bin, while the horizontal error bars
    show the width of the bin. We confirm high metallicities for the
    three re-analyzed systems, and find the two newly discovered
    systems also have high metallicities. This supports earlier
    findings that giant-planet-hosting M dwarfs have high metallicity,
    and that the metallicity--giant planet occurrence relation may be
    even stronger for M dwarfs than for hotter host stars.
\label{fig:stmassplmassstteffstfeh}
}
\ifthenelse{\boolean{emulateapj}}{
    \end{figure*}
}{
    \end{figure}
}

%
%
\ifthenelse{\boolean{emulateapj}}{
    \begin{figure*}[!ht]
}{
    \begin{figure}[!ht]
}
 {
 \centering
 \leavevmode
 \includegraphics[width={0.5\linewidth}]{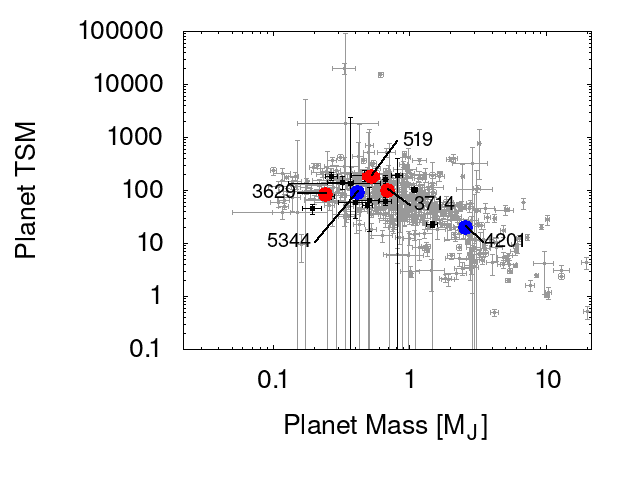}%
 \hfil
 \includegraphics[width={0.5\linewidth}]{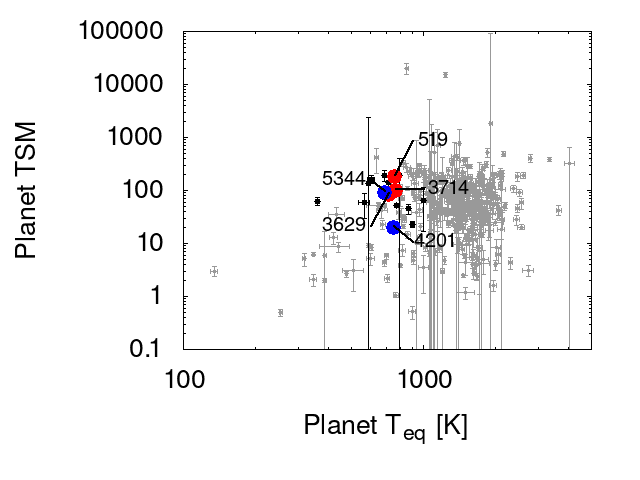}
 }
\caption{
    {\em Left:} Transmission Spectroscopy Metric (TSM) vs.\ planet mass. The two new planet
    discoveries and the three other systems that we re-analyze are
    indicated. Small gray points show all transiting planets with measured masses $> 0.1$\,\mjup\ and radii $> 0.5$\,\rjup\ from the NASA Exoplanet
    Archive, while small black points highlight those planets that orbit M dwarf stars. {\em Right:} TSM vs.\ planet T$_{\rm eq}$ computed assuming zero albedo and full redistribution of heat. The symbols are the same as in the left plot. \hatcurb{519} is the most favorable of the planets discussed in this paper for transmission spectroscopy. In general, giant planets transiting M dwarfs are among the highest TSM systems with T$_{\rm eq} < 1000$\,K.
\label{fig:tsm}
}
\ifthenelse{\boolean{emulateapj}}{
    \end{figure*}
}{
    \end{figure}
}



\acknowledgements 

JH, GB and ZC acknowledge funding from NASA grant 80NSSC22K0315. 
This work is based in part on observations made with the Keck-I telescope at Mauna Kea Observatory, HI. Time on this facility was awarded through NASA.
We would like to acknowledge the following individuals who contributed to gathering the Keck-I/HIRES observations presented in this paper: I.\ Angelo, A.\ Polanski, S.\ Yee, D.\ Tyler, C.\ Beard, J.\ Van Zandt, J.\ Akana Murphy, A.\ Mayo, E.\ Petigura, L.\ Weiss, G.\ Gilbert, L.\ Handley, M.\ MacDougall, F.\ Dai, A.\ Householder, M.\ Rice, N.\ Saunders, J.\ Zhang, C.\ Brinkman, M.\ He, A.\ Langford, D.\ Pidhorodetska, J.\ Lubin, S.\ Blunt, E.\ Turtelboom, E.\ Louden, S.\ Dulz, and D.\ Shaw.
The authors wish to recognize and acknowledge the very significant cultural role and reverence that the summit of Mauna Kea has always had within the indigenous Hawaiian community. We are most fortunate to have the opportunity to conduct observations from this mountain.
We acknowledge T.\ Gan for contributing to the data reduction of observations from the LCOGT facilities.
This work is partly supported by JSPS KAKENHI Grant Numbers JP17H04574, JP18H05439, JP21K20376, and JST CREST Grant Number JPMJCR1761.
E. P. acknowledges funding from the Spanish Ministry of Economics and Competitiveness through project PID2021-125627OB-C32.
E. E-B. acknowledges financial support from the European Union and the State Agency of Investigation of the Spanish Ministry of Science and Innovation (MICINN) under the grant PRE2020-093107 of the Pre-Doc Program for the Training of Doctors (FPI-SO) through FSE funds.
This article is based on observations made with the MuSCAT2 instrument, developed by ABC, at Telescopio Carlos Sánchez operated on the island of Tenerife by the IAC in the Spanish Observatorio del Teide.
This paper is based on observations made with the MuSCAT3 instrument, developed by the Astrobiology Center and under financial supports by JSPS KAKENHI (JP18H05439) and JST PRESTO (JPMJPR1775), at Faulkes Telescope North on Maui, HI, operated by the Las Cumbres Observatory.
A.J.\ acknowledges support from ANID -- Millennium  Science  Initiative -- ICN12\_009 and from FONDECYT project 1210718.
R.B.\ acknowledges support ANID -- Millennium  Science  Initiative -- ICN12\_009 and from FONDECYT project 11200751.
H.P. acknowledges support from the Spanish Ministry of Science and Innovation with the Ramon y Cajal fellowship number RYC2021-031798-I.
Funding from the University of La Laguna and the Spanish Ministry of Universities is acknowledged.
Based on data collected under the ExTrA project at the ESO La Silla Paranal Observatory. ExTrA is a project of Institut de Plan\'etologie et d'Astrophysique de Grenoble (IPAG/CNRS/UGA), funded by the European Research Council under the ERC Grant Agreement n. 337591-ExTrA. ExTrA has been supported by Labex OSUG@2020 (Investissements d'avenir -- ANR10 LABX56), the "Programme National de Physique Stellaire" (PNPS) and the “Programme National de Palnétologie of CNRS/INSU, co-funded by CEA and CNES.
We thank the Swiss National Science Foundation (SNSF) and the Geneva University for their continuous support to our planet search programs. This work has been in particular carried out in the frame of the National Centre for Competence in Research ‘PlanetS’ supported by the Swiss National Science Foundation (SNSF). 
This research has made use of the NASA Exoplanet Archive, which is operated by the California Institute of Technology, under contract with the National Aeronautics and Space Administration under the Exoplanet Exploration Program.
M.G. and A.T. acknowledge the support of M.V. Lomonosov Moscow State University Program of Development.
This work makes use of observations from the LCOGT network. Part of the LCOGT telescope time was granted by NOIRLab through the Mid-Scale Innovations Program (MSIP). MSIP is funded by NSF.
This research has made use of the Exoplanet Follow-up Observation Program (ExoFOP; DOI: 10.26134/ExoFOP5) website, which is operated by the California Institute of Technology, under contract with the National Aeronautics and Space Administration under the Exoplanet Exploration Program.
Funding for the TESS mission is provided by NASA's Science Mission Directorate. KAC acknowledges support from the TESS mission via subaward s3449 from MIT.
This publication makes use of data products from the TRAPPIST project. TRAPPIST is funded by the Belgian National Fund for Scientific Research (F.R.S.-FNRS) under grant PDRT.0120.21.
TRAPPIST-North is a project funded by the University of Liege (Belgium), in collaboration with Cadi Ayyad University of Marrakech (Morocco). EJ is F.R.S.-FNRS Senior Research Associate. MG is F.R.S.-FNRS Research Director. 
The postdoctoral fellowship of KB is funded by F.R.S.-FNRS grant T.0109.20 and by the Francqui Foundation.
The ULiege's contribution to SPECULOOS has received funding from the European Research Council under the European Union's Seventh Framework Programme (FP/2007-2013) (grant Agreement n$^\circ$ 336480/SPECULOOS), from the Balzan Prize and Francqui Foundations, from the Belgian Scientific Research Foundation (F.R.S.-FNRS; grant n$^\circ$ T.0109.20), from the University of Liege, and from the ARC grant for Concerted Research Actions financed by the Wallonia-Brussels Federation. SPECULOOS-North has received financial support from the Heising-Simons Foundation and from Dr. and Mrs. Colin Masson and Dr. Peter A. Gilman. 
This research received funding from the European Research Council (ERC) under the European Union's Horizon 2020 research and innovation programme (grant agreement n$^\circ$ 803193/BEBOP), and from the Science and Technology Facilities Council (STFC; grant n$^\circ$ ST/S00193X/1).
This
 publication benefits from the support of the French Community of Belgium in the context of the FRIA Doctoral Grant awarded to MT.
We acknowledge the use of public TESS data from pipelines at the TESS Science Office and at the TESS Science Processing Operations Center. 
Resources supporting this work were provided by the NASA High-End Computing (HEC) Program through the NASA Advanced Supercomputing (NAS) Division at Ames Research Center for the production of the SPOC data products.
This paper made use of data collected by the TESS mission and are publicly available from the Mikulski Archive for Space Telescopes (MAST) operated by the Space Telescope Science Institute (STScI). Funding for the TESS mission is provided by NASA’s Science Mission Directorate.

\facilities{TESS, Keck:I, LCOGT, Sanchez, FTN, TRAPPIST, Subaru, HET, WIYN, Gaia, Exoplanet Archive}

\software{VARTOOLS \citep{hartman:2016:vartools}, MWDUST \citep{bovy:2016}, Astropy \citep{astropy:2013,astropy:2018}, AstroImageJ \citep{collins:2017}, TAPIR \citep{jensen:2013}}


\bibliographystyle{aasjournal}
\bibliography{hatsbib}

\clearpage

%
%
\ifthenelse{\boolean{emulateapj}}{
    \begin{deluxetable*}{lccl}
}{
    \begin{deluxetable}{lccl}
}
\tablewidth{0pc}
\tabletypesize{\tiny}
\tablecaption{
    Astrometric, Spectroscopic and Photometric parameters for newly discovered systems \hatcur{4201} and \hatcur{5344}
    \label{tab:stellarobserved1}
}
\tablehead{
    \multicolumn{1}{c}{} &
    \multicolumn{1}{c}{\bf TOI~4201} &
    \multicolumn{1}{c}{\bf TOI~5344} &
    \multicolumn{1}{c}{} \\
    \multicolumn{1}{c}{~~~~~~~~Parameter~~~~~~~~} &
    \multicolumn{1}{c}{Value}                     &
    \multicolumn{1}{c}{Value}                     &
    \multicolumn{1}{c}{Source}
}
\startdata
\noalign{\vskip -3pt}
\sidehead{Astrometric properties and cross-identifications}
~~~~2MASS-ID\dotfill               & \hatcurCCtwomassshort{4201} & \hatcurCCtwomassshort{5344} & \\
~~~~TIC-ID\dotfill                 & \hatcurCCtic{4201} & \hatcurCCtic{5344} & \\
~~~~TOI-ID\dotfill                 & \hatcurCCtoi{4201} & \hatcurCCtoi{5344} & \\
~~~~GAIA~DR2-ID\dotfill                 & \hatcurCCgaiadrtwoshort{4201} & \hatcurCCgaiadrtwoshort{5344} & \\
~~~~R.A. (J2000)\dotfill            & \hatcurCCra{4201}    & \hatcurCCra{5344}    & GAIA DR3\\
~~~~Dec. (J2000)\dotfill            & \hatcurCCdec{4201}   & \hatcurCCdec{5344}   & GAIA DR3\\
~~~~$\mu_{\rm R.A.}$ (\masy)              & \hatcurCCpmra{4201} & \hatcurCCpmra{5344} & GAIA DR3\\
~~~~$\mu_{\rm Dec.}$ (\masy)              & \hatcurCCpmdec{4201} & \hatcurCCpmdec{5344} & GAIA DR3\\
~~~~parallax (mas)              & \hatcurCCparallax{4201} & \hatcurCCparallax{5344} & GAIA DR3\\
\sidehead{Spectroscopic properties}
~~~~$\teffstar$ (K)\dotfill         & \hatcurSMEiteff{4201} & \hatcurSMEteff{5344} & see Section~\ref{sec:atmosphericparams}\\
~~~~$\feh$\dotfill                  & \hatcurSMEizfeh{4201} & \hatcurSMEizfeh{5344} & see Section~\ref{sec:atmosphericparams}\\
\sidehead{Photometric properties\tablenotemark{a}}
~~~~$G$ (mag)\tablenotemark{b}\dotfill               & \hatcurCCgaiamG{4201} & \hatcurCCgaiamG{5344} & GAIA DR3 \\
~~~~$BP$ (mag)\tablenotemark{b}\dotfill               & \hatcurCCgaiamBP{4201} & \hatcurCCgaiamBP{5344} & GAIA DR3 \\
~~~~$RP$ (mag)\tablenotemark{b}\dotfill               & \hatcurCCgaiamRP{4201} & \hatcurCCgaiamRP{5344} & GAIA DR3 \\
~~~~$B$ (mag)\dotfill               & \hatcurCCtassmB{4201} & \hatcurCCtassmB{5344} & APASS\tablenotemark{c} \\
~~~~$V$ (mag)\dotfill               & \hatcurCCtassmv{4201} & \hatcurCCtassmv{5344} & APASS\tablenotemark{c} \\
~~~~$g$ (mag)\dotfill               & \hatcurCCtassmg{4201} & \hatcurCCtassmg{5344} & APASS\tablenotemark{c} \\
~~~~$r$ (mag)\dotfill               & \hatcurCCtassmr{4201} & \hatcurCCtassmr{5344} & APASS\tablenotemark{c} \\
~~~~$i$ (mag)\dotfill               & \hatcurCCtassmi{4201} & \hatcurCCtassmi{5344} & APASS\tablenotemark{c} \\
~~~~$J$ (mag)\tablenotemark{d}\dotfill               & \hatcurCCtwomassJmag{4201} & \hatcurCCtwomassJmag{5344} & 2MASS           \\
~~~~$H$ (mag)\tablenotemark{d}\dotfill               & \hatcurCCtwomassHmag{4201} & \hatcurCCtwomassHmag{5344} & 2MASS           \\
~~~~$K_s$ (mag)\tablenotemark{d}\dotfill             & \hatcurCCtwomassKmag{4201} & \hatcurCCtwomassKmag{5344} & 2MASS           \\
~~~~$W1$ (mag)\tablenotemark{e}\dotfill             & \hatcurCCWonemag{4201} & \hatcurCCWonemag{5344} & WISE           \\
~~~~$W2$ (mag)\tablenotemark{e}\dotfill             & \hatcurCCWtwomag{4201} & \hatcurCCWtwomag{5344} & WISE           \\
~~~~$W3$ (mag)\tablenotemark{e}\dotfill             & \hatcurCCWthreemag{4201} & \hatcurCCWthreemag{5344} & WISE           \\
\enddata
\tablenotetext{a}{
    We only include in the table catalog magnitudes that were included in our analysis of each system.
}
\tablenotetext{b}{
    The listed uncertainties for the Gaia DR3 photometry are taken from the catalog. For the analysis we assume an additional systematic uncertainty of 0.02\,mag for each bandpass.
}
\tablenotetext{c}{
    From APASS DR6 as
    listed in the UCAC 4 catalog \citep{zacharias:2013:ucac4}.  
}
\tablenotetext{d}{
    From the 2MASS catalog \citep{skrutskie:2006}.
}
\tablenotetext{e}{
    From the 2021 Feb 16 ALLWISE Data release of the WISE mission \citep{cutri:2021}.
}
\ifthenelse{\boolean{emulateapj}}{
    \end{deluxetable*}
}{
    \end{deluxetable}
}

%
%
\ifthenelse{\boolean{emulateapj}}{
    \begin{deluxetable*}{lcccl}
}{
    \begin{deluxetable}{lcccl}
}
\tablewidth{0pc}
\tabletypesize{\tiny}
\tablecaption{
    Astrometric, Spectroscopic and Photometric parameters for previously discovered systems \hatcur{519}, \hatcur{3629} and \hatcur{3714}
    \label{tab:stellarobserved2}
}
\tablehead{
    \multicolumn{1}{c}{} &
    \multicolumn{1}{c}{\bf TOI~519} &
    \multicolumn{1}{c}{\bf TOI~3629} &
    \multicolumn{1}{c}{\bf TOI~3714} &
    \multicolumn{1}{c}{} \\
    \multicolumn{1}{c}{~~~~~~~~Parameter~~~~~~~~} &
    \multicolumn{1}{c}{Value}                     &
    \multicolumn{1}{c}{Value}                     &
    \multicolumn{1}{c}{Value}                     &
    \multicolumn{1}{c}{Source}
}
\startdata
\noalign{\vskip -3pt}
\sidehead{Astrometric properties and cross-identifications}
~~~~2MASS-ID\dotfill               & \hatcurCCtwomassshort{519}  & \hatcurCCtwomassshort{3629} & \hatcurCCtwomassshort{3714} & \\
~~~~TIC-ID\dotfill                 & \hatcurCCtic{519} & \hatcurCCtic{3629} & \hatcurCCtic{3714} & \\
~~~~TOI-ID\dotfill                 & \hatcurCCtoi{519} & \hatcurCCtoi{3629} & \hatcurCCtoi{3714} & \\
~~~~GAIA~DR2-ID\dotfill                 & \hatcurCCgaiadrtwoshort{519}      & \hatcurCCgaiadrtwoshort{3629} & \hatcurCCgaiadrtwoshort{3714} & \\
~~~~R.A. (J2000)\dotfill            & \hatcurCCra{519}       & \hatcurCCra{3629}    & \hatcurCCra{3714}    & GAIA DR3\\
~~~~Dec. (J2000)\dotfill            & \hatcurCCdec{519}      & \hatcurCCdec{3629}   & \hatcurCCdec{3714}   & GAIA DR3\\
~~~~$\mu_{\rm R.A.}$ (\masy)              & \hatcurCCpmra{519}     & \hatcurCCpmra{3629} & \hatcurCCpmra{3714} & GAIA DR3\\
~~~~$\mu_{\rm Dec.}$ (\masy)              & \hatcurCCpmdec{519}    & \hatcurCCpmdec{3629} & \hatcurCCpmdec{3714} & GAIA DR3\\
~~~~parallax (mas)              & \hatcurCCparallax{519}    & \hatcurCCparallax{3629} & \hatcurCCparallax{3714} & GAIA DR3\\
\sidehead{Spectroscopic properties}
~~~~$\teffstar$ (K)\dotfill         &  \hatcurSMEiteff{519}   & \hatcurSMEiteff{3629} & \hatcurSMEteff{3714} & see Section~\ref{sec:atmosphericparams}\\
~~~~$\feh$\dotfill                  &  \hatcurSMEizfeh{519}   & \hatcurSMEizfeh{3629} & \hatcurSMEizfeh{3714} & see Section~\ref{sec:atmosphericparams}\\
\sidehead{Photometric properties\tablenotemark{a}}
~~~~$G$ (mag)\tablenotemark{b}\dotfill               &  \hatcurCCgaiamG{519}  & \hatcurCCgaiamG{3629} & \hatcurCCgaiamG{3714} & GAIA DR3 \\
~~~~$BP$ (mag)\tablenotemark{b}\dotfill               &  \hatcurCCgaiamBP{519}  & \hatcurCCgaiamBP{3629} & \hatcurCCgaiamBP{3714} & GAIA DR3 \\
~~~~$RP$ (mag)\tablenotemark{b}\dotfill               &  \hatcurCCgaiamRP{519}  & \hatcurCCgaiamRP{3629} & \hatcurCCgaiamRP{3714} & GAIA DR3 \\
~~~~$B$ (mag)\dotfill               &  $\cdots$  & \hatcurCCtassmB{3629} & \hatcurCCtassmB{3714} & APASS\tablenotemark{c} \\
~~~~$V$ (mag)\dotfill               &  $\cdots$  & \hatcurCCtassmv{3629} & \hatcurCCtassmv{3714} & APASS\tablenotemark{c} \\
~~~~$g$ (mag)\dotfill               &  $\cdots$  & \hatcurCCtassmg{3629} & \hatcurCCtassmg{3714} & APASS\tablenotemark{c} \\
~~~~$r$ (mag)\dotfill               &  $\cdots$  & \hatcurCCtassmr{3629} & \hatcurCCtassmr{3714} & APASS\tablenotemark{c} \\
~~~~$i$ (mag)\dotfill               &  $\cdots$  & \hatcurCCtassmi{3629} & \hatcurCCtassmi{3714} & APASS\tablenotemark{c} \\
~~~~$J$ (mag)\tablenotemark{d}\dotfill               &  \hatcurCCtwomassJmag{519} & \hatcurCCtwomassJmag{3629} & \hatcurCCtwomassJmag{3714} & 2MASS           \\
~~~~$H$ (mag)\tablenotemark{d}\dotfill               &  \hatcurCCtwomassHmag{519} & \hatcurCCtwomassHmag{3629} & \hatcurCCtwomassHmag{3714} & 2MASS           \\
~~~~$K_s$ (mag)\tablenotemark{d}\dotfill             &  \hatcurCCtwomassKmag{519} & \hatcurCCtwomassKmag{3629} & \hatcurCCtwomassKmag{3714} & 2MASS           \\
~~~~$W1$ (mag)\tablenotemark{e}\dotfill             &  \hatcurCCWonemag{519} & \hatcurCCWonemag{3629} & \hatcurCCWonemag{3714} & WISE           \\
~~~~$W2$ (mag)\tablenotemark{e}\dotfill             &  \hatcurCCWtwomag{519} & \hatcurCCWtwomag{3629} & \hatcurCCWtwomag{3714} & WISE           \\
~~~~$W3$ (mag)\tablenotemark{e}\dotfill             &  \hatcurCCWthreemag{519} & \hatcurCCWthreemag{3629} & \hatcurCCWthreemag{3714} & WISE           \\
\enddata
\tablenotetext{a}{
    We only include in the table catalog magnitudes that were included in our analysis of each system.
}
\tablenotetext{b}{
    The listed uncertainties for the Gaia DR3 photometry are taken from the catalog. For the analysis we assume an additional systematic uncertainties of 0.02\,mag for all bandpasses.
}
\tablenotetext{c}{
    From APASS DR6 as
    listed in the UCAC 4 catalog \citep{zacharias:2013:ucac4}.  
}
\tablenotetext{d}{
    From the 2MASS catalog \citep{skrutskie:2006}.
}
\tablenotetext{e}{
    From the 2021 Feb 16 ALLWISE Data release of the WISE mission \citep{cutri:2021}.
}
\ifthenelse{\boolean{emulateapj}}{
    \end{deluxetable*}
}{
    \end{deluxetable}
}

\startlongtable
\tabletypesize{\scriptsize}
\ifthenelse{\boolean{emulateapj}}{
    \begin{deluxetable*}{lrrrrrrrrl}
}{
    \begin{deluxetable}{lrrrrrrrrl}
}
\tablewidth{0pc}
\tablecaption{
    HIRES relative radial velocities and bisector spans for \hatcur{519}, \hatcur{3629}, \hatcur{3714}, \hatcur{4201} and \hatcur{5344}.
    \label{tab:rvs}
}
\tablehead{
    \colhead{System} &
    \colhead{BJD} &
    \colhead{RV\tablenotemark{a}} &
    \colhead{\ensuremath{\sigma_{\rm RV}}\tablenotemark{b}} &
    \colhead{BS} &
    \colhead{\ensuremath{\sigma_{\rm BS}}} &
    \colhead{Phase}\\
    \colhead{} &
    \colhead{\hbox{(2,450,000$+$)}} &
    \colhead{(\ms)} &
    \colhead{(\ms)} &
    \colhead{(\ms)} &
    \colhead{(\ms)} &
    \colhead{}
}
\startdata
TOI-519 & $ 9514.10954 $ & \nodata      & \nodata      & $  272.5 $ & $  717.8 $ & $   0.940 $ \\
TOI-519 & $ 9538.02709 $ & $   186.28 $ & $    19.67 $ & $  -13.1 $ & $  620.4 $ & $   0.844 $ \\
TOI-519 & $ 9543.01980 $ & $   208.48 $ & $    25.99 $ & $  152.7 $ & $  723.1 $ & $   0.790 $ \\
TOI-519 & $ 9545.01639 $ & $   -50.05 $ & $    18.86 $ & $ -289.7 $ & $  960.0 $ & $   0.368 $ \\
TOI-519 & $ 9546.03311 $ & $  -155.93 $ & $    21.90 $ & $ -558.1 $ & $ 1558.5 $ & $   0.172 $ \\
TOI-519 & $ 9547.08385 $ & $  -134.06 $ & $    21.11 $ & $  -57.3 $ & $ 1562.9 $ & $   0.002 $ \\
TOI-519 & $ 9565.95401 $ & $   119.69 $ & $    21.98 $ & $ 1498.4 $ & $  850.2 $ & $   0.917 $ \\
TOI-519 & $ 9587.87529 $ & $  -236.07 $ & $    24.08 $ & $  -10.4 $ & $ 1451.2 $ & $   0.244 $ \\
TOI-3629 & $ 9738.09644 $ & \nodata      & \nodata      & $    0.0 $ & $  147.1 $ & $   0.303 $ \\
TOI-3629 & $ 9803.11442 $ & $    61.65 $ & $     5.94 $ & $    0.0 $ & $  378.5 $ & $   0.820 $ \\
TOI-3629 & $ 9822.95907 $ & $    23.33 $ & $     4.06 $ & \nodata      & \nodata      & $   0.861 $ \\
TOI-3629 & $ 9827.06494 $ & $    12.29 $ & $     4.30 $ & $  -61.9 $ & $  148.5 $ & $   0.904 $ \\
TOI-3629 & $ 9831.96255 $ & $   -28.53 $ & $     4.59 $ & $    2.1 $ & $  259.3 $ & $   0.148 $ \\
TOI-3629 & $ 9833.99319 $ & $    24.30 $ & $     6.21 $ & $  416.8 $ & $  529.0 $ & $   0.664 $ \\
TOI-3629 & $ 9835.01084 $ & $    -7.34 $ & $     4.48 $ & $  116.3 $ & $  690.1 $ & $   0.922 $ \\
TOI-3629 & $ 9835.87585 $ & $   -31.10 $ & $     4.21 $ & $   10.6 $ & $  869.7 $ & $   0.142 $ \\
TOI-3629 & $ 9838.85563 $ & $    12.97 $ & $     3.86 $ & $  -75.2 $ & $  177.1 $ & $   0.899 $ \\
TOI-3629 & $ 9840.79082 $ & $   -14.30 $ & $     4.18 $ & $ -122.8 $ & $  228.9 $ & $   0.391 $ \\
TOI-3714 & $ 9479.00748 $ & \nodata      & \nodata      & $ -904.2 $ & $  649.4 $ & $   0.307 $ \\
TOI-3714 & $ 9481.98350 $ & $   126.75 $ & $     5.84 $ & $    0.0 $ & $  783.0 $ & $   0.688 $ \\
TOI-3714 & $ 9506.91898 $ & $  -162.97 $ & $     6.70 $ & $ 4173.0 $ & $ 1295.6 $ & $   0.260 $ \\
TOI-3714 & $ 9508.90928 $ & $  -171.54 $ & $     7.05 $ & $  825.9 $ & $ 1581.3 $ & $   0.184 $ \\
TOI-3714 & $ 9513.94311 $ & $    11.75 $ & $     6.59 $ & $ -2557.9 $ & $ 1283.5 $ & $   0.520 $ \\
TOI-3714 & $ 9537.93945 $ & $   159.89 $ & $     6.29 $ & $   18.6 $ & $  485.2 $ & $   0.656 $ \\
TOI-3714 & $ 9543.78921 $ & $  -103.68 $ & $     6.49 $ & $ 1551.1 $ & $  631.4 $ & $   0.371 $ \\
TOI-3714 & $ 9547.11122 $ & $    88.57 $ & $     7.75 $ & $    0.0 $ & $  794.6 $ & $   0.912 $ \\
TOI-4201 & $ 9830.11551 $ & \nodata      & \nodata      & $  262.9 $ & $  619.0 $ & $   0.448 $ \\
TOI-4201 & $ 9832.09351 $ & $    36.47 $ & $     8.40 $ & \nodata      & \nodata      & $   0.001 $ \\
TOI-4201 & $ 9835.11032 $ & $   413.78 $ & $     7.00 $ & \nodata      & \nodata      & $   0.843 $ \\
TOI-4201 & $ 9890.00349 $ & $  -333.41 $ & $    11.29 $ & \nodata      & \nodata      & $   0.168 $ \\
TOI-4201 & $ 9891.02938 $ & $  -174.50 $ & $     6.63 $ & $ -1048.0 $ & $ 1197.2 $ & $   0.454 $ \\
TOI-4201 & $ 9896.08512 $ & $   345.02 $ & $     7.33 $ & $ -733.1 $ & $  552.5 $ & $   0.866 $ \\
TOI-4201 & $ 9896.96372 $ & $  -290.74 $ & $    10.31 $ & $ -1051.1 $ & $ 1155.6 $ & $   0.111 $ \\
TOI-4201 & $ 9897.93348 $ & $  -385.68 $ & $     7.83 $ & $ -870.7 $ & $  831.8 $ & $   0.382 $ \\
TOI-4201 & $ 9928.05123 $ & $   468.67 $ & $     8.06 $ & $  259.0 $ & $  760.1 $ & $   0.790 $ \\
TOI-4201 & $ 9943.90169 $ & $  -436.74 $ & $     6.84 $ & $ -178.0 $ & $  281.4 $ & $   0.215 $ \\
TOI-4201 & $ 9944.89630 $ & $   -75.76 $ & $     7.46 $ & $  269.0 $ & $  289.7 $ & $   0.493 $ \\
TOI-4201 & $ 9953.89179 $ & $   -33.09 $ & $     6.64 $ & $   29.0 $ & $  514.8 $ & $   0.004 $ \\
TOI-4201 & $ 9954.99279 $ & $  -414.34 $ & $     7.11 $ & $    0.0 $ & $  202.4 $ & $   0.312 $ \\
TOI-5344 & $ 9824.02822 $ & $   -68.52 $ & $     5.52 $ & \nodata      & \nodata      & $   0.418 $ \\
TOI-5344 & $ 9830.05085 $ & \nodata      & \nodata      & $ -729.3 $ & $  494.3 $ & $   0.006 $ \\
TOI-5344 & $ 9832.00717 $ & $     6.71 $ & $     5.95 $ & \nodata      & \nodata      & $   0.522 $ \\
TOI-5344 & $ 9834.07972 $ & $   -29.60 $ & $     5.18 $ & \nodata      & \nodata      & $   0.069 $ \\
TOI-5344 & $ 9835.07406 $ & $   -45.75 $ & $     5.03 $ & $ -1709.3 $ & $ 1481.1 $ & $   0.331 $ \\
TOI-5344 & $ 9870.98245 $ & $    70.01 $ & $     5.06 $ & $  135.2 $ & $  277.5 $ & $   0.799 $ \\
TOI-5344 & $ 9889.95704 $ & $    65.59 $ & $     5.51 $ & \nodata      & \nodata      & $   0.802 $ \\
TOI-5344 & $ 9890.96999 $ & $   -27.01 $ & $     5.68 $ & $ -589.5 $ & $  320.9 $ & $   0.069 $ \\
TOI-5344 & $ 9897.81740 $ & $    68.08 $ & $     5.19 $ & $  277.1 $ & $  556.2 $ & $   0.874 $ \\
TOI-5344 & $ 9927.95984 $ & $    85.65 $ & $     6.03 $ & $  227.0 $ & $  361.7 $ & $   0.822 $ \\
TOI-5344 & $ 9943.85965 $ & $   -10.02 $ & $     5.42 $ & $  163.1 $ & $  387.4 $ & $   0.014 $ \\
TOI-5344 & $ 9944.91861 $ & $   -72.49 $ & $     6.29 $ & $   56.8 $ & $  602.7 $ & $   0.293 $ \\
TOI-5344 & $ 9953.84656 $ & $    32.62 $ & $     6.39 $ & $   60.4 $ & $  302.3 $ & $   0.647 $ \\
TOI-5344 & $ 9954.96159 $ & $    29.88 $ & $     8.25 $ & $  -27.8 $ & $ 1339.0 $ & $   0.941 $ \\
\enddata
\tablenotetext{a}{
    The zero-point of these velocities is arbitrary. An overall offset
    $\gamma_{\rm rel}$ fitted to the orbit has been subtracted for each system.
}
\tablenotetext{b}{
    Internal errors excluding the component of astrophysical jitter
    allowed to vary in the fit.
}
\ifthenelse{\boolean{emulateapj}}{
    \end{deluxetable*}
}{
    \end{deluxetable}
}

%
%
\ifthenelse{\boolean{emulateapj}}{
    \begin{deluxetable*}{lcc}
}{
    \begin{deluxetable}{lcc}
}
\tablewidth{0pc}
\tabletypesize{\footnotesize}
\tablecaption{
    Adopted derived stellar parameters for newly discovered systems \hatcur{4201}, and \hatcur{5344}.
    \label{tab:stellarderived1}
}
\tablehead{
    \multicolumn{1}{c}{} &
    \multicolumn{1}{c}{\bf TOI~4201} &
    \multicolumn{1}{c}{\bf TOI~5344} \\
    \multicolumn{1}{c}{~~~~~~~~Parameter~~~~~~~~} &
    \multicolumn{1}{c}{Value}                     &
    \multicolumn{1}{c}{Value}                     
}
\startdata
~~~~$\mstar$ ($\msun$)\dotfill      & \hatcurISOmlong{4201} & \hatcurISOmlong{5344} \\
~~~~$\rstar$ ($\rsun$)\dotfill      & \hatcurISOrlong{4201} & \hatcurISOrlong{5344} \\
~~~~$\loggstar$ (cgs)\dotfill       & \hatcurISOlogg{4201} & \hatcurISOlogg{5344} \\
~~~~$\rhostar$ (\gcmc)\dotfill       & \hatcurLCrho{4201} & \hatcurLCrho{5344} \\
~~~~$\lstar$ ($\lsun$)\dotfill      & \hatcurISOlum{4201} & \hatcurISOlum{5344} \\
~~~~$\teffstar$ (K)\dotfill      &  \hatcurISOteff{4201} &  \hatcurISOteff{5344} \\
~~~~$\feh$\dotfill      &  \hatcurISOzfeh{4201} &  \hatcurISOzfeh{5344} \\
~~~~Age (Gyr)\dotfill               & \hatcurISOage{4201} & \hatcurISOage{5344} \\
~~~~$A_{V}$ (mag)\dotfill               & \hatcurXAv{4201} & \hatcurXAv{5344} \\
~~~~Distance (pc)\dotfill           & \hatcurXdistred{4201} & \hatcurXdistred{5344} \\
\enddata
\tablecomments{
The listed parameters are those determined through the joint differential evolution Markov Chain analysis, including systematic errors in the stellar evolution models, described in Section~\ref{sec:transitmodel}. For all systems the RV observations are consistent with a circular orbit, and we assume a fixed circular orbit in generating the parameters listed here. 
}
\ifthenelse{\boolean{emulateapj}}{
    \end{deluxetable*}
}{
    \end{deluxetable}
}

%
%
\ifthenelse{\boolean{emulateapj}}{
    \begin{deluxetable*}{lccc}
}{
    \begin{deluxetable}{lccc}
}
\tablewidth{0pc}
\tabletypesize{\footnotesize}
\tablecaption{
    Adopted derived stellar parameters for previously discovered systems \hatcur{519}, \hatcur{3629}, and \hatcur{3714}.
    \label{tab:stellarderived2}
}
\tablehead{
    \multicolumn{1}{c}{} &
    \multicolumn{1}{c}{\bf TOI~519} &
    \multicolumn{1}{c}{\bf TOI~3629} &
    \multicolumn{1}{c}{\bf TOI~3714} \\
    \multicolumn{1}{c}{~~~~~~~~Parameter~~~~~~~~} &
    \multicolumn{1}{c}{Value}                     &
    \multicolumn{1}{c}{Value}                     &
    \multicolumn{1}{c}{Value}                     
}
\startdata
~~~~$\mstar$ ($\msun$)\dotfill      &  \hatcurISOmlong{519}   & \hatcurISOmlong{3629} & \hatcurISOmlong{3714} \\
~~~~$\rstar$ ($\rsun$)\dotfill      &  \hatcurISOrlong{519}   & \hatcurISOrlong{3629} & \hatcurISOrlong{3714} \\
~~~~$\loggstar$ (cgs)\dotfill       &  \hatcurISOlogg{519}    & \hatcurISOlogg{3629} & \hatcurISOlogg{3714} \\
~~~~$\rhostar$ (\gcmc)\dotfill       &  \hatcurLCrho{519}    & \hatcurLCrho{3629} & \hatcurLCrho{3714} \\
~~~~$\lstar$ ($\lsun$)\dotfill      &  \hatcurISOlum{519}     & \hatcurISOlum{3629} & \hatcurISOlum{3714} \\
~~~~$\teffstar$ (K)\dotfill      &  \hatcurISOteff{519} &  \hatcurISOteff{3629} &  \hatcurISOteff{3714} \\
~~~~$\feh$\dotfill      &  \hatcurISOzfeh{519} &  \hatcurISOzfeh{3629} &  \hatcurISOzfeh{3714} \\
~~~~Age (Gyr)\dotfill               &  \hatcurISOage{519}     & \hatcurISOage{3629} & \hatcurISOage{3714} \\
~~~~$A_{V}$ (mag)\dotfill               &  \hatcurXAv{519}     & \hatcurXAv{3629} & \hatcurXAv{3714} \\
~~~~Distance (pc)\dotfill           &  \hatcurXdistred{519}\phn  & \hatcurXdistred{3629} & \hatcurXdistred{3714} \\
\enddata
\tablecomments{
The listed parameters are those determined through the joint differential evolution Markov Chain analysis, including systematic errors in the stellar evolution models, described in Section~\ref{sec:transitmodel}. For all systems the RV observations are consistent with a circular orbit, and we assume a fixed circular orbit in generating the parameters listed here. 
}
\ifthenelse{\boolean{emulateapj}}{
    \end{deluxetable*}
}{
    \end{deluxetable}
}

%
\ifthenelse{\boolean{emulateapj}}{
    \begin{deluxetable*}{lcc}
}{
    \begin{deluxetable}{lcc}
}
\tabletypesize{\tiny}
\tablecaption{Adopted orbital and planetary parameters for newly discovered planets \hatcurb{4201}, and \hatcurb{5344}\label{tab:planetparam1}}
\tablehead{
    \multicolumn{1}{c}{} &
    \multicolumn{1}{c}{\bf TOI~4201b} &
    \multicolumn{1}{c}{\bf TOI~5344b} \\
    \multicolumn{1}{c}{~~~~~~~~~~~~~~~Parameter~~~~~~~~~~~~~~~} &
    \multicolumn{1}{c}{Value} &
    \multicolumn{1}{c}{Value}
}
\startdata
\noalign{\vskip -3pt}
\sidehead{\Lc{} parameters}
~~~$P$ (days)             \dotfill    & $\hatcurLCP{4201}$ & $\hatcurLCP{5344}$ \\
~~~$T_c$ (${\rm BJD\_{}TDB}$)    
      \tablenotemark{a}   \dotfill    & $\hatcurLCT{4201}$ & $\hatcurLCT{5344}$ \\
~~~$T_{14}$ (days)
      \tablenotemark{a}   \dotfill    & $\hatcurLCdur{4201}$ & $\hatcurLCdur{5344}$ \\
~~~$T_{12} = T_{34}$ (days)
      \tablenotemark{a}   \dotfill    & $\hatcurLCingdur{4201}$ & $\hatcurLCingdur{5344}$ \\
~~~$\arstar$              \dotfill    & $\hatcurPPar{4201}$ & $\hatcurPPar{5344}$ \\
~~~$\zrstar$ \tablenotemark{b}             \dotfill    & $\hatcurLCzeta{4201}$\phn& $\hatcurLCzeta{5344}$\phn\\
~~~$\rpl/\rstar$          \dotfill    & $\hatcurLCrprstar{4201}$& $\hatcurLCrprstar{5344}$\\
~~~$b^2$                  \dotfill    & $\hatcurLCbsq{4201}$& $\hatcurLCbsq{5344}$\\
~~~$b \equiv a \cos i/\rstar$
                          \dotfill    & $\hatcurLCimp{4201}$& $\hatcurLCimp{5344}$\\
~~~$i$ (deg)              \dotfill    & $\hatcurPPi{4201}$\phn& $\hatcurPPi{5344}$\phn\\
\sidehead{Limb-darkening coefficients \tablenotemark{c}}
~~~$c_1,g$                  \dotfill    & $\hatcurLBig{4201}$& $\hatcurLBig{5344}$\\
~~~$c_2,g$                  \dotfill    & $\hatcurLBiig{4201}$& $\hatcurLBiig{5344}$\\
~~~$c_1,r$                  \dotfill    & $\hatcurLBir{4201}$& $\hatcurLBir{5344}$\\
~~~$c_2,r$                  \dotfill    & $\hatcurLBiir{4201}$& $\hatcurLBiir{5344}$\\
~~~$c_1,i$                  \dotfill    & $\hatcurLBii{4201}$ & $\hatcurLBii{5344}$\\
~~~$c_2,i$                  \dotfill    & $\hatcurLBiii{4201}$ & $\hatcurLBiii{5344}$\\
~~~$c_1,zs$                  \dotfill   & $\hatcurLBiz{4201}$& $\hatcurLBiz{5344}$\\
~~~$c_2,zs$                  \dotfill   & $\hatcurLBiiz{4201}$& $\hatcurLBiiz{5344}$\\
~~~$c_1,I+z$                  \dotfill    & $\cdots$ & $\hatcurLBiI{5344}$\\
~~~$c_2,I+z$                  \dotfill    & $\cdots$ & $\hatcurLBiiI{5344}$\\
~~~$c_1,z$--$H$                  \dotfill    & $\hatcurLBiJ{4201}$ & $\hatcurLBiJ{5344}$\\
~~~$c_2,z$--$H$                  \dotfill    & $\hatcurLBiiJ{4201}$ & $\hatcurLBiiJ{5344}$\\
~~~$c_1,T$                  \dotfill    & $\hatcurLBiT{4201}$& $\hatcurLBiT{5344}$\\
~~~$c_2,T$                  \dotfill    & $\hatcurLBiiT{4201}$& $\hatcurLBiiT{5344}$\\
\sidehead{RV parameters}
~~~$K$ (\ms)              \dotfill    & $\hatcurRVK{4201}$\phn\phn& $\hatcurRVK{5344}$\phn\phn\\
~~~$e$ \tablenotemark{d}               \dotfill    & $\hatcurRVeccentwosiglimeccen{4201}$ & $\hatcurRVeccentwosiglimeccen{5344}$ \\
~~~RV jitter HIRES \tablenotemark{e} (\ms)        \dotfill    & $\hatcurRVjitter{4201}$& $\hatcurRVjitter{5344}$\\
\sidehead{Planetary parameters}
~~~$\mpl$ ($\mjup$)       \dotfill    & $\hatcurPPmlong{4201}$& $\hatcurPPmlong{5344}$\\
~~~$\rpl$ ($\rjup$)       \dotfill    & $\hatcurPPrlong{4201}$& $\hatcurPPrlong{5344}$\\
~~~$C(\mpl,\rpl)$
    \tablenotemark{f}     \dotfill    & $\hatcurPPmrcorr{4201}$& $\hatcurPPmrcorr{5344}$\\
~~~$\rhopl$ (\gcmc)       \dotfill    & $\hatcurPPrho{4201}$& $\hatcurPPrho{5344}$\\
~~~$\log g_p$ (cgs)       \dotfill    & $\hatcurPPlogg{4201}$& $\hatcurPPlogg{5344}$\\
~~~$a$ (AU)               \dotfill   & $\hatcurPParel{4201}$& $\hatcurPParel{5344}$\\
~~~$T_{\rm eq}$ (K)        \dotfill   & $\hatcurPPteff{4201}$& $\hatcurPPteff{5344}$\\
~~~$\Theta$ \tablenotemark{g} \dotfill & $\hatcurPPtheta{4201}$& $\hatcurPPtheta{5344}$\\
~~~$\log_{10}\langle F \rangle$ (cgs) \tablenotemark{h}
                          \dotfill    & $\hatcurPPfluxavglog{4201}$& $\hatcurPPfluxavglog{5344}$\\
\enddata
\tablecomments{
For all systems we adopt a model in which the orbit is assumed to be circular. See the discussion in Section~\ref{sec:transitmodel}.
}
\tablenotetext{a}{
    Times are in Barycentric Julian Date calculated on the Barycentric Dynamical Time (TDB) system.
    \ensuremath{T_c}: Reference epoch of
    mid transit that minimizes the correlation with the orbital
    period.
    \ensuremath{T_{12}}: total transit duration, time
    between first to last contact;
    \ensuremath{T_{12}=T_{34}}: ingress/egress time, time between first
    and second, or third and fourth contact.
}
\tablenotetext{b}{
   Reciprocal of the half duration of the transit used as a jump parameter in our MCMC analysis in place of $\arstar$. It is related to $\arstar$ by the expression $\zrstar = \arstar(2\pi(1+e\sin\omega))/(P\sqrt{1-b^2}\sqrt{1-e^2})$ \citep{bakos:2010:hat11}.
}
\tablenotetext{c}{
    Values for a quadratic law. The limb darkening parameters were
    directly varied in the fit, using the tabulations from
    \cite{claret:2012,claret:2013,claret:2018} to place Gaussian prior
    constraints on their values, assuming a prior uncertainty of $0.2$
    for each coefficient.
}
\tablenotetext{d}{
    The 95\% confidence upper limit on the eccentricity determined
    when $\sqrt{e}\cos\omega$ and $\sqrt{e}\sin\omega$ are allowed to
    vary in the fit.
}
\tablenotetext{e}{
    Term added in quadrature to the formal RV uncertainties for each
    instrument. This is treated as a free parameter in the fitting
    routine. 
}
\tablenotetext{f}{
    Correlation coefficient between the planetary mass \mpl\ and radius
    \rpl\ estimated from the posterior parameter distribution.
}
\tablenotetext{g}{
    The Safronov number is given by $\Theta = \frac{1}{2}(V_{\rm
    esc}/V_{\rm orb})^2 = (a/\rpl)(\mpl / \mstar )$
    \citep[see][]{hansen:2007}.
}
\tablenotetext{h}{
    Incoming flux per unit surface area, averaged over the orbit.
}
\ifthenelse{\boolean{emulateapj}}{
    \end{deluxetable*}
}{
    \end{deluxetable}
}

%
\ifthenelse{\boolean{emulateapj}}{
    \begin{deluxetable*}{lccc}
}{
    \begin{deluxetable}{lccc}
}
\tabletypesize{\tiny}
\tablecaption{Adopted orbital and planetary parameters for previously discovered planets \hatcurb{519}, \hatcurb{3629}, and \hatcurb{3714}\label{tab:planetparam2}}
\tablehead{
    \multicolumn{1}{c}{} &
    \multicolumn{1}{c}{\bf TOI~519b} &
    \multicolumn{1}{c}{\bf TOI~3629b} &
    \multicolumn{1}{c}{\bf TOI~3714b} \\
    \multicolumn{1}{c}{~~~~~~~~~~~~~~~Parameter~~~~~~~~~~~~~~~} &
    \multicolumn{1}{c}{Value} &
    \multicolumn{1}{c}{Value} &
    \multicolumn{1}{c}{Value}
}
\startdata
\noalign{\vskip -3pt}
\sidehead{\Lc{} parameters}
~~~$P$ (days)             \dotfill    & $\hatcurLCP{519}$ & $\hatcurLCP{3629}$ & $\hatcurLCP{3714}$ \\
~~~$T_c$ (${\rm BJD\_{}TDB}$)    
      \tablenotemark{a}   \dotfill    & $\hatcurLCT{519}$ & $\hatcurLCT{3629}$ & $\hatcurLCT{3714}$ \\
~~~$T_{14}$ (days)
      \tablenotemark{a}   \dotfill    & $\hatcurLCdur{519}$ & $\hatcurLCdur{3629}$ & $\hatcurLCdur{3714}$ \\
~~~$T_{12} = T_{34}$ (days)
      \tablenotemark{a}   \dotfill    & $\hatcurLCingdur{519}$ & $\hatcurLCingdur{3629}$ & $\hatcurLCingdur{3714}$ \\
~~~$\arstar$              \dotfill    & $\hatcurPPar{519}$ & $\hatcurPPar{3629}$ & $\hatcurPPar{3714}$ \\
~~~$\zrstar$ \tablenotemark{b}             \dotfill    & $\hatcurLCzeta{519}$\phn & $\hatcurLCzeta{3629}$\phn& $\hatcurLCzeta{3714}$\phn\\
~~~$\rpl/\rstar$          \dotfill    & $\hatcurLCrprstar{519}$ & $\hatcurLCrprstar{3629}$& $\hatcurLCrprstar{3714}$\\
~~~$b^2$                  \dotfill    & $\hatcurLCbsq{519}$ & $\hatcurLCbsq{3629}$& $\hatcurLCbsq{3714}$\\
~~~$b \equiv a \cos i/\rstar$
                          \dotfill    & $\hatcurLCimp{519}$ & $\hatcurLCimp{3629}$& $\hatcurLCimp{3714}$\\
~~~$i$ (deg)              \dotfill    & $\hatcurPPi{519}$\phn & $\hatcurPPi{3629}$\phn& $\hatcurPPi{3714}$\phn\\
\sidehead{Limb-darkening coefficients \tablenotemark{c}}
~~~$c_1,B$                  \dotfill    & $\hatcurLBiB{519}$ & $\cdots$ & $\cdots$\\
~~~$c_2,B$                  \dotfill    & $\hatcurLBiiB{519}$ & $\cdots$ & $\cdots$\\
~~~$c_1,g$                  \dotfill    & $\hatcurLBig{519}$ & $\hatcurLBig{3629}$& $\hatcurLBig{3714}$\\
~~~$c_2,g$                  \dotfill    & $\hatcurLBiig{519}$ & $\hatcurLBiig{3629}$& $\hatcurLBiig{3714}$\\
~~~$c_1,r$                  \dotfill    & $\hatcurLBir{519}$ & $\hatcurLBir{3629}$& $\hatcurLBir{3714}$\\
~~~$c_2,r$                  \dotfill    & $\hatcurLBiir{519}$ & $\hatcurLBiir{3629}$& $\hatcurLBiir{3714}$\\
~~~$c_1,i$                  \dotfill    & $\hatcurLBii{519}$ & $\hatcurLBii{3629}$ & $\hatcurLBii{3714}$\\
~~~$c_2,i$                  \dotfill    & $\hatcurLBiii{519}$ & $\hatcurLBiii{3629}$ & $\hatcurLBiii{3714}$\\
~~~$c_1,zs$                  \dotfill    & $\hatcurLBiz{519}$ & $\hatcurLBiz{3629}$& $\hatcurLBiz{3714}$\\
~~~$c_2,zs$                  \dotfill    & $\hatcurLBiiz{519}$ & $\hatcurLBiiz{3629}$& $\hatcurLBiiz{3714}$\\
~~~$c_1,I+z$                  \dotfill    & $\cdots$ & $\cdots$ & $\hatcurLBiI{3714}$\\
~~~$c_2,I+z$                  \dotfill    & $\cdots$ & $\cdots$ & $\hatcurLBiiI{3714}$\\
~~~$c_1,z$--$H$                  \dotfill    & $\hatcurLBiJ{519}$ & $\cdots$ & $\cdots$\\
~~~$c_2,z$--$H$                  \dotfill    & $\hatcurLBiiJ{519}$ & $\cdots$ & $\cdots$\\
~~~$c_1,T$                  \dotfill    & $\hatcurLBiT{519}$ & $\hatcurLBiT{3629}$& $\hatcurLBiT{3714}$\\
~~~$c_2,T$                  \dotfill    & $\hatcurLBiiT{519}$ & $\hatcurLBiiT{3629}$& $\hatcurLBiiT{3714}$\\
\sidehead{RV parameters}
~~~$K$ (\ms)              \dotfill    & $\hatcurRVK{519}$\phn\phn & $\hatcurRVK{3629}$\phn\phn& $\hatcurRVK{3714}$\phn\phn\\
~~~$e$ \tablenotemark{d}               \dotfill    & $\hatcurRVeccentwosiglimeccen{519}$ & $\hatcurRVeccentwosiglimeccen{3629}$ & $\hatcurRVeccentwosiglimeccen{3714}$ \\
~~~RV jitter HIRES \tablenotemark{e} (\ms)        \dotfill    & $\hatcurRVjitterA{519}$ & $\hatcurRVjitterA{3629}$& $\hatcurRVjitterA{3714}$\\
~~~RV jitter IRD \tablenotemark{e} (\ms)        \dotfill    & $\hatcurRVjitterB{519}$ & $\cdots$& $\cdots$\\
~~~RV jitter HPF \tablenotemark{e} (\ms)        \dotfill    & $\cdots$ & $\hatcurRVjittertwosiglimB{3629}$& $\hatcurRVjittertwosiglimB{3714}$\\
~~~RV jitter NEID \tablenotemark{e} (\ms)        \dotfill    & $\cdots$ & $\hatcurRVjittertwosiglimB{3629}$& $\hatcurRVjittertwosiglimC{3714}$\\
\sidehead{Planetary parameters}
~~~$\mpl$ ($\mjup$)       \dotfill    & $\hatcurPPmlong{519}$ & $\hatcurPPmlong{3629}$& $\hatcurPPmlong{3714}$\\
~~~$\rpl$ ($\rjup$)       \dotfill    & $\hatcurPPrlong{519}$ & $\hatcurPPrlong{3629}$& $\hatcurPPrlong{3714}$\\
~~~$C(\mpl,\rpl)$
    \tablenotemark{f}     \dotfill    & $\hatcurPPmrcorr{519}$ & $\hatcurPPmrcorr{3629}$& $\hatcurPPmrcorr{3714}$\\
~~~$\rhopl$ (\gcmc)       \dotfill    & $\hatcurPPrho{519}$ & $\hatcurPPrho{3629}$& $\hatcurPPrho{3714}$\\
~~~$\log g_p$ (cgs)       \dotfill    & $\hatcurPPlogg{519}$ & $\hatcurPPlogg{3629}$& $\hatcurPPlogg{3714}$\\
~~~$a$ (AU)               \dotfill    & $\hatcurPParel{519}$ & $\hatcurPParel{3629}$& $\hatcurPParel{3714}$\\
~~~$T_{\rm eq}$ (K)        \dotfill   & $\hatcurPPteff{519}$ & $\hatcurPPteff{3629}$& $\hatcurPPteff{3714}$\\
~~~$\Theta$ \tablenotemark{g} \dotfill & $\hatcurPPtheta{519}$ & $\hatcurPPtheta{3629}$& $\hatcurPPtheta{3714}$\\
~~~$\log_{10}\langle F \rangle$ (cgs) \tablenotemark{h}
                          \dotfill    & $\hatcurPPfluxavglog{519}$ & $\hatcurPPfluxavglog{3629}$& $\hatcurPPfluxavglog{3714}$\\
\enddata
\tablecomments{
For all systems we adopt a model in which the orbit is assumed to be circular. See the discussion in Section~\ref{sec:transitmodel}.
}
\tablenotetext{a}{
    Times are in Barycentric Julian Date calculated on the Barycentric Dynamical Time (TDB) system.
    \ensuremath{T_c}: Reference epoch of
    mid transit that minimizes the correlation with the orbital
    period.
    \ensuremath{T_{12}}: total transit duration, time
    between first to last contact;
    \ensuremath{T_{12}=T_{34}}: ingress/egress time, time between first
    and second, or third and fourth contact.
}
\tablenotetext{b}{
   Reciprocal of the half duration of the transit used as a jump parameter in our MCMC analysis in place of $\arstar$. It is related to $\arstar$ by the expression $\zrstar = \arstar(2\pi(1+e\sin\omega))/(P\sqrt{1-b^2}\sqrt{1-e^2})$ \citep{bakos:2010:hat11}.
}
\tablenotetext{c}{
    Values for a quadratic law. The limb darkening parameters were
    directly varied in the fit, using the tabulations from
    \cite{claret:2012,claret:2013,claret:2018} to place Gaussian prior
    constraints on their values, assuming a prior uncertainty of $0.2$
    for each coefficient.
}
\tablenotetext{d}{
    The 95\% confidence upper limit on the eccentricity determined
    when $\sqrt{e}\cos\omega$ and $\sqrt{e}\sin\omega$ are allowed to
    vary in the fit.
}
\tablenotetext{e}{
    Term added in quadrature to the formal RV uncertainties for each
    instrument. This is treated as a free parameter in the fitting
    routine. 
}
\tablenotetext{f}{
    Correlation coefficient between the planetary mass \mpl\ and radius
    \rpl\ estimated from the posterior parameter distribution.
}
\tablenotetext{g}{
    The Safronov number is given by $\Theta = \frac{1}{2}(V_{\rm
    esc}/V_{\rm orb})^2 = (a/\rpl)(\mpl / \mstar )$
    \citep[see][]{hansen:2007}.
}
\tablenotetext{h}{
    Incoming flux per unit surface area, averaged over the orbit.
}
\ifthenelse{\boolean{emulateapj}}{
    \end{deluxetable*}
}{
    \end{deluxetable}
}

%
%
\ifthenelse{\boolean{emulateapj}}{
    \begin{deluxetable*}{lrrrrrrr}
}{
    \begin{deluxetable}{lrrrrrrr}
}
\tablewidth{0pc}
\tabletypesize{\tiny}
\tablecaption{
    Comparison to Literature Parameters for \hatcur{519}, \hatcur{3629} and \hatcur{3714}
    \label{tab:paramcompare}
}
\tablehead{
    \multicolumn{1}{c}{} &
    \multicolumn{1}{c}{\bf TOI~519} &
    \multicolumn{1}{c}{\bf TOI~519} &
    \multicolumn{1}{c}{\bf TOI~519} &
    \multicolumn{1}{c}{\bf TOI~3629} &
    \multicolumn{1}{c}{\bf TOI~3629} &
    \multicolumn{1}{c}{\bf TOI~3714} &
    \multicolumn{1}{c}{\bf TOI~3714} \\
    \multicolumn{1}{c}{Parameter} &
    \multicolumn{1}{c}{This paper}                     &
    \multicolumn{1}{c}{\citep{kagetani:2023}}                     &
    \multicolumn{1}{c}{\citep{parviainen:2021}}                     &
    \multicolumn{1}{c}{This paper}                     &
    \multicolumn{1}{c}{\citep{canas:2022}}                     &
    \multicolumn{1}{c}{This paper}                     &
    \multicolumn{1}{c}{\citep{canas:2022}}                     
}
\startdata
~~~~$\mstar$ ($\msun$) & \hatcurISOm{519} & $0.335\pm0.008$ & $0.369^{+0.026}_{-0.097}$ & \hatcurISOm{3629} & $0.63 \pm 0.02$ & \hatcurISOm{3714} & $0.53 \pm 0.02$ \\
~~~~$\rstar$ ($\rsun$) & \hatcurISOr{519} & $0.350 \pm 0.010$ & $0.373^{+0.020}_{-0.088}$ & \hatcurISOr{3629} & $0.60^{+0.02}_{-0.01}$ & \hatcurISOr{3714} & $0.51 \pm 0.01$ \\
~~~~$\mpl$ ($\mjup$) & \hatcurPPm{519} & $0.463^{+0.082}_{-0.088}$ & $\cdots$ & \hatcurPPm{3629} & $0.26 \pm 0.02$ & \hatcurPPm{3714} & $0.70 \pm 0.03$ \\
~~~~$\rpl$ ($\rjup$) & \hatcurPPr{519} & $1.03\pm0.03$ & $1.06\pm0.17$ & \hatcurPPr{3629} & $0.74 \pm 0.02$ & \hatcurPPr{3714} & $1.01 \pm 0.03$ \\
\enddata
\ifthenelse{\boolean{emulateapj}}{
    \end{deluxetable*}
}{
    \end{deluxetable}
}

\end{document}

%% file: newcommand_gen.tex
\newcommand{\forloop}[5][1]%
{%
\setcounter{#2}{#3}%
\ifthenelse{#4}%
	{%
	#5%
	\addtocounter{#2}{#1}%
	\forloop[#1]{#2}{\value{#2}}{#4}{#5}%
	}%
	{%
	}%
}%

\newcommand{\ctbd}[1]{}

\newcommand{\Lc}{Light curve}

\newcommand{\masy}{\ensuremath{\rm mas\,yr^{-1}}}
\newcommand{\kms}{\ensuremath{\rm km\,s^{-1}}}
\newcommand{\ms}{\ensuremath{\rm m\,s^{-1}}}

\newcommand{\gcmc}{\ensuremath{\rm g\,cm^{-3}}}

\newcommand{\teff}{\ensuremath{T_{\rm eff}}}

\newcommand{\vsini}{\ensuremath{v \sin{i}}}
\newcommand{\feh}{\ensuremath{\rm [Fe/H]}}

\newcommand{\rsun}{\ensuremath{R_\sun}}
\newcommand{\msun}{\ensuremath{M_\sun}}
\newcommand{\lsun}{\ensuremath{L_\sun}}

\newcommand{\rstar}{\ensuremath{R_\star}}
\newcommand{\mstar}{\ensuremath{M_\star}}
\newcommand{\lstar}{\ensuremath{L_\star}}

\newcommand{\teffstar}{\ensuremath{T_{\rm eff\star}}}
\newcommand{\rhostar}{\ensuremath{\rho_\star}}
\newcommand{\loggstar}{\ensuremath{\log{g_{\star}}}}

\newcommand{\rpl}{\ensuremath{R_{p}}}
\newcommand{\mpl}{\ensuremath{M_{p}}}

\newcommand{\rhopl}{\ensuremath{\rho_{p}}}

\newcommand{\arstar}{\ensuremath{a/\rstar}}
\newcommand{\zrstar}{\ensuremath{\zeta/\rstar}}

\newcommand{\rjup}{\ensuremath{R_{\rm J}}}
\newcommand{\mjup}{\ensuremath{M_{\rm J}}}



%% file: newcommand_spec_multi.tex
\input{planetinfo_multi.tex}
\input{planetinfo_multi_eccen.tex}
\input{newcommand_spec_multi_TOI519.tex}
\input{newcommand_spec_multi_TOI3629.tex}
\input{newcommand_spec_multi_TOI3714.tex}
\input{newcommand_spec_multi_TOI4201.tex}
\input{newcommand_spec_multi_TOI5344.tex}
\newcommand{\hatcur}[1]{\ifnum#1=519 %
\hatcurxxxxxxA
\else
\ifnum#1=3629 %
\hatcurxxxxxxB
\else
\ifnum#1=3714 %
\hatcurxxxxxxC
\else
\ifnum#1=4201 %
\hatcurxxxxxxD
\else
\ifnum#1=5344 %
\hatcurxxxxxxE
\else
??????\fi
\fi
\fi
\fi
\fi
}
\newcommand{\hatcurb}[1]{\ifnum#1=519 %
\hatcurbxxxxxxA
\else
\ifnum#1=3629 %
\hatcurbxxxxxxB
\else
\ifnum#1=3714 %
\hatcurbxxxxxxC
\else
\ifnum#1=4201 %
\hatcurbxxxxxxD
\else
\ifnum#1=5344 %
\hatcurbxxxxxxE
\else
??????\fi
\fi
\fi
\fi
\fi
}
\newcommand{\hatcurc}[1]{\ifnum#1=519 %
\hatcurcxxxxxxA
\else
\ifnum#1=3629 %
\hatcurcxxxxxxB
\else
\ifnum#1=3714 %
\hatcurcxxxxxxC
\else
\ifnum#1=4201 %
\hatcurcxxxxxxD
\else
\ifnum#1=5344 %
\hatcurcxxxxxxE
\else
??????\fi
\fi
\fi
\fi
\fi
}
\newcommand{\hatcurCCgaiadrtwoshort}[1]{\ifnum#1=519 %
\hatcurCCgaiadrtwoshortxxxxxxA
\else
\ifnum#1=3629 %
\hatcurCCgaiadrtwoshortxxxxxxB
\else
\ifnum#1=3714 %
\hatcurCCgaiadrtwoshortxxxxxxC
\else
\ifnum#1=4201 %
\hatcurCCgaiadrtwoshortxxxxxxD
\else
\ifnum#1=5344 %
\hatcurCCgaiadrtwoshortxxxxxxE
\else
??????\fi
\fi
\fi
\fi
\fi
}
\newcommand{\hatcurCCtassvi}[1]{\ifnum#1=519 %
\hatcurCCtassvixxxxxxA
\else
\ifnum#1=3629 %
\hatcurCCtassvixxxxxxB
\else
\ifnum#1=3714 %
\hatcurCCtassvixxxxxxC
\else
\ifnum#1=4201 %
\hatcurCCtassvixxxxxxD
\else
\ifnum#1=5344 %
\hatcurCCtassvixxxxxxE
\else
??????\fi
\fi
\fi
\fi
\fi
}
\newcommand{\hatcurCCtic}[1]{\ifnum#1=519 %
\hatcurCCticxxxxxxA
\else
\ifnum#1=3629 %
\hatcurCCticxxxxxxB
\else
\ifnum#1=3714 %
\hatcurCCticxxxxxxC
\else
\ifnum#1=4201 %
\hatcurCCticxxxxxxD
\else
\ifnum#1=5344 %
\hatcurCCticxxxxxxE
\else
??????\fi
\fi
\fi
\fi
\fi
}
\newcommand{\hatcurCCtoi}[1]{\ifnum#1=519 %
\hatcurCCtoixxxxxxA
\else
\ifnum#1=3629 %
\hatcurCCtoixxxxxxB
\else
\ifnum#1=3714 %
\hatcurCCtoixxxxxxC
\else
\ifnum#1=4201 %
\hatcurCCtoixxxxxxD
\else
\ifnum#1=5344 %
\hatcurCCtoixxxxxxE
\else
??????\fi
\fi
\fi
\fi
\fi
}
\newcommand{\hatcurCCtwomassshort}[1]{\ifnum#1=519 %
\hatcurCCtwomassshortxxxxxxA
\else
\ifnum#1=3629 %
\hatcurCCtwomassshortxxxxxxB
\else
\ifnum#1=3714 %
\hatcurCCtwomassshortxxxxxxC
\else
\ifnum#1=4201 %
\hatcurCCtwomassshortxxxxxxD
\else
\ifnum#1=5344 %
\hatcurCCtwomassshortxxxxxxE
\else
??????\fi
\fi
\fi
\fi
\fi
}
\newcommand{\hatcurisocite}[1]{\ifnum#1=519 %
\hatcurisocitexxxxxxA
\else
\ifnum#1=3629 %
\hatcurisocitexxxxxxB
\else
\ifnum#1=3714 %
\hatcurisocitexxxxxxC
\else
\ifnum#1=4201 %
\hatcurisocitexxxxxxD
\else
\ifnum#1=5344 %
\hatcurisocitexxxxxxE
\else
??????\fi
\fi
\fi
\fi
\fi
}
\newcommand{\hatcurisofull}[1]{\ifnum#1=519 %
\hatcurisofullxxxxxxA
\else
\ifnum#1=3629 %
\hatcurisofullxxxxxxB
\else
\ifnum#1=3714 %
\hatcurisofullxxxxxxC
\else
\ifnum#1=4201 %
\hatcurisofullxxxxxxD
\else
\ifnum#1=5344 %
\hatcurisofullxxxxxxE
\else
??????\fi
\fi
\fi
\fi
\fi
}
\newcommand{\hatcurisoshort}[1]{\ifnum#1=519 %
\hatcurisoshortxxxxxxA
\else
\ifnum#1=3629 %
\hatcurisoshortxxxxxxB
\else
\ifnum#1=3714 %
\hatcurisoshortxxxxxxC
\else
\ifnum#1=4201 %
\hatcurisoshortxxxxxxD
\else
\ifnum#1=5344 %
\hatcurisoshortxxxxxxE
\else
??????\fi
\fi
\fi
\fi
\fi
}
\newcommand{\hatcurjhkfilset}[1]{\ifnum#1=519 %
\hatcurjhkfilsetxxxxxxA
\else
\ifnum#1=3629 %
\hatcurjhkfilsetxxxxxxB
\else
\ifnum#1=3714 %
\hatcurjhkfilsetxxxxxxC
\else
\ifnum#1=4201 %
\hatcurjhkfilsetxxxxxxD
\else
\ifnum#1=5344 %
\hatcurjhkfilsetxxxxxxE
\else
??????\fi
\fi
\fi
\fi
\fi
}
\newcommand{\hatcurlumind}[1]{\ifnum#1=519 %
\hatcurlumindxxxxxxA
\else
\ifnum#1=3629 %
\hatcurlumindxxxxxxB
\else
\ifnum#1=3714 %
\hatcurlumindxxxxxxC
\else
\ifnum#1=4201 %
\hatcurlumindxxxxxxD
\else
\ifnum#1=5344 %
\hatcurlumindxxxxxxE
\else
??????\fi
\fi
\fi
\fi
\fi
}
\newcommand{\hatcurplanetnum}[1]{\ifnum#1=519 %
\hatcurplanetnumxxxxxxA
\else
\ifnum#1=3629 %
\hatcurplanetnumxxxxxxB
\else
\ifnum#1=3714 %
\hatcurplanetnumxxxxxxC
\else
\ifnum#1=4201 %
\hatcurplanetnumxxxxxxD
\else
\ifnum#1=5344 %
\hatcurplanetnumxxxxxxE
\else
??????\fi
\fi
\fi
\fi
\fi
}
\newcommand{\hatcurRVgammaabs}[1]{\ifnum#1=519 %
\hatcurRVgammaabsxxxxxxA
\else
\ifnum#1=3629 %
\hatcurRVgammaabsxxxxxxB
\else
\ifnum#1=3714 %
\hatcurRVgammaabsxxxxxxC
\else
\ifnum#1=4201 %
\hatcurRVgammaabsxxxxxxD
\else
\ifnum#1=5344 %
\hatcurRVgammaabsxxxxxxE
\else
??????\fi
\fi
\fi
\fi
\fi
}
\newcommand{\hatcurRVgammarel}[1]{\ifnum#1=519 %
\hatcurRVgammarelxxxxxxA
\else
\ifnum#1=3629 %
\hatcurRVgammarelxxxxxxB
\else
\ifnum#1=3714 %
\hatcurRVgammarelxxxxxxC
\else
\ifnum#1=4201 %
\hatcurRVgammarelxxxxxxD
\else
\ifnum#1=5344 %
\hatcurRVgammarelxxxxxxE
\else
??????\fi
\fi
\fi
\fi
\fi
}
\newcommand{\hatcurSMElogg}[1]{\ifnum#1=519 %
\hatcurSMEloggxxxxxxA
\else
\ifnum#1=3629 %
\hatcurSMEloggxxxxxxB
\else
\ifnum#1=3714 %
\hatcurSMEloggxxxxxxC
\else
\ifnum#1=4201 %
\hatcurSMEloggxxxxxxD
\else
\ifnum#1=5344 %
\hatcurSMEloggxxxxxxE
\else
??????\fi
\fi
\fi
\fi
\fi
}
\newcommand{\hatcurSMEteff}[1]{\ifnum#1=519 %
\hatcurSMEteffxxxxxxA
\else
\ifnum#1=3629 %
\hatcurSMEteffxxxxxxB
\else
\ifnum#1=3714 %
\hatcurSMEteffxxxxxxC
\else
\ifnum#1=4201 %
\hatcurSMEteffxxxxxxD
\else
\ifnum#1=5344 %
\hatcurSMEteffxxxxxxE
\else
??????\fi
\fi
\fi
\fi
\fi
}
\newcommand{\hatcurSMEversion}[1]{\ifnum#1=519 %
\hatcurSMEversionxxxxxxA
\else
\ifnum#1=3629 %
\hatcurSMEversionxxxxxxB
\else
\ifnum#1=3714 %
\hatcurSMEversionxxxxxxC
\else
\ifnum#1=4201 %
\hatcurSMEversionxxxxxxD
\else
\ifnum#1=5344 %
\hatcurSMEversionxxxxxxE
\else
??????\fi
\fi
\fi
\fi
\fi
}
\newcommand{\hatcurSMEvmac}[1]{\ifnum#1=519 %
\hatcurSMEvmacxxxxxxA
\else
\ifnum#1=3629 %
\hatcurSMEvmacxxxxxxB
\else
\ifnum#1=3714 %
\hatcurSMEvmacxxxxxxC
\else
\ifnum#1=4201 %
\hatcurSMEvmacxxxxxxD
\else
\ifnum#1=5344 %
\hatcurSMEvmacxxxxxxE
\else
??????\fi
\fi
\fi
\fi
\fi
}
\newcommand{\hatcurSMEvmic}[1]{\ifnum#1=519 %
\hatcurSMEvmicxxxxxxA
\else
\ifnum#1=3629 %
\hatcurSMEvmicxxxxxxB
\else
\ifnum#1=3714 %
\hatcurSMEvmicxxxxxxC
\else
\ifnum#1=4201 %
\hatcurSMEvmicxxxxxxD
\else
\ifnum#1=5344 %
\hatcurSMEvmicxxxxxxE
\else
??????\fi
\fi
\fi
\fi
\fi
}
\newcommand{\hatcurSMEvsin}[1]{\ifnum#1=519 %
\hatcurSMEvsinxxxxxxA
\else
\ifnum#1=3629 %
\hatcurSMEvsinxxxxxxB
\else
\ifnum#1=3714 %
\hatcurSMEvsinxxxxxxC
\else
\ifnum#1=4201 %
\hatcurSMEvsinxxxxxxD
\else
\ifnum#1=5344 %
\hatcurSMEvsinxxxxxxE
\else
??????\fi
\fi
\fi
\fi
\fi
}
\newcommand{\hatcurSMEzfeh}[1]{\ifnum#1=519 %
\hatcurSMEzfehxxxxxxA
\else
\ifnum#1=3629 %
\hatcurSMEzfehxxxxxxB
\else
\ifnum#1=3714 %
\hatcurSMEzfehxxxxxxC
\else
\ifnum#1=4201 %
\hatcurSMEzfehxxxxxxD
\else
\ifnum#1=5344 %
\hatcurSMEzfehxxxxxxE
\else
??????\fi
\fi
\fi
\fi
\fi
}
\newcommand{\hatcurSMEzfehshort}[1]{\ifnum#1=519 %
\hatcurSMEzfehshortxxxxxxA
\else
\ifnum#1=3629 %
\hatcurSMEzfehshortxxxxxxB
\else
\ifnum#1=3714 %
\hatcurSMEzfehshortxxxxxxC
\else
\ifnum#1=4201 %
\hatcurSMEzfehshortxxxxxxD
\else
\ifnum#1=5344 %
\hatcurSMEzfehshortxxxxxxE
\else
??????\fi
\fi
\fi
\fi
\fi
}

%% file: planetinfo_multi.tex
\input{TOI519.tex}
\input{TOI3629.tex}
\input{TOI3714.tex}
\input{TOI4201.tex}
\input{TOI5344.tex}
\newcommand{\hatcurCCbbHmag}[1]{\ifnum#1=519 %
\hatcurCCbbHmagxxxxxxA
\else
\ifnum#1=3629 %
\hatcurCCbbHmagxxxxxxB
\else
\ifnum#1=3714 %
\hatcurCCbbHmagxxxxxxC
\else
\ifnum#1=4201 %
\hatcurCCbbHmagxxxxxxD
\else
\ifnum#1=5344 %
\hatcurCCbbHmagxxxxxxE
\else
??????\fi
\fi
\fi
\fi
\fi
}
\newcommand{\hatcurCCbbJmag}[1]{\ifnum#1=519 %
\hatcurCCbbJmagxxxxxxA
\else
\ifnum#1=3629 %
\hatcurCCbbJmagxxxxxxB
\else
\ifnum#1=3714 %
\hatcurCCbbJmagxxxxxxC
\else
\ifnum#1=4201 %
\hatcurCCbbJmagxxxxxxD
\else
\ifnum#1=5344 %
\hatcurCCbbJmagxxxxxxE
\else
??????\fi
\fi
\fi
\fi
\fi
}
\newcommand{\hatcurCCbbKmag}[1]{\ifnum#1=519 %
\hatcurCCbbKmagxxxxxxA
\else
\ifnum#1=3629 %
\hatcurCCbbKmagxxxxxxB
\else
\ifnum#1=3714 %
\hatcurCCbbKmagxxxxxxC
\else
\ifnum#1=4201 %
\hatcurCCbbKmagxxxxxxD
\else
\ifnum#1=5344 %
\hatcurCCbbKmagxxxxxxE
\else
??????\fi
\fi
\fi
\fi
\fi
}
\newcommand{\hatcurCCcitHmag}[1]{\ifnum#1=519 %
\hatcurCCcitHmagxxxxxxA
\else
\ifnum#1=3629 %
\hatcurCCcitHmagxxxxxxB
\else
\ifnum#1=3714 %
\hatcurCCcitHmagxxxxxxC
\else
\ifnum#1=4201 %
\hatcurCCcitHmagxxxxxxD
\else
\ifnum#1=5344 %
\hatcurCCcitHmagxxxxxxE
\else
??????\fi
\fi
\fi
\fi
\fi
}
\newcommand{\hatcurCCcitJmag}[1]{\ifnum#1=519 %
\hatcurCCcitJmagxxxxxxA
\else
\ifnum#1=3629 %
\hatcurCCcitJmagxxxxxxB
\else
\ifnum#1=3714 %
\hatcurCCcitJmagxxxxxxC
\else
\ifnum#1=4201 %
\hatcurCCcitJmagxxxxxxD
\else
\ifnum#1=5344 %
\hatcurCCcitJmagxxxxxxE
\else
??????\fi
\fi
\fi
\fi
\fi
}
\newcommand{\hatcurCCcitKmag}[1]{\ifnum#1=519 %
\hatcurCCcitKmagxxxxxxA
\else
\ifnum#1=3629 %
\hatcurCCcitKmagxxxxxxB
\else
\ifnum#1=3714 %
\hatcurCCcitKmagxxxxxxC
\else
\ifnum#1=4201 %
\hatcurCCcitKmagxxxxxxD
\else
\ifnum#1=5344 %
\hatcurCCcitKmagxxxxxxE
\else
??????\fi
\fi
\fi
\fi
\fi
}
\newcommand{\hatcurCCdec}[1]{\ifnum#1=519 %
\hatcurCCdecxxxxxxA
\else
\ifnum#1=3629 %
\hatcurCCdecxxxxxxB
\else
\ifnum#1=3714 %
\hatcurCCdecxxxxxxC
\else
\ifnum#1=4201 %
\hatcurCCdecxxxxxxD
\else
\ifnum#1=5344 %
\hatcurCCdecxxxxxxE
\else
??????\fi
\fi
\fi
\fi
\fi
}
\newcommand{\hatcurCCesoHKmag}[1]{\ifnum#1=519 %
\hatcurCCesoHKmagxxxxxxA
\else
\ifnum#1=3629 %
\hatcurCCesoHKmagxxxxxxB
\else
\ifnum#1=3714 %
\hatcurCCesoHKmagxxxxxxC
\else
\ifnum#1=4201 %
\hatcurCCesoHKmagxxxxxxD
\else
\ifnum#1=5344 %
\hatcurCCesoHKmagxxxxxxE
\else
??????\fi
\fi
\fi
\fi
\fi
}
\newcommand{\hatcurCCesoHmag}[1]{\ifnum#1=519 %
\hatcurCCesoHmagxxxxxxA
\else
\ifnum#1=3629 %
\hatcurCCesoHmagxxxxxxB
\else
\ifnum#1=3714 %
\hatcurCCesoHmagxxxxxxC
\else
\ifnum#1=4201 %
\hatcurCCesoHmagxxxxxxD
\else
\ifnum#1=5344 %
\hatcurCCesoHmagxxxxxxE
\else
??????\fi
\fi
\fi
\fi
\fi
}
\newcommand{\hatcurCCesoJHmag}[1]{\ifnum#1=519 %
\hatcurCCesoJHmagxxxxxxA
\else
\ifnum#1=3629 %
\hatcurCCesoJHmagxxxxxxB
\else
\ifnum#1=3714 %
\hatcurCCesoJHmagxxxxxxC
\else
\ifnum#1=4201 %
\hatcurCCesoJHmagxxxxxxD
\else
\ifnum#1=5344 %
\hatcurCCesoJHmagxxxxxxE
\else
??????\fi
\fi
\fi
\fi
\fi
}
\newcommand{\hatcurCCesoJKmag}[1]{\ifnum#1=519 %
\hatcurCCesoJKmagxxxxxxA
\else
\ifnum#1=3629 %
\hatcurCCesoJKmagxxxxxxB
\else
\ifnum#1=3714 %
\hatcurCCesoJKmagxxxxxxC
\else
\ifnum#1=4201 %
\hatcurCCesoJKmagxxxxxxD
\else
\ifnum#1=5344 %
\hatcurCCesoJKmagxxxxxxE
\else
??????\fi
\fi
\fi
\fi
\fi
}
\newcommand{\hatcurCCesoJmag}[1]{\ifnum#1=519 %
\hatcurCCesoJmagxxxxxxA
\else
\ifnum#1=3629 %
\hatcurCCesoJmagxxxxxxB
\else
\ifnum#1=3714 %
\hatcurCCesoJmagxxxxxxC
\else
\ifnum#1=4201 %
\hatcurCCesoJmagxxxxxxD
\else
\ifnum#1=5344 %
\hatcurCCesoJmagxxxxxxE
\else
??????\fi
\fi
\fi
\fi
\fi
}
\newcommand{\hatcurCCesoKmag}[1]{\ifnum#1=519 %
\hatcurCCesoKmagxxxxxxA
\else
\ifnum#1=3629 %
\hatcurCCesoKmagxxxxxxB
\else
\ifnum#1=3714 %
\hatcurCCesoKmagxxxxxxC
\else
\ifnum#1=4201 %
\hatcurCCesoKmagxxxxxxD
\else
\ifnum#1=5344 %
\hatcurCCesoKmagxxxxxxE
\else
??????\fi
\fi
\fi
\fi
\fi
}
\newcommand{\hatcurCCgaia}[1]{\ifnum#1=519 %
\hatcurCCgaiaxxxxxxA
\else
\ifnum#1=3629 %
\hatcurCCgaiaxxxxxxB
\else
\ifnum#1=3714 %
\hatcurCCgaiaxxxxxxC
\else
\ifnum#1=4201 %
\hatcurCCgaiaxxxxxxD
\else
\ifnum#1=5344 %
\hatcurCCgaiaxxxxxxE
\else
??????\fi
\fi
\fi
\fi
\fi
}
\newcommand{\hatcurCCgaiadrtwo}[1]{\ifnum#1=519 %
\hatcurCCgaiadrtwoxxxxxxA
\else
\ifnum#1=3629 %
\hatcurCCgaiadrtwoxxxxxxB
\else
\ifnum#1=3714 %
\hatcurCCgaiadrtwoxxxxxxC
\else
\ifnum#1=4201 %
\hatcurCCgaiadrtwoxxxxxxD
\else
\ifnum#1=5344 %
\hatcurCCgaiadrtwoxxxxxxE
\else
??????\fi
\fi
\fi
\fi
\fi
}
\newcommand{\hatcurCCgaiamBP}[1]{\ifnum#1=519 %
\hatcurCCgaiamBPxxxxxxA
\else
\ifnum#1=3629 %
\hatcurCCgaiamBPxxxxxxB
\else
\ifnum#1=3714 %
\hatcurCCgaiamBPxxxxxxC
\else
\ifnum#1=4201 %
\hatcurCCgaiamBPxxxxxxD
\else
\ifnum#1=5344 %
\hatcurCCgaiamBPxxxxxxE
\else
??????\fi
\fi
\fi
\fi
\fi
}
\newcommand{\hatcurCCgaiamG}[1]{\ifnum#1=519 %
\hatcurCCgaiamGxxxxxxA
\else
\ifnum#1=3629 %
\hatcurCCgaiamGxxxxxxB
\else
\ifnum#1=3714 %
\hatcurCCgaiamGxxxxxxC
\else
\ifnum#1=4201 %
\hatcurCCgaiamGxxxxxxD
\else
\ifnum#1=5344 %
\hatcurCCgaiamGxxxxxxE
\else
??????\fi
\fi
\fi
\fi
\fi
}
\newcommand{\hatcurCCgaiamRP}[1]{\ifnum#1=519 %
\hatcurCCgaiamRPxxxxxxA
\else
\ifnum#1=3629 %
\hatcurCCgaiamRPxxxxxxB
\else
\ifnum#1=3714 %
\hatcurCCgaiamRPxxxxxxC
\else
\ifnum#1=4201 %
\hatcurCCgaiamRPxxxxxxD
\else
\ifnum#1=5344 %
\hatcurCCgaiamRPxxxxxxE
\else
??????\fi
\fi
\fi
\fi
\fi
}
\newcommand{\hatcurCCgsc}[1]{\ifnum#1=519 %
\hatcurCCgscxxxxxxA
\else
\ifnum#1=3629 %
\hatcurCCgscxxxxxxB
\else
\ifnum#1=3714 %
\hatcurCCgscxxxxxxC
\else
\ifnum#1=4201 %
\hatcurCCgscxxxxxxD
\else
\ifnum#1=5344 %
\hatcurCCgscxxxxxxE
\else
??????\fi
\fi
\fi
\fi
\fi
}
\newcommand{\hatcurCCmag}[1]{\ifnum#1=519 %
\hatcurCCmagxxxxxxA
\else
\ifnum#1=3629 %
\hatcurCCmagxxxxxxB
\else
\ifnum#1=3714 %
\hatcurCCmagxxxxxxC
\else
\ifnum#1=4201 %
\hatcurCCmagxxxxxxD
\else
\ifnum#1=5344 %
\hatcurCCmagxxxxxxE
\else
??????\fi
\fi
\fi
\fi
\fi
}
\newcommand{\hatcurCCparallax}[1]{\ifnum#1=519 %
\hatcurCCparallaxxxxxxxA
\else
\ifnum#1=3629 %
\hatcurCCparallaxxxxxxxB
\else
\ifnum#1=3714 %
\hatcurCCparallaxxxxxxxC
\else
\ifnum#1=4201 %
\hatcurCCparallaxxxxxxxD
\else
\ifnum#1=5344 %
\hatcurCCparallaxxxxxxxE
\else
??????\fi
\fi
\fi
\fi
\fi
}
\newcommand{\hatcurCCpm}[1]{\ifnum#1=519 %
\hatcurCCpmxxxxxxA
\else
\ifnum#1=3629 %
\hatcurCCpmxxxxxxB
\else
\ifnum#1=3714 %
\hatcurCCpmxxxxxxC
\else
\ifnum#1=4201 %
\hatcurCCpmxxxxxxD
\else
\ifnum#1=5344 %
\hatcurCCpmxxxxxxE
\else
??????\fi
\fi
\fi
\fi
\fi
}
\newcommand{\hatcurCCpmdec}[1]{\ifnum#1=519 %
\hatcurCCpmdecxxxxxxA
\else
\ifnum#1=3629 %
\hatcurCCpmdecxxxxxxB
\else
\ifnum#1=3714 %
\hatcurCCpmdecxxxxxxC
\else
\ifnum#1=4201 %
\hatcurCCpmdecxxxxxxD
\else
\ifnum#1=5344 %
\hatcurCCpmdecxxxxxxE
\else
??????\fi
\fi
\fi
\fi
\fi
}
\newcommand{\hatcurCCpmra}[1]{\ifnum#1=519 %
\hatcurCCpmraxxxxxxA
\else
\ifnum#1=3629 %
\hatcurCCpmraxxxxxxB
\else
\ifnum#1=3714 %
\hatcurCCpmraxxxxxxC
\else
\ifnum#1=4201 %
\hatcurCCpmraxxxxxxD
\else
\ifnum#1=5344 %
\hatcurCCpmraxxxxxxE
\else
??????\fi
\fi
\fi
\fi
\fi
}
\newcommand{\hatcurCCra}[1]{\ifnum#1=519 %
\hatcurCCraxxxxxxA
\else
\ifnum#1=3629 %
\hatcurCCraxxxxxxB
\else
\ifnum#1=3714 %
\hatcurCCraxxxxxxC
\else
\ifnum#1=4201 %
\hatcurCCraxxxxxxD
\else
\ifnum#1=5344 %
\hatcurCCraxxxxxxE
\else
??????\fi
\fi
\fi
\fi
\fi
}
\newcommand{\hatcurCCtassmB}[1]{\ifnum#1=519 %
\hatcurCCtassmBxxxxxxA
\else
\ifnum#1=3629 %
\hatcurCCtassmBxxxxxxB
\else
\ifnum#1=3714 %
\hatcurCCtassmBxxxxxxC
\else
\ifnum#1=4201 %
\hatcurCCtassmBxxxxxxD
\else
\ifnum#1=5344 %
\hatcurCCtassmBxxxxxxE
\else
??????\fi
\fi
\fi
\fi
\fi
}
\newcommand{\hatcurCCtassmBshort}[1]{\ifnum#1=519 %
\hatcurCCtassmBshortxxxxxxA
\else
\ifnum#1=3629 %
\hatcurCCtassmBshortxxxxxxB
\else
\ifnum#1=3714 %
\hatcurCCtassmBshortxxxxxxC
\else
\ifnum#1=4201 %
\hatcurCCtassmBshortxxxxxxD
\else
\ifnum#1=5344 %
\hatcurCCtassmBshortxxxxxxE
\else
??????\fi
\fi
\fi
\fi
\fi
}
\newcommand{\hatcurCCtassmg}[1]{\ifnum#1=519 %
\hatcurCCtassmgxxxxxxA
\else
\ifnum#1=3629 %
\hatcurCCtassmgxxxxxxB
\else
\ifnum#1=3714 %
\hatcurCCtassmgxxxxxxC
\else
\ifnum#1=4201 %
\hatcurCCtassmgxxxxxxD
\else
\ifnum#1=5344 %
\hatcurCCtassmgxxxxxxE
\else
??????\fi
\fi
\fi
\fi
\fi
}
\newcommand{\hatcurCCtassmgshort}[1]{\ifnum#1=519 %
\hatcurCCtassmgshortxxxxxxA
\else
\ifnum#1=3629 %
\hatcurCCtassmgshortxxxxxxB
\else
\ifnum#1=3714 %
\hatcurCCtassmgshortxxxxxxC
\else
\ifnum#1=4201 %
\hatcurCCtassmgshortxxxxxxD
\else
\ifnum#1=5344 %
\hatcurCCtassmgshortxxxxxxE
\else
??????\fi
\fi
\fi
\fi
\fi
}
\newcommand{\hatcurCCtassmi}[1]{\ifnum#1=519 %
\hatcurCCtassmixxxxxxA
\else
\ifnum#1=3629 %
\hatcurCCtassmixxxxxxB
\else
\ifnum#1=3714 %
\hatcurCCtassmixxxxxxC
\else
\ifnum#1=4201 %
\hatcurCCtassmixxxxxxD
\else
\ifnum#1=5344 %
\hatcurCCtassmixxxxxxE
\else
??????\fi
\fi
\fi
\fi
\fi
}
\newcommand{\hatcurCCtassmI}[1]{\ifnum#1=519 %
\hatcurCCtassmIxxxxxxA
\else
\ifnum#1=3629 %
\hatcurCCtassmIxxxxxxB
\else
\ifnum#1=3714 %
\hatcurCCtassmIxxxxxxC
\else
\ifnum#1=4201 %
\hatcurCCtassmIxxxxxxD
\else
\ifnum#1=5344 %
\hatcurCCtassmIxxxxxxE
\else
??????\fi
\fi
\fi
\fi
\fi
}
\newcommand{\hatcurCCtassmishort}[1]{\ifnum#1=519 %
\hatcurCCtassmishortxxxxxxA
\else
\ifnum#1=3629 %
\hatcurCCtassmishortxxxxxxB
\else
\ifnum#1=3714 %
\hatcurCCtassmishortxxxxxxC
\else
\ifnum#1=4201 %
\hatcurCCtassmishortxxxxxxD
\else
\ifnum#1=5344 %
\hatcurCCtassmishortxxxxxxE
\else
??????\fi
\fi
\fi
\fi
\fi
}
\newcommand{\hatcurCCtassmIshort}[1]{\ifnum#1=519 %
\hatcurCCtassmIshortxxxxxxA
\else
\ifnum#1=3629 %
\hatcurCCtassmIshortxxxxxxB
\else
\ifnum#1=3714 %
\hatcurCCtassmIshortxxxxxxC
\else
\ifnum#1=4201 %
\hatcurCCtassmIshortxxxxxxD
\else
\ifnum#1=5344 %
\hatcurCCtassmIshortxxxxxxE
\else
??????\fi
\fi
\fi
\fi
\fi
}
\newcommand{\hatcurCCtassmr}[1]{\ifnum#1=519 %
\hatcurCCtassmrxxxxxxA
\else
\ifnum#1=3629 %
\hatcurCCtassmrxxxxxxB
\else
\ifnum#1=3714 %
\hatcurCCtassmrxxxxxxC
\else
\ifnum#1=4201 %
\hatcurCCtassmrxxxxxxD
\else
\ifnum#1=5344 %
\hatcurCCtassmrxxxxxxE
\else
??????\fi
\fi
\fi
\fi
\fi
}
\newcommand{\hatcurCCtassmrshort}[1]{\ifnum#1=519 %
\hatcurCCtassmrshortxxxxxxA
\else
\ifnum#1=3629 %
\hatcurCCtassmrshortxxxxxxB
\else
\ifnum#1=3714 %
\hatcurCCtassmrshortxxxxxxC
\else
\ifnum#1=4201 %
\hatcurCCtassmrshortxxxxxxD
\else
\ifnum#1=5344 %
\hatcurCCtassmrshortxxxxxxE
\else
??????\fi
\fi
\fi
\fi
\fi
}
\newcommand{\hatcurCCtassmv}[1]{\ifnum#1=519 %
\hatcurCCtassmvxxxxxxA
\else
\ifnum#1=3629 %
\hatcurCCtassmvxxxxxxB
\else
\ifnum#1=3714 %
\hatcurCCtassmvxxxxxxC
\else
\ifnum#1=4201 %
\hatcurCCtassmvxxxxxxD
\else
\ifnum#1=5344 %
\hatcurCCtassmvxxxxxxE
\else
??????\fi
\fi
\fi
\fi
\fi
}
\newcommand{\hatcurCCtassmvshort}[1]{\ifnum#1=519 %
\hatcurCCtassmvshortxxxxxxA
\else
\ifnum#1=3629 %
\hatcurCCtassmvshortxxxxxxB
\else
\ifnum#1=3714 %
\hatcurCCtassmvshortxxxxxxC
\else
\ifnum#1=4201 %
\hatcurCCtassmvshortxxxxxxD
\else
\ifnum#1=5344 %
\hatcurCCtassmvshortxxxxxxE
\else
??????\fi
\fi
\fi
\fi
\fi
}
\newcommand{\hatcurCCtwomass}[1]{\ifnum#1=519 %
\hatcurCCtwomassxxxxxxA
\else
\ifnum#1=3629 %
\hatcurCCtwomassxxxxxxB
\else
\ifnum#1=3714 %
\hatcurCCtwomassxxxxxxC
\else
\ifnum#1=4201 %
\hatcurCCtwomassxxxxxxD
\else
\ifnum#1=5344 %
\hatcurCCtwomassxxxxxxE
\else
??????\fi
\fi
\fi
\fi
\fi
}
\newcommand{\hatcurCCtwomassHmag}[1]{\ifnum#1=519 %
\hatcurCCtwomassHmagxxxxxxA
\else
\ifnum#1=3629 %
\hatcurCCtwomassHmagxxxxxxB
\else
\ifnum#1=3714 %
\hatcurCCtwomassHmagxxxxxxC
\else
\ifnum#1=4201 %
\hatcurCCtwomassHmagxxxxxxD
\else
\ifnum#1=5344 %
\hatcurCCtwomassHmagxxxxxxE
\else
??????\fi
\fi
\fi
\fi
\fi
}
\newcommand{\hatcurCCtwomassJmag}[1]{\ifnum#1=519 %
\hatcurCCtwomassJmagxxxxxxA
\else
\ifnum#1=3629 %
\hatcurCCtwomassJmagxxxxxxB
\else
\ifnum#1=3714 %
\hatcurCCtwomassJmagxxxxxxC
\else
\ifnum#1=4201 %
\hatcurCCtwomassJmagxxxxxxD
\else
\ifnum#1=5344 %
\hatcurCCtwomassJmagxxxxxxE
\else
??????\fi
\fi
\fi
\fi
\fi
}
\newcommand{\hatcurCCtwomassKmag}[1]{\ifnum#1=519 %
\hatcurCCtwomassKmagxxxxxxA
\else
\ifnum#1=3629 %
\hatcurCCtwomassKmagxxxxxxB
\else
\ifnum#1=3714 %
\hatcurCCtwomassKmagxxxxxxC
\else
\ifnum#1=4201 %
\hatcurCCtwomassKmagxxxxxxD
\else
\ifnum#1=5344 %
\hatcurCCtwomassKmagxxxxxxE
\else
??????\fi
\fi
\fi
\fi
\fi
}
\newcommand{\hatcurCCWfourmag}[1]{\ifnum#1=519 %
\hatcurCCWfourmagxxxxxxA
\else
\ifnum#1=3629 %
\hatcurCCWfourmagxxxxxxB
\else
\ifnum#1=3714 %
\hatcurCCWfourmagxxxxxxC
\else
\ifnum#1=4201 %
\hatcurCCWfourmagxxxxxxD
\else
\ifnum#1=5344 %
\hatcurCCWfourmagxxxxxxE
\else
??????\fi
\fi
\fi
\fi
\fi
}
\newcommand{\hatcurCCWonemag}[1]{\ifnum#1=519 %
\hatcurCCWonemagxxxxxxA
\else
\ifnum#1=3629 %
\hatcurCCWonemagxxxxxxB
\else
\ifnum#1=3714 %
\hatcurCCWonemagxxxxxxC
\else
\ifnum#1=4201 %
\hatcurCCWonemagxxxxxxD
\else
\ifnum#1=5344 %
\hatcurCCWonemagxxxxxxE
\else
??????\fi
\fi
\fi
\fi
\fi
}
\newcommand{\hatcurCCWthreemag}[1]{\ifnum#1=519 %
\hatcurCCWthreemagxxxxxxA
\else
\ifnum#1=3629 %
\hatcurCCWthreemagxxxxxxB
\else
\ifnum#1=3714 %
\hatcurCCWthreemagxxxxxxC
\else
\ifnum#1=4201 %
\hatcurCCWthreemagxxxxxxD
\else
\ifnum#1=5344 %
\hatcurCCWthreemagxxxxxxE
\else
??????\fi
\fi
\fi
\fi
\fi
}
\newcommand{\hatcurCCWtwomag}[1]{\ifnum#1=519 %
\hatcurCCWtwomagxxxxxxA
\else
\ifnum#1=3629 %
\hatcurCCWtwomagxxxxxxB
\else
\ifnum#1=3714 %
\hatcurCCWtwomagxxxxxxC
\else
\ifnum#1=4201 %
\hatcurCCWtwomagxxxxxxD
\else
\ifnum#1=5344 %
\hatcurCCWtwomagxxxxxxE
\else
??????\fi
\fi
\fi
\fi
\fi
}
\newcommand{\hatcurextraerrMg}[1]{\ifnum#1=3629 %
\hatcurextraerrMgxxxxxxB
\else
\ifnum#1=3714 %
\hatcurextraerrMgxxxxxxC
\else
\ifnum#1=4201 %
\hatcurextraerrMgxxxxxxD
\else
\ifnum#1=5344 %
\hatcurextraerrMgxxxxxxE
\else
??????\fi
\fi
\fi
\fi
}
\newcommand{\hatcurextraerrMG}[1]{\ifnum#1=519 %
\hatcurextraerrMGxxxxxxA
\else
\ifnum#1=3629 %
\hatcurextraerrMGxxxxxxB
\else
\ifnum#1=3714 %
\hatcurextraerrMGxxxxxxC
\else
\ifnum#1=4201 %
\hatcurextraerrMGxxxxxxD
\else
\ifnum#1=5344 %
\hatcurextraerrMGxxxxxxE
\else
??????\fi
\fi
\fi
\fi
\fi
}
\newcommand{\hatcurextraerrMgtwosiglim}[1]{\ifnum#1=3629 %
\hatcurextraerrMgtwosiglimxxxxxxB
\else
\ifnum#1=3714 %
\hatcurextraerrMgtwosiglimxxxxxxC
\else
\ifnum#1=4201 %
\hatcurextraerrMgtwosiglimxxxxxxD
\else
\ifnum#1=5344 %
\hatcurextraerrMgtwosiglimxxxxxxE
\else
??????\fi
\fi
\fi
\fi
}
\newcommand{\hatcurextraerrMGtwosiglim}[1]{\ifnum#1=519 %
\hatcurextraerrMGtwosiglimxxxxxxA
\else
\ifnum#1=3629 %
\hatcurextraerrMGtwosiglimxxxxxxB
\else
\ifnum#1=3714 %
\hatcurextraerrMGtwosiglimxxxxxxC
\else
\ifnum#1=4201 %
\hatcurextraerrMGtwosiglimxxxxxxD
\else
\ifnum#1=5344 %
\hatcurextraerrMGtwosiglimxxxxxxE
\else
??????\fi
\fi
\fi
\fi
\fi
}
\newcommand{\hatcurextraerrMH}[1]{\ifnum#1=519 %
\hatcurextraerrMHxxxxxxA
\else
\ifnum#1=3629 %
\hatcurextraerrMHxxxxxxB
\else
\ifnum#1=3714 %
\hatcurextraerrMHxxxxxxC
\else
\ifnum#1=4201 %
\hatcurextraerrMHxxxxxxD
\else
\ifnum#1=5344 %
\hatcurextraerrMHxxxxxxE
\else
??????\fi
\fi
\fi
\fi
\fi
}
\newcommand{\hatcurextraerrMHtwosiglim}[1]{\ifnum#1=519 %
\hatcurextraerrMHtwosiglimxxxxxxA
\else
\ifnum#1=3629 %
\hatcurextraerrMHtwosiglimxxxxxxB
\else
\ifnum#1=3714 %
\hatcurextraerrMHtwosiglimxxxxxxC
\else
\ifnum#1=4201 %
\hatcurextraerrMHtwosiglimxxxxxxD
\else
\ifnum#1=5344 %
\hatcurextraerrMHtwosiglimxxxxxxE
\else
??????\fi
\fi
\fi
\fi
\fi
}
\newcommand{\hatcurextraerrMi}[1]{\ifnum#1=3629 %
\hatcurextraerrMixxxxxxB
\else
\ifnum#1=3714 %
\hatcurextraerrMixxxxxxC
\else
\ifnum#1=4201 %
\hatcurextraerrMixxxxxxD
\else
\ifnum#1=5344 %
\hatcurextraerrMixxxxxxE
\else
??????\fi
\fi
\fi
\fi
}
\newcommand{\hatcurextraerrMitwosiglim}[1]{\ifnum#1=3629 %
\hatcurextraerrMitwosiglimxxxxxxB
\else
\ifnum#1=3714 %
\hatcurextraerrMitwosiglimxxxxxxC
\else
\ifnum#1=4201 %
\hatcurextraerrMitwosiglimxxxxxxD
\else
\ifnum#1=5344 %
\hatcurextraerrMitwosiglimxxxxxxE
\else
??????\fi
\fi
\fi
\fi
}
\newcommand{\hatcurextraerrMJ}[1]{\ifnum#1=519 %
\hatcurextraerrMJxxxxxxA
\else
\ifnum#1=3629 %
\hatcurextraerrMJxxxxxxB
\else
\ifnum#1=3714 %
\hatcurextraerrMJxxxxxxC
\else
\ifnum#1=4201 %
\hatcurextraerrMJxxxxxxD
\else
\ifnum#1=5344 %
\hatcurextraerrMJxxxxxxE
\else
??????\fi
\fi
\fi
\fi
\fi
}
\newcommand{\hatcurextraerrMJtwosiglim}[1]{\ifnum#1=519 %
\hatcurextraerrMJtwosiglimxxxxxxA
\else
\ifnum#1=3629 %
\hatcurextraerrMJtwosiglimxxxxxxB
\else
\ifnum#1=3714 %
\hatcurextraerrMJtwosiglimxxxxxxC
\else
\ifnum#1=4201 %
\hatcurextraerrMJtwosiglimxxxxxxD
\else
\ifnum#1=5344 %
\hatcurextraerrMJtwosiglimxxxxxxE
\else
??????\fi
\fi
\fi
\fi
\fi
}
\newcommand{\hatcurextraerrMKs}[1]{\ifnum#1=519 %
\hatcurextraerrMKsxxxxxxA
\else
\ifnum#1=3629 %
\hatcurextraerrMKsxxxxxxB
\else
\ifnum#1=3714 %
\hatcurextraerrMKsxxxxxxC
\else
\ifnum#1=4201 %
\hatcurextraerrMKsxxxxxxD
\else
\ifnum#1=5344 %
\hatcurextraerrMKsxxxxxxE
\else
??????\fi
\fi
\fi
\fi
\fi
}
\newcommand{\hatcurextraerrMKstwosiglim}[1]{\ifnum#1=519 %
\hatcurextraerrMKstwosiglimxxxxxxA
\else
\ifnum#1=3629 %
\hatcurextraerrMKstwosiglimxxxxxxB
\else
\ifnum#1=3714 %
\hatcurextraerrMKstwosiglimxxxxxxC
\else
\ifnum#1=4201 %
\hatcurextraerrMKstwosiglimxxxxxxD
\else
\ifnum#1=5344 %
\hatcurextraerrMKstwosiglimxxxxxxE
\else
??????\fi
\fi
\fi
\fi
\fi
}
\newcommand{\hatcurextraerrMr}[1]{\ifnum#1=3629 %
\hatcurextraerrMrxxxxxxB
\else
\ifnum#1=3714 %
\hatcurextraerrMrxxxxxxC
\else
\ifnum#1=4201 %
\hatcurextraerrMrxxxxxxD
\else
\ifnum#1=5344 %
\hatcurextraerrMrxxxxxxE
\else
??????\fi
\fi
\fi
\fi
}
\newcommand{\hatcurextraerrMRP}[1]{\ifnum#1=519 %
\hatcurextraerrMRPxxxxxxA
\else
\ifnum#1=3629 %
\hatcurextraerrMRPxxxxxxB
\else
\ifnum#1=3714 %
\hatcurextraerrMRPxxxxxxC
\else
\ifnum#1=4201 %
\hatcurextraerrMRPxxxxxxD
\else
\ifnum#1=5344 %
\hatcurextraerrMRPxxxxxxE
\else
??????\fi
\fi
\fi
\fi
\fi
}
\newcommand{\hatcurextraerrMRPtwosiglim}[1]{\ifnum#1=519 %
\hatcurextraerrMRPtwosiglimxxxxxxA
\else
\ifnum#1=3629 %
\hatcurextraerrMRPtwosiglimxxxxxxB
\else
\ifnum#1=3714 %
\hatcurextraerrMRPtwosiglimxxxxxxC
\else
\ifnum#1=4201 %
\hatcurextraerrMRPtwosiglimxxxxxxD
\else
\ifnum#1=5344 %
\hatcurextraerrMRPtwosiglimxxxxxxE
\else
??????\fi
\fi
\fi
\fi
\fi
}
\newcommand{\hatcurextraerrMrtwosiglim}[1]{\ifnum#1=3629 %
\hatcurextraerrMrtwosiglimxxxxxxB
\else
\ifnum#1=3714 %
\hatcurextraerrMrtwosiglimxxxxxxC
\else
\ifnum#1=4201 %
\hatcurextraerrMrtwosiglimxxxxxxD
\else
\ifnum#1=5344 %
\hatcurextraerrMrtwosiglimxxxxxxE
\else
??????\fi
\fi
\fi
\fi
}
\newcommand{\hatcurextraerrMWone}[1]{\ifnum#1=519 %
\hatcurextraerrMWonexxxxxxA
\else
\ifnum#1=3629 %
\hatcurextraerrMWonexxxxxxB
\else
\ifnum#1=3714 %
\hatcurextraerrMWonexxxxxxC
\else
\ifnum#1=4201 %
\hatcurextraerrMWonexxxxxxD
\else
\ifnum#1=5344 %
\hatcurextraerrMWonexxxxxxE
\else
??????\fi
\fi
\fi
\fi
\fi
}
\newcommand{\hatcurextraerrMWonetwosiglim}[1]{\ifnum#1=519 %
\hatcurextraerrMWonetwosiglimxxxxxxA
\else
\ifnum#1=3629 %
\hatcurextraerrMWonetwosiglimxxxxxxB
\else
\ifnum#1=3714 %
\hatcurextraerrMWonetwosiglimxxxxxxC
\else
\ifnum#1=4201 %
\hatcurextraerrMWonetwosiglimxxxxxxD
\else
\ifnum#1=5344 %
\hatcurextraerrMWonetwosiglimxxxxxxE
\else
??????\fi
\fi
\fi
\fi
\fi
}
\newcommand{\hatcurextraerrMWtwo}[1]{\ifnum#1=519 %
\hatcurextraerrMWtwoxxxxxxA
\else
\ifnum#1=3629 %
\hatcurextraerrMWtwoxxxxxxB
\else
\ifnum#1=3714 %
\hatcurextraerrMWtwoxxxxxxC
\else
\ifnum#1=4201 %
\hatcurextraerrMWtwoxxxxxxD
\else
\ifnum#1=5344 %
\hatcurextraerrMWtwoxxxxxxE
\else
??????\fi
\fi
\fi
\fi
\fi
}
\newcommand{\hatcurextraerrMWtwotwosiglim}[1]{\ifnum#1=519 %
\hatcurextraerrMWtwotwosiglimxxxxxxA
\else
\ifnum#1=3629 %
\hatcurextraerrMWtwotwosiglimxxxxxxB
\else
\ifnum#1=3714 %
\hatcurextraerrMWtwotwosiglimxxxxxxC
\else
\ifnum#1=4201 %
\hatcurextraerrMWtwotwosiglimxxxxxxD
\else
\ifnum#1=5344 %
\hatcurextraerrMWtwotwosiglimxxxxxxE
\else
??????\fi
\fi
\fi
\fi
\fi
}
\newcommand{\hatcurfield}[1]{\ifnum#1=519 %
\hatcurfieldxxxxxxA
\else
\ifnum#1=3629 %
\hatcurfieldxxxxxxB
\else
\ifnum#1=3714 %
\hatcurfieldxxxxxxC
\else
\ifnum#1=4201 %
\hatcurfieldxxxxxxD
\else
\ifnum#1=5344 %
\hatcurfieldxxxxxxE
\else
??????\fi
\fi
\fi
\fi
\fi
}
\newcommand{\hatcurhtr}[1]{\ifnum#1=519 %
\hatcurhtrxxxxxxA
\else
\ifnum#1=3629 %
\hatcurhtrxxxxxxB
\else
\ifnum#1=3714 %
\hatcurhtrxxxxxxC
\else
\ifnum#1=4201 %
\hatcurhtrxxxxxxD
\else
\ifnum#1=5344 %
\hatcurhtrxxxxxxE
\else
??????\fi
\fi
\fi
\fi
\fi
}
\newcommand{\hatcurISOage}[1]{\ifnum#1=519 %
\hatcurISOagexxxxxxA
\else
\ifnum#1=3629 %
\hatcurISOagexxxxxxB
\else
\ifnum#1=3714 %
\hatcurISOagexxxxxxC
\else
\ifnum#1=4201 %
\hatcurISOagexxxxxxD
\else
\ifnum#1=5344 %
\hatcurISOagexxxxxxE
\else
??????\fi
\fi
\fi
\fi
\fi
}
\newcommand{\hatcurISOlogg}[1]{\ifnum#1=519 %
\hatcurISOloggxxxxxxA
\else
\ifnum#1=3629 %
\hatcurISOloggxxxxxxB
\else
\ifnum#1=3714 %
\hatcurISOloggxxxxxxC
\else
\ifnum#1=4201 %
\hatcurISOloggxxxxxxD
\else
\ifnum#1=5344 %
\hatcurISOloggxxxxxxE
\else
??????\fi
\fi
\fi
\fi
\fi
}
\newcommand{\hatcurISOlum}[1]{\ifnum#1=519 %
\hatcurISOlumxxxxxxA
\else
\ifnum#1=3629 %
\hatcurISOlumxxxxxxB
\else
\ifnum#1=3714 %
\hatcurISOlumxxxxxxC
\else
\ifnum#1=4201 %
\hatcurISOlumxxxxxxD
\else
\ifnum#1=5344 %
\hatcurISOlumxxxxxxE
\else
??????\fi
\fi
\fi
\fi
\fi
}
\newcommand{\hatcurISOlumshort}[1]{\ifnum#1=519 %
\hatcurISOlumshortxxxxxxA
\else
\ifnum#1=3629 %
\hatcurISOlumshortxxxxxxB
\else
\ifnum#1=3714 %
\hatcurISOlumshortxxxxxxC
\else
\ifnum#1=4201 %
\hatcurISOlumshortxxxxxxD
\else
\ifnum#1=5344 %
\hatcurISOlumshortxxxxxxE
\else
??????\fi
\fi
\fi
\fi
\fi
}
\newcommand{\hatcurISOm}[1]{\ifnum#1=519 %
\hatcurISOmxxxxxxA
\else
\ifnum#1=3629 %
\hatcurISOmxxxxxxB
\else
\ifnum#1=3714 %
\hatcurISOmxxxxxxC
\else
\ifnum#1=4201 %
\hatcurISOmxxxxxxD
\else
\ifnum#1=5344 %
\hatcurISOmxxxxxxE
\else
??????\fi
\fi
\fi
\fi
\fi
}
\newcommand{\hatcurISOmlong}[1]{\ifnum#1=519 %
\hatcurISOmlongxxxxxxA
\else
\ifnum#1=3629 %
\hatcurISOmlongxxxxxxB
\else
\ifnum#1=3714 %
\hatcurISOmlongxxxxxxC
\else
\ifnum#1=4201 %
\hatcurISOmlongxxxxxxD
\else
\ifnum#1=5344 %
\hatcurISOmlongxxxxxxE
\else
??????\fi
\fi
\fi
\fi
\fi
}
\newcommand{\hatcurISOmshort}[1]{\ifnum#1=519 %
\hatcurISOmshortxxxxxxA
\else
\ifnum#1=3629 %
\hatcurISOmshortxxxxxxB
\else
\ifnum#1=3714 %
\hatcurISOmshortxxxxxxC
\else
\ifnum#1=4201 %
\hatcurISOmshortxxxxxxD
\else
\ifnum#1=5344 %
\hatcurISOmshortxxxxxxE
\else
??????\fi
\fi
\fi
\fi
\fi
}
\newcommand{\hatcurISOr}[1]{\ifnum#1=519 %
\hatcurISOrxxxxxxA
\else
\ifnum#1=3629 %
\hatcurISOrxxxxxxB
\else
\ifnum#1=3714 %
\hatcurISOrxxxxxxC
\else
\ifnum#1=4201 %
\hatcurISOrxxxxxxD
\else
\ifnum#1=5344 %
\hatcurISOrxxxxxxE
\else
??????\fi
\fi
\fi
\fi
\fi
}
\newcommand{\hatcurISOrho}[1]{\ifnum#1=519 %
\hatcurISOrhoxxxxxxA
\else
\ifnum#1=3629 %
\hatcurISOrhoxxxxxxB
\else
\ifnum#1=3714 %
\hatcurISOrhoxxxxxxC
\else
\ifnum#1=4201 %
\hatcurISOrhoxxxxxxD
\else
\ifnum#1=5344 %
\hatcurISOrhoxxxxxxE
\else
??????\fi
\fi
\fi
\fi
\fi
}
\newcommand{\hatcurISOrholong}[1]{\ifnum#1=519 %
\hatcurISOrholongxxxxxxA
\else
\ifnum#1=3629 %
\hatcurISOrholongxxxxxxB
\else
\ifnum#1=3714 %
\hatcurISOrholongxxxxxxC
\else
\ifnum#1=4201 %
\hatcurISOrholongxxxxxxD
\else
\ifnum#1=5344 %
\hatcurISOrholongxxxxxxE
\else
??????\fi
\fi
\fi
\fi
\fi
}
\newcommand{\hatcurISOrlong}[1]{\ifnum#1=519 %
\hatcurISOrlongxxxxxxA
\else
\ifnum#1=3629 %
\hatcurISOrlongxxxxxxB
\else
\ifnum#1=3714 %
\hatcurISOrlongxxxxxxC
\else
\ifnum#1=4201 %
\hatcurISOrlongxxxxxxD
\else
\ifnum#1=5344 %
\hatcurISOrlongxxxxxxE
\else
??????\fi
\fi
\fi
\fi
\fi
}
\newcommand{\hatcurISOrshort}[1]{\ifnum#1=519 %
\hatcurISOrshortxxxxxxA
\else
\ifnum#1=3629 %
\hatcurISOrshortxxxxxxB
\else
\ifnum#1=3714 %
\hatcurISOrshortxxxxxxC
\else
\ifnum#1=4201 %
\hatcurISOrshortxxxxxxD
\else
\ifnum#1=5344 %
\hatcurISOrshortxxxxxxE
\else
??????\fi
\fi
\fi
\fi
\fi
}
\newcommand{\hatcurISOspec}[1]{\ifnum#1=519 %
\hatcurISOspecxxxxxxA
\else
\ifnum#1=3629 %
\hatcurISOspecxxxxxxB
\else
\ifnum#1=3714 %
\hatcurISOspecxxxxxxC
\else
\ifnum#1=4201 %
\hatcurISOspecxxxxxxD
\else
\ifnum#1=5344 %
\hatcurISOspecxxxxxxE
\else
??????\fi
\fi
\fi
\fi
\fi
}
\newcommand{\hatcurISOteff}[1]{\ifnum#1=519 %
\hatcurISOteffxxxxxxA
\else
\ifnum#1=3629 %
\hatcurISOteffxxxxxxB
\else
\ifnum#1=3714 %
\hatcurISOteffxxxxxxC
\else
\ifnum#1=4201 %
\hatcurISOteffxxxxxxD
\else
\ifnum#1=5344 %
\hatcurISOteffxxxxxxE
\else
??????\fi
\fi
\fi
\fi
\fi
}
\newcommand{\hatcurISOzfeh}[1]{\ifnum#1=519 %
\hatcurISOzfehxxxxxxA
\else
\ifnum#1=3629 %
\hatcurISOzfehxxxxxxB
\else
\ifnum#1=3714 %
\hatcurISOzfehxxxxxxC
\else
\ifnum#1=4201 %
\hatcurISOzfehxxxxxxD
\else
\ifnum#1=5344 %
\hatcurISOzfehxxxxxxE
\else
??????\fi
\fi
\fi
\fi
\fi
}
\newcommand{\hatcurLBiB}[1]{\ifnum#1=519 %
\hatcurLBiBxxxxxxA
\else
\ifnum#1=3629 %
\hatcurLBiBxxxxxxB
\else
\ifnum#1=3714 %
\hatcurLBiBxxxxxxC
\else
\ifnum#1=4201 %
\hatcurLBiBxxxxxxD
\else
\ifnum#1=5344 %
\hatcurLBiBxxxxxxE
\else
??????\fi
\fi
\fi
\fi
\fi
}
\newcommand{\hatcurLBiC}[1]{\ifnum#1=519 %
\hatcurLBiCxxxxxxA
\else
\ifnum#1=3629 %
\hatcurLBiCxxxxxxB
\else
\ifnum#1=3714 %
\hatcurLBiCxxxxxxC
\else
\ifnum#1=4201 %
\hatcurLBiCxxxxxxD
\else
\ifnum#1=5344 %
\hatcurLBiCxxxxxxE
\else
??????\fi
\fi
\fi
\fi
\fi
}
\newcommand{\hatcurLBig}[1]{\ifnum#1=519 %
\hatcurLBigxxxxxxA
\else
\ifnum#1=3629 %
\hatcurLBigxxxxxxB
\else
\ifnum#1=3714 %
\hatcurLBigxxxxxxC
\else
\ifnum#1=4201 %
\hatcurLBigxxxxxxD
\else
\ifnum#1=5344 %
\hatcurLBigxxxxxxE
\else
??????\fi
\fi
\fi
\fi
\fi
}
\newcommand{\hatcurLBiH}[1]{\ifnum#1=519 %
\hatcurLBiHxxxxxxA
\else
\ifnum#1=3629 %
\hatcurLBiHxxxxxxB
\else
\ifnum#1=3714 %
\hatcurLBiHxxxxxxC
\else
\ifnum#1=4201 %
\hatcurLBiHxxxxxxD
\else
\ifnum#1=5344 %
\hatcurLBiHxxxxxxE
\else
??????\fi
\fi
\fi
\fi
\fi
}
\newcommand{\hatcurLBii}[1]{\ifnum#1=519 %
\hatcurLBiixxxxxxA
\else
\ifnum#1=3629 %
\hatcurLBiixxxxxxB
\else
\ifnum#1=3714 %
\hatcurLBiixxxxxxC
\else
\ifnum#1=4201 %
\hatcurLBiixxxxxxD
\else
\ifnum#1=5344 %
\hatcurLBiixxxxxxE
\else
??????\fi
\fi
\fi
\fi
\fi
}
\newcommand{\hatcurLBiI}[1]{\ifnum#1=519 %
\hatcurLBiIxxxxxxA
\else
\ifnum#1=3629 %
\hatcurLBiIxxxxxxB
\else
\ifnum#1=3714 %
\hatcurLBiIxxxxxxC
\else
\ifnum#1=4201 %
\hatcurLBiIxxxxxxD
\else
\ifnum#1=5344 %
\hatcurLBiIxxxxxxE
\else
??????\fi
\fi
\fi
\fi
\fi
}
\newcommand{\hatcurLBiiB}[1]{\ifnum#1=519 %
\hatcurLBiiBxxxxxxA
\else
\ifnum#1=3629 %
\hatcurLBiiBxxxxxxB
\else
\ifnum#1=3714 %
\hatcurLBiiBxxxxxxC
\else
\ifnum#1=4201 %
\hatcurLBiiBxxxxxxD
\else
\ifnum#1=5344 %
\hatcurLBiiBxxxxxxE
\else
??????\fi
\fi
\fi
\fi
\fi
}
\newcommand{\hatcurLBiiC}[1]{\ifnum#1=519 %
\hatcurLBiiCxxxxxxA
\else
\ifnum#1=3629 %
\hatcurLBiiCxxxxxxB
\else
\ifnum#1=3714 %
\hatcurLBiiCxxxxxxC
\else
\ifnum#1=4201 %
\hatcurLBiiCxxxxxxD
\else
\ifnum#1=5344 %
\hatcurLBiiCxxxxxxE
\else
??????\fi
\fi
\fi
\fi
\fi
}
\newcommand{\hatcurLBiig}[1]{\ifnum#1=519 %
\hatcurLBiigxxxxxxA
\else
\ifnum#1=3629 %
\hatcurLBiigxxxxxxB
\else
\ifnum#1=3714 %
\hatcurLBiigxxxxxxC
\else
\ifnum#1=4201 %
\hatcurLBiigxxxxxxD
\else
\ifnum#1=5344 %
\hatcurLBiigxxxxxxE
\else
??????\fi
\fi
\fi
\fi
\fi
}
\newcommand{\hatcurLBiiH}[1]{\ifnum#1=519 %
\hatcurLBiiHxxxxxxA
\else
\ifnum#1=3629 %
\hatcurLBiiHxxxxxxB
\else
\ifnum#1=3714 %
\hatcurLBiiHxxxxxxC
\else
\ifnum#1=4201 %
\hatcurLBiiHxxxxxxD
\else
\ifnum#1=5344 %
\hatcurLBiiHxxxxxxE
\else
??????\fi
\fi
\fi
\fi
\fi
}
\newcommand{\hatcurLBiii}[1]{\ifnum#1=519 %
\hatcurLBiiixxxxxxA
\else
\ifnum#1=3629 %
\hatcurLBiiixxxxxxB
\else
\ifnum#1=3714 %
\hatcurLBiiixxxxxxC
\else
\ifnum#1=4201 %
\hatcurLBiiixxxxxxD
\else
\ifnum#1=5344 %
\hatcurLBiiixxxxxxE
\else
??????\fi
\fi
\fi
\fi
\fi
}
\newcommand{\hatcurLBiiI}[1]{\ifnum#1=519 %
\hatcurLBiiIxxxxxxA
\else
\ifnum#1=3629 %
\hatcurLBiiIxxxxxxB
\else
\ifnum#1=3714 %
\hatcurLBiiIxxxxxxC
\else
\ifnum#1=4201 %
\hatcurLBiiIxxxxxxD
\else
\ifnum#1=5344 %
\hatcurLBiiIxxxxxxE
\else
??????\fi
\fi
\fi
\fi
\fi
}
\newcommand{\hatcurLBiiJ}[1]{\ifnum#1=519 %
\hatcurLBiiJxxxxxxA
\else
\ifnum#1=3629 %
\hatcurLBiiJxxxxxxB
\else
\ifnum#1=3714 %
\hatcurLBiiJxxxxxxC
\else
\ifnum#1=4201 %
\hatcurLBiiJxxxxxxD
\else
\ifnum#1=5344 %
\hatcurLBiiJxxxxxxE
\else
??????\fi
\fi
\fi
\fi
\fi
}
\newcommand{\hatcurLBiiK}[1]{\ifnum#1=519 %
\hatcurLBiiKxxxxxxA
\else
\ifnum#1=3629 %
\hatcurLBiiKxxxxxxB
\else
\ifnum#1=3714 %
\hatcurLBiiKxxxxxxC
\else
\ifnum#1=4201 %
\hatcurLBiiKxxxxxxD
\else
\ifnum#1=5344 %
\hatcurLBiiKxxxxxxE
\else
??????\fi
\fi
\fi
\fi
\fi
}
\newcommand{\hatcurLBiikep}[1]{\ifnum#1=519 %
\hatcurLBiikepxxxxxxA
\else
\ifnum#1=3629 %
\hatcurLBiikepxxxxxxB
\else
\ifnum#1=3714 %
\hatcurLBiikepxxxxxxC
\else
\ifnum#1=4201 %
\hatcurLBiikepxxxxxxD
\else
\ifnum#1=5344 %
\hatcurLBiikepxxxxxxE
\else
??????\fi
\fi
\fi
\fi
\fi
}
\newcommand{\hatcurLBiiM}[1]{\ifnum#1=519 %
\hatcurLBiiMxxxxxxA
\else
\ifnum#1=3629 %
\hatcurLBiiMxxxxxxB
\else
\ifnum#1=3714 %
\hatcurLBiiMxxxxxxC
\else
\ifnum#1=4201 %
\hatcurLBiiMxxxxxxD
\else
\ifnum#1=5344 %
\hatcurLBiiMxxxxxxE
\else
??????\fi
\fi
\fi
\fi
\fi
}
\newcommand{\hatcurLBiir}[1]{\ifnum#1=519 %
\hatcurLBiirxxxxxxA
\else
\ifnum#1=3629 %
\hatcurLBiirxxxxxxB
\else
\ifnum#1=3714 %
\hatcurLBiirxxxxxxC
\else
\ifnum#1=4201 %
\hatcurLBiirxxxxxxD
\else
\ifnum#1=5344 %
\hatcurLBiirxxxxxxE
\else
??????\fi
\fi
\fi
\fi
\fi
}
\newcommand{\hatcurLBiiR}[1]{\ifnum#1=519 %
\hatcurLBiiRxxxxxxA
\else
\ifnum#1=3629 %
\hatcurLBiiRxxxxxxB
\else
\ifnum#1=3714 %
\hatcurLBiiRxxxxxxC
\else
\ifnum#1=4201 %
\hatcurLBiiRxxxxxxD
\else
\ifnum#1=5344 %
\hatcurLBiiRxxxxxxE
\else
??????\fi
\fi
\fi
\fi
\fi
}
\newcommand{\hatcurLBiiSfour}[1]{\ifnum#1=519 %
\hatcurLBiiSfourxxxxxxA
\else
\ifnum#1=3629 %
\hatcurLBiiSfourxxxxxxB
\else
\ifnum#1=3714 %
\hatcurLBiiSfourxxxxxxC
\else
\ifnum#1=4201 %
\hatcurLBiiSfourxxxxxxD
\else
\ifnum#1=5344 %
\hatcurLBiiSfourxxxxxxE
\else
??????\fi
\fi
\fi
\fi
\fi
}
\newcommand{\hatcurLBiiSone}[1]{\ifnum#1=519 %
\hatcurLBiiSonexxxxxxA
\else
\ifnum#1=3629 %
\hatcurLBiiSonexxxxxxB
\else
\ifnum#1=3714 %
\hatcurLBiiSonexxxxxxC
\else
\ifnum#1=4201 %
\hatcurLBiiSonexxxxxxD
\else
\ifnum#1=5344 %
\hatcurLBiiSonexxxxxxE
\else
??????\fi
\fi
\fi
\fi
\fi
}
\newcommand{\hatcurLBiiSthree}[1]{\ifnum#1=519 %
\hatcurLBiiSthreexxxxxxA
\else
\ifnum#1=3629 %
\hatcurLBiiSthreexxxxxxB
\else
\ifnum#1=3714 %
\hatcurLBiiSthreexxxxxxC
\else
\ifnum#1=4201 %
\hatcurLBiiSthreexxxxxxD
\else
\ifnum#1=5344 %
\hatcurLBiiSthreexxxxxxE
\else
??????\fi
\fi
\fi
\fi
\fi
}
\newcommand{\hatcurLBiiStwo}[1]{\ifnum#1=519 %
\hatcurLBiiStwoxxxxxxA
\else
\ifnum#1=3629 %
\hatcurLBiiStwoxxxxxxB
\else
\ifnum#1=3714 %
\hatcurLBiiStwoxxxxxxC
\else
\ifnum#1=4201 %
\hatcurLBiiStwoxxxxxxD
\else
\ifnum#1=5344 %
\hatcurLBiiStwoxxxxxxE
\else
??????\fi
\fi
\fi
\fi
\fi
}
\newcommand{\hatcurLBiiT}[1]{\ifnum#1=519 %
\hatcurLBiiTxxxxxxA
\else
\ifnum#1=3629 %
\hatcurLBiiTxxxxxxB
\else
\ifnum#1=3714 %
\hatcurLBiiTxxxxxxC
\else
\ifnum#1=4201 %
\hatcurLBiiTxxxxxxD
\else
\ifnum#1=5344 %
\hatcurLBiiTxxxxxxE
\else
??????\fi
\fi
\fi
\fi
\fi
}
\newcommand{\hatcurLBiiu}[1]{\ifnum#1=519 %
\hatcurLBiiuxxxxxxA
\else
\ifnum#1=3629 %
\hatcurLBiiuxxxxxxB
\else
\ifnum#1=3714 %
\hatcurLBiiuxxxxxxC
\else
\ifnum#1=4201 %
\hatcurLBiiuxxxxxxD
\else
\ifnum#1=5344 %
\hatcurLBiiuxxxxxxE
\else
??????\fi
\fi
\fi
\fi
\fi
}
\newcommand{\hatcurLBiiV}[1]{\ifnum#1=519 %
\hatcurLBiiVxxxxxxA
\else
\ifnum#1=3629 %
\hatcurLBiiVxxxxxxB
\else
\ifnum#1=3714 %
\hatcurLBiiVxxxxxxC
\else
\ifnum#1=4201 %
\hatcurLBiiVxxxxxxD
\else
\ifnum#1=5344 %
\hatcurLBiiVxxxxxxE
\else
??????\fi
\fi
\fi
\fi
\fi
}
\newcommand{\hatcurLBiiz}[1]{\ifnum#1=519 %
\hatcurLBiizxxxxxxA
\else
\ifnum#1=3629 %
\hatcurLBiizxxxxxxB
\else
\ifnum#1=3714 %
\hatcurLBiizxxxxxxC
\else
\ifnum#1=4201 %
\hatcurLBiizxxxxxxD
\else
\ifnum#1=5344 %
\hatcurLBiizxxxxxxE
\else
??????\fi
\fi
\fi
\fi
\fi
}
\newcommand{\hatcurLBiJ}[1]{\ifnum#1=519 %
\hatcurLBiJxxxxxxA
\else
\ifnum#1=3629 %
\hatcurLBiJxxxxxxB
\else
\ifnum#1=3714 %
\hatcurLBiJxxxxxxC
\else
\ifnum#1=4201 %
\hatcurLBiJxxxxxxD
\else
\ifnum#1=5344 %
\hatcurLBiJxxxxxxE
\else
??????\fi
\fi
\fi
\fi
\fi
}
\newcommand{\hatcurLBiK}[1]{\ifnum#1=519 %
\hatcurLBiKxxxxxxA
\else
\ifnum#1=3629 %
\hatcurLBiKxxxxxxB
\else
\ifnum#1=3714 %
\hatcurLBiKxxxxxxC
\else
\ifnum#1=4201 %
\hatcurLBiKxxxxxxD
\else
\ifnum#1=5344 %
\hatcurLBiKxxxxxxE
\else
??????\fi
\fi
\fi
\fi
\fi
}
\newcommand{\hatcurLBikep}[1]{\ifnum#1=519 %
\hatcurLBikepxxxxxxA
\else
\ifnum#1=3629 %
\hatcurLBikepxxxxxxB
\else
\ifnum#1=3714 %
\hatcurLBikepxxxxxxC
\else
\ifnum#1=4201 %
\hatcurLBikepxxxxxxD
\else
\ifnum#1=5344 %
\hatcurLBikepxxxxxxE
\else
??????\fi
\fi
\fi
\fi
\fi
}
\newcommand{\hatcurLBiM}[1]{\ifnum#1=519 %
\hatcurLBiMxxxxxxA
\else
\ifnum#1=3629 %
\hatcurLBiMxxxxxxB
\else
\ifnum#1=3714 %
\hatcurLBiMxxxxxxC
\else
\ifnum#1=4201 %
\hatcurLBiMxxxxxxD
\else
\ifnum#1=5344 %
\hatcurLBiMxxxxxxE
\else
??????\fi
\fi
\fi
\fi
\fi
}
\newcommand{\hatcurLBir}[1]{\ifnum#1=519 %
\hatcurLBirxxxxxxA
\else
\ifnum#1=3629 %
\hatcurLBirxxxxxxB
\else
\ifnum#1=3714 %
\hatcurLBirxxxxxxC
\else
\ifnum#1=4201 %
\hatcurLBirxxxxxxD
\else
\ifnum#1=5344 %
\hatcurLBirxxxxxxE
\else
??????\fi
\fi
\fi
\fi
\fi
}
\newcommand{\hatcurLBiR}[1]{\ifnum#1=519 %
\hatcurLBiRxxxxxxA
\else
\ifnum#1=3629 %
\hatcurLBiRxxxxxxB
\else
\ifnum#1=3714 %
\hatcurLBiRxxxxxxC
\else
\ifnum#1=4201 %
\hatcurLBiRxxxxxxD
\else
\ifnum#1=5344 %
\hatcurLBiRxxxxxxE
\else
??????\fi
\fi
\fi
\fi
\fi
}
\newcommand{\hatcurLBiSfour}[1]{\ifnum#1=519 %
\hatcurLBiSfourxxxxxxA
\else
\ifnum#1=3629 %
\hatcurLBiSfourxxxxxxB
\else
\ifnum#1=3714 %
\hatcurLBiSfourxxxxxxC
\else
\ifnum#1=4201 %
\hatcurLBiSfourxxxxxxD
\else
\ifnum#1=5344 %
\hatcurLBiSfourxxxxxxE
\else
??????\fi
\fi
\fi
\fi
\fi
}
\newcommand{\hatcurLBiSone}[1]{\ifnum#1=519 %
\hatcurLBiSonexxxxxxA
\else
\ifnum#1=3629 %
\hatcurLBiSonexxxxxxB
\else
\ifnum#1=3714 %
\hatcurLBiSonexxxxxxC
\else
\ifnum#1=4201 %
\hatcurLBiSonexxxxxxD
\else
\ifnum#1=5344 %
\hatcurLBiSonexxxxxxE
\else
??????\fi
\fi
\fi
\fi
\fi
}
\newcommand{\hatcurLBiSthree}[1]{\ifnum#1=519 %
\hatcurLBiSthreexxxxxxA
\else
\ifnum#1=3629 %
\hatcurLBiSthreexxxxxxB
\else
\ifnum#1=3714 %
\hatcurLBiSthreexxxxxxC
\else
\ifnum#1=4201 %
\hatcurLBiSthreexxxxxxD
\else
\ifnum#1=5344 %
\hatcurLBiSthreexxxxxxE
\else
??????\fi
\fi
\fi
\fi
\fi
}
\newcommand{\hatcurLBiStwo}[1]{\ifnum#1=519 %
\hatcurLBiStwoxxxxxxA
\else
\ifnum#1=3629 %
\hatcurLBiStwoxxxxxxB
\else
\ifnum#1=3714 %
\hatcurLBiStwoxxxxxxC
\else
\ifnum#1=4201 %
\hatcurLBiStwoxxxxxxD
\else
\ifnum#1=5344 %
\hatcurLBiStwoxxxxxxE
\else
??????\fi
\fi
\fi
\fi
\fi
}
\newcommand{\hatcurLBiT}[1]{\ifnum#1=519 %
\hatcurLBiTxxxxxxA
\else
\ifnum#1=3629 %
\hatcurLBiTxxxxxxB
\else
\ifnum#1=3714 %
\hatcurLBiTxxxxxxC
\else
\ifnum#1=4201 %
\hatcurLBiTxxxxxxD
\else
\ifnum#1=5344 %
\hatcurLBiTxxxxxxE
\else
??????\fi
\fi
\fi
\fi
\fi
}
\newcommand{\hatcurLBiu}[1]{\ifnum#1=519 %
\hatcurLBiuxxxxxxA
\else
\ifnum#1=3629 %
\hatcurLBiuxxxxxxB
\else
\ifnum#1=3714 %
\hatcurLBiuxxxxxxC
\else
\ifnum#1=4201 %
\hatcurLBiuxxxxxxD
\else
\ifnum#1=5344 %
\hatcurLBiuxxxxxxE
\else
??????\fi
\fi
\fi
\fi
\fi
}
\newcommand{\hatcurLBiV}[1]{\ifnum#1=519 %
\hatcurLBiVxxxxxxA
\else
\ifnum#1=3629 %
\hatcurLBiVxxxxxxB
\else
\ifnum#1=3714 %
\hatcurLBiVxxxxxxC
\else
\ifnum#1=4201 %
\hatcurLBiVxxxxxxD
\else
\ifnum#1=5344 %
\hatcurLBiVxxxxxxE
\else
??????\fi
\fi
\fi
\fi
\fi
}
\newcommand{\hatcurLBiz}[1]{\ifnum#1=519 %
\hatcurLBizxxxxxxA
\else
\ifnum#1=3629 %
\hatcurLBizxxxxxxB
\else
\ifnum#1=3714 %
\hatcurLBizxxxxxxC
\else
\ifnum#1=4201 %
\hatcurLBizxxxxxxD
\else
\ifnum#1=5344 %
\hatcurLBizxxxxxxE
\else
??????\fi
\fi
\fi
\fi
\fi
}
\newcommand{\hatcurLCbsq}[1]{\ifnum#1=519 %
\hatcurLCbsqxxxxxxA
\else
\ifnum#1=3629 %
\hatcurLCbsqxxxxxxB
\else
\ifnum#1=3714 %
\hatcurLCbsqxxxxxxC
\else
\ifnum#1=4201 %
\hatcurLCbsqxxxxxxD
\else
\ifnum#1=5344 %
\hatcurLCbsqxxxxxxE
\else
??????\fi
\fi
\fi
\fi
\fi
}
\newcommand{\hatcurLCdip}[1]{\ifnum#1=519 %
\hatcurLCdipxxxxxxA
\else
\ifnum#1=3629 %
\hatcurLCdipxxxxxxB
\else
\ifnum#1=3714 %
\hatcurLCdipxxxxxxC
\else
\ifnum#1=4201 %
\hatcurLCdipxxxxxxD
\else
\ifnum#1=5344 %
\hatcurLCdipxxxxxxE
\else
??????\fi
\fi
\fi
\fi
\fi
}
\newcommand{\hatcurLCdur}[1]{\ifnum#1=519 %
\hatcurLCdurxxxxxxA
\else
\ifnum#1=3629 %
\hatcurLCdurxxxxxxB
\else
\ifnum#1=3714 %
\hatcurLCdurxxxxxxC
\else
\ifnum#1=4201 %
\hatcurLCdurxxxxxxD
\else
\ifnum#1=5344 %
\hatcurLCdurxxxxxxE
\else
??????\fi
\fi
\fi
\fi
\fi
}
\newcommand{\hatcurLCdurhr}[1]{\ifnum#1=519 %
\hatcurLCdurhrxxxxxxA
\else
\ifnum#1=3629 %
\hatcurLCdurhrxxxxxxB
\else
\ifnum#1=3714 %
\hatcurLCdurhrxxxxxxC
\else
\ifnum#1=4201 %
\hatcurLCdurhrxxxxxxD
\else
\ifnum#1=5344 %
\hatcurLCdurhrxxxxxxE
\else
??????\fi
\fi
\fi
\fi
\fi
}
\newcommand{\hatcurLCdurhrshort}[1]{\ifnum#1=519 %
\hatcurLCdurhrshortxxxxxxA
\else
\ifnum#1=3629 %
\hatcurLCdurhrshortxxxxxxB
\else
\ifnum#1=3714 %
\hatcurLCdurhrshortxxxxxxC
\else
\ifnum#1=4201 %
\hatcurLCdurhrshortxxxxxxD
\else
\ifnum#1=5344 %
\hatcurLCdurhrshortxxxxxxE
\else
??????\fi
\fi
\fi
\fi
\fi
}
\newcommand{\hatcurLCdurshort}[1]{\ifnum#1=519 %
\hatcurLCdurshortxxxxxxA
\else
\ifnum#1=3629 %
\hatcurLCdurshortxxxxxxB
\else
\ifnum#1=3714 %
\hatcurLCdurshortxxxxxxC
\else
\ifnum#1=4201 %
\hatcurLCdurshortxxxxxxD
\else
\ifnum#1=5344 %
\hatcurLCdurshortxxxxxxE
\else
??????\fi
\fi
\fi
\fi
\fi
}
\newcommand{\hatcurLChatnetm}[1]{\ifnum#1=3714 %
\hatcurLChatnetmxxxxxxC
\else
\ifnum#1=4201 %
\hatcurLChatnetmxxxxxxD
\else
??????\fi
\fi
}
\newcommand{\hatcurLChatnetmA}[1]{\ifnum#1=519 %
\hatcurLChatnetmAxxxxxxA
\else
\ifnum#1=3629 %
\hatcurLChatnetmAxxxxxxB
\else
\ifnum#1=5344 %
\hatcurLChatnetmAxxxxxxE
\else
??????\fi
\fi
\fi
}
\newcommand{\hatcurLChatnetmB}[1]{\ifnum#1=519 %
\hatcurLChatnetmBxxxxxxA
\else
\ifnum#1=3629 %
\hatcurLChatnetmBxxxxxxB
\else
\ifnum#1=5344 %
\hatcurLChatnetmBxxxxxxE
\else
??????\fi
\fi
\fi
}
\newcommand{\hatcurLChatnetmC}[1]{\ifnum#1=519 %
\hatcurLChatnetmCxxxxxxA
\else
??????\fi
}
\newcommand{\hatcurLChatnetmD}[1]{\ifnum#1=519 %
\hatcurLChatnetmDxxxxxxA
\else
??????\fi
}
\newcommand{\hatcurLCiblend}[1]{\ifnum#1=3714 %
\hatcurLCiblendxxxxxxC
\else
\ifnum#1=4201 %
\hatcurLCiblendxxxxxxD
\else
??????\fi
\fi
}
\newcommand{\hatcurLCiblendA}[1]{\ifnum#1=519 %
\hatcurLCiblendAxxxxxxA
\else
\ifnum#1=3629 %
\hatcurLCiblendAxxxxxxB
\else
\ifnum#1=5344 %
\hatcurLCiblendAxxxxxxE
\else
??????\fi
\fi
\fi
}
\newcommand{\hatcurLCiblendB}[1]{\ifnum#1=519 %
\hatcurLCiblendBxxxxxxA
\else
\ifnum#1=3629 %
\hatcurLCiblendBxxxxxxB
\else
\ifnum#1=5344 %
\hatcurLCiblendBxxxxxxE
\else
??????\fi
\fi
\fi
}
\newcommand{\hatcurLCiblendC}[1]{\ifnum#1=519 %
\hatcurLCiblendCxxxxxxA
\else
??????\fi
}
\newcommand{\hatcurLCiblendD}[1]{\ifnum#1=519 %
\hatcurLCiblendDxxxxxxA
\else
??????\fi
}
\newcommand{\hatcurLCimp}[1]{\ifnum#1=519 %
\hatcurLCimpxxxxxxA
\else
\ifnum#1=3629 %
\hatcurLCimpxxxxxxB
\else
\ifnum#1=3714 %
\hatcurLCimpxxxxxxC
\else
\ifnum#1=4201 %
\hatcurLCimpxxxxxxD
\else
\ifnum#1=5344 %
\hatcurLCimpxxxxxxE
\else
??????\fi
\fi
\fi
\fi
\fi
}
\newcommand{\hatcurLCingdur}[1]{\ifnum#1=519 %
\hatcurLCingdurxxxxxxA
\else
\ifnum#1=3629 %
\hatcurLCingdurxxxxxxB
\else
\ifnum#1=3714 %
\hatcurLCingdurxxxxxxC
\else
\ifnum#1=4201 %
\hatcurLCingdurxxxxxxD
\else
\ifnum#1=5344 %
\hatcurLCingdurxxxxxxE
\else
??????\fi
\fi
\fi
\fi
\fi
}
\newcommand{\hatcurLCP}[1]{\ifnum#1=519 %
\hatcurLCPxxxxxxA
\else
\ifnum#1=3629 %
\hatcurLCPxxxxxxB
\else
\ifnum#1=3714 %
\hatcurLCPxxxxxxC
\else
\ifnum#1=4201 %
\hatcurLCPxxxxxxD
\else
\ifnum#1=5344 %
\hatcurLCPxxxxxxE
\else
??????\fi
\fi
\fi
\fi
\fi
}
\newcommand{\hatcurLCPprec}[1]{\ifnum#1=519 %
\hatcurLCPprecxxxxxxA
\else
\ifnum#1=3629 %
\hatcurLCPprecxxxxxxB
\else
\ifnum#1=3714 %
\hatcurLCPprecxxxxxxC
\else
\ifnum#1=4201 %
\hatcurLCPprecxxxxxxD
\else
\ifnum#1=5344 %
\hatcurLCPprecxxxxxxE
\else
??????\fi
\fi
\fi
\fi
\fi
}
\newcommand{\hatcurLCPshort}[1]{\ifnum#1=519 %
\hatcurLCPshortxxxxxxA
\else
\ifnum#1=3629 %
\hatcurLCPshortxxxxxxB
\else
\ifnum#1=3714 %
\hatcurLCPshortxxxxxxC
\else
\ifnum#1=4201 %
\hatcurLCPshortxxxxxxD
\else
\ifnum#1=5344 %
\hatcurLCPshortxxxxxxE
\else
??????\fi
\fi
\fi
\fi
\fi
}
\newcommand{\hatcurLCq}[1]{\ifnum#1=519 %
\hatcurLCqxxxxxxA
\else
\ifnum#1=3629 %
\hatcurLCqxxxxxxB
\else
\ifnum#1=3714 %
\hatcurLCqxxxxxxC
\else
\ifnum#1=4201 %
\hatcurLCqxxxxxxD
\else
\ifnum#1=5344 %
\hatcurLCqxxxxxxE
\else
??????\fi
\fi
\fi
\fi
\fi
}
\newcommand{\hatcurLCqshort}[1]{\ifnum#1=519 %
\hatcurLCqshortxxxxxxA
\else
\ifnum#1=3629 %
\hatcurLCqshortxxxxxxB
\else
\ifnum#1=3714 %
\hatcurLCqshortxxxxxxC
\else
\ifnum#1=4201 %
\hatcurLCqshortxxxxxxD
\else
\ifnum#1=5344 %
\hatcurLCqshortxxxxxxE
\else
??????\fi
\fi
\fi
\fi
\fi
}
\newcommand{\hatcurLCrho}[1]{\ifnum#1=519 %
\hatcurLCrhoxxxxxxA
\else
\ifnum#1=3629 %
\hatcurLCrhoxxxxxxB
\else
\ifnum#1=3714 %
\hatcurLCrhoxxxxxxC
\else
\ifnum#1=4201 %
\hatcurLCrhoxxxxxxD
\else
\ifnum#1=5344 %
\hatcurLCrhoxxxxxxE
\else
??????\fi
\fi
\fi
\fi
\fi
}
\newcommand{\hatcurLCrprstar}[1]{\ifnum#1=519 %
\hatcurLCrprstarxxxxxxA
\else
\ifnum#1=3629 %
\hatcurLCrprstarxxxxxxB
\else
\ifnum#1=3714 %
\hatcurLCrprstarxxxxxxC
\else
\ifnum#1=4201 %
\hatcurLCrprstarxxxxxxD
\else
\ifnum#1=5344 %
\hatcurLCrprstarxxxxxxE
\else
??????\fi
\fi
\fi
\fi
\fi
}
\newcommand{\hatcurLCT}[1]{\ifnum#1=519 %
\hatcurLCTxxxxxxA
\else
\ifnum#1=3629 %
\hatcurLCTxxxxxxB
\else
\ifnum#1=3714 %
\hatcurLCTxxxxxxC
\else
\ifnum#1=4201 %
\hatcurLCTxxxxxxD
\else
\ifnum#1=5344 %
\hatcurLCTxxxxxxE
\else
??????\fi
\fi
\fi
\fi
\fi
}
\newcommand{\hatcurLCTA}[1]{\ifnum#1=519 %
\hatcurLCTAxxxxxxA
\else
\ifnum#1=3629 %
\hatcurLCTAxxxxxxB
\else
\ifnum#1=3714 %
\hatcurLCTAxxxxxxC
\else
\ifnum#1=4201 %
\hatcurLCTAxxxxxxD
\else
\ifnum#1=5344 %
\hatcurLCTAxxxxxxE
\else
??????\fi
\fi
\fi
\fi
\fi
}
\newcommand{\hatcurLCTB}[1]{\ifnum#1=519 %
\hatcurLCTBxxxxxxA
\else
\ifnum#1=3629 %
\hatcurLCTBxxxxxxB
\else
\ifnum#1=3714 %
\hatcurLCTBxxxxxxC
\else
\ifnum#1=4201 %
\hatcurLCTBxxxxxxD
\else
\ifnum#1=5344 %
\hatcurLCTBxxxxxxE
\else
??????\fi
\fi
\fi
\fi
\fi
}
\newcommand{\hatcurLCzeta}[1]{\ifnum#1=519 %
\hatcurLCzetaxxxxxxA
\else
\ifnum#1=3629 %
\hatcurLCzetaxxxxxxB
\else
\ifnum#1=3714 %
\hatcurLCzetaxxxxxxC
\else
\ifnum#1=4201 %
\hatcurLCzetaxxxxxxD
\else
\ifnum#1=5344 %
\hatcurLCzetaxxxxxxE
\else
??????\fi
\fi
\fi
\fi
\fi
}
\newcommand{\hatcurPPaequiv}[1]{\ifnum#1=519 %
\hatcurPPaequivxxxxxxA
\else
\ifnum#1=3629 %
\hatcurPPaequivxxxxxxB
\else
\ifnum#1=3714 %
\hatcurPPaequivxxxxxxC
\else
\ifnum#1=4201 %
\hatcurPPaequivxxxxxxD
\else
\ifnum#1=5344 %
\hatcurPPaequivxxxxxxE
\else
??????\fi
\fi
\fi
\fi
\fi
}
\newcommand{\hatcurPPar}[1]{\ifnum#1=519 %
\hatcurPParxxxxxxA
\else
\ifnum#1=3629 %
\hatcurPParxxxxxxB
\else
\ifnum#1=3714 %
\hatcurPParxxxxxxC
\else
\ifnum#1=4201 %
\hatcurPParxxxxxxD
\else
\ifnum#1=5344 %
\hatcurPParxxxxxxE
\else
??????\fi
\fi
\fi
\fi
\fi
}
\newcommand{\hatcurPParel}[1]{\ifnum#1=519 %
\hatcurPParelxxxxxxA
\else
\ifnum#1=3629 %
\hatcurPParelxxxxxxB
\else
\ifnum#1=3714 %
\hatcurPParelxxxxxxC
\else
\ifnum#1=4201 %
\hatcurPParelxxxxxxD
\else
\ifnum#1=5344 %
\hatcurPParelxxxxxxE
\else
??????\fi
\fi
\fi
\fi
\fi
}
\newcommand{\hatcurPPfluxap}[1]{\ifnum#1=519 %
\hatcurPPfluxapxxxxxxA
\else
\ifnum#1=3629 %
\hatcurPPfluxapxxxxxxB
\else
\ifnum#1=3714 %
\hatcurPPfluxapxxxxxxC
\else
\ifnum#1=4201 %
\hatcurPPfluxapxxxxxxD
\else
\ifnum#1=5344 %
\hatcurPPfluxapxxxxxxE
\else
??????\fi
\fi
\fi
\fi
\fi
}
\newcommand{\hatcurPPfluxapdim}[1]{\ifnum#1=519 %
\hatcurPPfluxapdimxxxxxxA
\else
\ifnum#1=3629 %
\hatcurPPfluxapdimxxxxxxB
\else
\ifnum#1=3714 %
\hatcurPPfluxapdimxxxxxxC
\else
\ifnum#1=4201 %
\hatcurPPfluxapdimxxxxxxD
\else
\ifnum#1=5344 %
\hatcurPPfluxapdimxxxxxxE
\else
??????\fi
\fi
\fi
\fi
\fi
}
\newcommand{\hatcurPPfluxavg}[1]{\ifnum#1=519 %
\hatcurPPfluxavgxxxxxxA
\else
\ifnum#1=3629 %
\hatcurPPfluxavgxxxxxxB
\else
\ifnum#1=3714 %
\hatcurPPfluxavgxxxxxxC
\else
\ifnum#1=4201 %
\hatcurPPfluxavgxxxxxxD
\else
\ifnum#1=5344 %
\hatcurPPfluxavgxxxxxxE
\else
??????\fi
\fi
\fi
\fi
\fi
}
\newcommand{\hatcurPPfluxavgdim}[1]{\ifnum#1=519 %
\hatcurPPfluxavgdimxxxxxxA
\else
\ifnum#1=3629 %
\hatcurPPfluxavgdimxxxxxxB
\else
\ifnum#1=3714 %
\hatcurPPfluxavgdimxxxxxxC
\else
\ifnum#1=4201 %
\hatcurPPfluxavgdimxxxxxxD
\else
\ifnum#1=5344 %
\hatcurPPfluxavgdimxxxxxxE
\else
??????\fi
\fi
\fi
\fi
\fi
}
\newcommand{\hatcurPPfluxavglog}[1]{\ifnum#1=519 %
\hatcurPPfluxavglogxxxxxxA
\else
\ifnum#1=3629 %
\hatcurPPfluxavglogxxxxxxB
\else
\ifnum#1=3714 %
\hatcurPPfluxavglogxxxxxxC
\else
\ifnum#1=4201 %
\hatcurPPfluxavglogxxxxxxD
\else
\ifnum#1=5344 %
\hatcurPPfluxavglogxxxxxxE
\else
??????\fi
\fi
\fi
\fi
\fi
}
\newcommand{\hatcurPPfluxperi}[1]{\ifnum#1=519 %
\hatcurPPfluxperixxxxxxA
\else
\ifnum#1=3629 %
\hatcurPPfluxperixxxxxxB
\else
\ifnum#1=3714 %
\hatcurPPfluxperixxxxxxC
\else
\ifnum#1=4201 %
\hatcurPPfluxperixxxxxxD
\else
\ifnum#1=5344 %
\hatcurPPfluxperixxxxxxE
\else
??????\fi
\fi
\fi
\fi
\fi
}
\newcommand{\hatcurPPfluxperidim}[1]{\ifnum#1=519 %
\hatcurPPfluxperidimxxxxxxA
\else
\ifnum#1=3629 %
\hatcurPPfluxperidimxxxxxxB
\else
\ifnum#1=3714 %
\hatcurPPfluxperidimxxxxxxC
\else
\ifnum#1=4201 %
\hatcurPPfluxperidimxxxxxxD
\else
\ifnum#1=5344 %
\hatcurPPfluxperidimxxxxxxE
\else
??????\fi
\fi
\fi
\fi
\fi
}
\newcommand{\hatcurPPg}[1]{\ifnum#1=519 %
\hatcurPPgxxxxxxA
\else
\ifnum#1=3629 %
\hatcurPPgxxxxxxB
\else
\ifnum#1=3714 %
\hatcurPPgxxxxxxC
\else
\ifnum#1=4201 %
\hatcurPPgxxxxxxD
\else
\ifnum#1=5344 %
\hatcurPPgxxxxxxE
\else
??????\fi
\fi
\fi
\fi
\fi
}
\newcommand{\hatcurPPi}[1]{\ifnum#1=519 %
\hatcurPPixxxxxxA
\else
\ifnum#1=3629 %
\hatcurPPixxxxxxB
\else
\ifnum#1=3714 %
\hatcurPPixxxxxxC
\else
\ifnum#1=4201 %
\hatcurPPixxxxxxD
\else
\ifnum#1=5344 %
\hatcurPPixxxxxxE
\else
??????\fi
\fi
\fi
\fi
\fi
}
\newcommand{\hatcurPPlogg}[1]{\ifnum#1=519 %
\hatcurPPloggxxxxxxA
\else
\ifnum#1=3629 %
\hatcurPPloggxxxxxxB
\else
\ifnum#1=3714 %
\hatcurPPloggxxxxxxC
\else
\ifnum#1=4201 %
\hatcurPPloggxxxxxxD
\else
\ifnum#1=5344 %
\hatcurPPloggxxxxxxE
\else
??????\fi
\fi
\fi
\fi
\fi
}
\newcommand{\hatcurPPm}[1]{\ifnum#1=519 %
\hatcurPPmxxxxxxA
\else
\ifnum#1=3629 %
\hatcurPPmxxxxxxB
\else
\ifnum#1=3714 %
\hatcurPPmxxxxxxC
\else
\ifnum#1=4201 %
\hatcurPPmxxxxxxD
\else
\ifnum#1=5344 %
\hatcurPPmxxxxxxE
\else
??????\fi
\fi
\fi
\fi
\fi
}
\newcommand{\hatcurPPme}[1]{\ifnum#1=519 %
\hatcurPPmexxxxxxA
\else
\ifnum#1=3629 %
\hatcurPPmexxxxxxB
\else
\ifnum#1=3714 %
\hatcurPPmexxxxxxC
\else
\ifnum#1=4201 %
\hatcurPPmexxxxxxD
\else
\ifnum#1=5344 %
\hatcurPPmexxxxxxE
\else
??????\fi
\fi
\fi
\fi
\fi
}
\newcommand{\hatcurPPmelong}[1]{\ifnum#1=519 %
\hatcurPPmelongxxxxxxA
\else
\ifnum#1=3629 %
\hatcurPPmelongxxxxxxB
\else
\ifnum#1=3714 %
\hatcurPPmelongxxxxxxC
\else
\ifnum#1=4201 %
\hatcurPPmelongxxxxxxD
\else
\ifnum#1=5344 %
\hatcurPPmelongxxxxxxE
\else
??????\fi
\fi
\fi
\fi
\fi
}
\newcommand{\hatcurPPmeshort}[1]{\ifnum#1=519 %
\hatcurPPmeshortxxxxxxA
\else
\ifnum#1=3629 %
\hatcurPPmeshortxxxxxxB
\else
\ifnum#1=3714 %
\hatcurPPmeshortxxxxxxC
\else
\ifnum#1=4201 %
\hatcurPPmeshortxxxxxxD
\else
\ifnum#1=5344 %
\hatcurPPmeshortxxxxxxE
\else
??????\fi
\fi
\fi
\fi
\fi
}
\newcommand{\hatcurPPmlong}[1]{\ifnum#1=519 %
\hatcurPPmlongxxxxxxA
\else
\ifnum#1=3629 %
\hatcurPPmlongxxxxxxB
\else
\ifnum#1=3714 %
\hatcurPPmlongxxxxxxC
\else
\ifnum#1=4201 %
\hatcurPPmlongxxxxxxD
\else
\ifnum#1=5344 %
\hatcurPPmlongxxxxxxE
\else
??????\fi
\fi
\fi
\fi
\fi
}
\newcommand{\hatcurPPmrcorr}[1]{\ifnum#1=519 %
\hatcurPPmrcorrxxxxxxA
\else
\ifnum#1=3629 %
\hatcurPPmrcorrxxxxxxB
\else
\ifnum#1=3714 %
\hatcurPPmrcorrxxxxxxC
\else
\ifnum#1=4201 %
\hatcurPPmrcorrxxxxxxD
\else
\ifnum#1=5344 %
\hatcurPPmrcorrxxxxxxE
\else
??????\fi
\fi
\fi
\fi
\fi
}
\newcommand{\hatcurPPmshort}[1]{\ifnum#1=519 %
\hatcurPPmshortxxxxxxA
\else
\ifnum#1=3629 %
\hatcurPPmshortxxxxxxB
\else
\ifnum#1=3714 %
\hatcurPPmshortxxxxxxC
\else
\ifnum#1=4201 %
\hatcurPPmshortxxxxxxD
\else
\ifnum#1=5344 %
\hatcurPPmshortxxxxxxE
\else
??????\fi
\fi
\fi
\fi
\fi
}
\newcommand{\hatcurPPmtwosiglim}[1]{\ifnum#1=519 %
\hatcurPPmtwosiglimxxxxxxA
\else
\ifnum#1=3629 %
\hatcurPPmtwosiglimxxxxxxB
\else
\ifnum#1=3714 %
\hatcurPPmtwosiglimxxxxxxC
\else
\ifnum#1=4201 %
\hatcurPPmtwosiglimxxxxxxD
\else
\ifnum#1=5344 %
\hatcurPPmtwosiglimxxxxxxE
\else
??????\fi
\fi
\fi
\fi
\fi
}
\newcommand{\hatcurPPperi}[1]{\ifnum#1=519 %
\hatcurPPperixxxxxxA
\else
\ifnum#1=3629 %
\hatcurPPperixxxxxxB
\else
\ifnum#1=3714 %
\hatcurPPperixxxxxxC
\else
\ifnum#1=4201 %
\hatcurPPperixxxxxxD
\else
\ifnum#1=5344 %
\hatcurPPperixxxxxxE
\else
??????\fi
\fi
\fi
\fi
\fi
}
\newcommand{\hatcurPPphiconj}[1]{\ifnum#1=519 %
\hatcurPPphiconjxxxxxxA
\else
\ifnum#1=3629 %
\hatcurPPphiconjxxxxxxB
\else
\ifnum#1=3714 %
\hatcurPPphiconjxxxxxxC
\else
\ifnum#1=4201 %
\hatcurPPphiconjxxxxxxD
\else
\ifnum#1=5344 %
\hatcurPPphiconjxxxxxxE
\else
??????\fi
\fi
\fi
\fi
\fi
}
\newcommand{\hatcurPPr}[1]{\ifnum#1=519 %
\hatcurPPrxxxxxxA
\else
\ifnum#1=3629 %
\hatcurPPrxxxxxxB
\else
\ifnum#1=3714 %
\hatcurPPrxxxxxxC
\else
\ifnum#1=4201 %
\hatcurPPrxxxxxxD
\else
\ifnum#1=5344 %
\hatcurPPrxxxxxxE
\else
??????\fi
\fi
\fi
\fi
\fi
}
\newcommand{\hatcurPPre}[1]{\ifnum#1=519 %
\hatcurPPrexxxxxxA
\else
\ifnum#1=3629 %
\hatcurPPrexxxxxxB
\else
\ifnum#1=3714 %
\hatcurPPrexxxxxxC
\else
\ifnum#1=4201 %
\hatcurPPrexxxxxxD
\else
\ifnum#1=5344 %
\hatcurPPrexxxxxxE
\else
??????\fi
\fi
\fi
\fi
\fi
}
\newcommand{\hatcurPPrelong}[1]{\ifnum#1=519 %
\hatcurPPrelongxxxxxxA
\else
\ifnum#1=3629 %
\hatcurPPrelongxxxxxxB
\else
\ifnum#1=3714 %
\hatcurPPrelongxxxxxxC
\else
\ifnum#1=4201 %
\hatcurPPrelongxxxxxxD
\else
\ifnum#1=5344 %
\hatcurPPrelongxxxxxxE
\else
??????\fi
\fi
\fi
\fi
\fi
}
\newcommand{\hatcurPPreshort}[1]{\ifnum#1=519 %
\hatcurPPreshortxxxxxxA
\else
\ifnum#1=3629 %
\hatcurPPreshortxxxxxxB
\else
\ifnum#1=3714 %
\hatcurPPreshortxxxxxxC
\else
\ifnum#1=4201 %
\hatcurPPreshortxxxxxxD
\else
\ifnum#1=5344 %
\hatcurPPreshortxxxxxxE
\else
??????\fi
\fi
\fi
\fi
\fi
}
\newcommand{\hatcurPPrho}[1]{\ifnum#1=519 %
\hatcurPPrhoxxxxxxA
\else
\ifnum#1=3629 %
\hatcurPPrhoxxxxxxB
\else
\ifnum#1=3714 %
\hatcurPPrhoxxxxxxC
\else
\ifnum#1=4201 %
\hatcurPPrhoxxxxxxD
\else
\ifnum#1=5344 %
\hatcurPPrhoxxxxxxE
\else
??????\fi
\fi
\fi
\fi
\fi
}
\newcommand{\hatcurPPrlong}[1]{\ifnum#1=519 %
\hatcurPPrlongxxxxxxA
\else
\ifnum#1=3629 %
\hatcurPPrlongxxxxxxB
\else
\ifnum#1=3714 %
\hatcurPPrlongxxxxxxC
\else
\ifnum#1=4201 %
\hatcurPPrlongxxxxxxD
\else
\ifnum#1=5344 %
\hatcurPPrlongxxxxxxE
\else
??????\fi
\fi
\fi
\fi
\fi
}
\newcommand{\hatcurPPrshort}[1]{\ifnum#1=519 %
\hatcurPPrshortxxxxxxA
\else
\ifnum#1=3629 %
\hatcurPPrshortxxxxxxB
\else
\ifnum#1=3714 %
\hatcurPPrshortxxxxxxC
\else
\ifnum#1=4201 %
\hatcurPPrshortxxxxxxD
\else
\ifnum#1=5344 %
\hatcurPPrshortxxxxxxE
\else
??????\fi
\fi
\fi
\fi
\fi
}
\newcommand{\hatcurPPtcirc}[1]{\ifnum#1=519 %
\hatcurPPtcircxxxxxxA
\else
\ifnum#1=3629 %
\hatcurPPtcircxxxxxxB
\else
\ifnum#1=3714 %
\hatcurPPtcircxxxxxxC
\else
\ifnum#1=4201 %
\hatcurPPtcircxxxxxxD
\else
\ifnum#1=5344 %
\hatcurPPtcircxxxxxxE
\else
??????\fi
\fi
\fi
\fi
\fi
}
\newcommand{\hatcurPPteff}[1]{\ifnum#1=519 %
\hatcurPPteffxxxxxxA
\else
\ifnum#1=3629 %
\hatcurPPteffxxxxxxB
\else
\ifnum#1=3714 %
\hatcurPPteffxxxxxxC
\else
\ifnum#1=4201 %
\hatcurPPteffxxxxxxD
\else
\ifnum#1=5344 %
\hatcurPPteffxxxxxxE
\else
??????\fi
\fi
\fi
\fi
\fi
}
\newcommand{\hatcurPPtheta}[1]{\ifnum#1=519 %
\hatcurPPthetaxxxxxxA
\else
\ifnum#1=3629 %
\hatcurPPthetaxxxxxxB
\else
\ifnum#1=3714 %
\hatcurPPthetaxxxxxxC
\else
\ifnum#1=4201 %
\hatcurPPthetaxxxxxxD
\else
\ifnum#1=5344 %
\hatcurPPthetaxxxxxxE
\else
??????\fi
\fi
\fi
\fi
\fi
}
\newcommand{\hatcurPPtinfall}[1]{\ifnum#1=519 %
\hatcurPPtinfallxxxxxxA
\else
\ifnum#1=3629 %
\hatcurPPtinfallxxxxxxB
\else
\ifnum#1=3714 %
\hatcurPPtinfallxxxxxxC
\else
\ifnum#1=4201 %
\hatcurPPtinfallxxxxxxD
\else
\ifnum#1=5344 %
\hatcurPPtinfallxxxxxxE
\else
??????\fi
\fi
\fi
\fi
\fi
}
\newcommand{\hatcurRVeccen}[1]{\ifnum#1=519 %
\hatcurRVeccenxxxxxxA
\else
\ifnum#1=3629 %
\hatcurRVeccenxxxxxxB
\else
\ifnum#1=3714 %
\hatcurRVeccenxxxxxxC
\else
\ifnum#1=4201 %
\hatcurRVeccenxxxxxxD
\else
\ifnum#1=5344 %
\hatcurRVeccenxxxxxxE
\else
??????\fi
\fi
\fi
\fi
\fi
}
\newcommand{\hatcurRVeccentwosiglim}[1]{\ifnum#1=519 %
\hatcurRVeccentwosiglimxxxxxxA
\else
\ifnum#1=3629 %
\hatcurRVeccentwosiglimxxxxxxB
\else
\ifnum#1=3714 %
\hatcurRVeccentwosiglimxxxxxxC
\else
\ifnum#1=4201 %
\hatcurRVeccentwosiglimxxxxxxD
\else
\ifnum#1=5344 %
\hatcurRVeccentwosiglimxxxxxxE
\else
??????\fi
\fi
\fi
\fi
\fi
}
\newcommand{\hatcurRVfitrms}[1]{\ifnum#1=4201 %
\hatcurRVfitrmsxxxxxxD
\else
\ifnum#1=5344 %
\hatcurRVfitrmsxxxxxxE
\else
??????\fi
\fi
}
\newcommand{\hatcurRVfitrmsA}[1]{\ifnum#1=519 %
\hatcurRVfitrmsAxxxxxxA
\else
\ifnum#1=3629 %
\hatcurRVfitrmsAxxxxxxB
\else
\ifnum#1=3714 %
\hatcurRVfitrmsAxxxxxxC
\else
??????\fi
\fi
\fi
}
\newcommand{\hatcurRVfitrmsB}[1]{\ifnum#1=519 %
\hatcurRVfitrmsBxxxxxxA
\else
\ifnum#1=3629 %
\hatcurRVfitrmsBxxxxxxB
\else
\ifnum#1=3714 %
\hatcurRVfitrmsBxxxxxxC
\else
??????\fi
\fi
\fi
}
\newcommand{\hatcurRVfitrmsC}[1]{\ifnum#1=3629 %
\hatcurRVfitrmsCxxxxxxB
\else
\ifnum#1=3714 %
\hatcurRVfitrmsCxxxxxxC
\else
??????\fi
\fi
}
\newcommand{\hatcurRVgamma}[1]{\ifnum#1=4201 %
\hatcurRVgammaxxxxxxD
\else
\ifnum#1=5344 %
\hatcurRVgammaxxxxxxE
\else
??????\fi
\fi
}
\newcommand{\hatcurRVgammaA}[1]{\ifnum#1=519 %
\hatcurRVgammaAxxxxxxA
\else
\ifnum#1=3629 %
\hatcurRVgammaAxxxxxxB
\else
\ifnum#1=3714 %
\hatcurRVgammaAxxxxxxC
\else
??????\fi
\fi
\fi
}
\newcommand{\hatcurRVgammaB}[1]{\ifnum#1=519 %
\hatcurRVgammaBxxxxxxA
\else
\ifnum#1=3629 %
\hatcurRVgammaBxxxxxxB
\else
\ifnum#1=3714 %
\hatcurRVgammaBxxxxxxC
\else
??????\fi
\fi
\fi
}
\newcommand{\hatcurRVgammaC}[1]{\ifnum#1=3629 %
\hatcurRVgammaCxxxxxxB
\else
\ifnum#1=3714 %
\hatcurRVgammaCxxxxxxC
\else
??????\fi
\fi
}
\newcommand{\hatcurRVh}[1]{\ifnum#1=519 %
\hatcurRVhxxxxxxA
\else
\ifnum#1=3629 %
\hatcurRVhxxxxxxB
\else
\ifnum#1=3714 %
\hatcurRVhxxxxxxC
\else
\ifnum#1=4201 %
\hatcurRVhxxxxxxD
\else
\ifnum#1=5344 %
\hatcurRVhxxxxxxE
\else
??????\fi
\fi
\fi
\fi
\fi
}
\newcommand{\hatcurRVjitter}[1]{\ifnum#1=4201 %
\hatcurRVjitterxxxxxxD
\else
\ifnum#1=5344 %
\hatcurRVjitterxxxxxxE
\else
??????\fi
\fi
}
\newcommand{\hatcurRVjitterA}[1]{\ifnum#1=519 %
\hatcurRVjitterAxxxxxxA
\else
\ifnum#1=3629 %
\hatcurRVjitterAxxxxxxB
\else
\ifnum#1=3714 %
\hatcurRVjitterAxxxxxxC
\else
??????\fi
\fi
\fi
}
\newcommand{\hatcurRVjitterB}[1]{\ifnum#1=519 %
\hatcurRVjitterBxxxxxxA
\else
\ifnum#1=3629 %
\hatcurRVjitterBxxxxxxB
\else
\ifnum#1=3714 %
\hatcurRVjitterBxxxxxxC
\else
??????\fi
\fi
\fi
}
\newcommand{\hatcurRVjitterC}[1]{\ifnum#1=3629 %
\hatcurRVjitterCxxxxxxB
\else
\ifnum#1=3714 %
\hatcurRVjitterCxxxxxxC
\else
??????\fi
\fi
}
\newcommand{\hatcurRVjittertwosiglim}[1]{\ifnum#1=4201 %
\hatcurRVjittertwosiglimxxxxxxD
\else
\ifnum#1=5344 %
\hatcurRVjittertwosiglimxxxxxxE
\else
??????\fi
\fi
}
\newcommand{\hatcurRVjittertwosiglimA}[1]{\ifnum#1=519 %
\hatcurRVjittertwosiglimAxxxxxxA
\else
\ifnum#1=3629 %
\hatcurRVjittertwosiglimAxxxxxxB
\else
\ifnum#1=3714 %
\hatcurRVjittertwosiglimAxxxxxxC
\else
??????\fi
\fi
\fi
}
\newcommand{\hatcurRVjittertwosiglimB}[1]{\ifnum#1=519 %
\hatcurRVjittertwosiglimBxxxxxxA
\else
\ifnum#1=3629 %
\hatcurRVjittertwosiglimBxxxxxxB
\else
\ifnum#1=3714 %
\hatcurRVjittertwosiglimBxxxxxxC
\else
??????\fi
\fi
\fi
}
\newcommand{\hatcurRVjittertwosiglimC}[1]{\ifnum#1=3629 %
\hatcurRVjittertwosiglimCxxxxxxB
\else
\ifnum#1=3714 %
\hatcurRVjittertwosiglimCxxxxxxC
\else
??????\fi
\fi
}
\newcommand{\hatcurRVk}[1]{\ifnum#1=519 %
\hatcurRVkxxxxxxA
\else
\ifnum#1=3629 %
\hatcurRVkxxxxxxB
\else
\ifnum#1=3714 %
\hatcurRVkxxxxxxC
\else
\ifnum#1=4201 %
\hatcurRVkxxxxxxD
\else
\ifnum#1=5344 %
\hatcurRVkxxxxxxE
\else
??????\fi
\fi
\fi
\fi
\fi
}
\newcommand{\hatcurRVK}[1]{\ifnum#1=519 %
\hatcurRVKxxxxxxA
\else
\ifnum#1=3629 %
\hatcurRVKxxxxxxB
\else
\ifnum#1=3714 %
\hatcurRVKxxxxxxC
\else
\ifnum#1=4201 %
\hatcurRVKxxxxxxD
\else
\ifnum#1=5344 %
\hatcurRVKxxxxxxE
\else
??????\fi
\fi
\fi
\fi
\fi
}
\newcommand{\hatcurRVKtwosiglim}[1]{\ifnum#1=519 %
\hatcurRVKtwosiglimxxxxxxA
\else
\ifnum#1=3629 %
\hatcurRVKtwosiglimxxxxxxB
\else
\ifnum#1=3714 %
\hatcurRVKtwosiglimxxxxxxC
\else
\ifnum#1=4201 %
\hatcurRVKtwosiglimxxxxxxD
\else
\ifnum#1=5344 %
\hatcurRVKtwosiglimxxxxxxE
\else
??????\fi
\fi
\fi
\fi
\fi
}
\newcommand{\hatcurRVomega}[1]{\ifnum#1=519 %
\hatcurRVomegaxxxxxxA
\else
\ifnum#1=3629 %
\hatcurRVomegaxxxxxxB
\else
\ifnum#1=3714 %
\hatcurRVomegaxxxxxxC
\else
\ifnum#1=4201 %
\hatcurRVomegaxxxxxxD
\else
\ifnum#1=5344 %
\hatcurRVomegaxxxxxxE
\else
??????\fi
\fi
\fi
\fi
\fi
}
\newcommand{\hatcurRVrh}[1]{\ifnum#1=519 %
\hatcurRVrhxxxxxxA
\else
\ifnum#1=3629 %
\hatcurRVrhxxxxxxB
\else
\ifnum#1=3714 %
\hatcurRVrhxxxxxxC
\else
\ifnum#1=4201 %
\hatcurRVrhxxxxxxD
\else
\ifnum#1=5344 %
\hatcurRVrhxxxxxxE
\else
??????\fi
\fi
\fi
\fi
\fi
}
\newcommand{\hatcurRVrk}[1]{\ifnum#1=519 %
\hatcurRVrkxxxxxxA
\else
\ifnum#1=3629 %
\hatcurRVrkxxxxxxB
\else
\ifnum#1=3714 %
\hatcurRVrkxxxxxxC
\else
\ifnum#1=4201 %
\hatcurRVrkxxxxxxD
\else
\ifnum#1=5344 %
\hatcurRVrkxxxxxxE
\else
??????\fi
\fi
\fi
\fi
\fi
}
\newcommand{\hatcurRVtrone}[1]{\ifnum#1=519 %
\hatcurRVtronexxxxxxA
\else
\ifnum#1=3629 %
\hatcurRVtronexxxxxxB
\else
\ifnum#1=3714 %
\hatcurRVtronexxxxxxC
\else
\ifnum#1=4201 %
\hatcurRVtronexxxxxxD
\else
\ifnum#1=5344 %
\hatcurRVtronexxxxxxE
\else
??????\fi
\fi
\fi
\fi
\fi
}
\newcommand{\hatcurRVtrtwo}[1]{\ifnum#1=519 %
\hatcurRVtrtwoxxxxxxA
\else
\ifnum#1=3629 %
\hatcurRVtrtwoxxxxxxB
\else
\ifnum#1=3714 %
\hatcurRVtrtwoxxxxxxC
\else
\ifnum#1=4201 %
\hatcurRVtrtwoxxxxxxD
\else
\ifnum#1=5344 %
\hatcurRVtrtwoxxxxxxE
\else
??????\fi
\fi
\fi
\fi
\fi
}
\newcommand{\hatcurSMEilogg}[1]{\ifnum#1=519 %
\hatcurSMEiloggxxxxxxA
\else
\ifnum#1=3629 %
\hatcurSMEiloggxxxxxxB
\else
\ifnum#1=3714 %
\hatcurSMEiloggxxxxxxC
\else
\ifnum#1=4201 %
\hatcurSMEiloggxxxxxxD
\else
\ifnum#1=5344 %
\hatcurSMEiloggxxxxxxE
\else
??????\fi
\fi
\fi
\fi
\fi
}
\newcommand{\hatcurSMEiteff}[1]{\ifnum#1=519 %
\hatcurSMEiteffxxxxxxA
\else
\ifnum#1=3629 %
\hatcurSMEiteffxxxxxxB
\else
\ifnum#1=3714 %
\hatcurSMEiteffxxxxxxC
\else
\ifnum#1=4201 %
\hatcurSMEiteffxxxxxxD
\else
\ifnum#1=5344 %
\hatcurSMEiteffxxxxxxE
\else
??????\fi
\fi
\fi
\fi
\fi
}
\newcommand{\hatcurSMEivmac}[1]{\ifnum#1=519 %
\hatcurSMEivmacxxxxxxA
\else
\ifnum#1=3629 %
\hatcurSMEivmacxxxxxxB
\else
\ifnum#1=3714 %
\hatcurSMEivmacxxxxxxC
\else
\ifnum#1=4201 %
\hatcurSMEivmacxxxxxxD
\else
\ifnum#1=5344 %
\hatcurSMEivmacxxxxxxE
\else
??????\fi
\fi
\fi
\fi
\fi
}
\newcommand{\hatcurSMEivmic}[1]{\ifnum#1=519 %
\hatcurSMEivmicxxxxxxA
\else
\ifnum#1=3629 %
\hatcurSMEivmicxxxxxxB
\else
\ifnum#1=3714 %
\hatcurSMEivmicxxxxxxC
\else
\ifnum#1=4201 %
\hatcurSMEivmicxxxxxxD
\else
\ifnum#1=5344 %
\hatcurSMEivmicxxxxxxE
\else
??????\fi
\fi
\fi
\fi
\fi
}
\newcommand{\hatcurSMEivsin}[1]{\ifnum#1=519 %
\hatcurSMEivsinxxxxxxA
\else
\ifnum#1=3629 %
\hatcurSMEivsinxxxxxxB
\else
\ifnum#1=3714 %
\hatcurSMEivsinxxxxxxC
\else
\ifnum#1=4201 %
\hatcurSMEivsinxxxxxxD
\else
\ifnum#1=5344 %
\hatcurSMEivsinxxxxxxE
\else
??????\fi
\fi
\fi
\fi
\fi
}
\newcommand{\hatcurSMEizfeh}[1]{\ifnum#1=519 %
\hatcurSMEizfehxxxxxxA
\else
\ifnum#1=3629 %
\hatcurSMEizfehxxxxxxB
\else
\ifnum#1=3714 %
\hatcurSMEizfehxxxxxxC
\else
\ifnum#1=4201 %
\hatcurSMEizfehxxxxxxD
\else
\ifnum#1=5344 %
\hatcurSMEizfehxxxxxxE
\else
??????\fi
\fi
\fi
\fi
\fi
}
\newcommand{\hatcurSMEizfehshort}[1]{\ifnum#1=519 %
\hatcurSMEizfehshortxxxxxxA
\else
\ifnum#1=3629 %
\hatcurSMEizfehshortxxxxxxB
\else
\ifnum#1=3714 %
\hatcurSMEizfehshortxxxxxxC
\else
\ifnum#1=4201 %
\hatcurSMEizfehshortxxxxxxD
\else
\ifnum#1=5344 %
\hatcurSMEizfehshortxxxxxxE
\else
??????\fi
\fi
\fi
\fi
\fi
}
\newcommand{\hatcurXAv}[1]{\ifnum#1=519 %
\hatcurXAvxxxxxxA
\else
\ifnum#1=3629 %
\hatcurXAvxxxxxxB
\else
\ifnum#1=3714 %
\hatcurXAvxxxxxxC
\else
\ifnum#1=4201 %
\hatcurXAvxxxxxxD
\else
\ifnum#1=5344 %
\hatcurXAvxxxxxxE
\else
??????\fi
\fi
\fi
\fi
\fi
}
\newcommand{\hatcurXdist}[1]{\ifnum#1=519 %
\hatcurXdistxxxxxxA
\else
\ifnum#1=3629 %
\hatcurXdistxxxxxxB
\else
\ifnum#1=3714 %
\hatcurXdistxxxxxxC
\else
\ifnum#1=4201 %
\hatcurXdistxxxxxxD
\else
\ifnum#1=5344 %
\hatcurXdistxxxxxxE
\else
??????\fi
\fi
\fi
\fi
\fi
}
\newcommand{\hatcurXdistred}[1]{\ifnum#1=519 %
\hatcurXdistredxxxxxxA
\else
\ifnum#1=3629 %
\hatcurXdistredxxxxxxB
\else
\ifnum#1=3714 %
\hatcurXdistredxxxxxxC
\else
\ifnum#1=4201 %
\hatcurXdistredxxxxxxD
\else
\ifnum#1=5344 %
\hatcurXdistredxxxxxxE
\else
??????\fi
\fi
\fi
\fi
\fi
}
\newcommand{\hatcurXEBV}[1]{\ifnum#1=519 %
\hatcurXEBVxxxxxxA
\else
\ifnum#1=3629 %
\hatcurXEBVxxxxxxB
\else
\ifnum#1=3714 %
\hatcurXEBVxxxxxxC
\else
\ifnum#1=4201 %
\hatcurXEBVxxxxxxD
\else
\ifnum#1=5344 %
\hatcurXEBVxxxxxxE
\else
??????\fi
\fi
\fi
\fi
\fi
}
\newcommand{\hatcurXsecdur}[1]{\ifnum#1=519 %
\hatcurXsecdurxxxxxxA
\else
\ifnum#1=3629 %
\hatcurXsecdurxxxxxxB
\else
\ifnum#1=3714 %
\hatcurXsecdurxxxxxxC
\else
\ifnum#1=4201 %
\hatcurXsecdurxxxxxxD
\else
\ifnum#1=5344 %
\hatcurXsecdurxxxxxxE
\else
??????\fi
\fi
\fi
\fi
\fi
}
\newcommand{\hatcurXsecingdur}[1]{\ifnum#1=519 %
\hatcurXsecingdurxxxxxxA
\else
\ifnum#1=3629 %
\hatcurXsecingdurxxxxxxB
\else
\ifnum#1=3714 %
\hatcurXsecingdurxxxxxxC
\else
\ifnum#1=4201 %
\hatcurXsecingdurxxxxxxD
\else
\ifnum#1=5344 %
\hatcurXsecingdurxxxxxxE
\else
??????\fi
\fi
\fi
\fi
\fi
}
\newcommand{\hatcurXsecondary}[1]{\ifnum#1=519 %
\hatcurXsecondaryxxxxxxA
\else
\ifnum#1=3629 %
\hatcurXsecondaryxxxxxxB
\else
\ifnum#1=3714 %
\hatcurXsecondaryxxxxxxC
\else
\ifnum#1=4201 %
\hatcurXsecondaryxxxxxxD
\else
\ifnum#1=5344 %
\hatcurXsecondaryxxxxxxE
\else
??????\fi
\fi
\fi
\fi
\fi
}
\newcommand{\hatcurXsecphase}[1]{\ifnum#1=519 %
\hatcurXsecphasexxxxxxA
\else
\ifnum#1=3629 %
\hatcurXsecphasexxxxxxB
\else
\ifnum#1=3714 %
\hatcurXsecphasexxxxxxC
\else
\ifnum#1=4201 %
\hatcurXsecphasexxxxxxD
\else
\ifnum#1=5344 %
\hatcurXsecphasexxxxxxE
\else
??????\fi
\fi
\fi
\fi
\fi
}

%% file: TOI519.tex
\newcommand{\hatcurhtrxxxxxxA}{TOI519}                           
\newcommand{\hatcurfieldxxxxxxA}{\ensuremath{string}}            
\newcommand{\hatcurCCraxxxxxxA}{\ensuremath{08^{\mathrm h}18^{\mathrm m}25.6680{\mathrm s}}}                   
\newcommand{\hatcurCCdecxxxxxxA}{\ensuremath{-19{\arcdeg}39{\arcmin}46.5010{\arcsec}}}                 
\newcommand{\hatcurCCmagxxxxxxA}{NULL}                           
\newcommand{\hatcurCCtwomassxxxxxxA}{2MASS~08182567-1939465}     
\newcommand{\hatcurCCgscxxxxxxA}{GSC~NULL}                       
\newcommand{\hatcurCCgaiaxxxxxxA}{GAIA~5707485523149430400}      
\newcommand{\hatcurCCgaiadrtwoxxxxxxA}{GAIA~DR2~5707485527450614656} 
\newcommand{\hatcurCCtassmvxxxxxxA}{\ensuremath{nff\pmnff}}      
\newcommand{\hatcurCCtassmvshortxxxxxxA}{\ensuremath{0.0}}       
\newcommand{\hatcurCCtassmBxxxxxxA}{\ensuremath{nff\pmnff}}      
\newcommand{\hatcurCCtassmBshortxxxxxxA}{\ensuremath{0.0}}       
\newcommand{\hatcurCCtassmIxxxxxxA}{\ensuremath{nff\pmnff}}      
\newcommand{\hatcurCCtassmIshortxxxxxxA}{\ensuremath{0.0}}       
\newcommand{\hatcurCCtassmgxxxxxxA}{\ensuremath{nff\pmnff}}      
\newcommand{\hatcurCCtassmgshortxxxxxxA}{\ensuremath{0.0}}       
\newcommand{\hatcurCCtassmrxxxxxxA}{\ensuremath{nff\pmnff}}      
\newcommand{\hatcurCCtassmrshortxxxxxxA}{\ensuremath{0.0}}       
\newcommand{\hatcurCCtassmixxxxxxA}{\ensuremath{nff\pmnff}}      
\newcommand{\hatcurCCtassmishortxxxxxxA}{\ensuremath{0.0}}       
\newcommand{\hatcurCCparallaxxxxxxxA}{\ensuremath{8.681\pm0.037}} 
\newcommand{\hatcurCCgaiamGxxxxxxA}{\ensuremath{15.6770\pm0.0028}} 
\newcommand{\hatcurCCgaiamBPxxxxxxA}{\ensuremath{17.1927\pm0.0057}} 
\newcommand{\hatcurCCgaiamRPxxxxxxA}{\ensuremath{14.4819\pm0.0039}} 
\newcommand{\hatcurCCtwomassJmagxxxxxxA}{\ensuremath{12.847\pm0.027}} 
\newcommand{\hatcurCCtwomassHmagxxxxxxA}{\ensuremath{12.226\pm0.027}} 
\newcommand{\hatcurCCtwomassKmagxxxxxxA}{\ensuremath{11.951\pm0.024}} 
\newcommand{\hatcurCCcitJmagxxxxxxA}{\ensuremath{12.836\pm0.028}} 
\newcommand{\hatcurCCcitHmagxxxxxxA}{\ensuremath{12.216\pm0.029}} 
\newcommand{\hatcurCCcitKmagxxxxxxA}{\ensuremath{11.975\pm0.024}} 
\newcommand{\hatcurCCbbJmagxxxxxxA}{\ensuremath{12.928\pm0.030}} 
\newcommand{\hatcurCCbbHmagxxxxxxA}{\ensuremath{12.243\pm0.029}} 
\newcommand{\hatcurCCbbKmagxxxxxxA}{\ensuremath{11.995\pm0.025}} 
\newcommand{\hatcurCCesoJmagxxxxxxA}{\ensuremath{12.936\pm0.035}} 
\newcommand{\hatcurCCesoHmagxxxxxxA}{\ensuremath{12.243\pm0.048}} 
\newcommand{\hatcurCCesoKmagxxxxxxA}{\ensuremath{11.991\pm0.026}} 
\newcommand{\hatcurCCesoJHmagxxxxxxA}{\ensuremath{0.692\pm0.040}} 
\newcommand{\hatcurCCesoJKmagxxxxxxA}{\ensuremath{0.946\pm0.041}} 
\newcommand{\hatcurCCesoHKmagxxxxxxA}{\ensuremath{0.252\pm0.053}} 
\newcommand{\hatcurCCWonemagxxxxxxA}{\ensuremath{11.790\pm0.024}} 
\newcommand{\hatcurCCWtwomagxxxxxxA}{\ensuremath{11.642\pm0.021}} 
\newcommand{\hatcurCCWthreemagxxxxxxA}{\ensuremath{11.59\pm0.17}} 
\newcommand{\hatcurCCWfourmagxxxxxxA}{\ensuremath{nff\pmnff}}    
\newcommand{\hatcurLCdipxxxxxxA}{\ensuremath{0.1}}               
\newcommand{\hatcurLCrprstarxxxxxxA}{\ensuremath{0.3040\pm0.0021}} 
\newcommand{\hatcurLCbsqxxxxxxA}{\ensuremath{0.098_{-0.014}^{+0.014}}} 
\newcommand{\hatcurLCimpxxxxxxA}{\ensuremath{0.313_{-0.024}^{+0.022}}} 
\newcommand{\hatcurLCzetaxxxxxxA}{\ensuremath{51.77\pm0.35}}     
\newcommand{\hatcurLCdurxxxxxxA}{\ensuremath{0.05149\pm0.00020}} 
\newcommand{\hatcurLCdurshortxxxxxxA}{\ensuremath{0.0515}}       
\newcommand{\hatcurLCdurhrxxxxxxA}{\ensuremath{1.2358\pm0.0048}} 
\newcommand{\hatcurLCdurhrshortxxxxxxA}{\ensuremath{1.236}}      
\newcommand{\hatcurLCqxxxxxxA}{\ensuremath{0.04070\pm0.00016}}   
\newcommand{\hatcurLCqshortxxxxxxA}{\ensuremath{0.041}}          
\newcommand{\hatcurLCingdurxxxxxxA}{\ensuremath{0.01310\pm0.00023}} 
\newcommand{\hatcurLCPxxxxxxA}{\ensuremath{1.26523248\pm0.00000012}} 
\newcommand{\hatcurLCPprecxxxxxxA}{\ensuremath{1.2652325}}       
\newcommand{\hatcurLCPshortxxxxxxA}{\ensuremath{1.2652}}         
\newcommand{\hatcurLCTxxxxxxA}{\ensuremath{2459013.152990\pm0.000036}} 
\newcommand{\hatcurLCTAxxxxxxA}{\ensuremath{2458491.877205\pm0.000059}} 
\newcommand{\hatcurLCTBxxxxxxA}{\ensuremath{2459658.421551\pm0.000074}} 
\newcommand{\hatcurLChatnetmAxxxxxxA}{\ensuremath{-5.80115\pm0.00025}} 
\newcommand{\hatcurLCiblendAxxxxxxA}{\ensuremath{0.9989\pm0.0016}} 
\newcommand{\hatcurLChatnetmBxxxxxxA}{\ensuremath{14.56898\pm0.00034}} 
\newcommand{\hatcurLCiblendBxxxxxxA}{\ensuremath{0.719\pm0.020}} 
\newcommand{\hatcurLChatnetmCxxxxxxA}{\ensuremath{-5.92499\pm0.00017}} 
\newcommand{\hatcurLCiblendCxxxxxxA}{\ensuremath{1.016\pm0.012}} 
\newcommand{\hatcurLChatnetmDxxxxxxA}{\ensuremath{-5.97861\pm0.00017}} 
\newcommand{\hatcurLCiblendDxxxxxxA}{\ensuremath{1.027\pm0.012}} 
\newcommand{\hatcurLCrhoxxxxxxA}{\ensuremath{11.45\pm0.18}}      
\newcommand{\hatcurSMEiteffxxxxxxA}{\ensuremath{3367\pm70}}      
\newcommand{\hatcurSMEizfehxxxxxxA}{\ensuremath{-0.010\pm0.090}} 
\newcommand{\hatcurSMEizfehshortxxxxxxA}{\ensuremath{-0.01}}     
\newcommand{\hatcurSMEiloggxxxxxxA}{\ensuremath{5.00\pm0.50}}    
\newcommand{\hatcurSMEivsinxxxxxxA}{\ensuremath{0\pm50}}         
\newcommand{\hatcurSMEivmacxxxxxxA}{\ensuremath{nff\pmnff}}      
\newcommand{\hatcurSMEivmicxxxxxxA}{\ensuremath{nff\pmnff}}      
\newcommand{\hatcurextraerrMJxxxxxxA}{\ensuremath{0\pm0}}        
\newcommand{\hatcurextraerrMJtwosiglimxxxxxxA}{\ensuremath{<0.0200}} 
\newcommand{\hatcurextraerrMHxxxxxxA}{\ensuremath{0\pm0}}        
\newcommand{\hatcurextraerrMHtwosiglimxxxxxxA}{\ensuremath{<0.0200}} 
\newcommand{\hatcurextraerrMKsxxxxxxA}{\ensuremath{0\pm0}}       
\newcommand{\hatcurextraerrMKstwosiglimxxxxxxA}{\ensuremath{<0.0200}} 
\newcommand{\hatcurextraerrMGxxxxxxA}{\ensuremath{0\pm0}}        
\newcommand{\hatcurextraerrMGtwosiglimxxxxxxA}{\ensuremath{<0.0200}} 
\newcommand{\hatcurextraerrMRPxxxxxxA}{\ensuremath{0\pm0}}       
\newcommand{\hatcurextraerrMRPtwosiglimxxxxxxA}{\ensuremath{<0.0200}} 
\newcommand{\hatcurextraerrMWonexxxxxxA}{\ensuremath{0\pm0}}     
\newcommand{\hatcurextraerrMWonetwosiglimxxxxxxA}{\ensuremath{<0.0200}} 
\newcommand{\hatcurextraerrMWtwoxxxxxxA}{\ensuremath{0\pm0}}     
\newcommand{\hatcurextraerrMWtwotwosiglimxxxxxxA}{\ensuremath{<0.0200}} 
\newcommand{\hatcurLBiBxxxxxxA}{\ensuremath{0.24\pm0.14}}        
\newcommand{\hatcurLBiiBxxxxxxA}{\ensuremath{0.21\pm0.18}}       
\newcommand{\hatcurLBiVxxxxxxA}{\ensuremath{0.4944}}             
\newcommand{\hatcurLBiiVxxxxxxA}{\ensuremath{0.3514}}            
\newcommand{\hatcurLBiRxxxxxxA}{\ensuremath{0.4204}}             
\newcommand{\hatcurLBiiRxxxxxxA}{\ensuremath{0.3353}}            
\newcommand{\hatcurLBiIxxxxxxA}{\ensuremath{0.2071}}             
\newcommand{\hatcurLBiiIxxxxxxA}{\ensuremath{0.3944}}            
\newcommand{\hatcurLBiuxxxxxxA}{\ensuremath{0.4340}}             
\newcommand{\hatcurLBiiuxxxxxxA}{\ensuremath{0.3422}}            
\newcommand{\hatcurLBigxxxxxxA}{\ensuremath{0.51\pm0.11}}        
\newcommand{\hatcurLBiigxxxxxxA}{\ensuremath{0.35\pm0.16}}       
\newcommand{\hatcurLBirxxxxxxA}{\ensuremath{0.53\pm0.10}}        
\newcommand{\hatcurLBiirxxxxxxA}{\ensuremath{0.22\pm0.15}}       
\newcommand{\hatcurLBiixxxxxxA}{\ensuremath{0.404\pm0.089}}      
\newcommand{\hatcurLBiiixxxxxxA}{\ensuremath{0.18\pm0.14}}       
\newcommand{\hatcurLBizxxxxxxA}{\ensuremath{0.329\pm0.081}}      
\newcommand{\hatcurLBiizxxxxxxA}{\ensuremath{0.11\pm0.14}}       
\newcommand{\hatcurLBiJxxxxxxA}{\ensuremath{0.141\pm0.080}}      
\newcommand{\hatcurLBiiJxxxxxxA}{\ensuremath{0.32\pm0.13}}       
\newcommand{\hatcurLBiHxxxxxxA}{\ensuremath{0.0905}}             
\newcommand{\hatcurLBiiHxxxxxxA}{\ensuremath{0.2823}}            
\newcommand{\hatcurLBiKxxxxxxA}{\ensuremath{0.0840}}             
\newcommand{\hatcurLBiiKxxxxxxA}{\ensuremath{0.2456}}            
\newcommand{\hatcurLBiTxxxxxxA}{\ensuremath{0.233\pm0.097}}      
\newcommand{\hatcurLBiiTxxxxxxA}{\ensuremath{0.46\pm0.15}}       
\newcommand{\hatcurLBikepxxxxxxA}{\ensuremath{0.2921}}           
\newcommand{\hatcurLBiikepxxxxxxA}{\ensuremath{0.4295}}          
\newcommand{\hatcurLBiCxxxxxxA}{\ensuremath{0.2393}}             
\newcommand{\hatcurLBiiCxxxxxxA}{\ensuremath{0.4349}}            
\newcommand{\hatcurLBiMxxxxxxA}{\ensuremath{0.4071}}             
\newcommand{\hatcurLBiiMxxxxxxA}{\ensuremath{0.4027}}            
\newcommand{\hatcurLBiSonexxxxxxA}{\ensuremath{0.0657}}            
\newcommand{\hatcurLBiiSonexxxxxxA}{\ensuremath{0.1832}}           
\newcommand{\hatcurLBiStwoxxxxxxA}{\ensuremath{0.0484}}            
\newcommand{\hatcurLBiiStwoxxxxxxA}{\ensuremath{0.1534}}           
\newcommand{\hatcurLBiSthreexxxxxxA}{\ensuremath{0.0531}}            
\newcommand{\hatcurLBiiSthreexxxxxxA}{\ensuremath{0.1359}}           
\newcommand{\hatcurLBiSfourxxxxxxA}{\ensuremath{0.0560}}            
\newcommand{\hatcurLBiiSfourxxxxxxA}{\ensuremath{0.1045}}           
\newcommand{\hatcurISOmxxxxxxA}{\ensuremath{0.372\pm0.018}}      
\newcommand{\hatcurISOmshortxxxxxxA}{\ensuremath{0.37}}          
\newcommand{\hatcurISOmlongxxxxxxA}{\ensuremath{0.372\pm0.018}}  
\newcommand{\hatcurISOrxxxxxxA}{\ensuremath{0.3578\pm0.0063}}    
\newcommand{\hatcurISOrshortxxxxxxA}{\ensuremath{0.36}}          
\newcommand{\hatcurISOrlongxxxxxxA}{\ensuremath{0.3578\pm0.0063}} 
\newcommand{\hatcurISOrhoxxxxxxA}{\ensuremath{11.45\pm0.18}}     
\newcommand{\hatcurISOrholongxxxxxxA}{\ensuremath{11.45\pm0.18}} 
\newcommand{\hatcurISOloggxxxxxxA}{\ensuremath{4.9016\pm0.0085}} 
\newcommand{\hatcurISOlumxxxxxxA}{\ensuremath{0.0146\pm0.0012}}  
\newcommand{\hatcurISOlumshortxxxxxxA}{\ensuremath{0.01}}        
\newcommand{\hatcurISOteffxxxxxxA}{\ensuremath{3354\pm63}}       
\newcommand{\hatcurISOzfehxxxxxxA}{\ensuremath{0.264\pm0.083}}   
\newcommand{\hatcurISOagexxxxxxA}{\ensuremath{0.55_{-0.19}^{+0.75}}} 
\newcommand{\hatcurISOspecxxxxxxA}{M}                            
\newcommand{\hatcurRVKxxxxxxA}{\ensuremath{191\pm22}}            
\newcommand{\hatcurRVKtwosiglimxxxxxxA}{\ensuremath{<226.2}}     
\newcommand{\hatcurRVrkxxxxxxA}{\ensuremath{0\pm0}}              
\newcommand{\hatcurRVrhxxxxxxA}{\ensuremath{0\pm0}}              
\newcommand{\hatcurRVkxxxxxxA}{\ensuremath{0\pm0}}               
\newcommand{\hatcurRVhxxxxxxA}{\ensuremath{0\pm0}}               
\newcommand{\hatcurRVtronexxxxxxA}{\ensuremath{0\pm0}}           
\newcommand{\hatcurRVtrtwoxxxxxxA}{\ensuremath{0\pm0}}           
\newcommand{\hatcurRVgammaAxxxxxxA}{\ensuremath{13\pm34}}        
\newcommand{\hatcurRVjitterAxxxxxxA}{\ensuremath{73\pm28}}       
\newcommand{\hatcurRVjittertwosiglimAxxxxxxA}{\ensuremath{<131.2}} 
\newcommand{\hatcurRVfitrmsAxxxxxxA}{\ensuremath{0.0}}           
\newcommand{\hatcurRVgammaBxxxxxxA}{\ensuremath{8\pm18}}         
\newcommand{\hatcurRVjitterBxxxxxxA}{\ensuremath{23.8\pm9.6}}    
\newcommand{\hatcurRVjittertwosiglimBxxxxxxA}{\ensuremath{<40.5}} 
\newcommand{\hatcurRVfitrmsBxxxxxxA}{\ensuremath{0.0}}           
\newcommand{\hatcurRVeccenxxxxxxA}{\ensuremath{0\pm0}}           
\newcommand{\hatcurRVeccentwosiglimxxxxxxA}{\ensuremath{<0.000}} 
\newcommand{\hatcurRVomegaxxxxxxA}{\ensuremath{0\pm0}}           
\newcommand{\hatcurPPixxxxxxA}{\ensuremath{88.19\pm0.14}}        
\newcommand{\hatcurPPgxxxxxxA}{\ensuremath{11.7\pm1.4}}          
\newcommand{\hatcurPPloggxxxxxxA}{\ensuremath{3.066\pm0.053}}    
\newcommand{\hatcurPParxxxxxxA}{\ensuremath{9.901\pm0.053}}      
\newcommand{\hatcurPParelxxxxxxA}{\ensuremath{0.01648\pm0.00027}} 
\newcommand{\hatcurPPrhoxxxxxxA}{\ensuremath{0.551\pm0.067}}     
\newcommand{\hatcurPPmxxxxxxA}{\ensuremath{0.525\pm0.064}}       
\newcommand{\hatcurPPmtwosiglimxxxxxxA}{\ensuremath{<0.63}}      
\newcommand{\hatcurPPmshortxxxxxxA}{\ensuremath{0.53}}           
\newcommand{\hatcurPPmlongxxxxxxA}{\ensuremath{0.525\pm0.064}}   
\newcommand{\hatcurPPmexxxxxxA}{\ensuremath{167\pm20}}           
\newcommand{\hatcurPPmeshortxxxxxxA}{\ensuremath{166.9}}         
\newcommand{\hatcurPPmelongxxxxxxA}{\ensuremath{167\pm20}}       
\newcommand{\hatcurPPrxxxxxxA}{\ensuremath{1.059\pm0.020}}       
\newcommand{\hatcurPPrshortxxxxxxA}{\ensuremath{1.06}}           
\newcommand{\hatcurPPrlongxxxxxxA}{\ensuremath{1.059\pm0.020}}   
\newcommand{\hatcurPPrexxxxxxA}{\ensuremath{11.87\pm0.23}}       
\newcommand{\hatcurPPreshortxxxxxxA}{\ensuremath{11.9}}          
\newcommand{\hatcurPPrelongxxxxxxA}{\ensuremath{11.87\pm0.23}}   
\newcommand{\hatcurPPmrcorrxxxxxxA}{\ensuremath{0.23}}           
\newcommand{\hatcurPPteffxxxxxxA}{\ensuremath{754\pm14}}         
\newcommand{\hatcurPPthetaxxxxxxA}{\ensuremath{0.0439\pm0.0052}} 
\newcommand{\hatcurPPfluxperixxxxxxA}{\ensuremath{7.33\pm0.56}}  
\newcommand{\hatcurPPfluxperidimxxxxxxA}{\ensuremath{7}}         
\newcommand{\hatcurPPfluxapxxxxxxA}{\ensuremath{7.33\pm0.56}}    
\newcommand{\hatcurPPfluxapdimxxxxxxA}{\ensuremath{7}}           
\newcommand{\hatcurPPfluxavgxxxxxxA}{\ensuremath{7.33\pm0.56}}   
\newcommand{\hatcurPPfluxavgdimxxxxxxA}{\ensuremath{7}}          
\newcommand{\hatcurPPfluxavglogxxxxxxA}{\ensuremath{7.865\pm0.033}} 
\newcommand{\hatcurXsecphasexxxxxxA}{\ensuremath{0\pm0}}         
\newcommand{\hatcurXsecondaryxxxxxxA}{\ensuremath{2459013.785600\pm0.000036}} 
\newcommand{\hatcurXsecdurxxxxxxA}{\ensuremath{0.05149\pm0.00020}} 
\newcommand{\hatcurXsecingdurxxxxxxA}{\ensuremath{0.01310\pm0.00023}} 
\newcommand{\hatcurPPphiconjxxxxxxA}{\ensuremath{0\pm0}}         
\newcommand{\hatcurPPperixxxxxxA}{\ensuremath{2459012.836679\pm0.000035}} 
\newcommand{\hatcurPPaequivxxxxxxA}{\ensuremath{0.1363\pm0.0052}} 
\newcommand{\hatcurPPtcircxxxxxxA}{\ensuremath{1.70\pm0.22}}     
\newcommand{\hatcurPPtinfallxxxxxxA}{\ensuremath{810\pm110}}     
\newcommand{\hatcurXdistxxxxxxA}{\ensuremath{114.90\pm0.48}}     
\newcommand{\hatcurXAvxxxxxxA}{\ensuremath{0.0110_{-0.0070}^{+0.0100}}} 
\newcommand{\hatcurXdistredxxxxxxA}{\ensuremath{114.88\pm0.48}}  
\newcommand{\hatcurXEBVxxxxxxA}{\ensuremath{0.0040\pm0.0027}}    
\newcommand{\hatcurCCpmraxxxxxxA}{\ensuremath{-41.959\pm0.029}}  
\newcommand{\hatcurCCpmdecxxxxxxA}{\ensuremath{29.074\pm0.027}}  
\newcommand{\hatcurCCpmxxxxxxA}{\ensuremath{51.048\pm0.040}}     

%% file: TOI3629.tex
\newcommand{\hatcurhtrxxxxxxB}{TOI3629}                          
\newcommand{\hatcurfieldxxxxxxB}{\ensuremath{string}}            
\newcommand{\hatcurCCraxxxxxxB}{\ensuremath{23^{\mathrm h}59^{\mathrm m}10.3200{\mathrm s}}}                   
\newcommand{\hatcurCCdecxxxxxxB}{\ensuremath{+39{\arcdeg}18{\arcmin}51.3200{\arcsec}}}                 
\newcommand{\hatcurCCmagxxxxxxB}{14.639}                         
\newcommand{\hatcurCCtwomassxxxxxxB}{2MASS~23591015+3918514}     
\newcommand{\hatcurCCgscxxxxxxB}{GSC~NULL}                       
\newcommand{\hatcurCCgaiaxxxxxxB}{GAIA~2881820319999000960}      
\newcommand{\hatcurCCgaiadrtwoxxxxxxB}{GAIA~DR2~2881820324294985856} 
\newcommand{\hatcurCCtassmvxxxxxxB}{\ensuremath{14.639\pm0.061}} 
\newcommand{\hatcurCCtassmvshortxxxxxxB}{\ensuremath{14.6}}      
\newcommand{\hatcurCCtassmBxxxxxxB}{\ensuremath{16.093\pm0.041}} 
\newcommand{\hatcurCCtassmBshortxxxxxxB}{\ensuremath{16.1}}      
\newcommand{\hatcurCCtassmIxxxxxxB}{\ensuremath{nff\pmnff}}      
\newcommand{\hatcurCCtassmIshortxxxxxxB}{\ensuremath{0.0}}       
\newcommand{\hatcurCCtassmgxxxxxxB}{\ensuremath{15.380\pm0.022}} 
\newcommand{\hatcurCCtassmgshortxxxxxxB}{\ensuremath{15.4}}      
\newcommand{\hatcurCCtassmrxxxxxxB}{\ensuremath{14.055\pm0.057}} 
\newcommand{\hatcurCCtassmrshortxxxxxxB}{\ensuremath{14.1}}      
\newcommand{\hatcurCCtassmixxxxxxB}{\ensuremath{13.111\pm0.032}} 
\newcommand{\hatcurCCtassmishortxxxxxxB}{\ensuremath{13.1}}      
\newcommand{\hatcurCCparallaxxxxxxxB}{\ensuremath{7.667\pm0.017}} 
\newcommand{\hatcurCCgaiamGxxxxxxB}{\ensuremath{13.8226\pm0.0028}} 
\newcommand{\hatcurCCgaiamBPxxxxxxB}{\ensuremath{14.8939\pm0.0031}} 
\newcommand{\hatcurCCgaiamRPxxxxxxB}{\ensuremath{12.7937\pm0.0038}} 
\newcommand{\hatcurCCtwomassJmagxxxxxxB}{\ensuremath{11.424\pm0.027}} 
\newcommand{\hatcurCCtwomassHmagxxxxxxB}{\ensuremath{10.733\pm0.030}} 
\newcommand{\hatcurCCtwomassKmagxxxxxxB}{\ensuremath{10.553\pm0.020}} 
\newcommand{\hatcurCCcitJmagxxxxxxB}{\ensuremath{11.414\pm0.028}} 
\newcommand{\hatcurCCcitHmagxxxxxxB}{\ensuremath{10.726\pm0.030}} 
\newcommand{\hatcurCCcitKmagxxxxxxB}{\ensuremath{10.577\pm0.021}} 
\newcommand{\hatcurCCbbJmagxxxxxxB}{\ensuremath{11.505\pm0.030}} 
\newcommand{\hatcurCCbbHmagxxxxxxB}{\ensuremath{10.749\pm0.032}} 
\newcommand{\hatcurCCbbKmagxxxxxxB}{\ensuremath{10.597\pm0.021}} 
\newcommand{\hatcurCCesoJmagxxxxxxB}{\ensuremath{11.513\pm0.035}} 
\newcommand{\hatcurCCesoHmagxxxxxxB}{\ensuremath{10.747\pm0.041}} 
\newcommand{\hatcurCCesoKmagxxxxxxB}{\ensuremath{10.593\pm0.023}} 
\newcommand{\hatcurCCesoJHmagxxxxxxB}{\ensuremath{0.765\pm0.050}} 
\newcommand{\hatcurCCesoJKmagxxxxxxB}{\ensuremath{0.920\pm0.039}} 
\newcommand{\hatcurCCesoHKmagxxxxxxB}{\ensuremath{0.154\pm0.045}} 
\newcommand{\hatcurCCWonemagxxxxxxB}{\ensuremath{10.484\pm0.022}} 
\newcommand{\hatcurCCWtwomagxxxxxxB}{\ensuremath{10.516\pm0.020}} 
\newcommand{\hatcurCCWthreemagxxxxxxB}{\ensuremath{10.375\pm0.067}} 
\newcommand{\hatcurCCWfourmagxxxxxxB}{\ensuremath{nff\pmnff}}    
\newcommand{\hatcurLCdipxxxxxxB}{\ensuremath{20.0}}              
\newcommand{\hatcurLCrprstarxxxxxxB}{\ensuremath{0.1247\pm0.0014}} 
\newcommand{\hatcurLCbsqxxxxxxB}{\ensuremath{0.106_{-0.024}^{+0.021}}} 
\newcommand{\hatcurLCimpxxxxxxB}{\ensuremath{0.325_{-0.039}^{+0.030}}} 
\newcommand{\hatcurLCzetaxxxxxxB}{\ensuremath{24.92\pm0.17}}     
\newcommand{\hatcurLCdurxxxxxxB}{\ensuremath{0.09139\pm0.00050}} 
\newcommand{\hatcurLCdurshortxxxxxxB}{\ensuremath{0.0914}}       
\newcommand{\hatcurLCdurhrxxxxxxB}{\ensuremath{2.193\pm0.012}}   
\newcommand{\hatcurLCdurhrshortxxxxxxB}{\ensuremath{2.193}}      
\newcommand{\hatcurLCqxxxxxxB}{\ensuremath{0.02320\pm0.00013}}   
\newcommand{\hatcurLCqshortxxxxxxB}{\ensuremath{0.023}}          
\newcommand{\hatcurLCingdurxxxxxxB}{\ensuremath{0.01119\pm0.00030}} 
\newcommand{\hatcurLCPxxxxxxB}{\ensuremath{3.9365578\pm0.0000043}} 
\newcommand{\hatcurLCPprecxxxxxxB}{\ensuremath{3.9365578}}       
\newcommand{\hatcurLCPshortxxxxxxB}{\ensuremath{3.9366}}         
\newcommand{\hatcurLCTxxxxxxB}{\ensuremath{2459662.10795\pm0.00019}} 
\newcommand{\hatcurLCTAxxxxxxB}{\ensuremath{2458768.50939\pm0.00087}} 
\newcommand{\hatcurLCTBxxxxxxB}{\ensuremath{2459878.61862\pm0.00038}} 
\newcommand{\hatcurLChatnetmAxxxxxxB}{\ensuremath{-0.000070\pm0.000079}} 
\newcommand{\hatcurLCiblendAxxxxxxB}{\ensuremath{0.862\pm0.035}} 
\newcommand{\hatcurLChatnetmBxxxxxxB}{\ensuremath{-7.707100\pm0.000048}} 
\newcommand{\hatcurLCiblendBxxxxxxB}{\ensuremath{0.970\pm0.022}} 
\newcommand{\hatcurLCrhoxxxxxxB}{\ensuremath{3.922_{-0.096}^{+0.138}}} 
\newcommand{\hatcurSMEiteffxxxxxxB}{\ensuremath{3870\pm90}}      
\newcommand{\hatcurSMEizfehxxxxxxB}{\ensuremath{0.40\pm0.10}}    
\newcommand{\hatcurSMEizfehshortxxxxxxB}{\ensuremath{0.40}}      
\newcommand{\hatcurSMEiloggxxxxxxB}{\ensuremath{4.670\pm0.050}}  
\newcommand{\hatcurSMEivsinxxxxxxB}{\ensuremath{0.0\pm2.0}}      
\newcommand{\hatcurSMEivmacxxxxxxB}{\ensuremath{nff\pmnff}}      
\newcommand{\hatcurSMEivmicxxxxxxB}{\ensuremath{nff\pmnff}}      
\newcommand{\hatcurextraerrMJxxxxxxB}{\ensuremath{0\pm0}}        
\newcommand{\hatcurextraerrMJtwosiglimxxxxxxB}{\ensuremath{<0.0200}} 
\newcommand{\hatcurextraerrMHxxxxxxB}{\ensuremath{0\pm0}}        
\newcommand{\hatcurextraerrMHtwosiglimxxxxxxB}{\ensuremath{<0.0200}} 
\newcommand{\hatcurextraerrMKsxxxxxxB}{\ensuremath{0\pm0}}       
\newcommand{\hatcurextraerrMKstwosiglimxxxxxxB}{\ensuremath{<0.0200}} 
\newcommand{\hatcurextraerrMGxxxxxxB}{\ensuremath{0\pm0}}        
\newcommand{\hatcurextraerrMGtwosiglimxxxxxxB}{\ensuremath{<0.0200}} 
\newcommand{\hatcurextraerrMRPxxxxxxB}{\ensuremath{0\pm0}}       
\newcommand{\hatcurextraerrMRPtwosiglimxxxxxxB}{\ensuremath{<0.0200}} 
\newcommand{\hatcurextraerrMgxxxxxxB}{\ensuremath{0\pm0}}        
\newcommand{\hatcurextraerrMgtwosiglimxxxxxxB}{\ensuremath{<0.0200}} 
\newcommand{\hatcurextraerrMrxxxxxxB}{\ensuremath{0\pm0}}        
\newcommand{\hatcurextraerrMrtwosiglimxxxxxxB}{\ensuremath{<0.0200}} 
\newcommand{\hatcurextraerrMixxxxxxB}{\ensuremath{0\pm0}}        
\newcommand{\hatcurextraerrMitwosiglimxxxxxxB}{\ensuremath{<0.0200}} 
\newcommand{\hatcurextraerrMWonexxxxxxB}{\ensuremath{0\pm0}}     
\newcommand{\hatcurextraerrMWonetwosiglimxxxxxxB}{\ensuremath{<0.0200}} 
\newcommand{\hatcurextraerrMWtwoxxxxxxB}{\ensuremath{0\pm0}}     
\newcommand{\hatcurextraerrMWtwotwosiglimxxxxxxB}{\ensuremath{<0.0200}} 
\newcommand{\hatcurLBiBxxxxxxB}{\ensuremath{0.5294}}             
\newcommand{\hatcurLBiiBxxxxxxB}{\ensuremath{0.2909}}            
\newcommand{\hatcurLBiVxxxxxxB}{\ensuremath{0.4674}}             
\newcommand{\hatcurLBiiVxxxxxxB}{\ensuremath{0.3330}}            
\newcommand{\hatcurLBiRxxxxxxB}{\ensuremath{0.4192}}             
\newcommand{\hatcurLBiiRxxxxxxB}{\ensuremath{0.3091}}            
\newcommand{\hatcurLBiIxxxxxxB}{\ensuremath{0.2292}}             
\newcommand{\hatcurLBiiIxxxxxxB}{\ensuremath{0.3468}}            
\newcommand{\hatcurLBiuxxxxxxB}{\ensuremath{0.5326}}             
\newcommand{\hatcurLBiiuxxxxxxB}{\ensuremath{0.2706}}            
\newcommand{\hatcurLBigxxxxxxB}{\ensuremath{0.39\pm0.12}}        
\newcommand{\hatcurLBiigxxxxxxB}{\ensuremath{0.35\pm0.15}}       
\newcommand{\hatcurLBirxxxxxxB}{\ensuremath{0.49\pm0.12}}        
\newcommand{\hatcurLBiirxxxxxxB}{\ensuremath{0.14\pm0.16}}       
\newcommand{\hatcurLBiixxxxxxB}{\ensuremath{0.370\pm0.093}}      
\newcommand{\hatcurLBiiixxxxxxB}{\ensuremath{0.26\pm0.15}}       
\newcommand{\hatcurLBizxxxxxxB}{\ensuremath{0.254\pm0.096}}      
\newcommand{\hatcurLBiizxxxxxxB}{\ensuremath{0.26\pm0.15}}       
\newcommand{\hatcurLBiJxxxxxxB}{\ensuremath{0.1266}}             
\newcommand{\hatcurLBiiJxxxxxxB}{\ensuremath{0.2533}}            
\newcommand{\hatcurLBiHxxxxxxB}{\ensuremath{0.0987}}             
\newcommand{\hatcurLBiiHxxxxxxB}{\ensuremath{0.2706}}            
\newcommand{\hatcurLBiKxxxxxxB}{\ensuremath{0.0784}}             
\newcommand{\hatcurLBiiKxxxxxxB}{\ensuremath{0.2508}}            
\newcommand{\hatcurLBiTxxxxxxB}{\ensuremath{0.29\pm0.11}}        
\newcommand{\hatcurLBiiTxxxxxxB}{\ensuremath{0.22\pm0.17}}       
\newcommand{\hatcurLBikepxxxxxxB}{\ensuremath{0.3414}}           
\newcommand{\hatcurLBiikepxxxxxxB}{\ensuremath{0.3695}}          
\newcommand{\hatcurLBiCxxxxxxB}{\ensuremath{0.3005}}             
\newcommand{\hatcurLBiiCxxxxxxB}{\ensuremath{0.3735}}            
\newcommand{\hatcurLBiMxxxxxxB}{\ensuremath{0.4294}}             
\newcommand{\hatcurLBiiMxxxxxxB}{\ensuremath{0.3538}}            
\newcommand{\hatcurLBiSonexxxxxxB}{\ensuremath{0.0646}}            
\newcommand{\hatcurLBiiSonexxxxxxB}{\ensuremath{0.1906}}           
\newcommand{\hatcurLBiStwoxxxxxxB}{\ensuremath{0.0527}}            
\newcommand{\hatcurLBiiStwoxxxxxxB}{\ensuremath{0.1496}}           
\newcommand{\hatcurLBiSthreexxxxxxB}{\ensuremath{0.0536}}            
\newcommand{\hatcurLBiiSthreexxxxxxB}{\ensuremath{0.1304}}           
\newcommand{\hatcurLBiSfourxxxxxxB}{\ensuremath{0.0638}}            
\newcommand{\hatcurLBiiSfourxxxxxxB}{\ensuremath{0.0992}}           
\newcommand{\hatcurISOmxxxxxxB}{\ensuremath{0.635\pm0.032}}      
\newcommand{\hatcurISOmshortxxxxxxB}{\ensuremath{0.63}}          
\newcommand{\hatcurISOmlongxxxxxxB}{\ensuremath{0.635\pm0.032}}  
\newcommand{\hatcurISOrxxxxxxB}{\ensuremath{0.6103\pm0.0099}}    
\newcommand{\hatcurISOrshortxxxxxxB}{\ensuremath{0.61}}          
\newcommand{\hatcurISOrlongxxxxxxB}{\ensuremath{0.6103\pm0.0099}} 
\newcommand{\hatcurISOrhoxxxxxxB}{\ensuremath{3.922_{-0.096}^{+0.138}}} 
\newcommand{\hatcurISOrholongxxxxxxB}{\ensuremath{3.922_{-0.096}^{+0.138}}} 
\newcommand{\hatcurISOloggxxxxxxB}{\ensuremath{4.670\pm0.013}}   
\newcommand{\hatcurISOlumxxxxxxB}{\ensuremath{0.0750\pm0.0068}}  
\newcommand{\hatcurISOlumshortxxxxxxB}{\ensuremath{0.07}}        
\newcommand{\hatcurISOteffxxxxxxB}{\ensuremath{3865\pm79}}       
\newcommand{\hatcurISOzfehxxxxxxB}{\ensuremath{0.549\pm0.093}}   
\newcommand{\hatcurISOagexxxxxxB}{\ensuremath{9.8_{-4.8}^{+6.6}}} 
\newcommand{\hatcurISOspecxxxxxxB}{M}                            
\newcommand{\hatcurRVKxxxxxxB}{\ensuremath{42.2\pm3.2}}          
\newcommand{\hatcurRVKtwosiglimxxxxxxB}{\ensuremath{<48.2}}      
\newcommand{\hatcurRVrkxxxxxxB}{\ensuremath{0\pm0}}              
\newcommand{\hatcurRVrhxxxxxxB}{\ensuremath{0\pm0}}              
\newcommand{\hatcurRVkxxxxxxB}{\ensuremath{0\pm0}}               
\newcommand{\hatcurRVhxxxxxxB}{\ensuremath{0\pm0}}               
\newcommand{\hatcurRVtronexxxxxxB}{\ensuremath{0\pm0}}           
\newcommand{\hatcurRVtrtwoxxxxxxB}{\ensuremath{0\pm0}}           
\newcommand{\hatcurRVgammaAxxxxxxB}{\ensuremath{-12.6\pm5.7}}    
\newcommand{\hatcurRVjitterAxxxxxxB}{\ensuremath{15.2\pm5.7}}    
\newcommand{\hatcurRVjittertwosiglimAxxxxxxB}{\ensuremath{<27.1}} 
\newcommand{\hatcurRVfitrmsAxxxxxxB}{\ensuremath{0.0}}           
\newcommand{\hatcurRVgammaBxxxxxxB}{\ensuremath{-16.0\pm3.8}}    
\newcommand{\hatcurRVjitterBxxxxxxB}{\ensuremath{0.0\pm3.0}}     
\newcommand{\hatcurRVjittertwosiglimBxxxxxxB}{\ensuremath{<7.2}} 
\newcommand{\hatcurRVfitrmsBxxxxxxB}{\ensuremath{0.0}}           
\newcommand{\hatcurRVgammaCxxxxxxB}{\ensuremath{-7.9\pm5.3}}     
\newcommand{\hatcurRVjitterCxxxxxxB}{\ensuremath{0\pm31}}        
\newcommand{\hatcurRVjittertwosiglimCxxxxxxB}{\ensuremath{<13.8}} 
\newcommand{\hatcurRVfitrmsCxxxxxxB}{\ensuremath{0.0}}           
\newcommand{\hatcurRVeccenxxxxxxB}{\ensuremath{0\pm0}}           
\newcommand{\hatcurRVeccentwosiglimxxxxxxB}{\ensuremath{<0.000}} 
\newcommand{\hatcurRVomegaxxxxxxB}{\ensuremath{0\pm0}}           
\newcommand{\hatcurPPixxxxxxB}{\ensuremath{88.74\pm0.14}}        
\newcommand{\hatcurPPgxxxxxxB}{\ensuremath{11.02\pm0.88}}        
\newcommand{\hatcurPPloggxxxxxxB}{\ensuremath{3.042\pm0.035}}    
\newcommand{\hatcurPParxxxxxxB}{\ensuremath{14.76_{-0.12}^{+0.17}}} 
\newcommand{\hatcurPParelxxxxxxB}{\ensuremath{0.04194\pm0.00071}} 
\newcommand{\hatcurPPrhoxxxxxxB}{\ensuremath{0.743\pm0.064}}     
\newcommand{\hatcurPPmxxxxxxB}{\ensuremath{0.243\pm0.020}}       
\newcommand{\hatcurPPmtwosiglimxxxxxxB}{\ensuremath{<0.28}}      
\newcommand{\hatcurPPmshortxxxxxxB}{\ensuremath{0.24}}           
\newcommand{\hatcurPPmlongxxxxxxB}{\ensuremath{0.243\pm0.020}}   
\newcommand{\hatcurPPmexxxxxxB}{\ensuremath{77.4\pm6.4}}         
\newcommand{\hatcurPPmeshortxxxxxxB}{\ensuremath{77.4}}          
\newcommand{\hatcurPPmelongxxxxxxB}{\ensuremath{77.4\pm6.4}}     
\newcommand{\hatcurPPrxxxxxxB}{\ensuremath{0.740\pm0.014}}       
\newcommand{\hatcurPPrshortxxxxxxB}{\ensuremath{0.74}}           
\newcommand{\hatcurPPrlongxxxxxxB}{\ensuremath{0.740\pm0.014}}   
\newcommand{\hatcurPPrexxxxxxB}{\ensuremath{8.30\pm0.16}}        
\newcommand{\hatcurPPreshortxxxxxxB}{\ensuremath{8.3}}           
\newcommand{\hatcurPPrelongxxxxxxB}{\ensuremath{8.30\pm0.16}}    
\newcommand{\hatcurPPmrcorrxxxxxxB}{\ensuremath{0.28}}           
\newcommand{\hatcurPPteffxxxxxxB}{\ensuremath{711\pm15}}         
\newcommand{\hatcurPPthetaxxxxxxB}{\ensuremath{0.0432\pm0.0033}} 
\newcommand{\hatcurPPfluxperixxxxxxB}{\ensuremath{5.80\pm0.48}}  
\newcommand{\hatcurPPfluxperidimxxxxxxB}{\ensuremath{7}}         
\newcommand{\hatcurPPfluxapxxxxxxB}{\ensuremath{5.80\pm0.48}}    
\newcommand{\hatcurPPfluxapdimxxxxxxB}{\ensuremath{7}}           
\newcommand{\hatcurPPfluxavgxxxxxxB}{\ensuremath{5.80\pm0.48}}   
\newcommand{\hatcurPPfluxavgdimxxxxxxB}{\ensuremath{7}}          
\newcommand{\hatcurPPfluxavglogxxxxxxB}{\ensuremath{7.763\pm0.036}} 
\newcommand{\hatcurXsecphasexxxxxxB}{\ensuremath{0\pm0}}         
\newcommand{\hatcurXsecondaryxxxxxxB}{\ensuremath{2459664.07622\pm0.00019}} 
\newcommand{\hatcurXsecdurxxxxxxB}{\ensuremath{0.09139\pm0.00050}} 
\newcommand{\hatcurXsecingdurxxxxxxB}{\ensuremath{0.01119\pm0.00030}} 
\newcommand{\hatcurPPphiconjxxxxxxB}{\ensuremath{0\pm0}}         
\newcommand{\hatcurPPperixxxxxxB}{\ensuremath{2459661.12381\pm0.00019}} 
\newcommand{\hatcurPPaequivxxxxxxB}{\ensuremath{0.1532\pm0.0064}} 
\newcommand{\hatcurPPtcircxxxxxxB}{\ensuremath{930\pm99}}        
\newcommand{\hatcurPPtinfallxxxxxxB}{\ensuremath{68200_{-5400}^{+7500}}} 
\newcommand{\hatcurXdistxxxxxxB}{\ensuremath{130.30_{-0.30}^{+0.40}}} 
\newcommand{\hatcurXAvxxxxxxB}{\ensuremath{0.163\pm0.029}}       
\newcommand{\hatcurXdistredxxxxxxB}{\ensuremath{130.34\pm0.30}}  
\newcommand{\hatcurXEBVxxxxxxB}{\ensuremath{0.0530\pm0.0094}}    
\newcommand{\hatcurCCpmraxxxxxxB}{\ensuremath{185.707\pm0.012}}  
\newcommand{\hatcurCCpmdecxxxxxxB}{\ensuremath{1.010\pm0.012}}   
\newcommand{\hatcurCCpmxxxxxxB}{\ensuremath{185.710\pm0.017}}    

%% file: TOI3714.tex
\newcommand{\hatcurhtrxxxxxxC}{TOI3714}                          
\newcommand{\hatcurfieldxxxxxxC}{\ensuremath{string}}            
\newcommand{\hatcurCCraxxxxxxC}{\ensuremath{04^{\mathrm h}38^{\mathrm m}12.5354{\mathrm s}}}                   
\newcommand{\hatcurCCdecxxxxxxC}{\ensuremath{+39{\arcdeg}27{\arcmin}29.9093{\arcsec}}}                 
\newcommand{\hatcurCCmagxxxxxxC}{15.278}                         
\newcommand{\hatcurCCtwomassxxxxxxC}{2MASS~04381253+3927299}     
\newcommand{\hatcurCCgscxxxxxxC}{GSC~NULL}                       
\newcommand{\hatcurCCgaiaxxxxxxC}{GAIA~178924386182591104}       
\newcommand{\hatcurCCgaiadrtwoxxxxxxC}{GAIA~DR2~178924390478792320} 
\newcommand{\hatcurCCtassmvxxxxxxC}{\ensuremath{15.278\pm0.020}} 
\newcommand{\hatcurCCtassmvshortxxxxxxC}{\ensuremath{15.3}}      
\newcommand{\hatcurCCtassmBxxxxxxC}{\ensuremath{16.86\pm0.13}}   
\newcommand{\hatcurCCtassmBshortxxxxxxC}{\ensuremath{16.9}}      
\newcommand{\hatcurCCtassmIxxxxxxC}{\ensuremath{nff\pmnff}}      
\newcommand{\hatcurCCtassmIshortxxxxxxC}{\ensuremath{0.0}}       
\newcommand{\hatcurCCtassmgxxxxxxC}{\ensuremath{15.99\pm0.18}}   
\newcommand{\hatcurCCtassmgshortxxxxxxC}{\ensuremath{16.0}}      
\newcommand{\hatcurCCtassmrxxxxxxC}{\ensuremath{14.737\pm0.011}} 
\newcommand{\hatcurCCtassmrshortxxxxxxC}{\ensuremath{14.7}}      
\newcommand{\hatcurCCtassmixxxxxxC}{\ensuremath{13.70\pm0.14}}   
\newcommand{\hatcurCCtassmishortxxxxxxC}{\ensuremath{13.7}}      
\newcommand{\hatcurCCparallaxxxxxxxC}{\ensuremath{8.838\pm0.022}} 
\newcommand{\hatcurCCgaiamGxxxxxxC}{\ensuremath{14.2908\pm0.0028}} 
\newcommand{\hatcurCCgaiamBPxxxxxxC}{\ensuremath{15.5114\pm0.0035}} 
\newcommand{\hatcurCCgaiamRPxxxxxxC}{\ensuremath{13.1948\pm0.0039}} 
\newcommand{\hatcurCCtwomassJmagxxxxxxC}{\ensuremath{11.788\pm0.020}} 
\newcommand{\hatcurCCtwomassHmagxxxxxxC}{\ensuremath{11.109\pm0.015}} 
\newcommand{\hatcurCCtwomassKmagxxxxxxC}{\ensuremath{10.906\pm0.017}} 
\newcommand{\hatcurCCcitJmagxxxxxxC}{\ensuremath{11.726\pm0.022}} 
\newcommand{\hatcurCCcitHmagxxxxxxC}{\ensuremath{11.048\pm0.018}} 
\newcommand{\hatcurCCcitKmagxxxxxxC}{\ensuremath{10.876\pm0.018}} 
\newcommand{\hatcurCCbbJmagxxxxxxC}{\ensuremath{11.817\pm0.024}} 
\newcommand{\hatcurCCbbHmagxxxxxxC}{\ensuremath{11.074\pm0.018}} 
\newcommand{\hatcurCCbbKmagxxxxxxC}{\ensuremath{10.896\pm0.018}} 
\newcommand{\hatcurCCesoJmagxxxxxxC}{\ensuremath{11.825\pm0.030}} 
\newcommand{\hatcurCCesoHmagxxxxxxC}{\ensuremath{11.072\pm0.033}} 
\newcommand{\hatcurCCesoKmagxxxxxxC}{\ensuremath{10.892\pm0.020}} 
\newcommand{\hatcurCCesoJHmagxxxxxxC}{\ensuremath{0.752\pm0.040}} 
\newcommand{\hatcurCCesoJKmagxxxxxxC}{\ensuremath{0.933\pm0.041}} 
\newcommand{\hatcurCCesoHKmagxxxxxxC}{\ensuremath{0.179\pm0.036}} 
\newcommand{\hatcurCCWonemagxxxxxxC}{\ensuremath{10.779\pm0.023}} 
\newcommand{\hatcurCCWtwomagxxxxxxC}{\ensuremath{10.744\pm0.019}} 
\newcommand{\hatcurCCWthreemagxxxxxxC}{\ensuremath{10.546\pm0.097}} 
\newcommand{\hatcurCCWfourmagxxxxxxC}{\ensuremath{nff\pmnff}}    
\newcommand{\hatcurLCdipxxxxxxC}{\ensuremath{51.0}}              
\newcommand{\hatcurLCrprstarxxxxxxC}{\ensuremath{0.2057\pm0.0013}} 
\newcommand{\hatcurLCbsqxxxxxxC}{\ensuremath{0.188_{-0.016}^{+0.015}}} 
\newcommand{\hatcurLCimpxxxxxxC}{\ensuremath{0.433_{-0.020}^{+0.017}}} 
\newcommand{\hatcurLCzetaxxxxxxC}{\ensuremath{36.85\pm0.23}}     
\newcommand{\hatcurLCdurxxxxxxC}{\ensuremath{0.06776\pm0.00029}} 
\newcommand{\hatcurLCdurshortxxxxxxC}{\ensuremath{0.0678}}       
\newcommand{\hatcurLCdurhrxxxxxxC}{\ensuremath{1.6262\pm0.0069}} 
\newcommand{\hatcurLCdurhrshortxxxxxxC}{\ensuremath{1.626}}      
\newcommand{\hatcurLCqxxxxxxC}{\ensuremath{0.03140\pm0.00014}}   
\newcommand{\hatcurLCqshortxxxxxxC}{\ensuremath{0.031}}          
\newcommand{\hatcurLCingdurxxxxxxC}{\ensuremath{0.01382\pm0.00028}} 
\newcommand{\hatcurLCPxxxxxxC}{\ensuremath{2.15484802\pm0.00000075}} 
\newcommand{\hatcurLCPprecxxxxxxC}{\ensuremath{2.1548480}}       
\newcommand{\hatcurLCPshortxxxxxxC}{\ensuremath{2.1548}}         
\newcommand{\hatcurLCTxxxxxxC}{\ensuremath{2459687.365240\pm0.000075}} 
\newcommand{\hatcurLCTAxxxxxxC}{\ensuremath{2458818.96148\pm0.00027}} 
\newcommand{\hatcurLCTBxxxxxxC}{\ensuremath{2459883.45641\pm0.00013}} 
\newcommand{\hatcurLChatnetmxxxxxxC}{\ensuremath{-0.00026\pm0.00014}} 
\newcommand{\hatcurLCiblendxxxxxxC}{\ensuremath{0.951\pm0.023}}  
\newcommand{\hatcurLCrhoxxxxxxC}{\ensuremath{6.01\pm0.12}}       
\newcommand{\hatcurSMEiteffxxxxxxC}{\ensuremath{3660\pm90}}      
\newcommand{\hatcurSMEizfehxxxxxxC}{\ensuremath{0.10\pm0.10}}    
\newcommand{\hatcurSMEizfehshortxxxxxxC}{\ensuremath{0.10}}      
\newcommand{\hatcurSMEiloggxxxxxxC}{\ensuremath{4.750\pm0.050}}  
\newcommand{\hatcurSMEivsinxxxxxxC}{\ensuremath{5.0\pm5.0}}      
\newcommand{\hatcurSMEivmacxxxxxxC}{\ensuremath{nff\pmnff}}      
\newcommand{\hatcurSMEivmicxxxxxxC}{\ensuremath{nff\pmnff}}      
\newcommand{\hatcurextraerrMJxxxxxxC}{\ensuremath{0\pm0}}        
\newcommand{\hatcurextraerrMJtwosiglimxxxxxxC}{\ensuremath{<0.0200}} 
\newcommand{\hatcurextraerrMHxxxxxxC}{\ensuremath{0\pm0}}        
\newcommand{\hatcurextraerrMHtwosiglimxxxxxxC}{\ensuremath{<0.0200}} 
\newcommand{\hatcurextraerrMKsxxxxxxC}{\ensuremath{0\pm0}}       
\newcommand{\hatcurextraerrMKstwosiglimxxxxxxC}{\ensuremath{<0.0200}} 
\newcommand{\hatcurextraerrMGxxxxxxC}{\ensuremath{0\pm0}}        
\newcommand{\hatcurextraerrMGtwosiglimxxxxxxC}{\ensuremath{<0.0200}} 
\newcommand{\hatcurextraerrMRPxxxxxxC}{\ensuremath{0\pm0}}       
\newcommand{\hatcurextraerrMRPtwosiglimxxxxxxC}{\ensuremath{<0.0200}} 
\newcommand{\hatcurextraerrMgxxxxxxC}{\ensuremath{0\pm0}}        
\newcommand{\hatcurextraerrMgtwosiglimxxxxxxC}{\ensuremath{<0.0200}} 
\newcommand{\hatcurextraerrMrxxxxxxC}{\ensuremath{0\pm0}}        
\newcommand{\hatcurextraerrMrtwosiglimxxxxxxC}{\ensuremath{<0.0200}} 
\newcommand{\hatcurextraerrMixxxxxxC}{\ensuremath{0\pm0}}        
\newcommand{\hatcurextraerrMitwosiglimxxxxxxC}{\ensuremath{<0.0200}} 
\newcommand{\hatcurextraerrMWonexxxxxxC}{\ensuremath{0\pm0}}     
\newcommand{\hatcurextraerrMWonetwosiglimxxxxxxC}{\ensuremath{<0.0200}} 
\newcommand{\hatcurextraerrMWtwoxxxxxxC}{\ensuremath{0\pm0}}     
\newcommand{\hatcurextraerrMWtwotwosiglimxxxxxxC}{\ensuremath{<0.0200}} 
\newcommand{\hatcurLBiBxxxxxxC}{\ensuremath{0.5256}}             
\newcommand{\hatcurLBiiBxxxxxxC}{\ensuremath{0.3086}}            
\newcommand{\hatcurLBiVxxxxxxC}{\ensuremath{0.5054}}             
\newcommand{\hatcurLBiiVxxxxxxC}{\ensuremath{0.3250}}            
\newcommand{\hatcurLBiRxxxxxxC}{\ensuremath{0.4394}}             
\newcommand{\hatcurLBiiRxxxxxxC}{\ensuremath{0.3031}}            
\newcommand{\hatcurLBiIxxxxxxC}{\ensuremath{0.38\pm0.11}}        
\newcommand{\hatcurLBiiIxxxxxxC}{\ensuremath{0.15\pm0.16}}       
\newcommand{\hatcurLBiuxxxxxxC}{\ensuremath{0.4650}}             
\newcommand{\hatcurLBiiuxxxxxxC}{\ensuremath{0.3204}}            
\newcommand{\hatcurLBigxxxxxxC}{\ensuremath{0.55\pm0.11}}        
\newcommand{\hatcurLBiigxxxxxxC}{\ensuremath{0.26\pm0.16}}       
\newcommand{\hatcurLBirxxxxxxC}{\ensuremath{0.55\pm0.10}}        
\newcommand{\hatcurLBiirxxxxxxC}{\ensuremath{0.16\pm0.15}}       
\newcommand{\hatcurLBiixxxxxxC}{\ensuremath{0.389\pm0.096}}      
\newcommand{\hatcurLBiiixxxxxxC}{\ensuremath{0.30\pm0.15}}       
\newcommand{\hatcurLBizxxxxxxC}{\ensuremath{0.288\pm0.081}}      
\newcommand{\hatcurLBiizxxxxxxC}{\ensuremath{0.23\pm0.13}}       
\newcommand{\hatcurLBiJxxxxxxC}{\ensuremath{0.1069}}             
\newcommand{\hatcurLBiiJxxxxxxC}{\ensuremath{0.2591}}            
\newcommand{\hatcurLBiHxxxxxxC}{\ensuremath{0.0865}}             
\newcommand{\hatcurLBiiHxxxxxxC}{\ensuremath{0.2899}}            
\newcommand{\hatcurLBiKxxxxxxC}{\ensuremath{0.0778}}             
\newcommand{\hatcurLBiiKxxxxxxC}{\ensuremath{0.2615}}            
\newcommand{\hatcurLBiTxxxxxxC}{\ensuremath{0.18\pm0.12}}        
\newcommand{\hatcurLBiiTxxxxxxC}{\ensuremath{0.16\pm0.16}}       
\newcommand{\hatcurLBikepxxxxxxC}{\ensuremath{0.3334}}           
\newcommand{\hatcurLBiikepxxxxxxC}{\ensuremath{0.3865}}          
\newcommand{\hatcurLBiCxxxxxxC}{\ensuremath{0.2805}}             
\newcommand{\hatcurLBiiCxxxxxxC}{\ensuremath{0.3966}}            
\newcommand{\hatcurLBiMxxxxxxC}{\ensuremath{0.4437}}             
\newcommand{\hatcurLBiiMxxxxxxC}{\ensuremath{0.3598}}            
\newcommand{\hatcurLBiSonexxxxxxC}{\ensuremath{0.0686}}            
\newcommand{\hatcurLBiiSonexxxxxxC}{\ensuremath{0.1876}}           
\newcommand{\hatcurLBiStwoxxxxxxC}{\ensuremath{0.0500}}            
\newcommand{\hatcurLBiiStwoxxxxxxC}{\ensuremath{0.1526}}           
\newcommand{\hatcurLBiSthreexxxxxxC}{\ensuremath{0.0535}}            
\newcommand{\hatcurLBiiSthreexxxxxxC}{\ensuremath{0.1327}}           
\newcommand{\hatcurLBiSfourxxxxxxC}{\ensuremath{0.0603}}            
\newcommand{\hatcurLBiiSfourxxxxxxC}{\ensuremath{0.0964}}           
\newcommand{\hatcurISOmxxxxxxC}{\ensuremath{0.522\pm0.028}}      
\newcommand{\hatcurISOmshortxxxxxxC}{\ensuremath{0.52}}          
\newcommand{\hatcurISOmlongxxxxxxC}{\ensuremath{0.522\pm0.028}}  
\newcommand{\hatcurISOrxxxxxxC}{\ensuremath{0.4958_{-0.0079}^{+0.0104}}} 
\newcommand{\hatcurISOrshortxxxxxxC}{\ensuremath{0.50}}          
\newcommand{\hatcurISOrlongxxxxxxC}{\ensuremath{0.4958_{-0.0079}^{+0.0104}}} 
\newcommand{\hatcurISOrhoxxxxxxC}{\ensuremath{6.01\pm0.12}}      
\newcommand{\hatcurISOrholongxxxxxxC}{\ensuremath{6.01\pm0.12}}  
\newcommand{\hatcurISOloggxxxxxxC}{\ensuremath{4.7638\pm0.0097}} 
\newcommand{\hatcurISOlumxxxxxxC}{\ensuremath{0.0392\pm0.0038}}  
\newcommand{\hatcurISOlumshortxxxxxxC}{\ensuremath{0.04}}        
\newcommand{\hatcurISOteffxxxxxxC}{\ensuremath{3652\pm81}}       
\newcommand{\hatcurISOzfehxxxxxxC}{\ensuremath{0.390\pm0.086}}   
\newcommand{\hatcurISOagexxxxxxC}{\ensuremath{12.5_{-6.5}^{+4.9}}} 
\newcommand{\hatcurISOspecxxxxxxC}{M}                            
\newcommand{\hatcurRVKxxxxxxC}{\ensuremath{167.1\pm4.1}}         
\newcommand{\hatcurRVKtwosiglimxxxxxxC}{\ensuremath{<172.9}}     
\newcommand{\hatcurRVrkxxxxxxC}{\ensuremath{0\pm0}}              
\newcommand{\hatcurRVrhxxxxxxC}{\ensuremath{0\pm0}}              
\newcommand{\hatcurRVkxxxxxxC}{\ensuremath{0\pm0}}               
\newcommand{\hatcurRVhxxxxxxC}{\ensuremath{0\pm0}}               
\newcommand{\hatcurRVtronexxxxxxC}{\ensuremath{0\pm0}}           
\newcommand{\hatcurRVtrtwoxxxxxxC}{\ensuremath{0\pm0}}           
\newcommand{\hatcurRVgammaAxxxxxxC}{\ensuremath{6.4\pm7.7}}      
\newcommand{\hatcurRVjitterAxxxxxxC}{\ensuremath{18.1\pm6.7}}    
\newcommand{\hatcurRVjittertwosiglimAxxxxxxC}{\ensuremath{<31.0}} 
\newcommand{\hatcurRVfitrmsAxxxxxxC}{\ensuremath{0.0}}           
\newcommand{\hatcurRVgammaBxxxxxxC}{\ensuremath{-1.2\pm7.8}}     
\newcommand{\hatcurRVjitterBxxxxxxC}{\ensuremath{0\pm12}}        
\newcommand{\hatcurRVjittertwosiglimBxxxxxxC}{\ensuremath{<30.2}} 
\newcommand{\hatcurRVfitrmsBxxxxxxC}{\ensuremath{0.0}}           
\newcommand{\hatcurRVgammaCxxxxxxC}{\ensuremath{-80.2\pm6.4}}    
\newcommand{\hatcurRVjitterCxxxxxxC}{\ensuremath{0.0\pm8.5}}     
\newcommand{\hatcurRVjittertwosiglimCxxxxxxC}{\ensuremath{<10.4}} 
\newcommand{\hatcurRVfitrmsCxxxxxxC}{\ensuremath{0.0}}           
\newcommand{\hatcurRVeccenxxxxxxC}{\ensuremath{0\pm0}}           
\newcommand{\hatcurRVeccentwosiglimxxxxxxC}{\ensuremath{<0.000}} 
\newcommand{\hatcurRVomegaxxxxxxC}{\ensuremath{0\pm0}}           
\newcommand{\hatcurPPixxxxxxC}{\ensuremath{87.830\pm0.098}}      
\newcommand{\hatcurPPgxxxxxxC}{\ensuremath{17.29\pm0.58}}        
\newcommand{\hatcurPPloggxxxxxxC}{\ensuremath{3.238\pm0.015}}    
\newcommand{\hatcurPParxxxxxxC}{\ensuremath{11.391\pm0.078}}     
\newcommand{\hatcurPParelxxxxxxC}{\ensuremath{0.02630\pm0.00046}} 
\newcommand{\hatcurPPrhoxxxxxxC}{\ensuremath{0.871\pm0.039}}     
\newcommand{\hatcurPPmxxxxxxC}{\ensuremath{0.689\pm0.030}}       
\newcommand{\hatcurPPmtwosiglimxxxxxxC}{\ensuremath{<0.74}}      
\newcommand{\hatcurPPmshortxxxxxxC}{\ensuremath{0.69}}           
\newcommand{\hatcurPPmlongxxxxxxC}{\ensuremath{0.689\pm0.030}}   
\newcommand{\hatcurPPmexxxxxxC}{\ensuremath{218.9\pm9.6}}        
\newcommand{\hatcurPPmeshortxxxxxxC}{\ensuremath{218.9}}         
\newcommand{\hatcurPPmelongxxxxxxC}{\ensuremath{218.9\pm9.6}}    
\newcommand{\hatcurPPrxxxxxxC}{\ensuremath{0.994\pm0.020}}       
\newcommand{\hatcurPPrshortxxxxxxC}{\ensuremath{0.99}}           
\newcommand{\hatcurPPrlongxxxxxxC}{\ensuremath{0.994\pm0.020}}   
\newcommand{\hatcurPPrexxxxxxC}{\ensuremath{11.14\pm0.23}}       
\newcommand{\hatcurPPreshortxxxxxxC}{\ensuremath{11.1}}          
\newcommand{\hatcurPPrelongxxxxxxC}{\ensuremath{11.14\pm0.23}}   
\newcommand{\hatcurPPmrcorrxxxxxxC}{\ensuremath{0.69}}           
\newcommand{\hatcurPPteffxxxxxxC}{\ensuremath{764\pm17}}         
\newcommand{\hatcurPPthetaxxxxxxC}{\ensuremath{0.0694\pm0.0023}} 
\newcommand{\hatcurPPfluxperixxxxxxC}{\ensuremath{7.73\pm0.71}}  
\newcommand{\hatcurPPfluxperidimxxxxxxC}{\ensuremath{7}}         
\newcommand{\hatcurPPfluxapxxxxxxC}{\ensuremath{7.73\pm0.71}}    
\newcommand{\hatcurPPfluxapdimxxxxxxC}{\ensuremath{7}}           
\newcommand{\hatcurPPfluxavgxxxxxxC}{\ensuremath{7.73\pm0.71}}   
\newcommand{\hatcurPPfluxavgdimxxxxxxC}{\ensuremath{7}}          
\newcommand{\hatcurPPfluxavglogxxxxxxC}{\ensuremath{7.888\pm0.040}} 
\newcommand{\hatcurXsecphasexxxxxxC}{\ensuremath{0\pm0}}         
\newcommand{\hatcurXsecondaryxxxxxxC}{\ensuremath{2459688.442660\pm0.000075}} 
\newcommand{\hatcurXsecdurxxxxxxC}{\ensuremath{0.06776\pm0.00029}} 
\newcommand{\hatcurXsecingdurxxxxxxC}{\ensuremath{0.01382\pm0.00028}} 
\newcommand{\hatcurPPphiconjxxxxxxC}{\ensuremath{0\pm0}}         
\newcommand{\hatcurPPperixxxxxxC}{\ensuremath{2459686.826525\pm0.000075}} 
\newcommand{\hatcurPPaequivxxxxxxC}{\ensuremath{0.1327\pm0.0061}} 
\newcommand{\hatcurPPtcircxxxxxxC}{\ensuremath{39.1\pm2.5}}      
\newcommand{\hatcurPPtinfallxxxxxxC}{\ensuremath{2990\pm130}}    
\newcommand{\hatcurXdistxxxxxxC}{\ensuremath{113.10_{-0.20}^{+0.30}}} 
\newcommand{\hatcurXAvxxxxxxC}{\ensuremath{0.046\pm0.015}}       
\newcommand{\hatcurXdistredxxxxxxC}{\ensuremath{113.12\pm0.30}}  
\newcommand{\hatcurXEBVxxxxxxC}{\ensuremath{0.0150\pm0.0048}}    
\newcommand{\hatcurCCpmraxxxxxxC}{\ensuremath{19.826\pm0.025}}   
\newcommand{\hatcurCCpmdecxxxxxxC}{\ensuremath{-70.762\pm0.020}} 
\newcommand{\hatcurCCpmxxxxxxC}{\ensuremath{73.487\pm0.032}}     

%% file: TOI4201.tex
\newcommand{\hatcurhtrxxxxxxD}{TOI4201}                                       
\newcommand{\hatcurfieldxxxxxxD}{\ensuremath{string}}                         
\newcommand{\hatcurCCraxxxxxxD}{\ensuremath{06^{\mathrm h}01^{\mathrm m}53.9212{\mathrm s}}}                                
\newcommand{\hatcurCCdecxxxxxxD}{\ensuremath{-13{\arcdeg}27{\arcmin}41.0292{\arcsec}}}                              
\newcommand{\hatcurCCmagxxxxxxD}{15.297}                                      
\newcommand{\hatcurCCtwomassxxxxxxD}{2MASS~06015391-1327410}                  
\newcommand{\hatcurCCgscxxxxxxD}{GSC~NULL}                                    
\newcommand{\hatcurCCgaiaxxxxxxD}{GAIA~2997312059307263616}                   
\newcommand{\hatcurCCgaiadrtwoxxxxxxD}{GAIA~DR2~2997312063605005056}          
\newcommand{\hatcurCCtassmvxxxxxxD}{\ensuremath{15.297\pm0.024}}              
\newcommand{\hatcurCCtassmvshortxxxxxxD}{\ensuremath{15.3}}                   
\newcommand{\hatcurCCtassmBxxxxxxD}{\ensuremath{16.70\pm0.14}}                
\newcommand{\hatcurCCtassmBshortxxxxxxD}{\ensuremath{16.7}}                   
\newcommand{\hatcurCCtassmIxxxxxxD}{\ensuremath{nff\pmnff}}                   
\newcommand{\hatcurCCtassmIshortxxxxxxD}{\ensuremath{0.0}}                    
\newcommand{\hatcurCCtassmgxxxxxxD}{\ensuremath{16.001\pm0.055}}              
\newcommand{\hatcurCCtassmgshortxxxxxxD}{\ensuremath{16.0}}                   
\newcommand{\hatcurCCtassmrxxxxxxD}{\ensuremath{14.655\pm0.064}}              
\newcommand{\hatcurCCtassmrshortxxxxxxD}{\ensuremath{14.7}}                   
\newcommand{\hatcurCCtassmixxxxxxD}{\ensuremath{13.911\pm0.084}}              
\newcommand{\hatcurCCtassmishortxxxxxxD}{\ensuremath{13.9}}                   
\newcommand{\hatcurCCparallaxxxxxxxD}{\ensuremath{5.291\pm0.019}}             
\newcommand{\hatcurCCgaiamGxxxxxxD}{\ensuremath{14.4805\pm0.0028}}            
\newcommand{\hatcurCCgaiamBPxxxxxxD}{\ensuremath{15.4724\pm0.0034}}           
\newcommand{\hatcurCCgaiamRPxxxxxxD}{\ensuremath{13.4948\pm0.0039}}           
\newcommand{\hatcurCCtwomassJmagxxxxxxD}{\ensuremath{12.258\pm0.021}}         
\newcommand{\hatcurCCtwomassHmagxxxxxxD}{\ensuremath{11.564\pm0.024}}         
\newcommand{\hatcurCCtwomassKmagxxxxxxD}{\ensuremath{11.368\pm0.025}}         
\newcommand{\hatcurCCcitJmagxxxxxxD}{\ensuremath{12.247\pm0.023}}             
\newcommand{\hatcurCCcitHmagxxxxxxD}{\ensuremath{11.556\pm0.025}}             
\newcommand{\hatcurCCcitKmagxxxxxxD}{\ensuremath{11.392\pm0.026}}             
\newcommand{\hatcurCCbbJmagxxxxxxD}{\ensuremath{12.339\pm0.025}}              
\newcommand{\hatcurCCbbHmagxxxxxxD}{\ensuremath{11.581\pm0.026}}              
\newcommand{\hatcurCCbbKmagxxxxxxD}{\ensuremath{11.412\pm0.026}}              
\newcommand{\hatcurCCesoJmagxxxxxxD}{\ensuremath{12.348\pm0.030}}             
\newcommand{\hatcurCCesoHmagxxxxxxD}{\ensuremath{11.579\pm0.038}}             
\newcommand{\hatcurCCesoKmagxxxxxxD}{\ensuremath{11.408\pm0.027}}             
\newcommand{\hatcurCCesoJHmagxxxxxxD}{\ensuremath{0.769\pm0.045}}             
\newcommand{\hatcurCCesoJKmagxxxxxxD}{\ensuremath{0.940\pm0.038}}             
\newcommand{\hatcurCCesoHKmagxxxxxxD}{\ensuremath{0.171\pm0.045}}             
\newcommand{\hatcurCCWonemagxxxxxxD}{\ensuremath{11.272\pm0.024}}             
\newcommand{\hatcurCCWtwomagxxxxxxD}{\ensuremath{11.301\pm0.021}}             
\newcommand{\hatcurCCWthreemagxxxxxxD}{\ensuremath{11.28\pm0.15}}             
\newcommand{\hatcurCCWfourmagxxxxxxD}{\ensuremath{nff\pmnff}}                 
\newcommand{\hatcurLCdipxxxxxxD}{\ensuremath{20.0}}                           
\newcommand{\hatcurLCrprstarxxxxxxD}{\ensuremath{0.1890\pm0.0014}}            
\newcommand{\hatcurLCbsqxxxxxxD}{\ensuremath{0.169_{-0.018}^{+0.016}}}        
\newcommand{\hatcurLCimpxxxxxxD}{\ensuremath{0.410_{-0.023}^{+0.020}}}        
\newcommand{\hatcurLCzetaxxxxxxD}{\ensuremath{26.33\pm0.18}}                  
\newcommand{\hatcurLCdurxxxxxxD}{\ensuremath{0.09296\pm0.00040}}              
\newcommand{\hatcurLCdurshortxxxxxxD}{\ensuremath{0.0930}}                    
\newcommand{\hatcurLCdurhrxxxxxxD}{\ensuremath{2.2311\pm0.0097}}              
\newcommand{\hatcurLCdurhrshortxxxxxxD}{\ensuremath{2.231}}                   
\newcommand{\hatcurLCqxxxxxxD}{\ensuremath{0.02600\pm0.00012}}                
\newcommand{\hatcurLCqshortxxxxxxD}{\ensuremath{0.026}}                       
\newcommand{\hatcurLCingdurxxxxxxD}{\ensuremath{0.01735\pm0.00038}}           
\newcommand{\hatcurLCPxxxxxxD}{\ensuremath{3.5819111\pm0.0000017}}            
\newcommand{\hatcurLCPprecxxxxxxD}{\ensuremath{3.5819111}}                    
\newcommand{\hatcurLCPshortxxxxxxD}{\ensuremath{3.5819}}                      
\newcommand{\hatcurLCTxxxxxxD}{\ensuremath{2459860.74636\pm0.00011}}          
\newcommand{\hatcurLCTAxxxxxxD}{\ensuremath{2458470.96486\pm0.00059}}         
\newcommand{\hatcurLCTBxxxxxxD}{\ensuremath{2460007.60471\pm0.00016}}         
\newcommand{\hatcurLChatnetmxxxxxxD}{\ensuremath{13.57618\pm0.00028}}         
\newcommand{\hatcurLCiblendxxxxxxD}{\ensuremath{0.979\pm0.024}}               
\newcommand{\hatcurLCrhoxxxxxxD}{\ensuremath{3.763\pm0.078}}                  
\newcommand{\hatcurSMEiteffxxxxxxD}{\ensuremath{3937\pm70}}                   
\newcommand{\hatcurSMEizfehxxxxxxD}{\ensuremath{0.390\pm0.090}}               
\newcommand{\hatcurSMEizfehshortxxxxxxD}{\ensuremath{0.39}}                   
\newcommand{\hatcurSMEiloggxxxxxxD}{\ensuremath{4.60\pm0.50}}                 
\newcommand{\hatcurSMEivsinxxxxxxD}{\ensuremath{5.4\pm1.0}}                   
\newcommand{\hatcurSMEivmacxxxxxxD}{\ensuremath{nff\pmnff}}                   
\newcommand{\hatcurSMEivmicxxxxxxD}{\ensuremath{nff\pmnff}}                   
\newcommand{\hatcurextraerrMJxxxxxxD}{\ensuremath{0\pm0}}                     
\newcommand{\hatcurextraerrMJtwosiglimxxxxxxD}{\ensuremath{<0.0200}}          
\newcommand{\hatcurextraerrMHxxxxxxD}{\ensuremath{0\pm0}}                     
\newcommand{\hatcurextraerrMHtwosiglimxxxxxxD}{\ensuremath{<0.0200}}          
\newcommand{\hatcurextraerrMKsxxxxxxD}{\ensuremath{0\pm0}}                    
\newcommand{\hatcurextraerrMKstwosiglimxxxxxxD}{\ensuremath{<0.0200}}         
\newcommand{\hatcurextraerrMGxxxxxxD}{\ensuremath{0\pm0}}                     
\newcommand{\hatcurextraerrMGtwosiglimxxxxxxD}{\ensuremath{<0.0200}}          
\newcommand{\hatcurextraerrMRPxxxxxxD}{\ensuremath{0\pm0}}                    
\newcommand{\hatcurextraerrMRPtwosiglimxxxxxxD}{\ensuremath{<0.0200}}         
\newcommand{\hatcurextraerrMgxxxxxxD}{\ensuremath{0\pm0}}                     
\newcommand{\hatcurextraerrMgtwosiglimxxxxxxD}{\ensuremath{<0.0200}}          
\newcommand{\hatcurextraerrMrxxxxxxD}{\ensuremath{0\pm0}}                     
\newcommand{\hatcurextraerrMrtwosiglimxxxxxxD}{\ensuremath{<0.0200}}          
\newcommand{\hatcurextraerrMixxxxxxD}{\ensuremath{0\pm0}}                     
\newcommand{\hatcurextraerrMitwosiglimxxxxxxD}{\ensuremath{<0.0200}}          
\newcommand{\hatcurextraerrMWonexxxxxxD}{\ensuremath{0\pm0}}                  
\newcommand{\hatcurextraerrMWonetwosiglimxxxxxxD}{\ensuremath{<0.0200}}       
\newcommand{\hatcurextraerrMWtwoxxxxxxD}{\ensuremath{0\pm0}}                  
\newcommand{\hatcurextraerrMWtwotwosiglimxxxxxxD}{\ensuremath{<0.0200}}       
\newcommand{\hatcurLBiBxxxxxxD}{\ensuremath{0.5515}}                          
\newcommand{\hatcurLBiiBxxxxxxD}{\ensuremath{0.2609}}                         
\newcommand{\hatcurLBiVxxxxxxD}{\ensuremath{0.4683}}                          
\newcommand{\hatcurLBiiVxxxxxxD}{\ensuremath{0.3143}}                         
\newcommand{\hatcurLBiRxxxxxxD}{\ensuremath{0.4169}}                          
\newcommand{\hatcurLBiiRxxxxxxD}{\ensuremath{0.3015}}                         
\newcommand{\hatcurLBiIxxxxxxD}{\ensuremath{0.2483}}                          
\newcommand{\hatcurLBiiIxxxxxxD}{\ensuremath{0.3232}}                         
\newcommand{\hatcurLBiuxxxxxxD}{\ensuremath{0.5807}}                          
\newcommand{\hatcurLBiiuxxxxxxD}{\ensuremath{0.2281}}                         
\newcommand{\hatcurLBigxxxxxxD}{\ensuremath{0.42\pm0.11}}                     
\newcommand{\hatcurLBiigxxxxxxD}{\ensuremath{0.35\pm0.15}}                    
\newcommand{\hatcurLBirxxxxxxD}{\ensuremath{0.533\pm0.091}}                   
\newcommand{\hatcurLBiirxxxxxxD}{\ensuremath{0.19\pm0.14}}                    
\newcommand{\hatcurLBiixxxxxxD}{\ensuremath{0.36\pm0.13}}                     
\newcommand{\hatcurLBiiixxxxxxD}{\ensuremath{0.31\pm0.17}}                    
\newcommand{\hatcurLBizxxxxxxD}{\ensuremath{0.129\pm0.073}}                   
\newcommand{\hatcurLBiizxxxxxxD}{\ensuremath{0.42\pm0.13}}                    
\newcommand{\hatcurLBiJxxxxxxD}{\ensuremath{0.179\pm0.091}}                   
\newcommand{\hatcurLBiiJxxxxxxD}{\ensuremath{0.17\pm0.14}}                    
\newcommand{\hatcurLBiHxxxxxxD}{\ensuremath{0.1161}}                          
\newcommand{\hatcurLBiiHxxxxxxD}{\ensuremath{0.2593}}                         
\newcommand{\hatcurLBiKxxxxxxD}{\ensuremath{0.0883}}                          
\newcommand{\hatcurLBiiKxxxxxxD}{\ensuremath{0.2403}}                         
\newcommand{\hatcurLBiTxxxxxxD}{\ensuremath{0.39\pm0.16}}                     
\newcommand{\hatcurLBiiTxxxxxxD}{\ensuremath{0.34\pm0.17}}                    
\newcommand{\hatcurLBikepxxxxxxD}{\ensuremath{0.3544}}                        
\newcommand{\hatcurLBiikepxxxxxxD}{\ensuremath{0.3567}}                       
\newcommand{\hatcurLBiCxxxxxxD}{\ensuremath{0.3188}}                          
\newcommand{\hatcurLBiiCxxxxxxD}{\ensuremath{0.3589}}                         
\newcommand{\hatcurLBiMxxxxxxD}{\ensuremath{0.4340}}                          
\newcommand{\hatcurLBiiMxxxxxxD}{\ensuremath{0.3442}}                         
\newcommand{\hatcurLBiSonexxxxxxD}{\ensuremath{0.0669}}                         
\newcommand{\hatcurLBiiSonexxxxxxD}{\ensuremath{0.1859}}                        
\newcommand{\hatcurLBiStwoxxxxxxD}{\ensuremath{0.0568}}                         
\newcommand{\hatcurLBiiStwoxxxxxxD}{\ensuremath{0.1468}}                        
\newcommand{\hatcurLBiSthreexxxxxxD}{\ensuremath{0.0550}}                         
\newcommand{\hatcurLBiiSthreexxxxxxD}{\ensuremath{0.1281}}                        
\newcommand{\hatcurLBiSfourxxxxxxD}{\ensuremath{0.0630}}                         
\newcommand{\hatcurLBiiSfourxxxxxxD}{\ensuremath{0.1031}}                        
\newcommand{\hatcurISOmxxxxxxD}{\ensuremath{0.625\pm0.033}}                   
\newcommand{\hatcurISOmshortxxxxxxD}{\ensuremath{0.63}}                       
\newcommand{\hatcurISOmlongxxxxxxD}{\ensuremath{0.625\pm0.033}}               
\newcommand{\hatcurISOrxxxxxxD}{\ensuremath{0.616\pm0.012}}                   
\newcommand{\hatcurISOrshortxxxxxxD}{\ensuremath{0.62}}                       
\newcommand{\hatcurISOrlongxxxxxxD}{\ensuremath{0.616\pm0.012}}               
\newcommand{\hatcurISOrhoxxxxxxD}{\ensuremath{3.763\pm0.078}}                 
\newcommand{\hatcurISOrholongxxxxxxD}{\ensuremath{3.763\pm0.078}}             
\newcommand{\hatcurISOloggxxxxxxD}{\ensuremath{4.6543\pm0.0099}}              
\newcommand{\hatcurISOlumxxxxxxD}{\ensuremath{0.0816\pm0.0063}}               
\newcommand{\hatcurISOlumshortxxxxxxD}{\ensuremath{0.08}}                     
\newcommand{\hatcurISOteffxxxxxxD}{\ensuremath{3932\pm65}}                    
\newcommand{\hatcurISOzfehxxxxxxD}{\ensuremath{0.432\pm0.085}}                
\newcommand{\hatcurISOagexxxxxxD}{\ensuremath{17.1_{-4.2}^{+2.0}}}            
\newcommand{\hatcurISOspecxxxxxxD}{M}                                         
\newcommand{\hatcurRVKxxxxxxD}{\ensuremath{466\pm21}}                         
\newcommand{\hatcurRVKtwosiglimxxxxxxD}{\ensuremath{<502.2}}                  
\newcommand{\hatcurRVrkxxxxxxD}{\ensuremath{0\pm0}}                           
\newcommand{\hatcurRVrhxxxxxxD}{\ensuremath{0\pm0}}                           
\newcommand{\hatcurRVkxxxxxxD}{\ensuremath{0\pm0}}                            
\newcommand{\hatcurRVhxxxxxxD}{\ensuremath{0\pm0}}                            
\newcommand{\hatcurRVtronexxxxxxD}{\ensuremath{0\pm0}}                        
\newcommand{\hatcurRVtrtwoxxxxxxD}{\ensuremath{0\pm0}}                        
\newcommand{\hatcurRVgammaxxxxxxD}{\ensuremath{74\pm14}}                      
\newcommand{\hatcurRVjitterxxxxxxD}{\ensuremath{43\pm12}}                     
\newcommand{\hatcurRVjittertwosiglimxxxxxxD}{\ensuremath{<68.5}}              
\newcommand{\hatcurRVfitrmsxxxxxxD}{\ensuremath{.1fym}}                       %
\newcommand{\hatcurRVeccenxxxxxxD}{\ensuremath{0\pm0}}                        
\newcommand{\hatcurRVeccentwosiglimxxxxxxD}{\ensuremath{<0.000}}              
\newcommand{\hatcurRVomegaxxxxxxD}{\ensuremath{0\pm0}}                        
\newcommand{\hatcurPPixxxxxxD}{\ensuremath{88.280\pm0.099}}                   
\newcommand{\hatcurPPgxxxxxxD}{\ensuremath{49.7\pm2.5}}                       
\newcommand{\hatcurPPloggxxxxxxD}{\ensuremath{3.696\pm0.022}}                 
\newcommand{\hatcurPParxxxxxxD}{\ensuremath{13.684\pm0.094}}                  
\newcommand{\hatcurPParelxxxxxxD}{\ensuremath{0.03923\pm0.00070}}             
\newcommand{\hatcurPPrhoxxxxxxD}{\ensuremath{2.19\pm0.13}}                    
\newcommand{\hatcurPPmxxxxxxD}{\ensuremath{2.57\pm0.15}}                      
\newcommand{\hatcurPPmtwosiglimxxxxxxD}{\ensuremath{<2.83}}                   
\newcommand{\hatcurPPmshortxxxxxxD}{\ensuremath{2.57}}                        
\newcommand{\hatcurPPmlongxxxxxxD}{\ensuremath{2.57\pm0.15}}                  
\newcommand{\hatcurPPmexxxxxxD}{\ensuremath{818\pm47}}                        
\newcommand{\hatcurPPmeshortxxxxxxD}{\ensuremath{818.2}}                      
\newcommand{\hatcurPPmelongxxxxxxD}{\ensuremath{818\pm47}}                    
\newcommand{\hatcurPPrxxxxxxD}{\ensuremath{1.133\pm0.024}}                    
\newcommand{\hatcurPPrshortxxxxxxD}{\ensuremath{1.13}}                        
\newcommand{\hatcurPPrlongxxxxxxD}{\ensuremath{1.133\pm0.024}}                
\newcommand{\hatcurPPrexxxxxxD}{\ensuremath{12.70\pm0.27}}                    
\newcommand{\hatcurPPreshortxxxxxxD}{\ensuremath{12.7}}                       
\newcommand{\hatcurPPrelongxxxxxxD}{\ensuremath{12.70\pm0.27}}                
\newcommand{\hatcurPPmrcorrxxxxxxD}{\ensuremath{0.51}}                        
\newcommand{\hatcurPPteffxxxxxxD}{\ensuremath{751\pm13}}                      
\newcommand{\hatcurPPthetaxxxxxxD}{\ensuremath{0.284\pm0.014}}                
\newcommand{\hatcurPPfluxperixxxxxxD}{\ensuremath{7.22\pm0.49}}               
\newcommand{\hatcurPPfluxperidimxxxxxxD}{\ensuremath{7}}                      
\newcommand{\hatcurPPfluxapxxxxxxD}{\ensuremath{7.22\pm0.49}}                 
\newcommand{\hatcurPPfluxapdimxxxxxxD}{\ensuremath{7}}                        
\newcommand{\hatcurPPfluxavgxxxxxxD}{\ensuremath{7.22\pm0.49}}                
\newcommand{\hatcurPPfluxavgdimxxxxxxD}{\ensuremath{7}}                       
\newcommand{\hatcurPPfluxavglogxxxxxxD}{\ensuremath{7.859\pm0.029}}           
\newcommand{\hatcurXsecphasexxxxxxD}{\ensuremath{0\pm0}}                      
\newcommand{\hatcurXsecondaryxxxxxxD}{\ensuremath{2459862.53731\pm0.00011}}   
\newcommand{\hatcurXsecdurxxxxxxD}{\ensuremath{0.09296\pm0.00040}}            
\newcommand{\hatcurXsecingdurxxxxxxD}{\ensuremath{0.01735\pm0.00038}}         
\newcommand{\hatcurPPphiconjxxxxxxD}{\ensuremath{0\pm0}}                      
\newcommand{\hatcurPPperixxxxxxD}{\ensuremath{2459859.85088\pm0.00011}}       
\newcommand{\hatcurPPaequivxxxxxxD}{\ensuremath{0.1373\pm0.0046}}             
\newcommand{\hatcurPPtcircxxxxxxD}{\ensuremath{777\pm60}}                     
\newcommand{\hatcurPPtinfallxxxxxxD}{\ensuremath{3970\pm240}}                 
\newcommand{\hatcurXdistxxxxxxD}{\ensuremath{189.20\pm0.65}}                  
\newcommand{\hatcurXAvxxxxxxD}{\ensuremath{0.053\pm0.018}}                    
\newcommand{\hatcurXdistredxxxxxxD}{\ensuremath{189.19\pm0.65}}               
\newcommand{\hatcurXEBVxxxxxxD}{\ensuremath{0.0170\pm0.0059}}                 
\newcommand{\hatcurCCpmraxxxxxxD}{\ensuremath{11.731\pm0.017}}                
\newcommand{\hatcurCCpmdecxxxxxxD}{\ensuremath{6.053\pm0.018}}                
\newcommand{\hatcurCCpmxxxxxxD}{\ensuremath{13.201\pm0.025}}                  

%% file: TOI5344.tex
\newcommand{\hatcurhtrxxxxxxE}{TOI5344}                          
\newcommand{\hatcurfieldxxxxxxE}{\ensuremath{string}}            
\newcommand{\hatcurCCraxxxxxxE}{\ensuremath{04^{\mathrm h}13^{\mathrm m}03.8474{\mathrm s}}}                   
\newcommand{\hatcurCCdecxxxxxxE}{\ensuremath{+20{\arcdeg}54{\arcmin}54.9086{\arcsec}}}                 
\newcommand{\hatcurCCmagxxxxxxE}{15.275}                         
\newcommand{\hatcurCCtwomassxxxxxxE}{2MASS~04130384+2054550}     
\newcommand{\hatcurCCgscxxxxxxE}{GSC~NULL}                       
\newcommand{\hatcurCCgaiaxxxxxxE}{GAIA~52359533989321472}        
\newcommand{\hatcurCCgaiadrtwoxxxxxxE}{GAIA~DR2~52359538285081728} 
\newcommand{\hatcurCCtassmvxxxxxxE}{\ensuremath{15.275\pm0.069}} 
\newcommand{\hatcurCCtassmvshortxxxxxxE}{\ensuremath{15.3}}      
\newcommand{\hatcurCCtassmBxxxxxxE}{\ensuremath{16.880\pm0.069}} 
\newcommand{\hatcurCCtassmBshortxxxxxxE}{\ensuremath{16.9}}      
\newcommand{\hatcurCCtassmIxxxxxxE}{\ensuremath{nff\pmnff}}      
\newcommand{\hatcurCCtassmIshortxxxxxxE}{\ensuremath{0.0}}       
\newcommand{\hatcurCCtassmgxxxxxxE}{\ensuremath{16.106\pm0.055}} 
\newcommand{\hatcurCCtassmgshortxxxxxxE}{\ensuremath{16.1}}      
\newcommand{\hatcurCCtassmrxxxxxxE}{\ensuremath{14.622\pm0.043}} 
\newcommand{\hatcurCCtassmrshortxxxxxxE}{\ensuremath{14.6}}      
\newcommand{\hatcurCCtassmixxxxxxE}{\ensuremath{13.603\pm0.062}} 
\newcommand{\hatcurCCtassmishortxxxxxxE}{\ensuremath{13.6}}      
\newcommand{\hatcurCCparallaxxxxxxxE}{\ensuremath{7.305\pm0.023}} 
\newcommand{\hatcurCCgaiamGxxxxxxE}{\ensuremath{14.3186\pm0.0028}} 
\newcommand{\hatcurCCgaiamBPxxxxxxE}{\ensuremath{15.5037\pm0.0037}} 
\newcommand{\hatcurCCgaiamRPxxxxxxE}{\ensuremath{13.2419\pm0.0039}} 
\newcommand{\hatcurCCtwomassJmagxxxxxxE}{\ensuremath{11.799\pm0.021}} 
\newcommand{\hatcurCCtwomassHmagxxxxxxE}{\ensuremath{11.087\pm0.022}} 
\newcommand{\hatcurCCtwomassKmagxxxxxxE}{\ensuremath{10.860\pm0.018}} 
\newcommand{\hatcurCCcitJmagxxxxxxE}{\ensuremath{11.786\pm0.023}} 
\newcommand{\hatcurCCcitHmagxxxxxxE}{\ensuremath{11.078\pm0.024}} 
\newcommand{\hatcurCCcitKmagxxxxxxE}{\ensuremath{10.884\pm0.019}} 
\newcommand{\hatcurCCbbJmagxxxxxxE}{\ensuremath{11.881\pm0.025}} 
\newcommand{\hatcurCCbbHmagxxxxxxE}{\ensuremath{11.104\pm0.025}} 
\newcommand{\hatcurCCbbKmagxxxxxxE}{\ensuremath{10.904\pm0.019}} 
\newcommand{\hatcurCCesoJmagxxxxxxE}{\ensuremath{11.891\pm0.031}} 
\newcommand{\hatcurCCesoHmagxxxxxxE}{\ensuremath{11.103\pm0.040}} 
\newcommand{\hatcurCCesoKmagxxxxxxE}{\ensuremath{10.900\pm0.021}} 
\newcommand{\hatcurCCesoJHmagxxxxxxE}{\ensuremath{0.788\pm0.046}} 
\newcommand{\hatcurCCesoJKmagxxxxxxE}{\ensuremath{0.991\pm0.034}} 
\newcommand{\hatcurCCesoHKmagxxxxxxE}{\ensuremath{0.203\pm0.043}} 
\newcommand{\hatcurCCWonemagxxxxxxE}{\ensuremath{10.739\pm0.024}} 
\newcommand{\hatcurCCWtwomagxxxxxxE}{\ensuremath{10.728\pm0.020}} 
\newcommand{\hatcurCCWthreemagxxxxxxE}{\ensuremath{10.55\pm0.11}} 
\newcommand{\hatcurCCWfourmagxxxxxxE}{\ensuremath{nff\pmnff}}    
\newcommand{\hatcurLCdipxxxxxxE}{\ensuremath{40.0}}              
\newcommand{\hatcurLCrprstarxxxxxxE}{\ensuremath{0.1653\pm0.0014}} 
\newcommand{\hatcurLCbsqxxxxxxE}{\ensuremath{0.540_{-0.017}^{+0.014}}} 
\newcommand{\hatcurLCimpxxxxxxE}{\ensuremath{0.7348_{-0.0113}^{+0.0094}}} 
\newcommand{\hatcurLCzetaxxxxxxE}{\ensuremath{36.10\pm0.42}}     
\newcommand{\hatcurLCdurxxxxxxE}{\ensuremath{0.07388\pm0.00062}} 
\newcommand{\hatcurLCdurshortxxxxxxE}{\ensuremath{0.0739}}       
\newcommand{\hatcurLCdurhrxxxxxxE}{\ensuremath{1.773\pm0.015}}   
\newcommand{\hatcurLCdurhrshortxxxxxxE}{\ensuremath{1.773}}      
\newcommand{\hatcurLCqxxxxxxE}{\ensuremath{0.01950\pm0.00017}}   
\newcommand{\hatcurLCqshortxxxxxxE}{\ensuremath{0.019}}          
\newcommand{\hatcurLCingdurxxxxxxE}{\ensuremath{0.02077\pm0.00072}} 
\newcommand{\hatcurLCPxxxxxxE}{\ensuremath{3.7926220\pm0.0000062}} 
\newcommand{\hatcurLCPprecxxxxxxE}{\ensuremath{3.7926220}}       
\newcommand{\hatcurLCPshortxxxxxxE}{\ensuremath{3.7926}}         
\newcommand{\hatcurLCTxxxxxxE}{\ensuremath{2459848.99030\pm0.00019}} 
\newcommand{\hatcurLCTAxxxxxxE}{\ensuremath{2459477.31335\pm0.00071}} 
\newcommand{\hatcurLCTBxxxxxxE}{\ensuremath{2459951.39109\pm0.00018}} 
\newcommand{\hatcurLChatnetmAxxxxxxE}{\ensuremath{-0.00025\pm0.00014}} 
\newcommand{\hatcurLCiblendAxxxxxxE}{\ensuremath{0.903\pm0.047}} 
\newcommand{\hatcurLChatnetmBxxxxxxE}{\ensuremath{-0.00029\pm0.00012}} 
\newcommand{\hatcurLCiblendBxxxxxxE}{\ensuremath{0.822\pm0.044}} 
\newcommand{\hatcurLCrhoxxxxxxE}{\ensuremath{4.24\pm0.14}}       
\newcommand{\hatcurSMEiteffxxxxxxE}{\ensuremath{3766\pm70}}      
\newcommand{\hatcurSMEizfehxxxxxxE}{\ensuremath{0.390\pm0.090}}  
\newcommand{\hatcurSMEizfehshortxxxxxxE}{\ensuremath{0.39}}      
\newcommand{\hatcurSMEiloggxxxxxxE}{\ensuremath{4.67\pm0.50}}    
\newcommand{\hatcurSMEivsinxxxxxxE}{\ensuremath{5.1\pm1.0}}      
\newcommand{\hatcurSMEivmacxxxxxxE}{\ensuremath{nff\pmnff}}      
\newcommand{\hatcurSMEivmicxxxxxxE}{\ensuremath{nff\pmnff}}      
\newcommand{\hatcurextraerrMJxxxxxxE}{\ensuremath{0\pm0}}        
\newcommand{\hatcurextraerrMJtwosiglimxxxxxxE}{\ensuremath{<0.0200}} 
\newcommand{\hatcurextraerrMHxxxxxxE}{\ensuremath{0\pm0}}        
\newcommand{\hatcurextraerrMHtwosiglimxxxxxxE}{\ensuremath{<0.0200}} 
\newcommand{\hatcurextraerrMKsxxxxxxE}{\ensuremath{0\pm0}}       
\newcommand{\hatcurextraerrMKstwosiglimxxxxxxE}{\ensuremath{<0.0200}} 
\newcommand{\hatcurextraerrMGxxxxxxE}{\ensuremath{0\pm0}}        
\newcommand{\hatcurextraerrMGtwosiglimxxxxxxE}{\ensuremath{<0.0200}} 
\newcommand{\hatcurextraerrMRPxxxxxxE}{\ensuremath{0\pm0}}       
\newcommand{\hatcurextraerrMRPtwosiglimxxxxxxE}{\ensuremath{<0.0200}} 
\newcommand{\hatcurextraerrMgxxxxxxE}{\ensuremath{0\pm0}}        
\newcommand{\hatcurextraerrMgtwosiglimxxxxxxE}{\ensuremath{<0.0200}} 
\newcommand{\hatcurextraerrMrxxxxxxE}{\ensuremath{0\pm0}}        
\newcommand{\hatcurextraerrMrtwosiglimxxxxxxE}{\ensuremath{<0.0200}} 
\newcommand{\hatcurextraerrMixxxxxxE}{\ensuremath{0\pm0}}        
\newcommand{\hatcurextraerrMitwosiglimxxxxxxE}{\ensuremath{<0.0200}} 
\newcommand{\hatcurextraerrMWonexxxxxxE}{\ensuremath{0\pm0}}     
\newcommand{\hatcurextraerrMWonetwosiglimxxxxxxE}{\ensuremath{<0.0200}} 
\newcommand{\hatcurextraerrMWtwoxxxxxxE}{\ensuremath{0\pm0}}     
\newcommand{\hatcurextraerrMWtwotwosiglimxxxxxxE}{\ensuremath{<0.0200}} 
\newcommand{\hatcurLBiBxxxxxxE}{\ensuremath{0.4527}}             
\newcommand{\hatcurLBiiBxxxxxxE}{\ensuremath{0.3417}}            
\newcommand{\hatcurLBiVxxxxxxE}{\ensuremath{0.4083}}             
\newcommand{\hatcurLBiiVxxxxxxE}{\ensuremath{0.3705}}            
\newcommand{\hatcurLBiRxxxxxxE}{\ensuremath{0.3642}}             
\newcommand{\hatcurLBiiRxxxxxxE}{\ensuremath{0.3397}}            
\newcommand{\hatcurLBiIxxxxxxE}{\ensuremath{0.31\pm0.14}}        
\newcommand{\hatcurLBiiIxxxxxxE}{\ensuremath{0.33\pm0.16}}       
\newcommand{\hatcurLBiuxxxxxxE}{\ensuremath{0.4343}}             
\newcommand{\hatcurLBiiuxxxxxxE}{\ensuremath{0.3245}}            
\newcommand{\hatcurLBigxxxxxxE}{\ensuremath{0.22\pm0.13}}        
\newcommand{\hatcurLBiigxxxxxxE}{\ensuremath{0.14\pm0.16}}       
\newcommand{\hatcurLBirxxxxxxE}{\ensuremath{0.25\pm0.13}}        
\newcommand{\hatcurLBiirxxxxxxE}{\ensuremath{0.29\pm0.15}}       
\newcommand{\hatcurLBiixxxxxxE}{\ensuremath{0.22\pm0.12}}        
\newcommand{\hatcurLBiiixxxxxxE}{\ensuremath{0.13\pm0.15}}       
\newcommand{\hatcurLBizxxxxxxE}{\ensuremath{0.18\pm0.12}}        
\newcommand{\hatcurLBiizxxxxxxE}{\ensuremath{0.21\pm0.15}}       
\newcommand{\hatcurLBiJxxxxxxE}{\ensuremath{0.18\pm0.12}}        
\newcommand{\hatcurLBiiJxxxxxxE}{\ensuremath{0.02\pm0.14}}       
\newcommand{\hatcurLBiHxxxxxxE}{\ensuremath{0.0809}}             
\newcommand{\hatcurLBiiHxxxxxxE}{\ensuremath{0.2652}}            
\newcommand{\hatcurLBiKxxxxxxE}{\ensuremath{0.0716}}             
\newcommand{\hatcurLBiiKxxxxxxE}{\ensuremath{0.2389}}            
\newcommand{\hatcurLBiTxxxxxxE}{\ensuremath{0.21\pm0.13}}        
\newcommand{\hatcurLBiiTxxxxxxE}{\ensuremath{0.17\pm0.17}}       
\newcommand{\hatcurLBikepxxxxxxE}{\ensuremath{0.2833}}           
\newcommand{\hatcurLBiikepxxxxxxE}{\ensuremath{0.4135}}          
\newcommand{\hatcurLBiCxxxxxxE}{\ensuremath{0.2423}}             
\newcommand{\hatcurLBiiCxxxxxxE}{\ensuremath{0.4143}}            
\newcommand{\hatcurLBiMxxxxxxE}{\ensuremath{0.3683}}             
\newcommand{\hatcurLBiiMxxxxxxE}{\ensuremath{0.4059}}            
\newcommand{\hatcurLBiSonexxxxxxE}{\ensuremath{0.0614}}            
\newcommand{\hatcurLBiiSonexxxxxxE}{\ensuremath{0.1756}}           
\newcommand{\hatcurLBiStwoxxxxxxE}{\ensuremath{0.0463}}            
\newcommand{\hatcurLBiiStwoxxxxxxE}{\ensuremath{0.1424}}           
\newcommand{\hatcurLBiSthreexxxxxxE}{\ensuremath{0.0506}}            
\newcommand{\hatcurLBiiSthreexxxxxxE}{\ensuremath{0.1245}}           
\newcommand{\hatcurLBiSfourxxxxxxE}{\ensuremath{0.0577}}            
\newcommand{\hatcurLBiiSfourxxxxxxE}{\ensuremath{0.0933}}           
\newcommand{\hatcurISOmxxxxxxE}{\ensuremath{0.612\pm0.034}}      
\newcommand{\hatcurISOmshortxxxxxxE}{\ensuremath{0.61}}          
\newcommand{\hatcurISOmlongxxxxxxE}{\ensuremath{0.612\pm0.034}}  
\newcommand{\hatcurISOrxxxxxxE}{\ensuremath{0.588\pm0.011}}      
\newcommand{\hatcurISOrshortxxxxxxE}{\ensuremath{0.59}}          
\newcommand{\hatcurISOrlongxxxxxxE}{\ensuremath{0.588\pm0.011}}  
\newcommand{\hatcurISOrhoxxxxxxE}{\ensuremath{4.24\pm0.14}}      
\newcommand{\hatcurISOrholongxxxxxxE}{\ensuremath{4.24\pm0.14}}  
\newcommand{\hatcurISOloggxxxxxxE}{\ensuremath{4.686\pm0.014}}   
\newcommand{\hatcurISOlumxxxxxxE}{\ensuremath{0.0613\pm0.0049}}  
\newcommand{\hatcurISOlumshortxxxxxxE}{\ensuremath{0.06}}        
\newcommand{\hatcurISOteffxxxxxxE}{\ensuremath{3747\pm64}}       
\newcommand{\hatcurISOzfehxxxxxxE}{\ensuremath{0.425\pm0.088}}   
\newcommand{\hatcurISOagexxxxxxE}{\ensuremath{10.9\pm5.8}}       
\newcommand{\hatcurISOspecxxxxxxE}{M}                            
\newcommand{\hatcurRVKxxxxxxE}{\ensuremath{74.4\pm6.8}}          
\newcommand{\hatcurRVKtwosiglimxxxxxxE}{\ensuremath{<85.7}}      
\newcommand{\hatcurRVrkxxxxxxE}{\ensuremath{0\pm0}}              
\newcommand{\hatcurRVrhxxxxxxE}{\ensuremath{0\pm0}}              
\newcommand{\hatcurRVkxxxxxxE}{\ensuremath{0\pm0}}               
\newcommand{\hatcurRVhxxxxxxE}{\ensuremath{0\pm0}}               
\newcommand{\hatcurRVtronexxxxxxE}{\ensuremath{0\pm0}}           
\newcommand{\hatcurRVtrtwoxxxxxxE}{\ensuremath{0\pm0}}           
\newcommand{\hatcurRVgammaxxxxxxE}{\ensuremath{-9.7\pm5.0}}      
\newcommand{\hatcurRVjitterxxxxxxE}{\ensuremath{15.2\pm4.6}}     
\newcommand{\hatcurRVjittertwosiglimxxxxxxE}{\ensuremath{<24.2}} 
\newcommand{\hatcurRVfitrmsxxxxxxE}{\ensuremath{.1fym}}          %
\newcommand{\hatcurRVeccenxxxxxxE}{\ensuremath{0\pm0}}           
\newcommand{\hatcurRVeccentwosiglimxxxxxxE}{\ensuremath{<0.000}} 
\newcommand{\hatcurRVomegaxxxxxxE}{\ensuremath{0\pm0}}           
\newcommand{\hatcurPPixxxxxxE}{\ensuremath{87.150\pm0.067}}      
\newcommand{\hatcurPPgxxxxxxE}{\ensuremath{11.4\pm1.1}}          
\newcommand{\hatcurPPloggxxxxxxE}{\ensuremath{3.058\pm0.042}}    
\newcommand{\hatcurPParxxxxxxE}{\ensuremath{14.78\pm0.16}}       
\newcommand{\hatcurPParelxxxxxxE}{\ensuremath{0.04041\pm0.00075}} 
\newcommand{\hatcurPPrhoxxxxxxE}{\ensuremath{0.604\pm0.063}}     
\newcommand{\hatcurPPmxxxxxxE}{\ensuremath{0.412\pm0.040}}       
\newcommand{\hatcurPPmtwosiglimxxxxxxE}{\ensuremath{<0.48}}      
\newcommand{\hatcurPPmshortxxxxxxE}{\ensuremath{0.41}}           
\newcommand{\hatcurPPmlongxxxxxxE}{\ensuremath{0.412\pm0.040}}   
\newcommand{\hatcurPPmexxxxxxE}{\ensuremath{131\pm13}}           
\newcommand{\hatcurPPmeshortxxxxxxE}{\ensuremath{130.9}}         
\newcommand{\hatcurPPmelongxxxxxxE}{\ensuremath{131\pm13}}       
\newcommand{\hatcurPPrxxxxxxE}{\ensuremath{0.946\pm0.021}}       
\newcommand{\hatcurPPrshortxxxxxxE}{\ensuremath{0.95}}           
\newcommand{\hatcurPPrlongxxxxxxE}{\ensuremath{0.946\pm0.021}}   
\newcommand{\hatcurPPrexxxxxxE}{\ensuremath{10.60\pm0.23}}       
\newcommand{\hatcurPPreshortxxxxxxE}{\ensuremath{10.6}}          
\newcommand{\hatcurPPrelongxxxxxxE}{\ensuremath{10.60\pm0.23}}   
\newcommand{\hatcurPPmrcorrxxxxxxE}{\ensuremath{0.25}}           
\newcommand{\hatcurPPteffxxxxxxE}{\ensuremath{689\pm12}}         
\newcommand{\hatcurPPthetaxxxxxxE}{\ensuremath{0.0574\pm0.0054}} 
\newcommand{\hatcurPPfluxperixxxxxxE}{\ensuremath{5.11\pm0.36}}  
\newcommand{\hatcurPPfluxperidimxxxxxxE}{\ensuremath{7}}         
\newcommand{\hatcurPPfluxapxxxxxxE}{\ensuremath{5.11\pm0.36}}    
\newcommand{\hatcurPPfluxapdimxxxxxxE}{\ensuremath{7}}           
\newcommand{\hatcurPPfluxavgxxxxxxE}{\ensuremath{5.11\pm0.36}}   
\newcommand{\hatcurPPfluxavgdimxxxxxxE}{\ensuremath{7}}          
\newcommand{\hatcurPPfluxavglogxxxxxxE}{\ensuremath{7.709\pm0.031}} 
\newcommand{\hatcurXsecphasexxxxxxE}{\ensuremath{0\pm0}}         
\newcommand{\hatcurXsecondaryxxxxxxE}{\ensuremath{2459850.88661\pm0.00019}} 
\newcommand{\hatcurXsecdurxxxxxxE}{\ensuremath{0.07388\pm0.00062}} 
\newcommand{\hatcurXsecingdurxxxxxxE}{\ensuremath{0.02077\pm0.00072}} 
\newcommand{\hatcurPPphiconjxxxxxxE}{\ensuremath{0\pm0}}         
\newcommand{\hatcurPPperixxxxxxE}{\ensuremath{2459848.04214\pm0.00019}} 
\newcommand{\hatcurPPaequivxxxxxxE}{\ensuremath{0.1632\pm0.0058}} 
\newcommand{\hatcurPPtcircxxxxxxE}{\ensuremath{385\pm47}}        
\newcommand{\hatcurPPtinfallxxxxxxE}{\ensuremath{37900\pm4400}}  
\newcommand{\hatcurXdistxxxxxxE}{\ensuremath{136.90\pm0.44}}     
\newcommand{\hatcurXAvxxxxxxE}{\ensuremath{0.536\pm0.068}}       
\newcommand{\hatcurXdistredxxxxxxE}{\ensuremath{136.88\pm0.44}}  
\newcommand{\hatcurXEBVxxxxxxE}{\ensuremath{0.173\pm0.022}}      
\newcommand{\hatcurCCpmraxxxxxxE}{\ensuremath{40.325\pm0.028}}   
\newcommand{\hatcurCCpmdecxxxxxxE}{\ensuremath{-22.194\pm0.020}} 
\newcommand{\hatcurCCpmxxxxxxE}{\ensuremath{46.029\pm0.034}}     

%% file: planetinfo_multi_eccen.tex
\input{TOI519.eccen.tex}
\input{TOI3629.eccen.tex}
\input{TOI3714.eccen.tex}
\input{TOI4201.eccen.tex}
\input{TOI5344.eccen.tex}
\newcommand{\hatcurCCbbHmageccen}[1]{\ifnum#1=519 %
\hatcurCCbbHmageccenxxxxxxA
\else
\ifnum#1=3629 %
\hatcurCCbbHmageccenxxxxxxB
\else
\ifnum#1=3714 %
\hatcurCCbbHmageccenxxxxxxC
\else
\ifnum#1=4201 %
\hatcurCCbbHmageccenxxxxxxD
\else
\ifnum#1=5344 %
\hatcurCCbbHmageccenxxxxxxE
\else
??????\fi
\fi
\fi
\fi
\fi
}
\newcommand{\hatcurCCbbJmageccen}[1]{\ifnum#1=519 %
\hatcurCCbbJmageccenxxxxxxA
\else
\ifnum#1=3629 %
\hatcurCCbbJmageccenxxxxxxB
\else
\ifnum#1=3714 %
\hatcurCCbbJmageccenxxxxxxC
\else
\ifnum#1=4201 %
\hatcurCCbbJmageccenxxxxxxD
\else
\ifnum#1=5344 %
\hatcurCCbbJmageccenxxxxxxE
\else
??????\fi
\fi
\fi
\fi
\fi
}
\newcommand{\hatcurCCbbKmageccen}[1]{\ifnum#1=519 %
\hatcurCCbbKmageccenxxxxxxA
\else
\ifnum#1=3629 %
\hatcurCCbbKmageccenxxxxxxB
\else
\ifnum#1=3714 %
\hatcurCCbbKmageccenxxxxxxC
\else
\ifnum#1=4201 %
\hatcurCCbbKmageccenxxxxxxD
\else
\ifnum#1=5344 %
\hatcurCCbbKmageccenxxxxxxE
\else
??????\fi
\fi
\fi
\fi
\fi
}
\newcommand{\hatcurCCcitHmageccen}[1]{\ifnum#1=519 %
\hatcurCCcitHmageccenxxxxxxA
\else
\ifnum#1=3629 %
\hatcurCCcitHmageccenxxxxxxB
\else
\ifnum#1=3714 %
\hatcurCCcitHmageccenxxxxxxC
\else
\ifnum#1=4201 %
\hatcurCCcitHmageccenxxxxxxD
\else
\ifnum#1=5344 %
\hatcurCCcitHmageccenxxxxxxE
\else
??????\fi
\fi
\fi
\fi
\fi
}
\newcommand{\hatcurCCcitJmageccen}[1]{\ifnum#1=519 %
\hatcurCCcitJmageccenxxxxxxA
\else
\ifnum#1=3629 %
\hatcurCCcitJmageccenxxxxxxB
\else
\ifnum#1=3714 %
\hatcurCCcitJmageccenxxxxxxC
\else
\ifnum#1=4201 %
\hatcurCCcitJmageccenxxxxxxD
\else
\ifnum#1=5344 %
\hatcurCCcitJmageccenxxxxxxE
\else
??????\fi
\fi
\fi
\fi
\fi
}
\newcommand{\hatcurCCcitKmageccen}[1]{\ifnum#1=519 %
\hatcurCCcitKmageccenxxxxxxA
\else
\ifnum#1=3629 %
\hatcurCCcitKmageccenxxxxxxB
\else
\ifnum#1=3714 %
\hatcurCCcitKmageccenxxxxxxC
\else
\ifnum#1=4201 %
\hatcurCCcitKmageccenxxxxxxD
\else
\ifnum#1=5344 %
\hatcurCCcitKmageccenxxxxxxE
\else
??????\fi
\fi
\fi
\fi
\fi
}
\newcommand{\hatcurCCdececcen}[1]{\ifnum#1=519 %
\hatcurCCdececcenxxxxxxA
\else
\ifnum#1=3629 %
\hatcurCCdececcenxxxxxxB
\else
\ifnum#1=3714 %
\hatcurCCdececcenxxxxxxC
\else
\ifnum#1=4201 %
\hatcurCCdececcenxxxxxxD
\else
\ifnum#1=5344 %
\hatcurCCdececcenxxxxxxE
\else
??????\fi
\fi
\fi
\fi
\fi
}
\newcommand{\hatcurCCesoHKmageccen}[1]{\ifnum#1=519 %
\hatcurCCesoHKmageccenxxxxxxA
\else
\ifnum#1=3629 %
\hatcurCCesoHKmageccenxxxxxxB
\else
\ifnum#1=3714 %
\hatcurCCesoHKmageccenxxxxxxC
\else
\ifnum#1=4201 %
\hatcurCCesoHKmageccenxxxxxxD
\else
\ifnum#1=5344 %
\hatcurCCesoHKmageccenxxxxxxE
\else
??????\fi
\fi
\fi
\fi
\fi
}
\newcommand{\hatcurCCesoHmageccen}[1]{\ifnum#1=519 %
\hatcurCCesoHmageccenxxxxxxA
\else
\ifnum#1=3629 %
\hatcurCCesoHmageccenxxxxxxB
\else
\ifnum#1=3714 %
\hatcurCCesoHmageccenxxxxxxC
\else
\ifnum#1=4201 %
\hatcurCCesoHmageccenxxxxxxD
\else
\ifnum#1=5344 %
\hatcurCCesoHmageccenxxxxxxE
\else
??????\fi
\fi
\fi
\fi
\fi
}
\newcommand{\hatcurCCesoJHmageccen}[1]{\ifnum#1=519 %
\hatcurCCesoJHmageccenxxxxxxA
\else
\ifnum#1=3629 %
\hatcurCCesoJHmageccenxxxxxxB
\else
\ifnum#1=3714 %
\hatcurCCesoJHmageccenxxxxxxC
\else
\ifnum#1=4201 %
\hatcurCCesoJHmageccenxxxxxxD
\else
\ifnum#1=5344 %
\hatcurCCesoJHmageccenxxxxxxE
\else
??????\fi
\fi
\fi
\fi
\fi
}
\newcommand{\hatcurCCesoJKmageccen}[1]{\ifnum#1=519 %
\hatcurCCesoJKmageccenxxxxxxA
\else
\ifnum#1=3629 %
\hatcurCCesoJKmageccenxxxxxxB
\else
\ifnum#1=3714 %
\hatcurCCesoJKmageccenxxxxxxC
\else
\ifnum#1=4201 %
\hatcurCCesoJKmageccenxxxxxxD
\else
\ifnum#1=5344 %
\hatcurCCesoJKmageccenxxxxxxE
\else
??????\fi
\fi
\fi
\fi
\fi
}
\newcommand{\hatcurCCesoJmageccen}[1]{\ifnum#1=519 %
\hatcurCCesoJmageccenxxxxxxA
\else
\ifnum#1=3629 %
\hatcurCCesoJmageccenxxxxxxB
\else
\ifnum#1=3714 %
\hatcurCCesoJmageccenxxxxxxC
\else
\ifnum#1=4201 %
\hatcurCCesoJmageccenxxxxxxD
\else
\ifnum#1=5344 %
\hatcurCCesoJmageccenxxxxxxE
\else
??????\fi
\fi
\fi
\fi
\fi
}
\newcommand{\hatcurCCesoKmageccen}[1]{\ifnum#1=519 %
\hatcurCCesoKmageccenxxxxxxA
\else
\ifnum#1=3629 %
\hatcurCCesoKmageccenxxxxxxB
\else
\ifnum#1=3714 %
\hatcurCCesoKmageccenxxxxxxC
\else
\ifnum#1=4201 %
\hatcurCCesoKmageccenxxxxxxD
\else
\ifnum#1=5344 %
\hatcurCCesoKmageccenxxxxxxE
\else
??????\fi
\fi
\fi
\fi
\fi
}
\newcommand{\hatcurCCgaiadrtwoeccen}[1]{\ifnum#1=519 %
\hatcurCCgaiadrtwoeccenxxxxxxA
\else
\ifnum#1=3629 %
\hatcurCCgaiadrtwoeccenxxxxxxB
\else
\ifnum#1=3714 %
\hatcurCCgaiadrtwoeccenxxxxxxC
\else
\ifnum#1=4201 %
\hatcurCCgaiadrtwoeccenxxxxxxD
\else
\ifnum#1=5344 %
\hatcurCCgaiadrtwoeccenxxxxxxE
\else
??????\fi
\fi
\fi
\fi
\fi
}
\newcommand{\hatcurCCgaiaeccen}[1]{\ifnum#1=519 %
\hatcurCCgaiaeccenxxxxxxA
\else
\ifnum#1=3629 %
\hatcurCCgaiaeccenxxxxxxB
\else
\ifnum#1=3714 %
\hatcurCCgaiaeccenxxxxxxC
\else
\ifnum#1=4201 %
\hatcurCCgaiaeccenxxxxxxD
\else
\ifnum#1=5344 %
\hatcurCCgaiaeccenxxxxxxE
\else
??????\fi
\fi
\fi
\fi
\fi
}
\newcommand{\hatcurCCgaiamBPeccen}[1]{\ifnum#1=519 %
\hatcurCCgaiamBPeccenxxxxxxA
\else
\ifnum#1=3629 %
\hatcurCCgaiamBPeccenxxxxxxB
\else
\ifnum#1=3714 %
\hatcurCCgaiamBPeccenxxxxxxC
\else
\ifnum#1=4201 %
\hatcurCCgaiamBPeccenxxxxxxD
\else
\ifnum#1=5344 %
\hatcurCCgaiamBPeccenxxxxxxE
\else
??????\fi
\fi
\fi
\fi
\fi
}
\newcommand{\hatcurCCgaiamGeccen}[1]{\ifnum#1=519 %
\hatcurCCgaiamGeccenxxxxxxA
\else
\ifnum#1=3629 %
\hatcurCCgaiamGeccenxxxxxxB
\else
\ifnum#1=3714 %
\hatcurCCgaiamGeccenxxxxxxC
\else
\ifnum#1=4201 %
\hatcurCCgaiamGeccenxxxxxxD
\else
\ifnum#1=5344 %
\hatcurCCgaiamGeccenxxxxxxE
\else
??????\fi
\fi
\fi
\fi
\fi
}
\newcommand{\hatcurCCgaiamRPeccen}[1]{\ifnum#1=519 %
\hatcurCCgaiamRPeccenxxxxxxA
\else
\ifnum#1=3629 %
\hatcurCCgaiamRPeccenxxxxxxB
\else
\ifnum#1=3714 %
\hatcurCCgaiamRPeccenxxxxxxC
\else
\ifnum#1=4201 %
\hatcurCCgaiamRPeccenxxxxxxD
\else
\ifnum#1=5344 %
\hatcurCCgaiamRPeccenxxxxxxE
\else
??????\fi
\fi
\fi
\fi
\fi
}
\newcommand{\hatcurCCgsceccen}[1]{\ifnum#1=519 %
\hatcurCCgsceccenxxxxxxA
\else
\ifnum#1=3629 %
\hatcurCCgsceccenxxxxxxB
\else
\ifnum#1=3714 %
\hatcurCCgsceccenxxxxxxC
\else
\ifnum#1=4201 %
\hatcurCCgsceccenxxxxxxD
\else
\ifnum#1=5344 %
\hatcurCCgsceccenxxxxxxE
\else
??????\fi
\fi
\fi
\fi
\fi
}
\newcommand{\hatcurCCmageccen}[1]{\ifnum#1=519 %
\hatcurCCmageccenxxxxxxA
\else
\ifnum#1=3629 %
\hatcurCCmageccenxxxxxxB
\else
\ifnum#1=3714 %
\hatcurCCmageccenxxxxxxC
\else
\ifnum#1=4201 %
\hatcurCCmageccenxxxxxxD
\else
\ifnum#1=5344 %
\hatcurCCmageccenxxxxxxE
\else
??????\fi
\fi
\fi
\fi
\fi
}
\newcommand{\hatcurCCparallaxeccen}[1]{\ifnum#1=519 %
\hatcurCCparallaxeccenxxxxxxA
\else
\ifnum#1=3629 %
\hatcurCCparallaxeccenxxxxxxB
\else
\ifnum#1=3714 %
\hatcurCCparallaxeccenxxxxxxC
\else
\ifnum#1=4201 %
\hatcurCCparallaxeccenxxxxxxD
\else
\ifnum#1=5344 %
\hatcurCCparallaxeccenxxxxxxE
\else
??????\fi
\fi
\fi
\fi
\fi
}
\newcommand{\hatcurCCpmdececcen}[1]{\ifnum#1=519 %
\hatcurCCpmdececcenxxxxxxA
\else
\ifnum#1=3629 %
\hatcurCCpmdececcenxxxxxxB
\else
\ifnum#1=3714 %
\hatcurCCpmdececcenxxxxxxC
\else
\ifnum#1=4201 %
\hatcurCCpmdececcenxxxxxxD
\else
\ifnum#1=5344 %
\hatcurCCpmdececcenxxxxxxE
\else
??????\fi
\fi
\fi
\fi
\fi
}
\newcommand{\hatcurCCpmeccen}[1]{\ifnum#1=519 %
\hatcurCCpmeccenxxxxxxA
\else
\ifnum#1=3629 %
\hatcurCCpmeccenxxxxxxB
\else
\ifnum#1=3714 %
\hatcurCCpmeccenxxxxxxC
\else
\ifnum#1=4201 %
\hatcurCCpmeccenxxxxxxD
\else
\ifnum#1=5344 %
\hatcurCCpmeccenxxxxxxE
\else
??????\fi
\fi
\fi
\fi
\fi
}
\newcommand{\hatcurCCpmraeccen}[1]{\ifnum#1=519 %
\hatcurCCpmraeccenxxxxxxA
\else
\ifnum#1=3629 %
\hatcurCCpmraeccenxxxxxxB
\else
\ifnum#1=3714 %
\hatcurCCpmraeccenxxxxxxC
\else
\ifnum#1=4201 %
\hatcurCCpmraeccenxxxxxxD
\else
\ifnum#1=5344 %
\hatcurCCpmraeccenxxxxxxE
\else
??????\fi
\fi
\fi
\fi
\fi
}
\newcommand{\hatcurCCraeccen}[1]{\ifnum#1=519 %
\hatcurCCraeccenxxxxxxA
\else
\ifnum#1=3629 %
\hatcurCCraeccenxxxxxxB
\else
\ifnum#1=3714 %
\hatcurCCraeccenxxxxxxC
\else
\ifnum#1=4201 %
\hatcurCCraeccenxxxxxxD
\else
\ifnum#1=5344 %
\hatcurCCraeccenxxxxxxE
\else
??????\fi
\fi
\fi
\fi
\fi
}
\newcommand{\hatcurCCtassmBeccen}[1]{\ifnum#1=519 %
\hatcurCCtassmBeccenxxxxxxA
\else
\ifnum#1=3629 %
\hatcurCCtassmBeccenxxxxxxB
\else
\ifnum#1=3714 %
\hatcurCCtassmBeccenxxxxxxC
\else
\ifnum#1=4201 %
\hatcurCCtassmBeccenxxxxxxD
\else
\ifnum#1=5344 %
\hatcurCCtassmBeccenxxxxxxE
\else
??????\fi
\fi
\fi
\fi
\fi
}
\newcommand{\hatcurCCtassmBshorteccen}[1]{\ifnum#1=519 %
\hatcurCCtassmBshorteccenxxxxxxA
\else
\ifnum#1=3629 %
\hatcurCCtassmBshorteccenxxxxxxB
\else
\ifnum#1=3714 %
\hatcurCCtassmBshorteccenxxxxxxC
\else
\ifnum#1=4201 %
\hatcurCCtassmBshorteccenxxxxxxD
\else
\ifnum#1=5344 %
\hatcurCCtassmBshorteccenxxxxxxE
\else
??????\fi
\fi
\fi
\fi
\fi
}
\newcommand{\hatcurCCtassmgeccen}[1]{\ifnum#1=519 %
\hatcurCCtassmgeccenxxxxxxA
\else
\ifnum#1=3629 %
\hatcurCCtassmgeccenxxxxxxB
\else
\ifnum#1=3714 %
\hatcurCCtassmgeccenxxxxxxC
\else
\ifnum#1=4201 %
\hatcurCCtassmgeccenxxxxxxD
\else
\ifnum#1=5344 %
\hatcurCCtassmgeccenxxxxxxE
\else
??????\fi
\fi
\fi
\fi
\fi
}
\newcommand{\hatcurCCtassmgshorteccen}[1]{\ifnum#1=519 %
\hatcurCCtassmgshorteccenxxxxxxA
\else
\ifnum#1=3629 %
\hatcurCCtassmgshorteccenxxxxxxB
\else
\ifnum#1=3714 %
\hatcurCCtassmgshorteccenxxxxxxC
\else
\ifnum#1=4201 %
\hatcurCCtassmgshorteccenxxxxxxD
\else
\ifnum#1=5344 %
\hatcurCCtassmgshorteccenxxxxxxE
\else
??????\fi
\fi
\fi
\fi
\fi
}
\newcommand{\hatcurCCtassmieccen}[1]{\ifnum#1=519 %
\hatcurCCtassmieccenxxxxxxA
\else
\ifnum#1=3629 %
\hatcurCCtassmieccenxxxxxxB
\else
\ifnum#1=3714 %
\hatcurCCtassmieccenxxxxxxC
\else
\ifnum#1=4201 %
\hatcurCCtassmieccenxxxxxxD
\else
\ifnum#1=5344 %
\hatcurCCtassmieccenxxxxxxE
\else
??????\fi
\fi
\fi
\fi
\fi
}
\newcommand{\hatcurCCtassmIeccen}[1]{\ifnum#1=519 %
\hatcurCCtassmIeccenxxxxxxA
\else
\ifnum#1=3629 %
\hatcurCCtassmIeccenxxxxxxB
\else
\ifnum#1=3714 %
\hatcurCCtassmIeccenxxxxxxC
\else
\ifnum#1=4201 %
\hatcurCCtassmIeccenxxxxxxD
\else
\ifnum#1=5344 %
\hatcurCCtassmIeccenxxxxxxE
\else
??????\fi
\fi
\fi
\fi
\fi
}
\newcommand{\hatcurCCtassmishorteccen}[1]{\ifnum#1=519 %
\hatcurCCtassmishorteccenxxxxxxA
\else
\ifnum#1=3629 %
\hatcurCCtassmishorteccenxxxxxxB
\else
\ifnum#1=3714 %
\hatcurCCtassmishorteccenxxxxxxC
\else
\ifnum#1=4201 %
\hatcurCCtassmishorteccenxxxxxxD
\else
\ifnum#1=5344 %
\hatcurCCtassmishorteccenxxxxxxE
\else
??????\fi
\fi
\fi
\fi
\fi
}
\newcommand{\hatcurCCtassmIshorteccen}[1]{\ifnum#1=519 %
\hatcurCCtassmIshorteccenxxxxxxA
\else
\ifnum#1=3629 %
\hatcurCCtassmIshorteccenxxxxxxB
\else
\ifnum#1=3714 %
\hatcurCCtassmIshorteccenxxxxxxC
\else
\ifnum#1=4201 %
\hatcurCCtassmIshorteccenxxxxxxD
\else
\ifnum#1=5344 %
\hatcurCCtassmIshorteccenxxxxxxE
\else
??????\fi
\fi
\fi
\fi
\fi
}
\newcommand{\hatcurCCtassmreccen}[1]{\ifnum#1=519 %
\hatcurCCtassmreccenxxxxxxA
\else
\ifnum#1=3629 %
\hatcurCCtassmreccenxxxxxxB
\else
\ifnum#1=3714 %
\hatcurCCtassmreccenxxxxxxC
\else
\ifnum#1=4201 %
\hatcurCCtassmreccenxxxxxxD
\else
\ifnum#1=5344 %
\hatcurCCtassmreccenxxxxxxE
\else
??????\fi
\fi
\fi
\fi
\fi
}
\newcommand{\hatcurCCtassmrshorteccen}[1]{\ifnum#1=519 %
\hatcurCCtassmrshorteccenxxxxxxA
\else
\ifnum#1=3629 %
\hatcurCCtassmrshorteccenxxxxxxB
\else
\ifnum#1=3714 %
\hatcurCCtassmrshorteccenxxxxxxC
\else
\ifnum#1=4201 %
\hatcurCCtassmrshorteccenxxxxxxD
\else
\ifnum#1=5344 %
\hatcurCCtassmrshorteccenxxxxxxE
\else
??????\fi
\fi
\fi
\fi
\fi
}
\newcommand{\hatcurCCtassmveccen}[1]{\ifnum#1=519 %
\hatcurCCtassmveccenxxxxxxA
\else
\ifnum#1=3629 %
\hatcurCCtassmveccenxxxxxxB
\else
\ifnum#1=3714 %
\hatcurCCtassmveccenxxxxxxC
\else
\ifnum#1=4201 %
\hatcurCCtassmveccenxxxxxxD
\else
\ifnum#1=5344 %
\hatcurCCtassmveccenxxxxxxE
\else
??????\fi
\fi
\fi
\fi
\fi
}
\newcommand{\hatcurCCtassmvshorteccen}[1]{\ifnum#1=519 %
\hatcurCCtassmvshorteccenxxxxxxA
\else
\ifnum#1=3629 %
\hatcurCCtassmvshorteccenxxxxxxB
\else
\ifnum#1=3714 %
\hatcurCCtassmvshorteccenxxxxxxC
\else
\ifnum#1=4201 %
\hatcurCCtassmvshorteccenxxxxxxD
\else
\ifnum#1=5344 %
\hatcurCCtassmvshorteccenxxxxxxE
\else
??????\fi
\fi
\fi
\fi
\fi
}
\newcommand{\hatcurCCtwomasseccen}[1]{\ifnum#1=519 %
\hatcurCCtwomasseccenxxxxxxA
\else
\ifnum#1=3629 %
\hatcurCCtwomasseccenxxxxxxB
\else
\ifnum#1=3714 %
\hatcurCCtwomasseccenxxxxxxC
\else
\ifnum#1=4201 %
\hatcurCCtwomasseccenxxxxxxD
\else
\ifnum#1=5344 %
\hatcurCCtwomasseccenxxxxxxE
\else
??????\fi
\fi
\fi
\fi
\fi
}
\newcommand{\hatcurCCtwomassHmageccen}[1]{\ifnum#1=519 %
\hatcurCCtwomassHmageccenxxxxxxA
\else
\ifnum#1=3629 %
\hatcurCCtwomassHmageccenxxxxxxB
\else
\ifnum#1=3714 %
\hatcurCCtwomassHmageccenxxxxxxC
\else
\ifnum#1=4201 %
\hatcurCCtwomassHmageccenxxxxxxD
\else
\ifnum#1=5344 %
\hatcurCCtwomassHmageccenxxxxxxE
\else
??????\fi
\fi
\fi
\fi
\fi
}
\newcommand{\hatcurCCtwomassJmageccen}[1]{\ifnum#1=519 %
\hatcurCCtwomassJmageccenxxxxxxA
\else
\ifnum#1=3629 %
\hatcurCCtwomassJmageccenxxxxxxB
\else
\ifnum#1=3714 %
\hatcurCCtwomassJmageccenxxxxxxC
\else
\ifnum#1=4201 %
\hatcurCCtwomassJmageccenxxxxxxD
\else
\ifnum#1=5344 %
\hatcurCCtwomassJmageccenxxxxxxE
\else
??????\fi
\fi
\fi
\fi
\fi
}
\newcommand{\hatcurCCtwomassKmageccen}[1]{\ifnum#1=519 %
\hatcurCCtwomassKmageccenxxxxxxA
\else
\ifnum#1=3629 %
\hatcurCCtwomassKmageccenxxxxxxB
\else
\ifnum#1=3714 %
\hatcurCCtwomassKmageccenxxxxxxC
\else
\ifnum#1=4201 %
\hatcurCCtwomassKmageccenxxxxxxD
\else
\ifnum#1=5344 %
\hatcurCCtwomassKmageccenxxxxxxE
\else
??????\fi
\fi
\fi
\fi
\fi
}
\newcommand{\hatcurCCWfourmageccen}[1]{\ifnum#1=519 %
\hatcurCCWfourmageccenxxxxxxA
\else
\ifnum#1=3629 %
\hatcurCCWfourmageccenxxxxxxB
\else
\ifnum#1=3714 %
\hatcurCCWfourmageccenxxxxxxC
\else
\ifnum#1=4201 %
\hatcurCCWfourmageccenxxxxxxD
\else
\ifnum#1=5344 %
\hatcurCCWfourmageccenxxxxxxE
\else
??????\fi
\fi
\fi
\fi
\fi
}
\newcommand{\hatcurCCWonemageccen}[1]{\ifnum#1=519 %
\hatcurCCWonemageccenxxxxxxA
\else
\ifnum#1=3629 %
\hatcurCCWonemageccenxxxxxxB
\else
\ifnum#1=3714 %
\hatcurCCWonemageccenxxxxxxC
\else
\ifnum#1=4201 %
\hatcurCCWonemageccenxxxxxxD
\else
\ifnum#1=5344 %
\hatcurCCWonemageccenxxxxxxE
\else
??????\fi
\fi
\fi
\fi
\fi
}
\newcommand{\hatcurCCWthreemageccen}[1]{\ifnum#1=519 %
\hatcurCCWthreemageccenxxxxxxA
\else
\ifnum#1=3629 %
\hatcurCCWthreemageccenxxxxxxB
\else
\ifnum#1=3714 %
\hatcurCCWthreemageccenxxxxxxC
\else
\ifnum#1=4201 %
\hatcurCCWthreemageccenxxxxxxD
\else
\ifnum#1=5344 %
\hatcurCCWthreemageccenxxxxxxE
\else
??????\fi
\fi
\fi
\fi
\fi
}
\newcommand{\hatcurCCWtwomageccen}[1]{\ifnum#1=519 %
\hatcurCCWtwomageccenxxxxxxA
\else
\ifnum#1=3629 %
\hatcurCCWtwomageccenxxxxxxB
\else
\ifnum#1=3714 %
\hatcurCCWtwomageccenxxxxxxC
\else
\ifnum#1=4201 %
\hatcurCCWtwomageccenxxxxxxD
\else
\ifnum#1=5344 %
\hatcurCCWtwomageccenxxxxxxE
\else
??????\fi
\fi
\fi
\fi
\fi
}
\newcommand{\hatcurextraerrMgeccen}[1]{\ifnum#1=3629 %
\hatcurextraerrMgeccenxxxxxxB
\else
\ifnum#1=3714 %
\hatcurextraerrMgeccenxxxxxxC
\else
\ifnum#1=4201 %
\hatcurextraerrMgeccenxxxxxxD
\else
\ifnum#1=5344 %
\hatcurextraerrMgeccenxxxxxxE
\else
??????\fi
\fi
\fi
\fi
}
\newcommand{\hatcurextraerrMGeccen}[1]{\ifnum#1=519 %
\hatcurextraerrMGeccenxxxxxxA
\else
\ifnum#1=3629 %
\hatcurextraerrMGeccenxxxxxxB
\else
\ifnum#1=3714 %
\hatcurextraerrMGeccenxxxxxxC
\else
\ifnum#1=4201 %
\hatcurextraerrMGeccenxxxxxxD
\else
\ifnum#1=5344 %
\hatcurextraerrMGeccenxxxxxxE
\else
??????\fi
\fi
\fi
\fi
\fi
}
\newcommand{\hatcurextraerrMgtwosiglimeccen}[1]{\ifnum#1=3629 %
\hatcurextraerrMgtwosiglimeccenxxxxxxB
\else
\ifnum#1=3714 %
\hatcurextraerrMgtwosiglimeccenxxxxxxC
\else
\ifnum#1=4201 %
\hatcurextraerrMgtwosiglimeccenxxxxxxD
\else
\ifnum#1=5344 %
\hatcurextraerrMgtwosiglimeccenxxxxxxE
\else
??????\fi
\fi
\fi
\fi
}
\newcommand{\hatcurextraerrMGtwosiglimeccen}[1]{\ifnum#1=519 %
\hatcurextraerrMGtwosiglimeccenxxxxxxA
\else
\ifnum#1=3629 %
\hatcurextraerrMGtwosiglimeccenxxxxxxB
\else
\ifnum#1=3714 %
\hatcurextraerrMGtwosiglimeccenxxxxxxC
\else
\ifnum#1=4201 %
\hatcurextraerrMGtwosiglimeccenxxxxxxD
\else
\ifnum#1=5344 %
\hatcurextraerrMGtwosiglimeccenxxxxxxE
\else
??????\fi
\fi
\fi
\fi
\fi
}
\newcommand{\hatcurextraerrMHeccen}[1]{\ifnum#1=519 %
\hatcurextraerrMHeccenxxxxxxA
\else
\ifnum#1=3629 %
\hatcurextraerrMHeccenxxxxxxB
\else
\ifnum#1=3714 %
\hatcurextraerrMHeccenxxxxxxC
\else
\ifnum#1=4201 %
\hatcurextraerrMHeccenxxxxxxD
\else
\ifnum#1=5344 %
\hatcurextraerrMHeccenxxxxxxE
\else
??????\fi
\fi
\fi
\fi
\fi
}
\newcommand{\hatcurextraerrMHtwosiglimeccen}[1]{\ifnum#1=519 %
\hatcurextraerrMHtwosiglimeccenxxxxxxA
\else
\ifnum#1=3629 %
\hatcurextraerrMHtwosiglimeccenxxxxxxB
\else
\ifnum#1=3714 %
\hatcurextraerrMHtwosiglimeccenxxxxxxC
\else
\ifnum#1=4201 %
\hatcurextraerrMHtwosiglimeccenxxxxxxD
\else
\ifnum#1=5344 %
\hatcurextraerrMHtwosiglimeccenxxxxxxE
\else
??????\fi
\fi
\fi
\fi
\fi
}
\newcommand{\hatcurextraerrMieccen}[1]{\ifnum#1=3629 %
\hatcurextraerrMieccenxxxxxxB
\else
\ifnum#1=3714 %
\hatcurextraerrMieccenxxxxxxC
\else
\ifnum#1=4201 %
\hatcurextraerrMieccenxxxxxxD
\else
\ifnum#1=5344 %
\hatcurextraerrMieccenxxxxxxE
\else
??????\fi
\fi
\fi
\fi
}
\newcommand{\hatcurextraerrMitwosiglimeccen}[1]{\ifnum#1=3629 %
\hatcurextraerrMitwosiglimeccenxxxxxxB
\else
\ifnum#1=3714 %
\hatcurextraerrMitwosiglimeccenxxxxxxC
\else
\ifnum#1=4201 %
\hatcurextraerrMitwosiglimeccenxxxxxxD
\else
\ifnum#1=5344 %
\hatcurextraerrMitwosiglimeccenxxxxxxE
\else
??????\fi
\fi
\fi
\fi
}
\newcommand{\hatcurextraerrMJeccen}[1]{\ifnum#1=519 %
\hatcurextraerrMJeccenxxxxxxA
\else
\ifnum#1=3629 %
\hatcurextraerrMJeccenxxxxxxB
\else
\ifnum#1=3714 %
\hatcurextraerrMJeccenxxxxxxC
\else
\ifnum#1=4201 %
\hatcurextraerrMJeccenxxxxxxD
\else
\ifnum#1=5344 %
\hatcurextraerrMJeccenxxxxxxE
\else
??????\fi
\fi
\fi
\fi
\fi
}
\newcommand{\hatcurextraerrMJtwosiglimeccen}[1]{\ifnum#1=519 %
\hatcurextraerrMJtwosiglimeccenxxxxxxA
\else
\ifnum#1=3629 %
\hatcurextraerrMJtwosiglimeccenxxxxxxB
\else
\ifnum#1=3714 %
\hatcurextraerrMJtwosiglimeccenxxxxxxC
\else
\ifnum#1=4201 %
\hatcurextraerrMJtwosiglimeccenxxxxxxD
\else
\ifnum#1=5344 %
\hatcurextraerrMJtwosiglimeccenxxxxxxE
\else
??????\fi
\fi
\fi
\fi
\fi
}
\newcommand{\hatcurextraerrMKseccen}[1]{\ifnum#1=519 %
\hatcurextraerrMKseccenxxxxxxA
\else
\ifnum#1=3629 %
\hatcurextraerrMKseccenxxxxxxB
\else
\ifnum#1=3714 %
\hatcurextraerrMKseccenxxxxxxC
\else
\ifnum#1=4201 %
\hatcurextraerrMKseccenxxxxxxD
\else
\ifnum#1=5344 %
\hatcurextraerrMKseccenxxxxxxE
\else
??????\fi
\fi
\fi
\fi
\fi
}
\newcommand{\hatcurextraerrMKstwosiglimeccen}[1]{\ifnum#1=519 %
\hatcurextraerrMKstwosiglimeccenxxxxxxA
\else
\ifnum#1=3629 %
\hatcurextraerrMKstwosiglimeccenxxxxxxB
\else
\ifnum#1=3714 %
\hatcurextraerrMKstwosiglimeccenxxxxxxC
\else
\ifnum#1=4201 %
\hatcurextraerrMKstwosiglimeccenxxxxxxD
\else
\ifnum#1=5344 %
\hatcurextraerrMKstwosiglimeccenxxxxxxE
\else
??????\fi
\fi
\fi
\fi
\fi
}
\newcommand{\hatcurextraerrMreccen}[1]{\ifnum#1=3629 %
\hatcurextraerrMreccenxxxxxxB
\else
\ifnum#1=3714 %
\hatcurextraerrMreccenxxxxxxC
\else
\ifnum#1=4201 %
\hatcurextraerrMreccenxxxxxxD
\else
\ifnum#1=5344 %
\hatcurextraerrMreccenxxxxxxE
\else
??????\fi
\fi
\fi
\fi
}
\newcommand{\hatcurextraerrMRPeccen}[1]{\ifnum#1=519 %
\hatcurextraerrMRPeccenxxxxxxA
\else
\ifnum#1=3629 %
\hatcurextraerrMRPeccenxxxxxxB
\else
\ifnum#1=3714 %
\hatcurextraerrMRPeccenxxxxxxC
\else
\ifnum#1=4201 %
\hatcurextraerrMRPeccenxxxxxxD
\else
\ifnum#1=5344 %
\hatcurextraerrMRPeccenxxxxxxE
\else
??????\fi
\fi
\fi
\fi
\fi
}
\newcommand{\hatcurextraerrMRPtwosiglimeccen}[1]{\ifnum#1=519 %
\hatcurextraerrMRPtwosiglimeccenxxxxxxA
\else
\ifnum#1=3629 %
\hatcurextraerrMRPtwosiglimeccenxxxxxxB
\else
\ifnum#1=3714 %
\hatcurextraerrMRPtwosiglimeccenxxxxxxC
\else
\ifnum#1=4201 %
\hatcurextraerrMRPtwosiglimeccenxxxxxxD
\else
\ifnum#1=5344 %
\hatcurextraerrMRPtwosiglimeccenxxxxxxE
\else
??????\fi
\fi
\fi
\fi
\fi
}
\newcommand{\hatcurextraerrMrtwosiglimeccen}[1]{\ifnum#1=3629 %
\hatcurextraerrMrtwosiglimeccenxxxxxxB
\else
\ifnum#1=3714 %
\hatcurextraerrMrtwosiglimeccenxxxxxxC
\else
\ifnum#1=4201 %
\hatcurextraerrMrtwosiglimeccenxxxxxxD
\else
\ifnum#1=5344 %
\hatcurextraerrMrtwosiglimeccenxxxxxxE
\else
??????\fi
\fi
\fi
\fi
}
\newcommand{\hatcurextraerrMWoneeccen}[1]{\ifnum#1=519 %
\hatcurextraerrMWoneeccenxxxxxxA
\else
\ifnum#1=3629 %
\hatcurextraerrMWoneeccenxxxxxxB
\else
\ifnum#1=3714 %
\hatcurextraerrMWoneeccenxxxxxxC
\else
\ifnum#1=4201 %
\hatcurextraerrMWoneeccenxxxxxxD
\else
\ifnum#1=5344 %
\hatcurextraerrMWoneeccenxxxxxxE
\else
??????\fi
\fi
\fi
\fi
\fi
}
\newcommand{\hatcurextraerrMWonetwosiglimeccen}[1]{\ifnum#1=519 %
\hatcurextraerrMWonetwosiglimeccenxxxxxxA
\else
\ifnum#1=3629 %
\hatcurextraerrMWonetwosiglimeccenxxxxxxB
\else
\ifnum#1=3714 %
\hatcurextraerrMWonetwosiglimeccenxxxxxxC
\else
\ifnum#1=4201 %
\hatcurextraerrMWonetwosiglimeccenxxxxxxD
\else
\ifnum#1=5344 %
\hatcurextraerrMWonetwosiglimeccenxxxxxxE
\else
??????\fi
\fi
\fi
\fi
\fi
}
\newcommand{\hatcurextraerrMWtwoeccen}[1]{\ifnum#1=519 %
\hatcurextraerrMWtwoeccenxxxxxxA
\else
\ifnum#1=3629 %
\hatcurextraerrMWtwoeccenxxxxxxB
\else
\ifnum#1=3714 %
\hatcurextraerrMWtwoeccenxxxxxxC
\else
\ifnum#1=4201 %
\hatcurextraerrMWtwoeccenxxxxxxD
\else
\ifnum#1=5344 %
\hatcurextraerrMWtwoeccenxxxxxxE
\else
??????\fi
\fi
\fi
\fi
\fi
}
\newcommand{\hatcurextraerrMWtwotwosiglimeccen}[1]{\ifnum#1=519 %
\hatcurextraerrMWtwotwosiglimeccenxxxxxxA
\else
\ifnum#1=3629 %
\hatcurextraerrMWtwotwosiglimeccenxxxxxxB
\else
\ifnum#1=3714 %
\hatcurextraerrMWtwotwosiglimeccenxxxxxxC
\else
\ifnum#1=4201 %
\hatcurextraerrMWtwotwosiglimeccenxxxxxxD
\else
\ifnum#1=5344 %
\hatcurextraerrMWtwotwosiglimeccenxxxxxxE
\else
??????\fi
\fi
\fi
\fi
\fi
}
\newcommand{\hatcurfieldeccen}[1]{\ifnum#1=519 %
\hatcurfieldeccenxxxxxxA
\else
\ifnum#1=3629 %
\hatcurfieldeccenxxxxxxB
\else
\ifnum#1=3714 %
\hatcurfieldeccenxxxxxxC
\else
\ifnum#1=4201 %
\hatcurfieldeccenxxxxxxD
\else
\ifnum#1=5344 %
\hatcurfieldeccenxxxxxxE
\else
??????\fi
\fi
\fi
\fi
\fi
}
\newcommand{\hatcurhtreccen}[1]{\ifnum#1=519 %
\hatcurhtreccenxxxxxxA
\else
\ifnum#1=3629 %
\hatcurhtreccenxxxxxxB
\else
\ifnum#1=3714 %
\hatcurhtreccenxxxxxxC
\else
\ifnum#1=4201 %
\hatcurhtreccenxxxxxxD
\else
\ifnum#1=5344 %
\hatcurhtreccenxxxxxxE
\else
??????\fi
\fi
\fi
\fi
\fi
}
\newcommand{\hatcurISOageeccen}[1]{\ifnum#1=519 %
\hatcurISOageeccenxxxxxxA
\else
\ifnum#1=3629 %
\hatcurISOageeccenxxxxxxB
\else
\ifnum#1=3714 %
\hatcurISOageeccenxxxxxxC
\else
\ifnum#1=4201 %
\hatcurISOageeccenxxxxxxD
\else
\ifnum#1=5344 %
\hatcurISOageeccenxxxxxxE
\else
??????\fi
\fi
\fi
\fi
\fi
}
\newcommand{\hatcurISOloggeccen}[1]{\ifnum#1=519 %
\hatcurISOloggeccenxxxxxxA
\else
\ifnum#1=3629 %
\hatcurISOloggeccenxxxxxxB
\else
\ifnum#1=3714 %
\hatcurISOloggeccenxxxxxxC
\else
\ifnum#1=4201 %
\hatcurISOloggeccenxxxxxxD
\else
\ifnum#1=5344 %
\hatcurISOloggeccenxxxxxxE
\else
??????\fi
\fi
\fi
\fi
\fi
}
\newcommand{\hatcurISOlumeccen}[1]{\ifnum#1=519 %
\hatcurISOlumeccenxxxxxxA
\else
\ifnum#1=3629 %
\hatcurISOlumeccenxxxxxxB
\else
\ifnum#1=3714 %
\hatcurISOlumeccenxxxxxxC
\else
\ifnum#1=4201 %
\hatcurISOlumeccenxxxxxxD
\else
\ifnum#1=5344 %
\hatcurISOlumeccenxxxxxxE
\else
??????\fi
\fi
\fi
\fi
\fi
}
\newcommand{\hatcurISOlumshorteccen}[1]{\ifnum#1=519 %
\hatcurISOlumshorteccenxxxxxxA
\else
\ifnum#1=3629 %
\hatcurISOlumshorteccenxxxxxxB
\else
\ifnum#1=3714 %
\hatcurISOlumshorteccenxxxxxxC
\else
\ifnum#1=4201 %
\hatcurISOlumshorteccenxxxxxxD
\else
\ifnum#1=5344 %
\hatcurISOlumshorteccenxxxxxxE
\else
??????\fi
\fi
\fi
\fi
\fi
}
\newcommand{\hatcurISOmeccen}[1]{\ifnum#1=519 %
\hatcurISOmeccenxxxxxxA
\else
\ifnum#1=3629 %
\hatcurISOmeccenxxxxxxB
\else
\ifnum#1=3714 %
\hatcurISOmeccenxxxxxxC
\else
\ifnum#1=4201 %
\hatcurISOmeccenxxxxxxD
\else
\ifnum#1=5344 %
\hatcurISOmeccenxxxxxxE
\else
??????\fi
\fi
\fi
\fi
\fi
}
\newcommand{\hatcurISOmlongeccen}[1]{\ifnum#1=519 %
\hatcurISOmlongeccenxxxxxxA
\else
\ifnum#1=3629 %
\hatcurISOmlongeccenxxxxxxB
\else
\ifnum#1=3714 %
\hatcurISOmlongeccenxxxxxxC
\else
\ifnum#1=4201 %
\hatcurISOmlongeccenxxxxxxD
\else
\ifnum#1=5344 %
\hatcurISOmlongeccenxxxxxxE
\else
??????\fi
\fi
\fi
\fi
\fi
}
\newcommand{\hatcurISOmshorteccen}[1]{\ifnum#1=519 %
\hatcurISOmshorteccenxxxxxxA
\else
\ifnum#1=3629 %
\hatcurISOmshorteccenxxxxxxB
\else
\ifnum#1=3714 %
\hatcurISOmshorteccenxxxxxxC
\else
\ifnum#1=4201 %
\hatcurISOmshorteccenxxxxxxD
\else
\ifnum#1=5344 %
\hatcurISOmshorteccenxxxxxxE
\else
??????\fi
\fi
\fi
\fi
\fi
}
\newcommand{\hatcurISOreccen}[1]{\ifnum#1=519 %
\hatcurISOreccenxxxxxxA
\else
\ifnum#1=3629 %
\hatcurISOreccenxxxxxxB
\else
\ifnum#1=3714 %
\hatcurISOreccenxxxxxxC
\else
\ifnum#1=4201 %
\hatcurISOreccenxxxxxxD
\else
\ifnum#1=5344 %
\hatcurISOreccenxxxxxxE
\else
??????\fi
\fi
\fi
\fi
\fi
}
\newcommand{\hatcurISOrhoeccen}[1]{\ifnum#1=519 %
\hatcurISOrhoeccenxxxxxxA
\else
\ifnum#1=3629 %
\hatcurISOrhoeccenxxxxxxB
\else
\ifnum#1=3714 %
\hatcurISOrhoeccenxxxxxxC
\else
\ifnum#1=4201 %
\hatcurISOrhoeccenxxxxxxD
\else
\ifnum#1=5344 %
\hatcurISOrhoeccenxxxxxxE
\else
??????\fi
\fi
\fi
\fi
\fi
}
\newcommand{\hatcurISOrholongeccen}[1]{\ifnum#1=519 %
\hatcurISOrholongeccenxxxxxxA
\else
\ifnum#1=3629 %
\hatcurISOrholongeccenxxxxxxB
\else
\ifnum#1=3714 %
\hatcurISOrholongeccenxxxxxxC
\else
\ifnum#1=4201 %
\hatcurISOrholongeccenxxxxxxD
\else
\ifnum#1=5344 %
\hatcurISOrholongeccenxxxxxxE
\else
??????\fi
\fi
\fi
\fi
\fi
}
\newcommand{\hatcurISOrlongeccen}[1]{\ifnum#1=519 %
\hatcurISOrlongeccenxxxxxxA
\else
\ifnum#1=3629 %
\hatcurISOrlongeccenxxxxxxB
\else
\ifnum#1=3714 %
\hatcurISOrlongeccenxxxxxxC
\else
\ifnum#1=4201 %
\hatcurISOrlongeccenxxxxxxD
\else
\ifnum#1=5344 %
\hatcurISOrlongeccenxxxxxxE
\else
??????\fi
\fi
\fi
\fi
\fi
}
\newcommand{\hatcurISOrshorteccen}[1]{\ifnum#1=519 %
\hatcurISOrshorteccenxxxxxxA
\else
\ifnum#1=3629 %
\hatcurISOrshorteccenxxxxxxB
\else
\ifnum#1=3714 %
\hatcurISOrshorteccenxxxxxxC
\else
\ifnum#1=4201 %
\hatcurISOrshorteccenxxxxxxD
\else
\ifnum#1=5344 %
\hatcurISOrshorteccenxxxxxxE
\else
??????\fi
\fi
\fi
\fi
\fi
}
\newcommand{\hatcurISOspececcen}[1]{\ifnum#1=519 %
\hatcurISOspececcenxxxxxxA
\else
\ifnum#1=3629 %
\hatcurISOspececcenxxxxxxB
\else
\ifnum#1=3714 %
\hatcurISOspececcenxxxxxxC
\else
\ifnum#1=4201 %
\hatcurISOspececcenxxxxxxD
\else
\ifnum#1=5344 %
\hatcurISOspececcenxxxxxxE
\else
??????\fi
\fi
\fi
\fi
\fi
}
\newcommand{\hatcurISOteffeccen}[1]{\ifnum#1=519 %
\hatcurISOteffeccenxxxxxxA
\else
\ifnum#1=3629 %
\hatcurISOteffeccenxxxxxxB
\else
\ifnum#1=3714 %
\hatcurISOteffeccenxxxxxxC
\else
\ifnum#1=4201 %
\hatcurISOteffeccenxxxxxxD
\else
\ifnum#1=5344 %
\hatcurISOteffeccenxxxxxxE
\else
??????\fi
\fi
\fi
\fi
\fi
}
\newcommand{\hatcurISOzfeheccen}[1]{\ifnum#1=519 %
\hatcurISOzfeheccenxxxxxxA
\else
\ifnum#1=3629 %
\hatcurISOzfeheccenxxxxxxB
\else
\ifnum#1=3714 %
\hatcurISOzfeheccenxxxxxxC
\else
\ifnum#1=4201 %
\hatcurISOzfeheccenxxxxxxD
\else
\ifnum#1=5344 %
\hatcurISOzfeheccenxxxxxxE
\else
??????\fi
\fi
\fi
\fi
\fi
}
\newcommand{\hatcurLBiBeccen}[1]{\ifnum#1=519 %
\hatcurLBiBeccenxxxxxxA
\else
\ifnum#1=3629 %
\hatcurLBiBeccenxxxxxxB
\else
\ifnum#1=3714 %
\hatcurLBiBeccenxxxxxxC
\else
\ifnum#1=4201 %
\hatcurLBiBeccenxxxxxxD
\else
\ifnum#1=5344 %
\hatcurLBiBeccenxxxxxxE
\else
??????\fi
\fi
\fi
\fi
\fi
}
\newcommand{\hatcurLBiCeccen}[1]{\ifnum#1=519 %
\hatcurLBiCeccenxxxxxxA
\else
\ifnum#1=3629 %
\hatcurLBiCeccenxxxxxxB
\else
\ifnum#1=3714 %
\hatcurLBiCeccenxxxxxxC
\else
\ifnum#1=4201 %
\hatcurLBiCeccenxxxxxxD
\else
\ifnum#1=5344 %
\hatcurLBiCeccenxxxxxxE
\else
??????\fi
\fi
\fi
\fi
\fi
}
\newcommand{\hatcurLBigeccen}[1]{\ifnum#1=519 %
\hatcurLBigeccenxxxxxxA
\else
\ifnum#1=3629 %
\hatcurLBigeccenxxxxxxB
\else
\ifnum#1=3714 %
\hatcurLBigeccenxxxxxxC
\else
\ifnum#1=4201 %
\hatcurLBigeccenxxxxxxD
\else
\ifnum#1=5344 %
\hatcurLBigeccenxxxxxxE
\else
??????\fi
\fi
\fi
\fi
\fi
}
\newcommand{\hatcurLBiHeccen}[1]{\ifnum#1=519 %
\hatcurLBiHeccenxxxxxxA
\else
\ifnum#1=3629 %
\hatcurLBiHeccenxxxxxxB
\else
\ifnum#1=3714 %
\hatcurLBiHeccenxxxxxxC
\else
\ifnum#1=4201 %
\hatcurLBiHeccenxxxxxxD
\else
\ifnum#1=5344 %
\hatcurLBiHeccenxxxxxxE
\else
??????\fi
\fi
\fi
\fi
\fi
}
\newcommand{\hatcurLBiiBeccen}[1]{\ifnum#1=519 %
\hatcurLBiiBeccenxxxxxxA
\else
\ifnum#1=3629 %
\hatcurLBiiBeccenxxxxxxB
\else
\ifnum#1=3714 %
\hatcurLBiiBeccenxxxxxxC
\else
\ifnum#1=4201 %
\hatcurLBiiBeccenxxxxxxD
\else
\ifnum#1=5344 %
\hatcurLBiiBeccenxxxxxxE
\else
??????\fi
\fi
\fi
\fi
\fi
}
\newcommand{\hatcurLBiiCeccen}[1]{\ifnum#1=519 %
\hatcurLBiiCeccenxxxxxxA
\else
\ifnum#1=3629 %
\hatcurLBiiCeccenxxxxxxB
\else
\ifnum#1=3714 %
\hatcurLBiiCeccenxxxxxxC
\else
\ifnum#1=4201 %
\hatcurLBiiCeccenxxxxxxD
\else
\ifnum#1=5344 %
\hatcurLBiiCeccenxxxxxxE
\else
??????\fi
\fi
\fi
\fi
\fi
}
\newcommand{\hatcurLBiieccen}[1]{\ifnum#1=519 %
\hatcurLBiieccenxxxxxxA
\else
\ifnum#1=3629 %
\hatcurLBiieccenxxxxxxB
\else
\ifnum#1=3714 %
\hatcurLBiieccenxxxxxxC
\else
\ifnum#1=4201 %
\hatcurLBiieccenxxxxxxD
\else
\ifnum#1=5344 %
\hatcurLBiieccenxxxxxxE
\else
??????\fi
\fi
\fi
\fi
\fi
}
\newcommand{\hatcurLBiIeccen}[1]{\ifnum#1=519 %
\hatcurLBiIeccenxxxxxxA
\else
\ifnum#1=3629 %
\hatcurLBiIeccenxxxxxxB
\else
\ifnum#1=3714 %
\hatcurLBiIeccenxxxxxxC
\else
\ifnum#1=4201 %
\hatcurLBiIeccenxxxxxxD
\else
\ifnum#1=5344 %
\hatcurLBiIeccenxxxxxxE
\else
??????\fi
\fi
\fi
\fi
\fi
}
\newcommand{\hatcurLBiigeccen}[1]{\ifnum#1=519 %
\hatcurLBiigeccenxxxxxxA
\else
\ifnum#1=3629 %
\hatcurLBiigeccenxxxxxxB
\else
\ifnum#1=3714 %
\hatcurLBiigeccenxxxxxxC
\else
\ifnum#1=4201 %
\hatcurLBiigeccenxxxxxxD
\else
\ifnum#1=5344 %
\hatcurLBiigeccenxxxxxxE
\else
??????\fi
\fi
\fi
\fi
\fi
}
\newcommand{\hatcurLBiiHeccen}[1]{\ifnum#1=519 %
\hatcurLBiiHeccenxxxxxxA
\else
\ifnum#1=3629 %
\hatcurLBiiHeccenxxxxxxB
\else
\ifnum#1=3714 %
\hatcurLBiiHeccenxxxxxxC
\else
\ifnum#1=4201 %
\hatcurLBiiHeccenxxxxxxD
\else
\ifnum#1=5344 %
\hatcurLBiiHeccenxxxxxxE
\else
??????\fi
\fi
\fi
\fi
\fi
}
\newcommand{\hatcurLBiiieccen}[1]{\ifnum#1=519 %
\hatcurLBiiieccenxxxxxxA
\else
\ifnum#1=3629 %
\hatcurLBiiieccenxxxxxxB
\else
\ifnum#1=3714 %
\hatcurLBiiieccenxxxxxxC
\else
\ifnum#1=4201 %
\hatcurLBiiieccenxxxxxxD
\else
\ifnum#1=5344 %
\hatcurLBiiieccenxxxxxxE
\else
??????\fi
\fi
\fi
\fi
\fi
}
\newcommand{\hatcurLBiiIeccen}[1]{\ifnum#1=519 %
\hatcurLBiiIeccenxxxxxxA
\else
\ifnum#1=3629 %
\hatcurLBiiIeccenxxxxxxB
\else
\ifnum#1=3714 %
\hatcurLBiiIeccenxxxxxxC
\else
\ifnum#1=4201 %
\hatcurLBiiIeccenxxxxxxD
\else
\ifnum#1=5344 %
\hatcurLBiiIeccenxxxxxxE
\else
??????\fi
\fi
\fi
\fi
\fi
}
\newcommand{\hatcurLBiiJeccen}[1]{\ifnum#1=519 %
\hatcurLBiiJeccenxxxxxxA
\else
\ifnum#1=3629 %
\hatcurLBiiJeccenxxxxxxB
\else
\ifnum#1=3714 %
\hatcurLBiiJeccenxxxxxxC
\else
\ifnum#1=4201 %
\hatcurLBiiJeccenxxxxxxD
\else
\ifnum#1=5344 %
\hatcurLBiiJeccenxxxxxxE
\else
??????\fi
\fi
\fi
\fi
\fi
}
\newcommand{\hatcurLBiiKeccen}[1]{\ifnum#1=519 %
\hatcurLBiiKeccenxxxxxxA
\else
\ifnum#1=3629 %
\hatcurLBiiKeccenxxxxxxB
\else
\ifnum#1=3714 %
\hatcurLBiiKeccenxxxxxxC
\else
\ifnum#1=4201 %
\hatcurLBiiKeccenxxxxxxD
\else
\ifnum#1=5344 %
\hatcurLBiiKeccenxxxxxxE
\else
??????\fi
\fi
\fi
\fi
\fi
}
\newcommand{\hatcurLBiikepeccen}[1]{\ifnum#1=519 %
\hatcurLBiikepeccenxxxxxxA
\else
\ifnum#1=3629 %
\hatcurLBiikepeccenxxxxxxB
\else
\ifnum#1=3714 %
\hatcurLBiikepeccenxxxxxxC
\else
\ifnum#1=4201 %
\hatcurLBiikepeccenxxxxxxD
\else
\ifnum#1=5344 %
\hatcurLBiikepeccenxxxxxxE
\else
??????\fi
\fi
\fi
\fi
\fi
}
\newcommand{\hatcurLBiiMeccen}[1]{\ifnum#1=519 %
\hatcurLBiiMeccenxxxxxxA
\else
\ifnum#1=3629 %
\hatcurLBiiMeccenxxxxxxB
\else
\ifnum#1=3714 %
\hatcurLBiiMeccenxxxxxxC
\else
\ifnum#1=4201 %
\hatcurLBiiMeccenxxxxxxD
\else
\ifnum#1=5344 %
\hatcurLBiiMeccenxxxxxxE
\else
??????\fi
\fi
\fi
\fi
\fi
}
\newcommand{\hatcurLBiireccen}[1]{\ifnum#1=519 %
\hatcurLBiireccenxxxxxxA
\else
\ifnum#1=3629 %
\hatcurLBiireccenxxxxxxB
\else
\ifnum#1=3714 %
\hatcurLBiireccenxxxxxxC
\else
\ifnum#1=4201 %
\hatcurLBiireccenxxxxxxD
\else
\ifnum#1=5344 %
\hatcurLBiireccenxxxxxxE
\else
??????\fi
\fi
\fi
\fi
\fi
}
\newcommand{\hatcurLBiiReccen}[1]{\ifnum#1=519 %
\hatcurLBiiReccenxxxxxxA
\else
\ifnum#1=3629 %
\hatcurLBiiReccenxxxxxxB
\else
\ifnum#1=3714 %
\hatcurLBiiReccenxxxxxxC
\else
\ifnum#1=4201 %
\hatcurLBiiReccenxxxxxxD
\else
\ifnum#1=5344 %
\hatcurLBiiReccenxxxxxxE
\else
??????\fi
\fi
\fi
\fi
\fi
}
\newcommand{\hatcurLBiiSfoureccen}[1]{\ifnum#1=519 %
\hatcurLBiiSfoureccenxxxxxxA
\else
\ifnum#1=3629 %
\hatcurLBiiSfoureccenxxxxxxB
\else
\ifnum#1=3714 %
\hatcurLBiiSfoureccenxxxxxxC
\else
\ifnum#1=4201 %
\hatcurLBiiSfoureccenxxxxxxD
\else
\ifnum#1=5344 %
\hatcurLBiiSfoureccenxxxxxxE
\else
??????\fi
\fi
\fi
\fi
\fi
}
\newcommand{\hatcurLBiiSoneeccen}[1]{\ifnum#1=519 %
\hatcurLBiiSoneeccenxxxxxxA
\else
\ifnum#1=3629 %
\hatcurLBiiSoneeccenxxxxxxB
\else
\ifnum#1=3714 %
\hatcurLBiiSoneeccenxxxxxxC
\else
\ifnum#1=4201 %
\hatcurLBiiSoneeccenxxxxxxD
\else
\ifnum#1=5344 %
\hatcurLBiiSoneeccenxxxxxxE
\else
??????\fi
\fi
\fi
\fi
\fi
}
\newcommand{\hatcurLBiiSthreeeccen}[1]{\ifnum#1=519 %
\hatcurLBiiSthreeeccenxxxxxxA
\else
\ifnum#1=3629 %
\hatcurLBiiSthreeeccenxxxxxxB
\else
\ifnum#1=3714 %
\hatcurLBiiSthreeeccenxxxxxxC
\else
\ifnum#1=4201 %
\hatcurLBiiSthreeeccenxxxxxxD
\else
\ifnum#1=5344 %
\hatcurLBiiSthreeeccenxxxxxxE
\else
??????\fi
\fi
\fi
\fi
\fi
}
\newcommand{\hatcurLBiiStwoeccen}[1]{\ifnum#1=519 %
\hatcurLBiiStwoeccenxxxxxxA
\else
\ifnum#1=3629 %
\hatcurLBiiStwoeccenxxxxxxB
\else
\ifnum#1=3714 %
\hatcurLBiiStwoeccenxxxxxxC
\else
\ifnum#1=4201 %
\hatcurLBiiStwoeccenxxxxxxD
\else
\ifnum#1=5344 %
\hatcurLBiiStwoeccenxxxxxxE
\else
??????\fi
\fi
\fi
\fi
\fi
}
\newcommand{\hatcurLBiiTeccen}[1]{\ifnum#1=519 %
\hatcurLBiiTeccenxxxxxxA
\else
\ifnum#1=3629 %
\hatcurLBiiTeccenxxxxxxB
\else
\ifnum#1=3714 %
\hatcurLBiiTeccenxxxxxxC
\else
\ifnum#1=4201 %
\hatcurLBiiTeccenxxxxxxD
\else
\ifnum#1=5344 %
\hatcurLBiiTeccenxxxxxxE
\else
??????\fi
\fi
\fi
\fi
\fi
}
\newcommand{\hatcurLBiiueccen}[1]{\ifnum#1=519 %
\hatcurLBiiueccenxxxxxxA
\else
\ifnum#1=3629 %
\hatcurLBiiueccenxxxxxxB
\else
\ifnum#1=3714 %
\hatcurLBiiueccenxxxxxxC
\else
\ifnum#1=4201 %
\hatcurLBiiueccenxxxxxxD
\else
\ifnum#1=5344 %
\hatcurLBiiueccenxxxxxxE
\else
??????\fi
\fi
\fi
\fi
\fi
}
\newcommand{\hatcurLBiiVeccen}[1]{\ifnum#1=519 %
\hatcurLBiiVeccenxxxxxxA
\else
\ifnum#1=3629 %
\hatcurLBiiVeccenxxxxxxB
\else
\ifnum#1=3714 %
\hatcurLBiiVeccenxxxxxxC
\else
\ifnum#1=4201 %
\hatcurLBiiVeccenxxxxxxD
\else
\ifnum#1=5344 %
\hatcurLBiiVeccenxxxxxxE
\else
??????\fi
\fi
\fi
\fi
\fi
}
\newcommand{\hatcurLBiizeccen}[1]{\ifnum#1=519 %
\hatcurLBiizeccenxxxxxxA
\else
\ifnum#1=3629 %
\hatcurLBiizeccenxxxxxxB
\else
\ifnum#1=3714 %
\hatcurLBiizeccenxxxxxxC
\else
\ifnum#1=4201 %
\hatcurLBiizeccenxxxxxxD
\else
\ifnum#1=5344 %
\hatcurLBiizeccenxxxxxxE
\else
??????\fi
\fi
\fi
\fi
\fi
}
\newcommand{\hatcurLBiJeccen}[1]{\ifnum#1=519 %
\hatcurLBiJeccenxxxxxxA
\else
\ifnum#1=3629 %
\hatcurLBiJeccenxxxxxxB
\else
\ifnum#1=3714 %
\hatcurLBiJeccenxxxxxxC
\else
\ifnum#1=4201 %
\hatcurLBiJeccenxxxxxxD
\else
\ifnum#1=5344 %
\hatcurLBiJeccenxxxxxxE
\else
??????\fi
\fi
\fi
\fi
\fi
}
\newcommand{\hatcurLBiKeccen}[1]{\ifnum#1=519 %
\hatcurLBiKeccenxxxxxxA
\else
\ifnum#1=3629 %
\hatcurLBiKeccenxxxxxxB
\else
\ifnum#1=3714 %
\hatcurLBiKeccenxxxxxxC
\else
\ifnum#1=4201 %
\hatcurLBiKeccenxxxxxxD
\else
\ifnum#1=5344 %
\hatcurLBiKeccenxxxxxxE
\else
??????\fi
\fi
\fi
\fi
\fi
}
\newcommand{\hatcurLBikepeccen}[1]{\ifnum#1=519 %
\hatcurLBikepeccenxxxxxxA
\else
\ifnum#1=3629 %
\hatcurLBikepeccenxxxxxxB
\else
\ifnum#1=3714 %
\hatcurLBikepeccenxxxxxxC
\else
\ifnum#1=4201 %
\hatcurLBikepeccenxxxxxxD
\else
\ifnum#1=5344 %
\hatcurLBikepeccenxxxxxxE
\else
??????\fi
\fi
\fi
\fi
\fi
}
\newcommand{\hatcurLBiMeccen}[1]{\ifnum#1=519 %
\hatcurLBiMeccenxxxxxxA
\else
\ifnum#1=3629 %
\hatcurLBiMeccenxxxxxxB
\else
\ifnum#1=3714 %
\hatcurLBiMeccenxxxxxxC
\else
\ifnum#1=4201 %
\hatcurLBiMeccenxxxxxxD
\else
\ifnum#1=5344 %
\hatcurLBiMeccenxxxxxxE
\else
??????\fi
\fi
\fi
\fi
\fi
}
\newcommand{\hatcurLBireccen}[1]{\ifnum#1=519 %
\hatcurLBireccenxxxxxxA
\else
\ifnum#1=3629 %
\hatcurLBireccenxxxxxxB
\else
\ifnum#1=3714 %
\hatcurLBireccenxxxxxxC
\else
\ifnum#1=4201 %
\hatcurLBireccenxxxxxxD
\else
\ifnum#1=5344 %
\hatcurLBireccenxxxxxxE
\else
??????\fi
\fi
\fi
\fi
\fi
}
\newcommand{\hatcurLBiReccen}[1]{\ifnum#1=519 %
\hatcurLBiReccenxxxxxxA
\else
\ifnum#1=3629 %
\hatcurLBiReccenxxxxxxB
\else
\ifnum#1=3714 %
\hatcurLBiReccenxxxxxxC
\else
\ifnum#1=4201 %
\hatcurLBiReccenxxxxxxD
\else
\ifnum#1=5344 %
\hatcurLBiReccenxxxxxxE
\else
??????\fi
\fi
\fi
\fi
\fi
}
\newcommand{\hatcurLBiSfoureccen}[1]{\ifnum#1=519 %
\hatcurLBiSfoureccenxxxxxxA
\else
\ifnum#1=3629 %
\hatcurLBiSfoureccenxxxxxxB
\else
\ifnum#1=3714 %
\hatcurLBiSfoureccenxxxxxxC
\else
\ifnum#1=4201 %
\hatcurLBiSfoureccenxxxxxxD
\else
\ifnum#1=5344 %
\hatcurLBiSfoureccenxxxxxxE
\else
??????\fi
\fi
\fi
\fi
\fi
}
\newcommand{\hatcurLBiSoneeccen}[1]{\ifnum#1=519 %
\hatcurLBiSoneeccenxxxxxxA
\else
\ifnum#1=3629 %
\hatcurLBiSoneeccenxxxxxxB
\else
\ifnum#1=3714 %
\hatcurLBiSoneeccenxxxxxxC
\else
\ifnum#1=4201 %
\hatcurLBiSoneeccenxxxxxxD
\else
\ifnum#1=5344 %
\hatcurLBiSoneeccenxxxxxxE
\else
??????\fi
\fi
\fi
\fi
\fi
}
\newcommand{\hatcurLBiSthreeeccen}[1]{\ifnum#1=519 %
\hatcurLBiSthreeeccenxxxxxxA
\else
\ifnum#1=3629 %
\hatcurLBiSthreeeccenxxxxxxB
\else
\ifnum#1=3714 %
\hatcurLBiSthreeeccenxxxxxxC
\else
\ifnum#1=4201 %
\hatcurLBiSthreeeccenxxxxxxD
\else
\ifnum#1=5344 %
\hatcurLBiSthreeeccenxxxxxxE
\else
??????\fi
\fi
\fi
\fi
\fi
}
\newcommand{\hatcurLBiStwoeccen}[1]{\ifnum#1=519 %
\hatcurLBiStwoeccenxxxxxxA
\else
\ifnum#1=3629 %
\hatcurLBiStwoeccenxxxxxxB
\else
\ifnum#1=3714 %
\hatcurLBiStwoeccenxxxxxxC
\else
\ifnum#1=4201 %
\hatcurLBiStwoeccenxxxxxxD
\else
\ifnum#1=5344 %
\hatcurLBiStwoeccenxxxxxxE
\else
??????\fi
\fi
\fi
\fi
\fi
}
\newcommand{\hatcurLBiTeccen}[1]{\ifnum#1=519 %
\hatcurLBiTeccenxxxxxxA
\else
\ifnum#1=3629 %
\hatcurLBiTeccenxxxxxxB
\else
\ifnum#1=3714 %
\hatcurLBiTeccenxxxxxxC
\else
\ifnum#1=4201 %
\hatcurLBiTeccenxxxxxxD
\else
\ifnum#1=5344 %
\hatcurLBiTeccenxxxxxxE
\else
??????\fi
\fi
\fi
\fi
\fi
}
\newcommand{\hatcurLBiueccen}[1]{\ifnum#1=519 %
\hatcurLBiueccenxxxxxxA
\else
\ifnum#1=3629 %
\hatcurLBiueccenxxxxxxB
\else
\ifnum#1=3714 %
\hatcurLBiueccenxxxxxxC
\else
\ifnum#1=4201 %
\hatcurLBiueccenxxxxxxD
\else
\ifnum#1=5344 %
\hatcurLBiueccenxxxxxxE
\else
??????\fi
\fi
\fi
\fi
\fi
}
\newcommand{\hatcurLBiVeccen}[1]{\ifnum#1=519 %
\hatcurLBiVeccenxxxxxxA
\else
\ifnum#1=3629 %
\hatcurLBiVeccenxxxxxxB
\else
\ifnum#1=3714 %
\hatcurLBiVeccenxxxxxxC
\else
\ifnum#1=4201 %
\hatcurLBiVeccenxxxxxxD
\else
\ifnum#1=5344 %
\hatcurLBiVeccenxxxxxxE
\else
??????\fi
\fi
\fi
\fi
\fi
}
\newcommand{\hatcurLBizeccen}[1]{\ifnum#1=519 %
\hatcurLBizeccenxxxxxxA
\else
\ifnum#1=3629 %
\hatcurLBizeccenxxxxxxB
\else
\ifnum#1=3714 %
\hatcurLBizeccenxxxxxxC
\else
\ifnum#1=4201 %
\hatcurLBizeccenxxxxxxD
\else
\ifnum#1=5344 %
\hatcurLBizeccenxxxxxxE
\else
??????\fi
\fi
\fi
\fi
\fi
}
\newcommand{\hatcurLCbsqeccen}[1]{\ifnum#1=519 %
\hatcurLCbsqeccenxxxxxxA
\else
\ifnum#1=3629 %
\hatcurLCbsqeccenxxxxxxB
\else
\ifnum#1=3714 %
\hatcurLCbsqeccenxxxxxxC
\else
\ifnum#1=4201 %
\hatcurLCbsqeccenxxxxxxD
\else
\ifnum#1=5344 %
\hatcurLCbsqeccenxxxxxxE
\else
??????\fi
\fi
\fi
\fi
\fi
}
\newcommand{\hatcurLCdipeccen}[1]{\ifnum#1=519 %
\hatcurLCdipeccenxxxxxxA
\else
\ifnum#1=3629 %
\hatcurLCdipeccenxxxxxxB
\else
\ifnum#1=3714 %
\hatcurLCdipeccenxxxxxxC
\else
\ifnum#1=4201 %
\hatcurLCdipeccenxxxxxxD
\else
\ifnum#1=5344 %
\hatcurLCdipeccenxxxxxxE
\else
??????\fi
\fi
\fi
\fi
\fi
}
\newcommand{\hatcurLCdureccen}[1]{\ifnum#1=519 %
\hatcurLCdureccenxxxxxxA
\else
\ifnum#1=3629 %
\hatcurLCdureccenxxxxxxB
\else
\ifnum#1=3714 %
\hatcurLCdureccenxxxxxxC
\else
\ifnum#1=4201 %
\hatcurLCdureccenxxxxxxD
\else
\ifnum#1=5344 %
\hatcurLCdureccenxxxxxxE
\else
??????\fi
\fi
\fi
\fi
\fi
}
\newcommand{\hatcurLCdurhreccen}[1]{\ifnum#1=519 %
\hatcurLCdurhreccenxxxxxxA
\else
\ifnum#1=3629 %
\hatcurLCdurhreccenxxxxxxB
\else
\ifnum#1=3714 %
\hatcurLCdurhreccenxxxxxxC
\else
\ifnum#1=4201 %
\hatcurLCdurhreccenxxxxxxD
\else
\ifnum#1=5344 %
\hatcurLCdurhreccenxxxxxxE
\else
??????\fi
\fi
\fi
\fi
\fi
}
\newcommand{\hatcurLCdurhrshorteccen}[1]{\ifnum#1=519 %
\hatcurLCdurhrshorteccenxxxxxxA
\else
\ifnum#1=3629 %
\hatcurLCdurhrshorteccenxxxxxxB
\else
\ifnum#1=3714 %
\hatcurLCdurhrshorteccenxxxxxxC
\else
\ifnum#1=4201 %
\hatcurLCdurhrshorteccenxxxxxxD
\else
\ifnum#1=5344 %
\hatcurLCdurhrshorteccenxxxxxxE
\else
??????\fi
\fi
\fi
\fi
\fi
}
\newcommand{\hatcurLCdurshorteccen}[1]{\ifnum#1=519 %
\hatcurLCdurshorteccenxxxxxxA
\else
\ifnum#1=3629 %
\hatcurLCdurshorteccenxxxxxxB
\else
\ifnum#1=3714 %
\hatcurLCdurshorteccenxxxxxxC
\else
\ifnum#1=4201 %
\hatcurLCdurshorteccenxxxxxxD
\else
\ifnum#1=5344 %
\hatcurLCdurshorteccenxxxxxxE
\else
??????\fi
\fi
\fi
\fi
\fi
}
\newcommand{\hatcurLChatnetmAeccen}[1]{\ifnum#1=519 %
\hatcurLChatnetmAeccenxxxxxxA
\else
\ifnum#1=3629 %
\hatcurLChatnetmAeccenxxxxxxB
\else
\ifnum#1=5344 %
\hatcurLChatnetmAeccenxxxxxxE
\else
??????\fi
\fi
\fi
}
\newcommand{\hatcurLChatnetmBeccen}[1]{\ifnum#1=519 %
\hatcurLChatnetmBeccenxxxxxxA
\else
\ifnum#1=3629 %
\hatcurLChatnetmBeccenxxxxxxB
\else
\ifnum#1=5344 %
\hatcurLChatnetmBeccenxxxxxxE
\else
??????\fi
\fi
\fi
}
\newcommand{\hatcurLChatnetmCeccen}[1]{\ifnum#1=519 %
\hatcurLChatnetmCeccenxxxxxxA
\else
??????\fi
}
\newcommand{\hatcurLChatnetmDeccen}[1]{\ifnum#1=519 %
\hatcurLChatnetmDeccenxxxxxxA
\else
??????\fi
}
\newcommand{\hatcurLChatnetmeccen}[1]{\ifnum#1=3714 %
\hatcurLChatnetmeccenxxxxxxC
\else
\ifnum#1=4201 %
\hatcurLChatnetmeccenxxxxxxD
\else
??????\fi
\fi
}
\newcommand{\hatcurLCiblendAeccen}[1]{\ifnum#1=519 %
\hatcurLCiblendAeccenxxxxxxA
\else
\ifnum#1=3629 %
\hatcurLCiblendAeccenxxxxxxB
\else
\ifnum#1=5344 %
\hatcurLCiblendAeccenxxxxxxE
\else
??????\fi
\fi
\fi
}
\newcommand{\hatcurLCiblendBeccen}[1]{\ifnum#1=519 %
\hatcurLCiblendBeccenxxxxxxA
\else
\ifnum#1=3629 %
\hatcurLCiblendBeccenxxxxxxB
\else
\ifnum#1=5344 %
\hatcurLCiblendBeccenxxxxxxE
\else
??????\fi
\fi
\fi
}
\newcommand{\hatcurLCiblendCeccen}[1]{\ifnum#1=519 %
\hatcurLCiblendCeccenxxxxxxA
\else
??????\fi
}
\newcommand{\hatcurLCiblendDeccen}[1]{\ifnum#1=519 %
\hatcurLCiblendDeccenxxxxxxA
\else
??????\fi
}
\newcommand{\hatcurLCiblendeccen}[1]{\ifnum#1=3714 %
\hatcurLCiblendeccenxxxxxxC
\else
\ifnum#1=4201 %
\hatcurLCiblendeccenxxxxxxD
\else
??????\fi
\fi
}
\newcommand{\hatcurLCimpeccen}[1]{\ifnum#1=519 %
\hatcurLCimpeccenxxxxxxA
\else
\ifnum#1=3629 %
\hatcurLCimpeccenxxxxxxB
\else
\ifnum#1=3714 %
\hatcurLCimpeccenxxxxxxC
\else
\ifnum#1=4201 %
\hatcurLCimpeccenxxxxxxD
\else
\ifnum#1=5344 %
\hatcurLCimpeccenxxxxxxE
\else
??????\fi
\fi
\fi
\fi
\fi
}
\newcommand{\hatcurLCingdureccen}[1]{\ifnum#1=519 %
\hatcurLCingdureccenxxxxxxA
\else
\ifnum#1=3629 %
\hatcurLCingdureccenxxxxxxB
\else
\ifnum#1=3714 %
\hatcurLCingdureccenxxxxxxC
\else
\ifnum#1=4201 %
\hatcurLCingdureccenxxxxxxD
\else
\ifnum#1=5344 %
\hatcurLCingdureccenxxxxxxE
\else
??????\fi
\fi
\fi
\fi
\fi
}
\newcommand{\hatcurLCPeccen}[1]{\ifnum#1=519 %
\hatcurLCPeccenxxxxxxA
\else
\ifnum#1=3629 %
\hatcurLCPeccenxxxxxxB
\else
\ifnum#1=3714 %
\hatcurLCPeccenxxxxxxC
\else
\ifnum#1=4201 %
\hatcurLCPeccenxxxxxxD
\else
\ifnum#1=5344 %
\hatcurLCPeccenxxxxxxE
\else
??????\fi
\fi
\fi
\fi
\fi
}
\newcommand{\hatcurLCPprececcen}[1]{\ifnum#1=519 %
\hatcurLCPprececcenxxxxxxA
\else
\ifnum#1=3629 %
\hatcurLCPprececcenxxxxxxB
\else
\ifnum#1=3714 %
\hatcurLCPprececcenxxxxxxC
\else
\ifnum#1=4201 %
\hatcurLCPprececcenxxxxxxD
\else
\ifnum#1=5344 %
\hatcurLCPprececcenxxxxxxE
\else
??????\fi
\fi
\fi
\fi
\fi
}
\newcommand{\hatcurLCPshorteccen}[1]{\ifnum#1=519 %
\hatcurLCPshorteccenxxxxxxA
\else
\ifnum#1=3629 %
\hatcurLCPshorteccenxxxxxxB
\else
\ifnum#1=3714 %
\hatcurLCPshorteccenxxxxxxC
\else
\ifnum#1=4201 %
\hatcurLCPshorteccenxxxxxxD
\else
\ifnum#1=5344 %
\hatcurLCPshorteccenxxxxxxE
\else
??????\fi
\fi
\fi
\fi
\fi
}
\newcommand{\hatcurLCqeccen}[1]{\ifnum#1=519 %
\hatcurLCqeccenxxxxxxA
\else
\ifnum#1=3629 %
\hatcurLCqeccenxxxxxxB
\else
\ifnum#1=3714 %
\hatcurLCqeccenxxxxxxC
\else
\ifnum#1=4201 %
\hatcurLCqeccenxxxxxxD
\else
\ifnum#1=5344 %
\hatcurLCqeccenxxxxxxE
\else
??????\fi
\fi
\fi
\fi
\fi
}
\newcommand{\hatcurLCqshorteccen}[1]{\ifnum#1=519 %
\hatcurLCqshorteccenxxxxxxA
\else
\ifnum#1=3629 %
\hatcurLCqshorteccenxxxxxxB
\else
\ifnum#1=3714 %
\hatcurLCqshorteccenxxxxxxC
\else
\ifnum#1=4201 %
\hatcurLCqshorteccenxxxxxxD
\else
\ifnum#1=5344 %
\hatcurLCqshorteccenxxxxxxE
\else
??????\fi
\fi
\fi
\fi
\fi
}
\newcommand{\hatcurLCrhoeccen}[1]{\ifnum#1=519 %
\hatcurLCrhoeccenxxxxxxA
\else
\ifnum#1=3629 %
\hatcurLCrhoeccenxxxxxxB
\else
\ifnum#1=3714 %
\hatcurLCrhoeccenxxxxxxC
\else
\ifnum#1=4201 %
\hatcurLCrhoeccenxxxxxxD
\else
\ifnum#1=5344 %
\hatcurLCrhoeccenxxxxxxE
\else
??????\fi
\fi
\fi
\fi
\fi
}
\newcommand{\hatcurLCrprstareccen}[1]{\ifnum#1=519 %
\hatcurLCrprstareccenxxxxxxA
\else
\ifnum#1=3629 %
\hatcurLCrprstareccenxxxxxxB
\else
\ifnum#1=3714 %
\hatcurLCrprstareccenxxxxxxC
\else
\ifnum#1=4201 %
\hatcurLCrprstareccenxxxxxxD
\else
\ifnum#1=5344 %
\hatcurLCrprstareccenxxxxxxE
\else
??????\fi
\fi
\fi
\fi
\fi
}
\newcommand{\hatcurLCTAeccen}[1]{\ifnum#1=519 %
\hatcurLCTAeccenxxxxxxA
\else
\ifnum#1=3629 %
\hatcurLCTAeccenxxxxxxB
\else
\ifnum#1=3714 %
\hatcurLCTAeccenxxxxxxC
\else
\ifnum#1=4201 %
\hatcurLCTAeccenxxxxxxD
\else
\ifnum#1=5344 %
\hatcurLCTAeccenxxxxxxE
\else
??????\fi
\fi
\fi
\fi
\fi
}
\newcommand{\hatcurLCTBeccen}[1]{\ifnum#1=519 %
\hatcurLCTBeccenxxxxxxA
\else
\ifnum#1=3629 %
\hatcurLCTBeccenxxxxxxB
\else
\ifnum#1=3714 %
\hatcurLCTBeccenxxxxxxC
\else
\ifnum#1=4201 %
\hatcurLCTBeccenxxxxxxD
\else
\ifnum#1=5344 %
\hatcurLCTBeccenxxxxxxE
\else
??????\fi
\fi
\fi
\fi
\fi
}
\newcommand{\hatcurLCTeccen}[1]{\ifnum#1=519 %
\hatcurLCTeccenxxxxxxA
\else
\ifnum#1=3629 %
\hatcurLCTeccenxxxxxxB
\else
\ifnum#1=3714 %
\hatcurLCTeccenxxxxxxC
\else
\ifnum#1=4201 %
\hatcurLCTeccenxxxxxxD
\else
\ifnum#1=5344 %
\hatcurLCTeccenxxxxxxE
\else
??????\fi
\fi
\fi
\fi
\fi
}
\newcommand{\hatcurLCzetaeccen}[1]{\ifnum#1=519 %
\hatcurLCzetaeccenxxxxxxA
\else
\ifnum#1=3629 %
\hatcurLCzetaeccenxxxxxxB
\else
\ifnum#1=3714 %
\hatcurLCzetaeccenxxxxxxC
\else
\ifnum#1=4201 %
\hatcurLCzetaeccenxxxxxxD
\else
\ifnum#1=5344 %
\hatcurLCzetaeccenxxxxxxE
\else
??????\fi
\fi
\fi
\fi
\fi
}
\newcommand{\hatcurPPaequiveccen}[1]{\ifnum#1=519 %
\hatcurPPaequiveccenxxxxxxA
\else
\ifnum#1=3629 %
\hatcurPPaequiveccenxxxxxxB
\else
\ifnum#1=3714 %
\hatcurPPaequiveccenxxxxxxC
\else
\ifnum#1=4201 %
\hatcurPPaequiveccenxxxxxxD
\else
\ifnum#1=5344 %
\hatcurPPaequiveccenxxxxxxE
\else
??????\fi
\fi
\fi
\fi
\fi
}
\newcommand{\hatcurPPareccen}[1]{\ifnum#1=519 %
\hatcurPPareccenxxxxxxA
\else
\ifnum#1=3629 %
\hatcurPPareccenxxxxxxB
\else
\ifnum#1=3714 %
\hatcurPPareccenxxxxxxC
\else
\ifnum#1=4201 %
\hatcurPPareccenxxxxxxD
\else
\ifnum#1=5344 %
\hatcurPPareccenxxxxxxE
\else
??????\fi
\fi
\fi
\fi
\fi
}
\newcommand{\hatcurPPareleccen}[1]{\ifnum#1=519 %
\hatcurPPareleccenxxxxxxA
\else
\ifnum#1=3629 %
\hatcurPPareleccenxxxxxxB
\else
\ifnum#1=3714 %
\hatcurPPareleccenxxxxxxC
\else
\ifnum#1=4201 %
\hatcurPPareleccenxxxxxxD
\else
\ifnum#1=5344 %
\hatcurPPareleccenxxxxxxE
\else
??????\fi
\fi
\fi
\fi
\fi
}
\newcommand{\hatcurPPfluxapdimeccen}[1]{\ifnum#1=519 %
\hatcurPPfluxapdimeccenxxxxxxA
\else
\ifnum#1=3629 %
\hatcurPPfluxapdimeccenxxxxxxB
\else
\ifnum#1=3714 %
\hatcurPPfluxapdimeccenxxxxxxC
\else
\ifnum#1=4201 %
\hatcurPPfluxapdimeccenxxxxxxD
\else
\ifnum#1=5344 %
\hatcurPPfluxapdimeccenxxxxxxE
\else
??????\fi
\fi
\fi
\fi
\fi
}
\newcommand{\hatcurPPfluxapeccen}[1]{\ifnum#1=519 %
\hatcurPPfluxapeccenxxxxxxA
\else
\ifnum#1=3629 %
\hatcurPPfluxapeccenxxxxxxB
\else
\ifnum#1=3714 %
\hatcurPPfluxapeccenxxxxxxC
\else
\ifnum#1=4201 %
\hatcurPPfluxapeccenxxxxxxD
\else
\ifnum#1=5344 %
\hatcurPPfluxapeccenxxxxxxE
\else
??????\fi
\fi
\fi
\fi
\fi
}
\newcommand{\hatcurPPfluxavgdimeccen}[1]{\ifnum#1=519 %
\hatcurPPfluxavgdimeccenxxxxxxA
\else
\ifnum#1=3629 %
\hatcurPPfluxavgdimeccenxxxxxxB
\else
\ifnum#1=3714 %
\hatcurPPfluxavgdimeccenxxxxxxC
\else
\ifnum#1=4201 %
\hatcurPPfluxavgdimeccenxxxxxxD
\else
\ifnum#1=5344 %
\hatcurPPfluxavgdimeccenxxxxxxE
\else
??????\fi
\fi
\fi
\fi
\fi
}
\newcommand{\hatcurPPfluxavgeccen}[1]{\ifnum#1=519 %
\hatcurPPfluxavgeccenxxxxxxA
\else
\ifnum#1=3629 %
\hatcurPPfluxavgeccenxxxxxxB
\else
\ifnum#1=3714 %
\hatcurPPfluxavgeccenxxxxxxC
\else
\ifnum#1=4201 %
\hatcurPPfluxavgeccenxxxxxxD
\else
\ifnum#1=5344 %
\hatcurPPfluxavgeccenxxxxxxE
\else
??????\fi
\fi
\fi
\fi
\fi
}
\newcommand{\hatcurPPfluxavglogeccen}[1]{\ifnum#1=519 %
\hatcurPPfluxavglogeccenxxxxxxA
\else
\ifnum#1=3629 %
\hatcurPPfluxavglogeccenxxxxxxB
\else
\ifnum#1=3714 %
\hatcurPPfluxavglogeccenxxxxxxC
\else
\ifnum#1=4201 %
\hatcurPPfluxavglogeccenxxxxxxD
\else
\ifnum#1=5344 %
\hatcurPPfluxavglogeccenxxxxxxE
\else
??????\fi
\fi
\fi
\fi
\fi
}
\newcommand{\hatcurPPfluxperidimeccen}[1]{\ifnum#1=519 %
\hatcurPPfluxperidimeccenxxxxxxA
\else
\ifnum#1=3629 %
\hatcurPPfluxperidimeccenxxxxxxB
\else
\ifnum#1=3714 %
\hatcurPPfluxperidimeccenxxxxxxC
\else
\ifnum#1=4201 %
\hatcurPPfluxperidimeccenxxxxxxD
\else
\ifnum#1=5344 %
\hatcurPPfluxperidimeccenxxxxxxE
\else
??????\fi
\fi
\fi
\fi
\fi
}
\newcommand{\hatcurPPfluxperieccen}[1]{\ifnum#1=519 %
\hatcurPPfluxperieccenxxxxxxA
\else
\ifnum#1=3629 %
\hatcurPPfluxperieccenxxxxxxB
\else
\ifnum#1=3714 %
\hatcurPPfluxperieccenxxxxxxC
\else
\ifnum#1=4201 %
\hatcurPPfluxperieccenxxxxxxD
\else
\ifnum#1=5344 %
\hatcurPPfluxperieccenxxxxxxE
\else
??????\fi
\fi
\fi
\fi
\fi
}
\newcommand{\hatcurPPgeccen}[1]{\ifnum#1=519 %
\hatcurPPgeccenxxxxxxA
\else
\ifnum#1=3629 %
\hatcurPPgeccenxxxxxxB
\else
\ifnum#1=3714 %
\hatcurPPgeccenxxxxxxC
\else
\ifnum#1=4201 %
\hatcurPPgeccenxxxxxxD
\else
\ifnum#1=5344 %
\hatcurPPgeccenxxxxxxE
\else
??????\fi
\fi
\fi
\fi
\fi
}
\newcommand{\hatcurPPieccen}[1]{\ifnum#1=519 %
\hatcurPPieccenxxxxxxA
\else
\ifnum#1=3629 %
\hatcurPPieccenxxxxxxB
\else
\ifnum#1=3714 %
\hatcurPPieccenxxxxxxC
\else
\ifnum#1=4201 %
\hatcurPPieccenxxxxxxD
\else
\ifnum#1=5344 %
\hatcurPPieccenxxxxxxE
\else
??????\fi
\fi
\fi
\fi
\fi
}
\newcommand{\hatcurPPloggeccen}[1]{\ifnum#1=519 %
\hatcurPPloggeccenxxxxxxA
\else
\ifnum#1=3629 %
\hatcurPPloggeccenxxxxxxB
\else
\ifnum#1=3714 %
\hatcurPPloggeccenxxxxxxC
\else
\ifnum#1=4201 %
\hatcurPPloggeccenxxxxxxD
\else
\ifnum#1=5344 %
\hatcurPPloggeccenxxxxxxE
\else
??????\fi
\fi
\fi
\fi
\fi
}
\newcommand{\hatcurPPmeccen}[1]{\ifnum#1=519 %
\hatcurPPmeccenxxxxxxA
\else
\ifnum#1=3629 %
\hatcurPPmeccenxxxxxxB
\else
\ifnum#1=3714 %
\hatcurPPmeccenxxxxxxC
\else
\ifnum#1=4201 %
\hatcurPPmeccenxxxxxxD
\else
\ifnum#1=5344 %
\hatcurPPmeccenxxxxxxE
\else
??????\fi
\fi
\fi
\fi
\fi
}
\newcommand{\hatcurPPmeeccen}[1]{\ifnum#1=519 %
\hatcurPPmeeccenxxxxxxA
\else
\ifnum#1=3629 %
\hatcurPPmeeccenxxxxxxB
\else
\ifnum#1=3714 %
\hatcurPPmeeccenxxxxxxC
\else
\ifnum#1=4201 %
\hatcurPPmeeccenxxxxxxD
\else
\ifnum#1=5344 %
\hatcurPPmeeccenxxxxxxE
\else
??????\fi
\fi
\fi
\fi
\fi
}
\newcommand{\hatcurPPmelongeccen}[1]{\ifnum#1=519 %
\hatcurPPmelongeccenxxxxxxA
\else
\ifnum#1=3629 %
\hatcurPPmelongeccenxxxxxxB
\else
\ifnum#1=3714 %
\hatcurPPmelongeccenxxxxxxC
\else
\ifnum#1=4201 %
\hatcurPPmelongeccenxxxxxxD
\else
\ifnum#1=5344 %
\hatcurPPmelongeccenxxxxxxE
\else
??????\fi
\fi
\fi
\fi
\fi
}
\newcommand{\hatcurPPmeshorteccen}[1]{\ifnum#1=519 %
\hatcurPPmeshorteccenxxxxxxA
\else
\ifnum#1=3629 %
\hatcurPPmeshorteccenxxxxxxB
\else
\ifnum#1=3714 %
\hatcurPPmeshorteccenxxxxxxC
\else
\ifnum#1=4201 %
\hatcurPPmeshorteccenxxxxxxD
\else
\ifnum#1=5344 %
\hatcurPPmeshorteccenxxxxxxE
\else
??????\fi
\fi
\fi
\fi
\fi
}
\newcommand{\hatcurPPmlongeccen}[1]{\ifnum#1=519 %
\hatcurPPmlongeccenxxxxxxA
\else
\ifnum#1=3629 %
\hatcurPPmlongeccenxxxxxxB
\else
\ifnum#1=3714 %
\hatcurPPmlongeccenxxxxxxC
\else
\ifnum#1=4201 %
\hatcurPPmlongeccenxxxxxxD
\else
\ifnum#1=5344 %
\hatcurPPmlongeccenxxxxxxE
\else
??????\fi
\fi
\fi
\fi
\fi
}
\newcommand{\hatcurPPmrcorreccen}[1]{\ifnum#1=519 %
\hatcurPPmrcorreccenxxxxxxA
\else
\ifnum#1=3629 %
\hatcurPPmrcorreccenxxxxxxB
\else
\ifnum#1=3714 %
\hatcurPPmrcorreccenxxxxxxC
\else
\ifnum#1=4201 %
\hatcurPPmrcorreccenxxxxxxD
\else
\ifnum#1=5344 %
\hatcurPPmrcorreccenxxxxxxE
\else
??????\fi
\fi
\fi
\fi
\fi
}
\newcommand{\hatcurPPmshorteccen}[1]{\ifnum#1=519 %
\hatcurPPmshorteccenxxxxxxA
\else
\ifnum#1=3629 %
\hatcurPPmshorteccenxxxxxxB
\else
\ifnum#1=3714 %
\hatcurPPmshorteccenxxxxxxC
\else
\ifnum#1=4201 %
\hatcurPPmshorteccenxxxxxxD
\else
\ifnum#1=5344 %
\hatcurPPmshorteccenxxxxxxE
\else
??????\fi
\fi
\fi
\fi
\fi
}
\newcommand{\hatcurPPmtwosiglimeccen}[1]{\ifnum#1=519 %
\hatcurPPmtwosiglimeccenxxxxxxA
\else
\ifnum#1=3629 %
\hatcurPPmtwosiglimeccenxxxxxxB
\else
\ifnum#1=3714 %
\hatcurPPmtwosiglimeccenxxxxxxC
\else
\ifnum#1=4201 %
\hatcurPPmtwosiglimeccenxxxxxxD
\else
\ifnum#1=5344 %
\hatcurPPmtwosiglimeccenxxxxxxE
\else
??????\fi
\fi
\fi
\fi
\fi
}
\newcommand{\hatcurPPperieccen}[1]{\ifnum#1=519 %
\hatcurPPperieccenxxxxxxA
\else
\ifnum#1=3629 %
\hatcurPPperieccenxxxxxxB
\else
\ifnum#1=3714 %
\hatcurPPperieccenxxxxxxC
\else
\ifnum#1=4201 %
\hatcurPPperieccenxxxxxxD
\else
\ifnum#1=5344 %
\hatcurPPperieccenxxxxxxE
\else
??????\fi
\fi
\fi
\fi
\fi
}
\newcommand{\hatcurPPphiconjeccen}[1]{\ifnum#1=519 %
\hatcurPPphiconjeccenxxxxxxA
\else
\ifnum#1=3629 %
\hatcurPPphiconjeccenxxxxxxB
\else
\ifnum#1=3714 %
\hatcurPPphiconjeccenxxxxxxC
\else
\ifnum#1=4201 %
\hatcurPPphiconjeccenxxxxxxD
\else
\ifnum#1=5344 %
\hatcurPPphiconjeccenxxxxxxE
\else
??????\fi
\fi
\fi
\fi
\fi
}
\newcommand{\hatcurPPreccen}[1]{\ifnum#1=519 %
\hatcurPPreccenxxxxxxA
\else
\ifnum#1=3629 %
\hatcurPPreccenxxxxxxB
\else
\ifnum#1=3714 %
\hatcurPPreccenxxxxxxC
\else
\ifnum#1=4201 %
\hatcurPPreccenxxxxxxD
\else
\ifnum#1=5344 %
\hatcurPPreccenxxxxxxE
\else
??????\fi
\fi
\fi
\fi
\fi
}
\newcommand{\hatcurPPreeccen}[1]{\ifnum#1=519 %
\hatcurPPreeccenxxxxxxA
\else
\ifnum#1=3629 %
\hatcurPPreeccenxxxxxxB
\else
\ifnum#1=3714 %
\hatcurPPreeccenxxxxxxC
\else
\ifnum#1=4201 %
\hatcurPPreeccenxxxxxxD
\else
\ifnum#1=5344 %
\hatcurPPreeccenxxxxxxE
\else
??????\fi
\fi
\fi
\fi
\fi
}
\newcommand{\hatcurPPrelongeccen}[1]{\ifnum#1=519 %
\hatcurPPrelongeccenxxxxxxA
\else
\ifnum#1=3629 %
\hatcurPPrelongeccenxxxxxxB
\else
\ifnum#1=3714 %
\hatcurPPrelongeccenxxxxxxC
\else
\ifnum#1=4201 %
\hatcurPPrelongeccenxxxxxxD
\else
\ifnum#1=5344 %
\hatcurPPrelongeccenxxxxxxE
\else
??????\fi
\fi
\fi
\fi
\fi
}
\newcommand{\hatcurPPreshorteccen}[1]{\ifnum#1=519 %
\hatcurPPreshorteccenxxxxxxA
\else
\ifnum#1=3629 %
\hatcurPPreshorteccenxxxxxxB
\else
\ifnum#1=3714 %
\hatcurPPreshorteccenxxxxxxC
\else
\ifnum#1=4201 %
\hatcurPPreshorteccenxxxxxxD
\else
\ifnum#1=5344 %
\hatcurPPreshorteccenxxxxxxE
\else
??????\fi
\fi
\fi
\fi
\fi
}
\newcommand{\hatcurPPrhoeccen}[1]{\ifnum#1=519 %
\hatcurPPrhoeccenxxxxxxA
\else
\ifnum#1=3629 %
\hatcurPPrhoeccenxxxxxxB
\else
\ifnum#1=3714 %
\hatcurPPrhoeccenxxxxxxC
\else
\ifnum#1=4201 %
\hatcurPPrhoeccenxxxxxxD
\else
\ifnum#1=5344 %
\hatcurPPrhoeccenxxxxxxE
\else
??????\fi
\fi
\fi
\fi
\fi
}
\newcommand{\hatcurPPrlongeccen}[1]{\ifnum#1=519 %
\hatcurPPrlongeccenxxxxxxA
\else
\ifnum#1=3629 %
\hatcurPPrlongeccenxxxxxxB
\else
\ifnum#1=3714 %
\hatcurPPrlongeccenxxxxxxC
\else
\ifnum#1=4201 %
\hatcurPPrlongeccenxxxxxxD
\else
\ifnum#1=5344 %
\hatcurPPrlongeccenxxxxxxE
\else
??????\fi
\fi
\fi
\fi
\fi
}
\newcommand{\hatcurPPrshorteccen}[1]{\ifnum#1=519 %
\hatcurPPrshorteccenxxxxxxA
\else
\ifnum#1=3629 %
\hatcurPPrshorteccenxxxxxxB
\else
\ifnum#1=3714 %
\hatcurPPrshorteccenxxxxxxC
\else
\ifnum#1=4201 %
\hatcurPPrshorteccenxxxxxxD
\else
\ifnum#1=5344 %
\hatcurPPrshorteccenxxxxxxE
\else
??????\fi
\fi
\fi
\fi
\fi
}
\newcommand{\hatcurPPtcirceccen}[1]{\ifnum#1=519 %
\hatcurPPtcirceccenxxxxxxA
\else
\ifnum#1=3629 %
\hatcurPPtcirceccenxxxxxxB
\else
\ifnum#1=3714 %
\hatcurPPtcirceccenxxxxxxC
\else
\ifnum#1=4201 %
\hatcurPPtcirceccenxxxxxxD
\else
\ifnum#1=5344 %
\hatcurPPtcirceccenxxxxxxE
\else
??????\fi
\fi
\fi
\fi
\fi
}
\newcommand{\hatcurPPteffeccen}[1]{\ifnum#1=519 %
\hatcurPPteffeccenxxxxxxA
\else
\ifnum#1=3629 %
\hatcurPPteffeccenxxxxxxB
\else
\ifnum#1=3714 %
\hatcurPPteffeccenxxxxxxC
\else
\ifnum#1=4201 %
\hatcurPPteffeccenxxxxxxD
\else
\ifnum#1=5344 %
\hatcurPPteffeccenxxxxxxE
\else
??????\fi
\fi
\fi
\fi
\fi
}
\newcommand{\hatcurPPthetaeccen}[1]{\ifnum#1=519 %
\hatcurPPthetaeccenxxxxxxA
\else
\ifnum#1=3629 %
\hatcurPPthetaeccenxxxxxxB
\else
\ifnum#1=3714 %
\hatcurPPthetaeccenxxxxxxC
\else
\ifnum#1=4201 %
\hatcurPPthetaeccenxxxxxxD
\else
\ifnum#1=5344 %
\hatcurPPthetaeccenxxxxxxE
\else
??????\fi
\fi
\fi
\fi
\fi
}
\newcommand{\hatcurPPtinfalleccen}[1]{\ifnum#1=519 %
\hatcurPPtinfalleccenxxxxxxA
\else
\ifnum#1=3629 %
\hatcurPPtinfalleccenxxxxxxB
\else
\ifnum#1=3714 %
\hatcurPPtinfalleccenxxxxxxC
\else
\ifnum#1=4201 %
\hatcurPPtinfalleccenxxxxxxD
\else
\ifnum#1=5344 %
\hatcurPPtinfalleccenxxxxxxE
\else
??????\fi
\fi
\fi
\fi
\fi
}
\newcommand{\hatcurRVecceneccen}[1]{\ifnum#1=519 %
\hatcurRVecceneccenxxxxxxA
\else
\ifnum#1=3629 %
\hatcurRVecceneccenxxxxxxB
\else
\ifnum#1=3714 %
\hatcurRVecceneccenxxxxxxC
\else
\ifnum#1=4201 %
\hatcurRVecceneccenxxxxxxD
\else
\ifnum#1=5344 %
\hatcurRVecceneccenxxxxxxE
\else
??????\fi
\fi
\fi
\fi
\fi
}
\newcommand{\hatcurRVeccentwosiglimeccen}[1]{\ifnum#1=519 %
\hatcurRVeccentwosiglimeccenxxxxxxA
\else
\ifnum#1=3629 %
\hatcurRVeccentwosiglimeccenxxxxxxB
\else
\ifnum#1=3714 %
\hatcurRVeccentwosiglimeccenxxxxxxC
\else
\ifnum#1=4201 %
\hatcurRVeccentwosiglimeccenxxxxxxD
\else
\ifnum#1=5344 %
\hatcurRVeccentwosiglimeccenxxxxxxE
\else
??????\fi
\fi
\fi
\fi
\fi
}
\newcommand{\hatcurRVfitrmsAeccen}[1]{\ifnum#1=519 %
\hatcurRVfitrmsAeccenxxxxxxA
\else
\ifnum#1=3629 %
\hatcurRVfitrmsAeccenxxxxxxB
\else
\ifnum#1=3714 %
\hatcurRVfitrmsAeccenxxxxxxC
\else
??????\fi
\fi
\fi
}
\newcommand{\hatcurRVfitrmsBeccen}[1]{\ifnum#1=519 %
\hatcurRVfitrmsBeccenxxxxxxA
\else
\ifnum#1=3629 %
\hatcurRVfitrmsBeccenxxxxxxB
\else
\ifnum#1=3714 %
\hatcurRVfitrmsBeccenxxxxxxC
\else
??????\fi
\fi
\fi
}
\newcommand{\hatcurRVfitrmsCeccen}[1]{\ifnum#1=3629 %
\hatcurRVfitrmsCeccenxxxxxxB
\else
\ifnum#1=3714 %
\hatcurRVfitrmsCeccenxxxxxxC
\else
??????\fi
\fi
}
\newcommand{\hatcurRVfitrmseccen}[1]{\ifnum#1=4201 %
\hatcurRVfitrmseccenxxxxxxD
\else
\ifnum#1=5344 %
\hatcurRVfitrmseccenxxxxxxE
\else
??????\fi
\fi
}
\newcommand{\hatcurRVgammaAeccen}[1]{\ifnum#1=519 %
\hatcurRVgammaAeccenxxxxxxA
\else
\ifnum#1=3629 %
\hatcurRVgammaAeccenxxxxxxB
\else
\ifnum#1=3714 %
\hatcurRVgammaAeccenxxxxxxC
\else
??????\fi
\fi
\fi
}
\newcommand{\hatcurRVgammaBeccen}[1]{\ifnum#1=519 %
\hatcurRVgammaBeccenxxxxxxA
\else
\ifnum#1=3629 %
\hatcurRVgammaBeccenxxxxxxB
\else
\ifnum#1=3714 %
\hatcurRVgammaBeccenxxxxxxC
\else
??????\fi
\fi
\fi
}
\newcommand{\hatcurRVgammaCeccen}[1]{\ifnum#1=3629 %
\hatcurRVgammaCeccenxxxxxxB
\else
\ifnum#1=3714 %
\hatcurRVgammaCeccenxxxxxxC
\else
??????\fi
\fi
}
\newcommand{\hatcurRVgammaeccen}[1]{\ifnum#1=4201 %
\hatcurRVgammaeccenxxxxxxD
\else
\ifnum#1=5344 %
\hatcurRVgammaeccenxxxxxxE
\else
??????\fi
\fi
}
\newcommand{\hatcurRVheccen}[1]{\ifnum#1=519 %
\hatcurRVheccenxxxxxxA
\else
\ifnum#1=3629 %
\hatcurRVheccenxxxxxxB
\else
\ifnum#1=3714 %
\hatcurRVheccenxxxxxxC
\else
\ifnum#1=4201 %
\hatcurRVheccenxxxxxxD
\else
\ifnum#1=5344 %
\hatcurRVheccenxxxxxxE
\else
??????\fi
\fi
\fi
\fi
\fi
}
\newcommand{\hatcurRVjitterAeccen}[1]{\ifnum#1=519 %
\hatcurRVjitterAeccenxxxxxxA
\else
\ifnum#1=3629 %
\hatcurRVjitterAeccenxxxxxxB
\else
\ifnum#1=3714 %
\hatcurRVjitterAeccenxxxxxxC
\else
??????\fi
\fi
\fi
}
\newcommand{\hatcurRVjitterBeccen}[1]{\ifnum#1=519 %
\hatcurRVjitterBeccenxxxxxxA
\else
\ifnum#1=3629 %
\hatcurRVjitterBeccenxxxxxxB
\else
\ifnum#1=3714 %
\hatcurRVjitterBeccenxxxxxxC
\else
??????\fi
\fi
\fi
}
\newcommand{\hatcurRVjitterCeccen}[1]{\ifnum#1=3629 %
\hatcurRVjitterCeccenxxxxxxB
\else
\ifnum#1=3714 %
\hatcurRVjitterCeccenxxxxxxC
\else
??????\fi
\fi
}
\newcommand{\hatcurRVjittereccen}[1]{\ifnum#1=4201 %
\hatcurRVjittereccenxxxxxxD
\else
\ifnum#1=5344 %
\hatcurRVjittereccenxxxxxxE
\else
??????\fi
\fi
}
\newcommand{\hatcurRVjittertwosiglimAeccen}[1]{\ifnum#1=519 %
\hatcurRVjittertwosiglimAeccenxxxxxxA
\else
\ifnum#1=3629 %
\hatcurRVjittertwosiglimAeccenxxxxxxB
\else
\ifnum#1=3714 %
\hatcurRVjittertwosiglimAeccenxxxxxxC
\else
??????\fi
\fi
\fi
}
\newcommand{\hatcurRVjittertwosiglimBeccen}[1]{\ifnum#1=519 %
\hatcurRVjittertwosiglimBeccenxxxxxxA
\else
\ifnum#1=3629 %
\hatcurRVjittertwosiglimBeccenxxxxxxB
\else
\ifnum#1=3714 %
\hatcurRVjittertwosiglimBeccenxxxxxxC
\else
??????\fi
\fi
\fi
}
\newcommand{\hatcurRVjittertwosiglimCeccen}[1]{\ifnum#1=3629 %
\hatcurRVjittertwosiglimCeccenxxxxxxB
\else
\ifnum#1=3714 %
\hatcurRVjittertwosiglimCeccenxxxxxxC
\else
??????\fi
\fi
}
\newcommand{\hatcurRVjittertwosiglimeccen}[1]{\ifnum#1=4201 %
\hatcurRVjittertwosiglimeccenxxxxxxD
\else
\ifnum#1=5344 %
\hatcurRVjittertwosiglimeccenxxxxxxE
\else
??????\fi
\fi
}
\newcommand{\hatcurRVkeccen}[1]{\ifnum#1=519 %
\hatcurRVkeccenxxxxxxA
\else
\ifnum#1=3629 %
\hatcurRVkeccenxxxxxxB
\else
\ifnum#1=3714 %
\hatcurRVkeccenxxxxxxC
\else
\ifnum#1=4201 %
\hatcurRVkeccenxxxxxxD
\else
\ifnum#1=5344 %
\hatcurRVkeccenxxxxxxE
\else
??????\fi
\fi
\fi
\fi
\fi
}
\newcommand{\hatcurRVKeccen}[1]{\ifnum#1=519 %
\hatcurRVKeccenxxxxxxA
\else
\ifnum#1=3629 %
\hatcurRVKeccenxxxxxxB
\else
\ifnum#1=3714 %
\hatcurRVKeccenxxxxxxC
\else
\ifnum#1=4201 %
\hatcurRVKeccenxxxxxxD
\else
\ifnum#1=5344 %
\hatcurRVKeccenxxxxxxE
\else
??????\fi
\fi
\fi
\fi
\fi
}
\newcommand{\hatcurRVKtwosiglimeccen}[1]{\ifnum#1=519 %
\hatcurRVKtwosiglimeccenxxxxxxA
\else
\ifnum#1=3629 %
\hatcurRVKtwosiglimeccenxxxxxxB
\else
\ifnum#1=3714 %
\hatcurRVKtwosiglimeccenxxxxxxC
\else
\ifnum#1=4201 %
\hatcurRVKtwosiglimeccenxxxxxxD
\else
\ifnum#1=5344 %
\hatcurRVKtwosiglimeccenxxxxxxE
\else
??????\fi
\fi
\fi
\fi
\fi
}
\newcommand{\hatcurRVomegaeccen}[1]{\ifnum#1=519 %
\hatcurRVomegaeccenxxxxxxA
\else
\ifnum#1=3629 %
\hatcurRVomegaeccenxxxxxxB
\else
\ifnum#1=3714 %
\hatcurRVomegaeccenxxxxxxC
\else
\ifnum#1=4201 %
\hatcurRVomegaeccenxxxxxxD
\else
\ifnum#1=5344 %
\hatcurRVomegaeccenxxxxxxE
\else
??????\fi
\fi
\fi
\fi
\fi
}
\newcommand{\hatcurRVrheccen}[1]{\ifnum#1=519 %
\hatcurRVrheccenxxxxxxA
\else
\ifnum#1=3629 %
\hatcurRVrheccenxxxxxxB
\else
\ifnum#1=3714 %
\hatcurRVrheccenxxxxxxC
\else
\ifnum#1=4201 %
\hatcurRVrheccenxxxxxxD
\else
\ifnum#1=5344 %
\hatcurRVrheccenxxxxxxE
\else
??????\fi
\fi
\fi
\fi
\fi
}
\newcommand{\hatcurRVrkeccen}[1]{\ifnum#1=519 %
\hatcurRVrkeccenxxxxxxA
\else
\ifnum#1=3629 %
\hatcurRVrkeccenxxxxxxB
\else
\ifnum#1=3714 %
\hatcurRVrkeccenxxxxxxC
\else
\ifnum#1=4201 %
\hatcurRVrkeccenxxxxxxD
\else
\ifnum#1=5344 %
\hatcurRVrkeccenxxxxxxE
\else
??????\fi
\fi
\fi
\fi
\fi
}
\newcommand{\hatcurRVtroneeccen}[1]{\ifnum#1=519 %
\hatcurRVtroneeccenxxxxxxA
\else
\ifnum#1=3629 %
\hatcurRVtroneeccenxxxxxxB
\else
\ifnum#1=3714 %
\hatcurRVtroneeccenxxxxxxC
\else
\ifnum#1=4201 %
\hatcurRVtroneeccenxxxxxxD
\else
\ifnum#1=5344 %
\hatcurRVtroneeccenxxxxxxE
\else
??????\fi
\fi
\fi
\fi
\fi
}
\newcommand{\hatcurRVtrtwoeccen}[1]{\ifnum#1=519 %
\hatcurRVtrtwoeccenxxxxxxA
\else
\ifnum#1=3629 %
\hatcurRVtrtwoeccenxxxxxxB
\else
\ifnum#1=3714 %
\hatcurRVtrtwoeccenxxxxxxC
\else
\ifnum#1=4201 %
\hatcurRVtrtwoeccenxxxxxxD
\else
\ifnum#1=5344 %
\hatcurRVtrtwoeccenxxxxxxE
\else
??????\fi
\fi
\fi
\fi
\fi
}
\newcommand{\hatcurSMEiloggeccen}[1]{\ifnum#1=519 %
\hatcurSMEiloggeccenxxxxxxA
\else
\ifnum#1=3629 %
\hatcurSMEiloggeccenxxxxxxB
\else
\ifnum#1=3714 %
\hatcurSMEiloggeccenxxxxxxC
\else
\ifnum#1=4201 %
\hatcurSMEiloggeccenxxxxxxD
\else
\ifnum#1=5344 %
\hatcurSMEiloggeccenxxxxxxE
\else
??????\fi
\fi
\fi
\fi
\fi
}
\newcommand{\hatcurSMEiteffeccen}[1]{\ifnum#1=519 %
\hatcurSMEiteffeccenxxxxxxA
\else
\ifnum#1=3629 %
\hatcurSMEiteffeccenxxxxxxB
\else
\ifnum#1=3714 %
\hatcurSMEiteffeccenxxxxxxC
\else
\ifnum#1=4201 %
\hatcurSMEiteffeccenxxxxxxD
\else
\ifnum#1=5344 %
\hatcurSMEiteffeccenxxxxxxE
\else
??????\fi
\fi
\fi
\fi
\fi
}
\newcommand{\hatcurSMEivmaceccen}[1]{\ifnum#1=519 %
\hatcurSMEivmaceccenxxxxxxA
\else
\ifnum#1=3629 %
\hatcurSMEivmaceccenxxxxxxB
\else
\ifnum#1=3714 %
\hatcurSMEivmaceccenxxxxxxC
\else
\ifnum#1=4201 %
\hatcurSMEivmaceccenxxxxxxD
\else
\ifnum#1=5344 %
\hatcurSMEivmaceccenxxxxxxE
\else
??????\fi
\fi
\fi
\fi
\fi
}
\newcommand{\hatcurSMEivmiceccen}[1]{\ifnum#1=519 %
\hatcurSMEivmiceccenxxxxxxA
\else
\ifnum#1=3629 %
\hatcurSMEivmiceccenxxxxxxB
\else
\ifnum#1=3714 %
\hatcurSMEivmiceccenxxxxxxC
\else
\ifnum#1=4201 %
\hatcurSMEivmiceccenxxxxxxD
\else
\ifnum#1=5344 %
\hatcurSMEivmiceccenxxxxxxE
\else
??????\fi
\fi
\fi
\fi
\fi
}
\newcommand{\hatcurSMEivsineccen}[1]{\ifnum#1=519 %
\hatcurSMEivsineccenxxxxxxA
\else
\ifnum#1=3629 %
\hatcurSMEivsineccenxxxxxxB
\else
\ifnum#1=3714 %
\hatcurSMEivsineccenxxxxxxC
\else
\ifnum#1=4201 %
\hatcurSMEivsineccenxxxxxxD
\else
\ifnum#1=5344 %
\hatcurSMEivsineccenxxxxxxE
\else
??????\fi
\fi
\fi
\fi
\fi
}
\newcommand{\hatcurSMEizfeheccen}[1]{\ifnum#1=519 %
\hatcurSMEizfeheccenxxxxxxA
\else
\ifnum#1=3629 %
\hatcurSMEizfeheccenxxxxxxB
\else
\ifnum#1=3714 %
\hatcurSMEizfeheccenxxxxxxC
\else
\ifnum#1=4201 %
\hatcurSMEizfeheccenxxxxxxD
\else
\ifnum#1=5344 %
\hatcurSMEizfeheccenxxxxxxE
\else
??????\fi
\fi
\fi
\fi
\fi
}
\newcommand{\hatcurSMEizfehshorteccen}[1]{\ifnum#1=519 %
\hatcurSMEizfehshorteccenxxxxxxA
\else
\ifnum#1=3629 %
\hatcurSMEizfehshorteccenxxxxxxB
\else
\ifnum#1=3714 %
\hatcurSMEizfehshorteccenxxxxxxC
\else
\ifnum#1=4201 %
\hatcurSMEizfehshorteccenxxxxxxD
\else
\ifnum#1=5344 %
\hatcurSMEizfehshorteccenxxxxxxE
\else
??????\fi
\fi
\fi
\fi
\fi
}
\newcommand{\hatcurXAveccen}[1]{\ifnum#1=519 %
\hatcurXAveccenxxxxxxA
\else
\ifnum#1=3629 %
\hatcurXAveccenxxxxxxB
\else
\ifnum#1=3714 %
\hatcurXAveccenxxxxxxC
\else
\ifnum#1=4201 %
\hatcurXAveccenxxxxxxD
\else
\ifnum#1=5344 %
\hatcurXAveccenxxxxxxE
\else
??????\fi
\fi
\fi
\fi
\fi
}
\newcommand{\hatcurXdisteccen}[1]{\ifnum#1=519 %
\hatcurXdisteccenxxxxxxA
\else
\ifnum#1=3629 %
\hatcurXdisteccenxxxxxxB
\else
\ifnum#1=3714 %
\hatcurXdisteccenxxxxxxC
\else
\ifnum#1=4201 %
\hatcurXdisteccenxxxxxxD
\else
\ifnum#1=5344 %
\hatcurXdisteccenxxxxxxE
\else
??????\fi
\fi
\fi
\fi
\fi
}
\newcommand{\hatcurXdistredeccen}[1]{\ifnum#1=519 %
\hatcurXdistredeccenxxxxxxA
\else
\ifnum#1=3629 %
\hatcurXdistredeccenxxxxxxB
\else
\ifnum#1=3714 %
\hatcurXdistredeccenxxxxxxC
\else
\ifnum#1=4201 %
\hatcurXdistredeccenxxxxxxD
\else
\ifnum#1=5344 %
\hatcurXdistredeccenxxxxxxE
\else
??????\fi
\fi
\fi
\fi
\fi
}
\newcommand{\hatcurXEBVeccen}[1]{\ifnum#1=519 %
\hatcurXEBVeccenxxxxxxA
\else
\ifnum#1=3629 %
\hatcurXEBVeccenxxxxxxB
\else
\ifnum#1=3714 %
\hatcurXEBVeccenxxxxxxC
\else
\ifnum#1=4201 %
\hatcurXEBVeccenxxxxxxD
\else
\ifnum#1=5344 %
\hatcurXEBVeccenxxxxxxE
\else
??????\fi
\fi
\fi
\fi
\fi
}
\newcommand{\hatcurXsecdureccen}[1]{\ifnum#1=519 %
\hatcurXsecdureccenxxxxxxA
\else
\ifnum#1=3629 %
\hatcurXsecdureccenxxxxxxB
\else
\ifnum#1=3714 %
\hatcurXsecdureccenxxxxxxC
\else
\ifnum#1=4201 %
\hatcurXsecdureccenxxxxxxD
\else
\ifnum#1=5344 %
\hatcurXsecdureccenxxxxxxE
\else
??????\fi
\fi
\fi
\fi
\fi
}
\newcommand{\hatcurXsecingdureccen}[1]{\ifnum#1=519 %
\hatcurXsecingdureccenxxxxxxA
\else
\ifnum#1=3629 %
\hatcurXsecingdureccenxxxxxxB
\else
\ifnum#1=3714 %
\hatcurXsecingdureccenxxxxxxC
\else
\ifnum#1=4201 %
\hatcurXsecingdureccenxxxxxxD
\else
\ifnum#1=5344 %
\hatcurXsecingdureccenxxxxxxE
\else
??????\fi
\fi
\fi
\fi
\fi
}
\newcommand{\hatcurXsecondaryeccen}[1]{\ifnum#1=519 %
\hatcurXsecondaryeccenxxxxxxA
\else
\ifnum#1=3629 %
\hatcurXsecondaryeccenxxxxxxB
\else
\ifnum#1=3714 %
\hatcurXsecondaryeccenxxxxxxC
\else
\ifnum#1=4201 %
\hatcurXsecondaryeccenxxxxxxD
\else
\ifnum#1=5344 %
\hatcurXsecondaryeccenxxxxxxE
\else
??????\fi
\fi
\fi
\fi
\fi
}
\newcommand{\hatcurXsecphaseeccen}[1]{\ifnum#1=519 %
\hatcurXsecphaseeccenxxxxxxA
\else
\ifnum#1=3629 %
\hatcurXsecphaseeccenxxxxxxB
\else
\ifnum#1=3714 %
\hatcurXsecphaseeccenxxxxxxC
\else
\ifnum#1=4201 %
\hatcurXsecphaseeccenxxxxxxD
\else
\ifnum#1=5344 %
\hatcurXsecphaseeccenxxxxxxE
\else
??????\fi
\fi
\fi
\fi
\fi
}

%% file: TOI519.eccen.tex
\newcommand{\hatcurhtreccenxxxxxxA}{TOI519}                                        
\newcommand{\hatcurfieldeccenxxxxxxA}{\ensuremath{string}}                         
\newcommand{\hatcurCCraeccenxxxxxxA}{\ensuremath{08^{\mathrm h}18^{\mathrm m}25.6680{\mathrm s}}}                                
\newcommand{\hatcurCCdececcenxxxxxxA}{\ensuremath{-19{\arcdeg}39{\arcmin}46.5010{\arcsec}}}                              
\newcommand{\hatcurCCmageccenxxxxxxA}{NULL}                                        
\newcommand{\hatcurCCtwomasseccenxxxxxxA}{2MASS~08182567-1939465}                  
\newcommand{\hatcurCCgsceccenxxxxxxA}{GSC~NULL}                                    
\newcommand{\hatcurCCgaiaeccenxxxxxxA}{GAIA~5707485523149430400}                   
\newcommand{\hatcurCCgaiadrtwoeccenxxxxxxA}{GAIA~DR2~5707485527450614656}          
\newcommand{\hatcurCCtassmveccenxxxxxxA}{\ensuremath{nff\pmnff}}                   
\newcommand{\hatcurCCtassmvshorteccenxxxxxxA}{\ensuremath{0.0}}                    
\newcommand{\hatcurCCtassmBeccenxxxxxxA}{\ensuremath{nff\pmnff}}                   
\newcommand{\hatcurCCtassmBshorteccenxxxxxxA}{\ensuremath{0.0}}                    
\newcommand{\hatcurCCtassmIeccenxxxxxxA}{\ensuremath{nff\pmnff}}                   
\newcommand{\hatcurCCtassmIshorteccenxxxxxxA}{\ensuremath{0.0}}                    
\newcommand{\hatcurCCtassmgeccenxxxxxxA}{\ensuremath{nff\pmnff}}                   
\newcommand{\hatcurCCtassmgshorteccenxxxxxxA}{\ensuremath{0.0}}                    
\newcommand{\hatcurCCtassmreccenxxxxxxA}{\ensuremath{nff\pmnff}}                   
\newcommand{\hatcurCCtassmrshorteccenxxxxxxA}{\ensuremath{0.0}}                    
\newcommand{\hatcurCCtassmieccenxxxxxxA}{\ensuremath{nff\pmnff}}                   
\newcommand{\hatcurCCtassmishorteccenxxxxxxA}{\ensuremath{0.0}}                    
\newcommand{\hatcurCCparallaxeccenxxxxxxA}{\ensuremath{8.681\pm0.037}}             
\newcommand{\hatcurCCgaiamGeccenxxxxxxA}{\ensuremath{15.6770\pm0.0028}}            
\newcommand{\hatcurCCgaiamBPeccenxxxxxxA}{\ensuremath{17.1927\pm0.0057}}           
\newcommand{\hatcurCCgaiamRPeccenxxxxxxA}{\ensuremath{14.4819\pm0.0039}}           
\newcommand{\hatcurCCtwomassJmageccenxxxxxxA}{\ensuremath{12.847\pm0.027}}         
\newcommand{\hatcurCCtwomassHmageccenxxxxxxA}{\ensuremath{12.226\pm0.027}}         
\newcommand{\hatcurCCtwomassKmageccenxxxxxxA}{\ensuremath{11.951\pm0.024}}         
\newcommand{\hatcurCCcitJmageccenxxxxxxA}{\ensuremath{12.836\pm0.028}}             
\newcommand{\hatcurCCcitHmageccenxxxxxxA}{\ensuremath{12.216\pm0.029}}             
\newcommand{\hatcurCCcitKmageccenxxxxxxA}{\ensuremath{11.975\pm0.024}}             
\newcommand{\hatcurCCbbJmageccenxxxxxxA}{\ensuremath{12.928\pm0.030}}              
\newcommand{\hatcurCCbbHmageccenxxxxxxA}{\ensuremath{12.243\pm0.029}}              
\newcommand{\hatcurCCbbKmageccenxxxxxxA}{\ensuremath{11.995\pm0.025}}              
\newcommand{\hatcurCCesoJmageccenxxxxxxA}{\ensuremath{12.936\pm0.035}}             
\newcommand{\hatcurCCesoHmageccenxxxxxxA}{\ensuremath{12.243\pm0.048}}             
\newcommand{\hatcurCCesoKmageccenxxxxxxA}{\ensuremath{11.991\pm0.026}}             
\newcommand{\hatcurCCesoJHmageccenxxxxxxA}{\ensuremath{0.692\pm0.040}}             
\newcommand{\hatcurCCesoJKmageccenxxxxxxA}{\ensuremath{0.946\pm0.041}}             
\newcommand{\hatcurCCesoHKmageccenxxxxxxA}{\ensuremath{0.252\pm0.053}}             
\newcommand{\hatcurCCWonemageccenxxxxxxA}{\ensuremath{11.790\pm0.024}}             
\newcommand{\hatcurCCWtwomageccenxxxxxxA}{\ensuremath{11.642\pm0.021}}             
\newcommand{\hatcurCCWthreemageccenxxxxxxA}{\ensuremath{11.59\pm0.17}}             
\newcommand{\hatcurCCWfourmageccenxxxxxxA}{\ensuremath{nff\pmnff}}                 
\newcommand{\hatcurLCdipeccenxxxxxxA}{\ensuremath{0.1}}                            
\newcommand{\hatcurLCrprstareccenxxxxxxA}{\ensuremath{0.3039\pm0.0020}}            
\newcommand{\hatcurLCbsqeccenxxxxxxA}{\ensuremath{0.088_{-0.016}^{+0.015}}}        
\newcommand{\hatcurLCimpeccenxxxxxxA}{\ensuremath{0.297_{-0.028}^{+0.025}}}        
\newcommand{\hatcurLCzetaeccenxxxxxxA}{\ensuremath{51.86\pm0.34}}                  
\newcommand{\hatcurLCdureccenxxxxxxA}{\ensuremath{0.05129\pm0.00023}}              
\newcommand{\hatcurLCdurshorteccenxxxxxxA}{\ensuremath{0.0513}}                    
\newcommand{\hatcurLCdurhreccenxxxxxxA}{\ensuremath{1.2309\pm0.0056}}              
\newcommand{\hatcurLCdurhrshorteccenxxxxxxA}{\ensuremath{1.231}}                   
\newcommand{\hatcurLCqeccenxxxxxxA}{\ensuremath{0.04050\pm0.00019}}                
\newcommand{\hatcurLCqshorteccenxxxxxxA}{\ensuremath{0.041}}                       
\newcommand{\hatcurLCingdureccenxxxxxxA}{\ensuremath{0.01293\pm0.00025}}           
\newcommand{\hatcurLCPeccenxxxxxxA}{\ensuremath{1.26523247\pm0.00000012}}          
\newcommand{\hatcurLCPprececcenxxxxxxA}{\ensuremath{1.2652325}}                    
\newcommand{\hatcurLCPshorteccenxxxxxxA}{\ensuremath{1.2652}}                      
\newcommand{\hatcurLCTeccenxxxxxxA}{\ensuremath{2458920.791020\pm0.000034}}        
\newcommand{\hatcurLCTAeccenxxxxxxA}{\ensuremath{2458491.877206\pm0.000057}}       
\newcommand{\hatcurLCTBeccenxxxxxxA}{\ensuremath{2459658.421549\pm0.000070}}       
\newcommand{\hatcurLChatnetmAeccenxxxxxxA}{\ensuremath{-5.80115\pm0.00025}}        
\newcommand{\hatcurLCiblendAeccenxxxxxxA}{\ensuremath{0.9988\pm0.0021}}            
\newcommand{\hatcurLChatnetmBeccenxxxxxxA}{\ensuremath{14.56900\pm0.00033}}        
\newcommand{\hatcurLCiblendBeccenxxxxxxA}{\ensuremath{0.720\pm0.021}}              
\newcommand{\hatcurLChatnetmCeccenxxxxxxA}{\ensuremath{-5.92498\pm0.00017}}        
\newcommand{\hatcurLCiblendCeccenxxxxxxA}{\ensuremath{1.016\pm0.011}}              
\newcommand{\hatcurLChatnetmDeccenxxxxxxA}{\ensuremath{-5.97859\pm0.00018}}        
\newcommand{\hatcurLCiblendDeccenxxxxxxA}{\ensuremath{1.027\pm0.013}}              
\newcommand{\hatcurLCrhoeccenxxxxxxA}{\ensuremath{10.91\pm0.30}}                   
\newcommand{\hatcurSMEiteffeccenxxxxxxA}{\ensuremath{3367\pm70}}                   
\newcommand{\hatcurSMEizfeheccenxxxxxxA}{\ensuremath{-0.010\pm0.090}}              
\newcommand{\hatcurSMEizfehshorteccenxxxxxxA}{\ensuremath{-0.01}}                  
\newcommand{\hatcurSMEiloggeccenxxxxxxA}{\ensuremath{5.00\pm0.50}}                 
\newcommand{\hatcurSMEivsineccenxxxxxxA}{\ensuremath{0\pm50}}                      
\newcommand{\hatcurSMEivmaceccenxxxxxxA}{\ensuremath{nff\pmnff}}                   
\newcommand{\hatcurSMEivmiceccenxxxxxxA}{\ensuremath{nff\pmnff}}                   
\newcommand{\hatcurextraerrMJeccenxxxxxxA}{\ensuremath{0\pm0}}                     
\newcommand{\hatcurextraerrMJtwosiglimeccenxxxxxxA}{\ensuremath{<0.0200}}          
\newcommand{\hatcurextraerrMHeccenxxxxxxA}{\ensuremath{0\pm0}}                     
\newcommand{\hatcurextraerrMHtwosiglimeccenxxxxxxA}{\ensuremath{<0.0200}}          
\newcommand{\hatcurextraerrMKseccenxxxxxxA}{\ensuremath{0\pm0}}                    
\newcommand{\hatcurextraerrMKstwosiglimeccenxxxxxxA}{\ensuremath{<0.0200}}         
\newcommand{\hatcurextraerrMGeccenxxxxxxA}{\ensuremath{0\pm0}}                     
\newcommand{\hatcurextraerrMGtwosiglimeccenxxxxxxA}{\ensuremath{<0.0200}}          
\newcommand{\hatcurextraerrMRPeccenxxxxxxA}{\ensuremath{0\pm0}}                    
\newcommand{\hatcurextraerrMRPtwosiglimeccenxxxxxxA}{\ensuremath{<0.0200}}         
\newcommand{\hatcurextraerrMWoneeccenxxxxxxA}{\ensuremath{0\pm0}}                  
\newcommand{\hatcurextraerrMWonetwosiglimeccenxxxxxxA}{\ensuremath{<0.0200}}       
\newcommand{\hatcurextraerrMWtwoeccenxxxxxxA}{\ensuremath{0\pm0}}                  
\newcommand{\hatcurextraerrMWtwotwosiglimeccenxxxxxxA}{\ensuremath{<0.0200}}       
\newcommand{\hatcurLBiBeccenxxxxxxA}{\ensuremath{0.25\pm0.14}}                     
\newcommand{\hatcurLBiiBeccenxxxxxxA}{\ensuremath{0.18\pm0.18}}                    
\newcommand{\hatcurLBiVeccenxxxxxxA}{\ensuremath{0.4944}}                          
\newcommand{\hatcurLBiiVeccenxxxxxxA}{\ensuremath{0.3514}}                         
\newcommand{\hatcurLBiReccenxxxxxxA}{\ensuremath{0.4204}}                          
\newcommand{\hatcurLBiiReccenxxxxxxA}{\ensuremath{0.3353}}                         
\newcommand{\hatcurLBiIeccenxxxxxxA}{\ensuremath{0.2071}}                          
\newcommand{\hatcurLBiiIeccenxxxxxxA}{\ensuremath{0.3944}}                         
\newcommand{\hatcurLBiueccenxxxxxxA}{\ensuremath{0.4340}}                          
\newcommand{\hatcurLBiiueccenxxxxxxA}{\ensuremath{0.3422}}                         
\newcommand{\hatcurLBigeccenxxxxxxA}{\ensuremath{0.51\pm0.11}}                     
\newcommand{\hatcurLBiigeccenxxxxxxA}{\ensuremath{0.33\pm0.16}}                    
\newcommand{\hatcurLBireccenxxxxxxA}{\ensuremath{0.53\pm0.10}}                     
\newcommand{\hatcurLBiireccenxxxxxxA}{\ensuremath{0.21\pm0.15}}                    
\newcommand{\hatcurLBiieccenxxxxxxA}{\ensuremath{0.410\pm0.093}}                   
\newcommand{\hatcurLBiiieccenxxxxxxA}{\ensuremath{0.16\pm0.15}}                    
\newcommand{\hatcurLBizeccenxxxxxxA}{\ensuremath{0.343\pm0.081}}                   
\newcommand{\hatcurLBiizeccenxxxxxxA}{\ensuremath{0.08\pm0.14}}                    
\newcommand{\hatcurLBiJeccenxxxxxxA}{\ensuremath{0.161\pm0.086}}                   
\newcommand{\hatcurLBiiJeccenxxxxxxA}{\ensuremath{0.28\pm0.14}}                    
\newcommand{\hatcurLBiHeccenxxxxxxA}{\ensuremath{0.0905}}                          
\newcommand{\hatcurLBiiHeccenxxxxxxA}{\ensuremath{0.2823}}                         
\newcommand{\hatcurLBiKeccenxxxxxxA}{\ensuremath{0.0840}}                          
\newcommand{\hatcurLBiiKeccenxxxxxxA}{\ensuremath{0.2456}}                         
\newcommand{\hatcurLBiTeccenxxxxxxA}{\ensuremath{0.239\pm0.096}}                   
\newcommand{\hatcurLBiiTeccenxxxxxxA}{\ensuremath{0.44\pm0.15}}                    
\newcommand{\hatcurLBikepeccenxxxxxxA}{\ensuremath{0.2921}}                        
\newcommand{\hatcurLBiikepeccenxxxxxxA}{\ensuremath{0.4295}}                       
\newcommand{\hatcurLBiCeccenxxxxxxA}{\ensuremath{0.2393}}                          
\newcommand{\hatcurLBiiCeccenxxxxxxA}{\ensuremath{0.4349}}                         
\newcommand{\hatcurLBiMeccenxxxxxxA}{\ensuremath{0.4071}}                          
\newcommand{\hatcurLBiiMeccenxxxxxxA}{\ensuremath{0.4027}}                         
\newcommand{\hatcurLBiSoneeccenxxxxxxA}{\ensuremath{0.0657}}                         
\newcommand{\hatcurLBiiSoneeccenxxxxxxA}{\ensuremath{0.1832}}                        
\newcommand{\hatcurLBiStwoeccenxxxxxxA}{\ensuremath{0.0484}}                         
\newcommand{\hatcurLBiiStwoeccenxxxxxxA}{\ensuremath{0.1534}}                        
\newcommand{\hatcurLBiSthreeeccenxxxxxxA}{\ensuremath{0.0531}}                         
\newcommand{\hatcurLBiiSthreeeccenxxxxxxA}{\ensuremath{0.1359}}                        
\newcommand{\hatcurLBiSfoureccenxxxxxxA}{\ensuremath{0.0560}}                         
\newcommand{\hatcurLBiiSfoureccenxxxxxxA}{\ensuremath{0.1045}}                        
\newcommand{\hatcurISOmeccenxxxxxxA}{\ensuremath{0.368\pm0.019}}                   
\newcommand{\hatcurISOmshorteccenxxxxxxA}{\ensuremath{0.37}}                       
\newcommand{\hatcurISOmlongeccenxxxxxxA}{\ensuremath{0.368\pm0.019}}               
\newcommand{\hatcurISOreccenxxxxxxA}{\ensuremath{0.3620\pm0.0070}}                 
\newcommand{\hatcurISOrshorteccenxxxxxxA}{\ensuremath{0.36}}                       
\newcommand{\hatcurISOrlongeccenxxxxxxA}{\ensuremath{0.3620\pm0.0070}}             
\newcommand{\hatcurISOrhoeccenxxxxxxA}{\ensuremath{10.91\pm0.30}}                  
\newcommand{\hatcurISOrholongeccenxxxxxxA}{\ensuremath{10.91\pm0.30}}              
\newcommand{\hatcurISOloggeccenxxxxxxA}{\ensuremath{4.886\pm0.011}}                
\newcommand{\hatcurISOlumeccenxxxxxxA}{\ensuremath{0.0149\pm0.0013}}               
\newcommand{\hatcurISOlumshorteccenxxxxxxA}{\ensuremath{0.01}}                     
\newcommand{\hatcurISOteffeccenxxxxxxA}{\ensuremath{3351\pm66}}                    
\newcommand{\hatcurISOzfeheccenxxxxxxA}{\ensuremath{0.261\pm0.079}}                
\newcommand{\hatcurISOageeccenxxxxxxA}{\ensuremath{3.1_{-2.2}^{+3.8}}}             
\newcommand{\hatcurISOspececcenxxxxxxA}{M}                                         
\newcommand{\hatcurRVKeccenxxxxxxA}{\ensuremath{190\pm21}}                         
\newcommand{\hatcurRVKtwosiglimeccenxxxxxxA}{\ensuremath{<224.2}}                  
\newcommand{\hatcurRVrkeccenxxxxxxA}{\ensuremath{0.028\pm0.079}}                   
\newcommand{\hatcurRVrheccenxxxxxxA}{\ensuremath{0.141_{-0.059}^{+0.037}}}         
\newcommand{\hatcurRVkeccenxxxxxxA}{\ensuremath{0.004_{-0.011}^{+0.018}}}          
\newcommand{\hatcurRVheccenxxxxxxA}{\ensuremath{0.023\pm0.012}}                    
\newcommand{\hatcurRVtroneeccenxxxxxxA}{\ensuremath{0\pm0}}                        
\newcommand{\hatcurRVtrtwoeccenxxxxxxA}{\ensuremath{0\pm0}}                        
\newcommand{\hatcurRVgammaAeccenxxxxxxA}{\ensuremath{7\pm32}}                      
\newcommand{\hatcurRVjitterAeccenxxxxxxA}{\ensuremath{73\pm30}}                    
\newcommand{\hatcurRVjittertwosiglimAeccenxxxxxxA}{\ensuremath{<131.6}}            
\newcommand{\hatcurRVfitrmsAeccenxxxxxxA}{\ensuremath{0.0}}                        
\newcommand{\hatcurRVgammaBeccenxxxxxxA}{\ensuremath{7\pm17}}                      
\newcommand{\hatcurRVjitterBeccenxxxxxxA}{\ensuremath{23\pm10}}                    
\newcommand{\hatcurRVjittertwosiglimBeccenxxxxxxA}{\ensuremath{<39.9}}             
\newcommand{\hatcurRVfitrmsBeccenxxxxxxA}{\ensuremath{0.0}}                        
\newcommand{\hatcurRVecceneccenxxxxxxA}{\ensuremath{0.027\pm0.013}}                
\newcommand{\hatcurRVeccentwosiglimeccenxxxxxxA}{\ensuremath{<0.048}}              
\newcommand{\hatcurRVomegaeccenxxxxxxA}{\ensuremath{80\pm42}}                      
\newcommand{\hatcurPPieccenxxxxxxA}{\ensuremath{88.21\pm0.15}}                     
\newcommand{\hatcurPPgeccenxxxxxxA}{\ensuremath{11.2\pm1.3}}                       
\newcommand{\hatcurPPloggeccenxxxxxxA}{\ensuremath{3.051\pm0.051}}                 
\newcommand{\hatcurPPareccenxxxxxxA}{\ensuremath{9.741\pm0.088}}                   
\newcommand{\hatcurPPareleccenxxxxxxA}{\ensuremath{0.01642\pm0.00029}}             
\newcommand{\hatcurPPrhoeccenxxxxxxA}{\ensuremath{0.525\pm0.062}}                  
\newcommand{\hatcurPPmeccenxxxxxxA}{\ensuremath{0.520\pm0.062}}                    
\newcommand{\hatcurPPmtwosiglimeccenxxxxxxA}{\ensuremath{<0.62}}                   
\newcommand{\hatcurPPmshorteccenxxxxxxA}{\ensuremath{0.52}}                        
\newcommand{\hatcurPPmlongeccenxxxxxxA}{\ensuremath{0.520\pm0.062}}                
\newcommand{\hatcurPPmeeccenxxxxxxA}{\ensuremath{165\pm20}}                        
\newcommand{\hatcurPPmeshorteccenxxxxxxA}{\ensuremath{165.3}}                      
\newcommand{\hatcurPPmelongeccenxxxxxxA}{\ensuremath{165\pm20}}                    
\newcommand{\hatcurPPreccenxxxxxxA}{\ensuremath{1.071\pm0.022}}                    
\newcommand{\hatcurPPrshorteccenxxxxxxA}{\ensuremath{1.07}}                        
\newcommand{\hatcurPPrlongeccenxxxxxxA}{\ensuremath{1.071\pm0.022}}                
\newcommand{\hatcurPPreeccenxxxxxxA}{\ensuremath{12.00\pm0.24}}                    
\newcommand{\hatcurPPreshorteccenxxxxxxA}{\ensuremath{12.0}}                       
\newcommand{\hatcurPPrelongeccenxxxxxxA}{\ensuremath{12.00\pm0.24}}                
\newcommand{\hatcurPPmrcorreccenxxxxxxA}{\ensuremath{0.27}}                        
\newcommand{\hatcurPPteffeccenxxxxxxA}{\ensuremath{759\pm15}}                      
\newcommand{\hatcurPPthetaeccenxxxxxxA}{\ensuremath{0.0432\pm0.0049}}              
\newcommand{\hatcurPPfluxperieccenxxxxxxA}{\ensuremath{7.93\pm0.74}}               
\newcommand{\hatcurPPfluxperidimeccenxxxxxxA}{\ensuremath{7}}                      
\newcommand{\hatcurPPfluxapeccenxxxxxxA}{\ensuremath{7.14\pm0.57}}                 
\newcommand{\hatcurPPfluxapdimeccenxxxxxxA}{\ensuremath{7}}                        
\newcommand{\hatcurPPfluxavgeccenxxxxxxA}{\ensuremath{7.52\pm0.62}}                
\newcommand{\hatcurPPfluxavgdimeccenxxxxxxA}{\ensuremath{7}}                       
\newcommand{\hatcurPPfluxavglogeccenxxxxxxA}{\ensuremath{7.877\pm0.035}}           
\newcommand{\hatcurXsecphaseeccenxxxxxxA}{\ensuremath{0.5025\pm0.0094}}            
\newcommand{\hatcurXsecondaryeccenxxxxxxA}{\ensuremath{2458921.427\pm0.012}}       
\newcommand{\hatcurXsecdureccenxxxxxxA}{\ensuremath{0.0535\pm0.0011}}              
\newcommand{\hatcurXsecingdureccenxxxxxxA}{\ensuremath{0.01360\pm0.00033}}         
\newcommand{\hatcurPPphiconjeccenxxxxxxA}{\ensuremath{0.03\pm0.10}}                
\newcommand{\hatcurPPperieccenxxxxxxA}{\ensuremath{2458920.75\pm0.13}}             
\newcommand{\hatcurPPaequiveccenxxxxxxA}{\ensuremath{0.1345\pm0.0055}}             
\newcommand{\hatcurPPtcirceccenxxxxxxA}{\ensuremath{1.60\pm0.21}}                  
\newcommand{\hatcurPPtinfalleccenxxxxxxA}{\ensuremath{749\pm99}}                   
\newcommand{\hatcurXdisteccenxxxxxxA}{\ensuremath{115.00\pm0.50}}                  
\newcommand{\hatcurXAveccenxxxxxxA}{\ensuremath{0.0110\pm0.0079}}                  
\newcommand{\hatcurXdistredeccenxxxxxxA}{\ensuremath{115.04\pm0.49}}               
\newcommand{\hatcurXEBVeccenxxxxxxA}{\ensuremath{0.0030_{-0.0020}^{+0.0030}}}      
\newcommand{\hatcurCCpmraeccenxxxxxxA}{\ensuremath{-41.959\pm0.029}}               
\newcommand{\hatcurCCpmdececcenxxxxxxA}{\ensuremath{29.074\pm0.027}}               
\newcommand{\hatcurCCpmeccenxxxxxxA}{\ensuremath{51.048\pm0.040}}                  

%% file: TOI3629.eccen.tex
\newcommand{\hatcurhtreccenxxxxxxB}{TOI3629}                          
\newcommand{\hatcurfieldeccenxxxxxxB}{\ensuremath{string}}            
\newcommand{\hatcurCCraeccenxxxxxxB}{\ensuremath{23^{\mathrm h}59^{\mathrm m}10.3200{\mathrm s}}}                   
\newcommand{\hatcurCCdececcenxxxxxxB}{\ensuremath{+39{\arcdeg}18{\arcmin}51.3200{\arcsec}}}                 
\newcommand{\hatcurCCmageccenxxxxxxB}{14.639}                         
\newcommand{\hatcurCCtwomasseccenxxxxxxB}{2MASS~23591015+3918514}     
\newcommand{\hatcurCCgsceccenxxxxxxB}{GSC~NULL}                       
\newcommand{\hatcurCCgaiaeccenxxxxxxB}{GAIA~2881820319999000960}      
\newcommand{\hatcurCCgaiadrtwoeccenxxxxxxB}{GAIA~DR2~2881820324294985856} 
\newcommand{\hatcurCCtassmveccenxxxxxxB}{\ensuremath{14.639\pm0.061}} 
\newcommand{\hatcurCCtassmvshorteccenxxxxxxB}{\ensuremath{14.6}}      
\newcommand{\hatcurCCtassmBeccenxxxxxxB}{\ensuremath{16.093\pm0.041}} 
\newcommand{\hatcurCCtassmBshorteccenxxxxxxB}{\ensuremath{16.1}}      
\newcommand{\hatcurCCtassmIeccenxxxxxxB}{\ensuremath{nff\pmnff}}      
\newcommand{\hatcurCCtassmIshorteccenxxxxxxB}{\ensuremath{0.0}}       
\newcommand{\hatcurCCtassmgeccenxxxxxxB}{\ensuremath{15.380\pm0.022}} 
\newcommand{\hatcurCCtassmgshorteccenxxxxxxB}{\ensuremath{15.4}}      
\newcommand{\hatcurCCtassmreccenxxxxxxB}{\ensuremath{14.055\pm0.057}} 
\newcommand{\hatcurCCtassmrshorteccenxxxxxxB}{\ensuremath{14.1}}      
\newcommand{\hatcurCCtassmieccenxxxxxxB}{\ensuremath{13.111\pm0.032}} 
\newcommand{\hatcurCCtassmishorteccenxxxxxxB}{\ensuremath{13.1}}      
\newcommand{\hatcurCCparallaxeccenxxxxxxB}{\ensuremath{7.667\pm0.017}} 
\newcommand{\hatcurCCgaiamGeccenxxxxxxB}{\ensuremath{13.8226\pm0.0028}} 
\newcommand{\hatcurCCgaiamBPeccenxxxxxxB}{\ensuremath{14.8939\pm0.0031}} 
\newcommand{\hatcurCCgaiamRPeccenxxxxxxB}{\ensuremath{12.7937\pm0.0038}} 
\newcommand{\hatcurCCtwomassJmageccenxxxxxxB}{\ensuremath{11.424\pm0.027}} 
\newcommand{\hatcurCCtwomassHmageccenxxxxxxB}{\ensuremath{10.733\pm0.030}} 
\newcommand{\hatcurCCtwomassKmageccenxxxxxxB}{\ensuremath{10.553\pm0.020}} 
\newcommand{\hatcurCCcitJmageccenxxxxxxB}{\ensuremath{11.414\pm0.028}} 
\newcommand{\hatcurCCcitHmageccenxxxxxxB}{\ensuremath{10.726\pm0.030}} 
\newcommand{\hatcurCCcitKmageccenxxxxxxB}{\ensuremath{10.577\pm0.021}} 
\newcommand{\hatcurCCbbJmageccenxxxxxxB}{\ensuremath{11.505\pm0.030}} 
\newcommand{\hatcurCCbbHmageccenxxxxxxB}{\ensuremath{10.749\pm0.032}} 
\newcommand{\hatcurCCbbKmageccenxxxxxxB}{\ensuremath{10.597\pm0.021}} 
\newcommand{\hatcurCCesoJmageccenxxxxxxB}{\ensuremath{11.513\pm0.035}} 
\newcommand{\hatcurCCesoHmageccenxxxxxxB}{\ensuremath{10.747\pm0.041}} 
\newcommand{\hatcurCCesoKmageccenxxxxxxB}{\ensuremath{10.593\pm0.023}} 
\newcommand{\hatcurCCesoJHmageccenxxxxxxB}{\ensuremath{0.765\pm0.050}} 
\newcommand{\hatcurCCesoJKmageccenxxxxxxB}{\ensuremath{0.920\pm0.039}} 
\newcommand{\hatcurCCesoHKmageccenxxxxxxB}{\ensuremath{0.154\pm0.045}} 
\newcommand{\hatcurCCWonemageccenxxxxxxB}{\ensuremath{10.484\pm0.022}} 
\newcommand{\hatcurCCWtwomageccenxxxxxxB}{\ensuremath{10.516\pm0.020}} 
\newcommand{\hatcurCCWthreemageccenxxxxxxB}{\ensuremath{10.375\pm0.067}} 
\newcommand{\hatcurCCWfourmageccenxxxxxxB}{\ensuremath{nff\pmnff}}    
\newcommand{\hatcurLCdipeccenxxxxxxB}{\ensuremath{20.0}}              
\newcommand{\hatcurLCrprstareccenxxxxxxB}{\ensuremath{0.1239\pm0.0013}} 
\newcommand{\hatcurLCbsqeccenxxxxxxB}{\ensuremath{0.053_{-0.021}^{+0.025}}} 
\newcommand{\hatcurLCimpeccenxxxxxxB}{\ensuremath{0.231_{-0.051}^{+0.050}}} 
\newcommand{\hatcurLCzetaeccenxxxxxxB}{\ensuremath{24.88\pm0.15}}     
\newcommand{\hatcurLCdureccenxxxxxxB}{\ensuremath{0.09087\pm0.00058}} 
\newcommand{\hatcurLCdurshorteccenxxxxxxB}{\ensuremath{0.0909}}       
\newcommand{\hatcurLCdurhreccenxxxxxxB}{\ensuremath{2.181\pm0.014}}   
\newcommand{\hatcurLCdurhrshorteccenxxxxxxB}{\ensuremath{2.181}}      
\newcommand{\hatcurLCqeccenxxxxxxB}{\ensuremath{0.02310\pm0.00015}}   
\newcommand{\hatcurLCqshorteccenxxxxxxB}{\ensuremath{0.023}}          
\newcommand{\hatcurLCingdureccenxxxxxxB}{\ensuremath{0.01051\pm0.00032}} 
\newcommand{\hatcurLCPeccenxxxxxxB}{\ensuremath{3.9365578\pm0.0000038}} 
\newcommand{\hatcurLCPprececcenxxxxxxB}{\ensuremath{3.9365578}}       
\newcommand{\hatcurLCPshorteccenxxxxxxB}{\ensuremath{3.9366}}         
\newcommand{\hatcurLCTeccenxxxxxxB}{\ensuremath{2459579.44012\pm0.00016}} 
\newcommand{\hatcurLCTAeccenxxxxxxB}{\ensuremath{2458768.50919\pm0.00079}} 
\newcommand{\hatcurLCTBeccenxxxxxxB}{\ensuremath{2459878.61852\pm0.00034}} 
\newcommand{\hatcurLChatnetmAeccenxxxxxxB}{\ensuremath{-0.000060\pm0.000072}} 
\newcommand{\hatcurLCiblendAeccenxxxxxxB}{\ensuremath{0.847\pm0.038}} 
\newcommand{\hatcurLChatnetmBeccenxxxxxxB}{\ensuremath{-7.707110\pm0.000045}} 
\newcommand{\hatcurLCiblendBeccenxxxxxxB}{\ensuremath{0.953\pm0.025}} 
\newcommand{\hatcurLCrhoeccenxxxxxxB}{\ensuremath{3.92\pm0.12}}       
\newcommand{\hatcurSMEiteffeccenxxxxxxB}{\ensuremath{3870\pm90}}      
\newcommand{\hatcurSMEizfeheccenxxxxxxB}{\ensuremath{0.40\pm0.10}}    
\newcommand{\hatcurSMEizfehshorteccenxxxxxxB}{\ensuremath{0.40}}      
\newcommand{\hatcurSMEiloggeccenxxxxxxB}{\ensuremath{4.670\pm0.050}}  
\newcommand{\hatcurSMEivsineccenxxxxxxB}{\ensuremath{0.0\pm2.0}}      
\newcommand{\hatcurSMEivmaceccenxxxxxxB}{\ensuremath{nff\pmnff}}      
\newcommand{\hatcurSMEivmiceccenxxxxxxB}{\ensuremath{nff\pmnff}}      
\newcommand{\hatcurextraerrMJeccenxxxxxxB}{\ensuremath{0\pm0}}        
\newcommand{\hatcurextraerrMJtwosiglimeccenxxxxxxB}{\ensuremath{<0.0200}} 
\newcommand{\hatcurextraerrMHeccenxxxxxxB}{\ensuremath{0\pm0}}        
\newcommand{\hatcurextraerrMHtwosiglimeccenxxxxxxB}{\ensuremath{<0.0200}} 
\newcommand{\hatcurextraerrMKseccenxxxxxxB}{\ensuremath{0\pm0}}       
\newcommand{\hatcurextraerrMKstwosiglimeccenxxxxxxB}{\ensuremath{<0.0200}} 
\newcommand{\hatcurextraerrMGeccenxxxxxxB}{\ensuremath{0\pm0}}        
\newcommand{\hatcurextraerrMGtwosiglimeccenxxxxxxB}{\ensuremath{<0.0200}} 
\newcommand{\hatcurextraerrMRPeccenxxxxxxB}{\ensuremath{0\pm0}}       
\newcommand{\hatcurextraerrMRPtwosiglimeccenxxxxxxB}{\ensuremath{<0.0200}} 
\newcommand{\hatcurextraerrMgeccenxxxxxxB}{\ensuremath{0\pm0}}        
\newcommand{\hatcurextraerrMgtwosiglimeccenxxxxxxB}{\ensuremath{<0.0200}} 
\newcommand{\hatcurextraerrMreccenxxxxxxB}{\ensuremath{0\pm0}}        
\newcommand{\hatcurextraerrMrtwosiglimeccenxxxxxxB}{\ensuremath{<0.0200}} 
\newcommand{\hatcurextraerrMieccenxxxxxxB}{\ensuremath{0\pm0}}        
\newcommand{\hatcurextraerrMitwosiglimeccenxxxxxxB}{\ensuremath{<0.0200}} 
\newcommand{\hatcurextraerrMWoneeccenxxxxxxB}{\ensuremath{0\pm0}}     
\newcommand{\hatcurextraerrMWonetwosiglimeccenxxxxxxB}{\ensuremath{<0.0200}} 
\newcommand{\hatcurextraerrMWtwoeccenxxxxxxB}{\ensuremath{0\pm0}}     
\newcommand{\hatcurextraerrMWtwotwosiglimeccenxxxxxxB}{\ensuremath{<0.0200}} 
\newcommand{\hatcurLBiBeccenxxxxxxB}{\ensuremath{0.5294}}             
\newcommand{\hatcurLBiiBeccenxxxxxxB}{\ensuremath{0.2909}}            
\newcommand{\hatcurLBiVeccenxxxxxxB}{\ensuremath{0.4674}}             
\newcommand{\hatcurLBiiVeccenxxxxxxB}{\ensuremath{0.3330}}            
\newcommand{\hatcurLBiReccenxxxxxxB}{\ensuremath{0.4192}}             
\newcommand{\hatcurLBiiReccenxxxxxxB}{\ensuremath{0.3091}}            
\newcommand{\hatcurLBiIeccenxxxxxxB}{\ensuremath{0.2292}}             
\newcommand{\hatcurLBiiIeccenxxxxxxB}{\ensuremath{0.3468}}            
\newcommand{\hatcurLBiueccenxxxxxxB}{\ensuremath{0.5326}}             
\newcommand{\hatcurLBiiueccenxxxxxxB}{\ensuremath{0.2706}}            
\newcommand{\hatcurLBigeccenxxxxxxB}{\ensuremath{0.39\pm0.12}}        
\newcommand{\hatcurLBiigeccenxxxxxxB}{\ensuremath{0.32\pm0.16}}       
\newcommand{\hatcurLBireccenxxxxxxB}{\ensuremath{0.43\pm0.11}}        
\newcommand{\hatcurLBiireccenxxxxxxB}{\ensuremath{0.25\pm0.15}}       
\newcommand{\hatcurLBiieccenxxxxxxB}{\ensuremath{0.353\pm0.089}}      
\newcommand{\hatcurLBiiieccenxxxxxxB}{\ensuremath{0.27\pm0.14}}       
\newcommand{\hatcurLBizeccenxxxxxxB}{\ensuremath{0.225\pm0.092}}      
\newcommand{\hatcurLBiizeccenxxxxxxB}{\ensuremath{0.34\pm0.15}}       
\newcommand{\hatcurLBiJeccenxxxxxxB}{\ensuremath{0.1266}}             
\newcommand{\hatcurLBiiJeccenxxxxxxB}{\ensuremath{0.2533}}            
\newcommand{\hatcurLBiHeccenxxxxxxB}{\ensuremath{0.0987}}             
\newcommand{\hatcurLBiiHeccenxxxxxxB}{\ensuremath{0.2706}}            
\newcommand{\hatcurLBiKeccenxxxxxxB}{\ensuremath{0.0784}}             
\newcommand{\hatcurLBiiKeccenxxxxxxB}{\ensuremath{0.2508}}            
\newcommand{\hatcurLBiTeccenxxxxxxB}{\ensuremath{0.27\pm0.10}}        
\newcommand{\hatcurLBiiTeccenxxxxxxB}{\ensuremath{0.31\pm0.15}}       
\newcommand{\hatcurLBikepeccenxxxxxxB}{\ensuremath{0.3414}}           
\newcommand{\hatcurLBiikepeccenxxxxxxB}{\ensuremath{0.3695}}          
\newcommand{\hatcurLBiCeccenxxxxxxB}{\ensuremath{0.3005}}             
\newcommand{\hatcurLBiiCeccenxxxxxxB}{\ensuremath{0.3735}}            
\newcommand{\hatcurLBiMeccenxxxxxxB}{\ensuremath{0.4294}}             
\newcommand{\hatcurLBiiMeccenxxxxxxB}{\ensuremath{0.3538}}            
\newcommand{\hatcurLBiSoneeccenxxxxxxB}{\ensuremath{0.0646}}            
\newcommand{\hatcurLBiiSoneeccenxxxxxxB}{\ensuremath{0.1906}}           
\newcommand{\hatcurLBiStwoeccenxxxxxxB}{\ensuremath{0.0527}}            
\newcommand{\hatcurLBiiStwoeccenxxxxxxB}{\ensuremath{0.1496}}           
\newcommand{\hatcurLBiSthreeeccenxxxxxxB}{\ensuremath{0.0536}}            
\newcommand{\hatcurLBiiSthreeeccenxxxxxxB}{\ensuremath{0.1304}}           
\newcommand{\hatcurLBiSfoureccenxxxxxxB}{\ensuremath{0.0638}}            
\newcommand{\hatcurLBiiSfoureccenxxxxxxB}{\ensuremath{0.0992}}           
\newcommand{\hatcurISOmeccenxxxxxxB}{\ensuremath{0.640\pm0.030}}      
\newcommand{\hatcurISOmshorteccenxxxxxxB}{\ensuremath{0.64}}          
\newcommand{\hatcurISOmlongeccenxxxxxxB}{\ensuremath{0.640\pm0.030}}  
\newcommand{\hatcurISOreccenxxxxxxB}{\ensuremath{0.6118\pm0.0094}}    
\newcommand{\hatcurISOrshorteccenxxxxxxB}{\ensuremath{0.61}}          
\newcommand{\hatcurISOrlongeccenxxxxxxB}{\ensuremath{0.6118\pm0.0094}} 
\newcommand{\hatcurISOrhoeccenxxxxxxB}{\ensuremath{3.92\pm0.12}}      
\newcommand{\hatcurISOrholongeccenxxxxxxB}{\ensuremath{3.92\pm0.12}}  
\newcommand{\hatcurISOloggeccenxxxxxxB}{\ensuremath{4.670\pm0.013}}   
\newcommand{\hatcurISOlumeccenxxxxxxB}{\ensuremath{0.0745\pm0.0063}}  
\newcommand{\hatcurISOlumshorteccenxxxxxxB}{\ensuremath{0.07}}        
\newcommand{\hatcurISOteffeccenxxxxxxB}{\ensuremath{3853\pm72}}       
\newcommand{\hatcurISOzfeheccenxxxxxxB}{\ensuremath{0.519\pm0.083}}   
\newcommand{\hatcurISOageeccenxxxxxxB}{\ensuremath{9.4_{-5.1}^{+6.6}}} 
\newcommand{\hatcurISOspececcenxxxxxxB}{M}                            
\newcommand{\hatcurRVKeccenxxxxxxB}{\ensuremath{39.1\pm3.2}}          
\newcommand{\hatcurRVKtwosiglimeccenxxxxxxB}{\ensuremath{<45.0}}      
\newcommand{\hatcurRVrkeccenxxxxxxB}{\ensuremath{0.052\pm0.085}}      
\newcommand{\hatcurRVrheccenxxxxxxB}{\ensuremath{0.146_{-0.059}^{+0.045}}} 
\newcommand{\hatcurRVkeccenxxxxxxB}{\ensuremath{0.008_{-0.015}^{+0.019}}} 
\newcommand{\hatcurRVheccenxxxxxxB}{\ensuremath{0.026\pm0.014}}       
\newcommand{\hatcurRVtroneeccenxxxxxxB}{\ensuremath{0\pm0}}           
\newcommand{\hatcurRVtrtwoeccenxxxxxxB}{\ensuremath{0\pm0}}           
\newcommand{\hatcurRVgammaAeccenxxxxxxB}{\ensuremath{-8.9\pm2.2}}     
\newcommand{\hatcurRVjitterAeccenxxxxxxB}{\ensuremath{13.6\pm4.2}}    
\newcommand{\hatcurRVjittertwosiglimAeccenxxxxxxB}{\ensuremath{<22.3}} 
\newcommand{\hatcurRVfitrmsAeccenxxxxxxB}{\ensuremath{0.0}}           
\newcommand{\hatcurRVgammaBeccenxxxxxxB}{\ensuremath{-16.1\pm3.8}}    
\newcommand{\hatcurRVjitterBeccenxxxxxxB}{\ensuremath{0.3\pm5.5}}     
\newcommand{\hatcurRVjittertwosiglimBeccenxxxxxxB}{\ensuremath{<13.6}} 
\newcommand{\hatcurRVfitrmsBeccenxxxxxxB}{\ensuremath{0.0}}           
\newcommand{\hatcurRVgammaCeccenxxxxxxB}{\ensuremath{-5.6\pm5.8}}     
\newcommand{\hatcurRVjitterCeccenxxxxxxB}{\ensuremath{0\pm14}}        
\newcommand{\hatcurRVjittertwosiglimCeccenxxxxxxB}{\ensuremath{<22.3}} 
\newcommand{\hatcurRVfitrmsCeccenxxxxxxB}{\ensuremath{0.0}}           
\newcommand{\hatcurRVecceneccenxxxxxxB}{\ensuremath{0.031\pm0.015}}   
\newcommand{\hatcurRVeccentwosiglimeccenxxxxxxB}{\ensuremath{<0.056}} 
\newcommand{\hatcurRVomegaeccenxxxxxxB}{\ensuremath{71\pm41}}         
\newcommand{\hatcurPPieccenxxxxxxB}{\ensuremath{89.09\pm0.21}}        
\newcommand{\hatcurPPgeccenxxxxxxB}{\ensuremath{10.28\pm0.94}}        
\newcommand{\hatcurPPloggeccenxxxxxxB}{\ensuremath{3.012\pm0.039}}    
\newcommand{\hatcurPPareccenxxxxxxB}{\ensuremath{14.76\pm0.15}}       
\newcommand{\hatcurPPareleccenxxxxxxB}{\ensuremath{0.04204\pm0.00065}} 
\newcommand{\hatcurPPrhoeccenxxxxxxB}{\ensuremath{0.698\pm0.068}}     
\newcommand{\hatcurPPmeccenxxxxxxB}{\ensuremath{0.224_{-0.017}^{+0.023}}} 
\newcommand{\hatcurPPmtwosiglimeccenxxxxxxB}{\ensuremath{<0.26}}      
\newcommand{\hatcurPPmshorteccenxxxxxxB}{\ensuremath{0.22}}           
\newcommand{\hatcurPPmlongeccenxxxxxxB}{\ensuremath{0.224_{-0.017}^{+0.023}}} 
\newcommand{\hatcurPPmeeccenxxxxxxB}{\ensuremath{71.1_{-5.4}^{+7.2}}} 
\newcommand{\hatcurPPmeshorteccenxxxxxxB}{\ensuremath{71.1}}          
\newcommand{\hatcurPPmelongeccenxxxxxxB}{\ensuremath{71.1_{-5.4}^{+7.2}}} 
\newcommand{\hatcurPPreccenxxxxxxB}{\ensuremath{0.738\pm0.013}}       
\newcommand{\hatcurPPrshorteccenxxxxxxB}{\ensuremath{0.74}}           
\newcommand{\hatcurPPrlongeccenxxxxxxB}{\ensuremath{0.738\pm0.013}}   
\newcommand{\hatcurPPreeccenxxxxxxB}{\ensuremath{8.27\pm0.14}}        
\newcommand{\hatcurPPreshorteccenxxxxxxB}{\ensuremath{8.3}}           
\newcommand{\hatcurPPrelongeccenxxxxxxB}{\ensuremath{8.27\pm0.14}}    
\newcommand{\hatcurPPmrcorreccenxxxxxxB}{\ensuremath{0.14}}           
\newcommand{\hatcurPPteffeccenxxxxxxB}{\ensuremath{709\pm14}}         
\newcommand{\hatcurPPthetaeccenxxxxxxB}{\ensuremath{0.0402\pm0.0034}} 
\newcommand{\hatcurPPfluxperieccenxxxxxxB}{\ensuremath{6.13\pm0.53}}  
\newcommand{\hatcurPPfluxperidimeccenxxxxxxB}{\ensuremath{7}}         
\newcommand{\hatcurPPfluxapeccenxxxxxxB}{\ensuremath{5.39\pm0.42}}    
\newcommand{\hatcurPPfluxapdimeccenxxxxxxB}{\ensuremath{7}}           
\newcommand{\hatcurPPfluxavgeccenxxxxxxB}{\ensuremath{5.73\pm0.44}}   
\newcommand{\hatcurPPfluxavgdimeccenxxxxxxB}{\ensuremath{7}}          
\newcommand{\hatcurPPfluxavglogeccenxxxxxxB}{\ensuremath{7.758\pm0.034}} 
\newcommand{\hatcurXsecphaseeccenxxxxxxB}{\ensuremath{0.505\pm0.011}} 
\newcommand{\hatcurXsecondaryeccenxxxxxxB}{\ensuremath{2459581.429\pm0.042}} 
\newcommand{\hatcurXsecdureccenxxxxxxB}{\ensuremath{0.0953\pm0.0023}} 
\newcommand{\hatcurXsecingdureccenxxxxxxB}{\ensuremath{0.01114\pm0.00029}} 
\newcommand{\hatcurPPphiconjeccenxxxxxxB}{\ensuremath{0.051_{-0.105}^{+0.072}}} 
\newcommand{\hatcurPPperieccenxxxxxxB}{\ensuremath{2459579.24\pm0.37}} 
\newcommand{\hatcurPPaequiveccenxxxxxxB}{\ensuremath{0.1541\pm0.0060}} 
\newcommand{\hatcurPPtcirceccenxxxxxxB}{\ensuremath{880\pm100}}       
\newcommand{\hatcurPPtinfalleccenxxxxxxB}{\ensuremath{74700\pm7300}}  
\newcommand{\hatcurXdisteccenxxxxxxB}{\ensuremath{130.50\pm0.25}}     
\newcommand{\hatcurXAveccenxxxxxxB}{\ensuremath{0.1700_{-0.0020}^{+0.0050}}} 
\newcommand{\hatcurXdistredeccenxxxxxxB}{\ensuremath{130.51\pm0.25}}  
\newcommand{\hatcurXEBVeccenxxxxxxB}{\ensuremath{0.0550\pm0.0013}}    
\newcommand{\hatcurCCpmraeccenxxxxxxB}{\ensuremath{185.707\pm0.012}}  
\newcommand{\hatcurCCpmdececcenxxxxxxB}{\ensuremath{1.010\pm0.012}}   
\newcommand{\hatcurCCpmeccenxxxxxxB}{\ensuremath{185.710\pm0.017}}    

%% file: TOI3714.eccen.tex
\newcommand{\hatcurhtreccenxxxxxxC}{TOI3714}                          
\newcommand{\hatcurfieldeccenxxxxxxC}{\ensuremath{string}}            
\newcommand{\hatcurCCraeccenxxxxxxC}{\ensuremath{04^{\mathrm h}38^{\mathrm m}12.5354{\mathrm s}}}                   
\newcommand{\hatcurCCdececcenxxxxxxC}{\ensuremath{+39{\arcdeg}27{\arcmin}29.9093{\arcsec}}}                 
\newcommand{\hatcurCCmageccenxxxxxxC}{15.278}                         
\newcommand{\hatcurCCtwomasseccenxxxxxxC}{2MASS~04381253+3927299}     
\newcommand{\hatcurCCgsceccenxxxxxxC}{GSC~NULL}                       
\newcommand{\hatcurCCgaiaeccenxxxxxxC}{GAIA~178924386182591104}       
\newcommand{\hatcurCCgaiadrtwoeccenxxxxxxC}{GAIA~DR2~178924390478792320} 
\newcommand{\hatcurCCtassmveccenxxxxxxC}{\ensuremath{15.278\pm0.020}} 
\newcommand{\hatcurCCtassmvshorteccenxxxxxxC}{\ensuremath{15.3}}      
\newcommand{\hatcurCCtassmBeccenxxxxxxC}{\ensuremath{16.86\pm0.13}}   
\newcommand{\hatcurCCtassmBshorteccenxxxxxxC}{\ensuremath{16.9}}      
\newcommand{\hatcurCCtassmIeccenxxxxxxC}{\ensuremath{nff\pmnff}}      
\newcommand{\hatcurCCtassmIshorteccenxxxxxxC}{\ensuremath{0.0}}       
\newcommand{\hatcurCCtassmgeccenxxxxxxC}{\ensuremath{15.99\pm0.18}}   
\newcommand{\hatcurCCtassmgshorteccenxxxxxxC}{\ensuremath{16.0}}      
\newcommand{\hatcurCCtassmreccenxxxxxxC}{\ensuremath{14.737\pm0.011}} 
\newcommand{\hatcurCCtassmrshorteccenxxxxxxC}{\ensuremath{14.7}}      
\newcommand{\hatcurCCtassmieccenxxxxxxC}{\ensuremath{13.70\pm0.14}}   
\newcommand{\hatcurCCtassmishorteccenxxxxxxC}{\ensuremath{13.7}}      
\newcommand{\hatcurCCparallaxeccenxxxxxxC}{\ensuremath{8.838\pm0.022}} 
\newcommand{\hatcurCCgaiamGeccenxxxxxxC}{\ensuremath{14.2908\pm0.0028}} 
\newcommand{\hatcurCCgaiamBPeccenxxxxxxC}{\ensuremath{15.5114\pm0.0035}} 
\newcommand{\hatcurCCgaiamRPeccenxxxxxxC}{\ensuremath{13.1948\pm0.0039}} 
\newcommand{\hatcurCCtwomassJmageccenxxxxxxC}{\ensuremath{11.788\pm0.020}} 
\newcommand{\hatcurCCtwomassHmageccenxxxxxxC}{\ensuremath{11.109\pm0.015}} 
\newcommand{\hatcurCCtwomassKmageccenxxxxxxC}{\ensuremath{10.906\pm0.017}} 
\newcommand{\hatcurCCcitJmageccenxxxxxxC}{\ensuremath{11.726\pm0.022}} 
\newcommand{\hatcurCCcitHmageccenxxxxxxC}{\ensuremath{11.048\pm0.018}} 
\newcommand{\hatcurCCcitKmageccenxxxxxxC}{\ensuremath{10.876\pm0.018}} 
\newcommand{\hatcurCCbbJmageccenxxxxxxC}{\ensuremath{11.817\pm0.024}} 
\newcommand{\hatcurCCbbHmageccenxxxxxxC}{\ensuremath{11.074\pm0.018}} 
\newcommand{\hatcurCCbbKmageccenxxxxxxC}{\ensuremath{10.896\pm0.018}} 
\newcommand{\hatcurCCesoJmageccenxxxxxxC}{\ensuremath{11.825\pm0.030}} 
\newcommand{\hatcurCCesoHmageccenxxxxxxC}{\ensuremath{11.072\pm0.033}} 
\newcommand{\hatcurCCesoKmageccenxxxxxxC}{\ensuremath{10.892\pm0.020}} 
\newcommand{\hatcurCCesoJHmageccenxxxxxxC}{\ensuremath{0.752\pm0.040}} 
\newcommand{\hatcurCCesoJKmageccenxxxxxxC}{\ensuremath{0.933\pm0.041}} 
\newcommand{\hatcurCCesoHKmageccenxxxxxxC}{\ensuremath{0.179\pm0.036}} 
\newcommand{\hatcurCCWonemageccenxxxxxxC}{\ensuremath{10.779\pm0.023}} 
\newcommand{\hatcurCCWtwomageccenxxxxxxC}{\ensuremath{10.744\pm0.019}} 
\newcommand{\hatcurCCWthreemageccenxxxxxxC}{\ensuremath{10.546\pm0.097}} 
\newcommand{\hatcurCCWfourmageccenxxxxxxC}{\ensuremath{nff\pmnff}}    
\newcommand{\hatcurLCdipeccenxxxxxxC}{\ensuremath{51.0}}              
\newcommand{\hatcurLCrprstareccenxxxxxxC}{\ensuremath{0.2064\pm0.0013}} 
\newcommand{\hatcurLCbsqeccenxxxxxxC}{\ensuremath{0.192_{-0.018}^{+0.016}}} 
\newcommand{\hatcurLCimpeccenxxxxxxC}{\ensuremath{0.438_{-0.021}^{+0.018}}} 
\newcommand{\hatcurLCzetaeccenxxxxxxC}{\ensuremath{36.85\pm0.23}}     
\newcommand{\hatcurLCdureccenxxxxxxC}{\ensuremath{0.06788\pm0.00035}} 
\newcommand{\hatcurLCdurshorteccenxxxxxxC}{\ensuremath{0.0679}}       
\newcommand{\hatcurLCdurhreccenxxxxxxC}{\ensuremath{1.6290\pm0.0083}} 
\newcommand{\hatcurLCdurhrshorteccenxxxxxxC}{\ensuremath{1.629}}      
\newcommand{\hatcurLCqeccenxxxxxxC}{\ensuremath{0.03150\pm0.00016}}   
\newcommand{\hatcurLCqshorteccenxxxxxxC}{\ensuremath{0.032}}          
\newcommand{\hatcurLCingdureccenxxxxxxC}{\ensuremath{0.01394\pm0.00034}} 
\newcommand{\hatcurLCPeccenxxxxxxC}{\ensuremath{2.15484828\pm0.00000075}} 
\newcommand{\hatcurLCPprececcenxxxxxxC}{\ensuremath{2.1548483}}       
\newcommand{\hatcurLCPshorteccenxxxxxxC}{\ensuremath{2.1548}}         
\newcommand{\hatcurLCTeccenxxxxxxC}{\ensuremath{2459670.126470\pm0.000075}} 
\newcommand{\hatcurLCTAeccenxxxxxxC}{\ensuremath{2458818.96140\pm0.00026}} 
\newcommand{\hatcurLCTBeccenxxxxxxC}{\ensuremath{2459883.45645\pm0.00013}} 
\newcommand{\hatcurLChatnetmeccenxxxxxxC}{\ensuremath{-0.00027\pm0.00015}} 
\newcommand{\hatcurLCiblendeccenxxxxxxC}{\ensuremath{0.947\pm0.025}}  
\newcommand{\hatcurLCrhoeccenxxxxxxC}{\ensuremath{6.07\pm0.17}}       
\newcommand{\hatcurSMEiteffeccenxxxxxxC}{\ensuremath{3660\pm90}}      
\newcommand{\hatcurSMEizfeheccenxxxxxxC}{\ensuremath{0.10\pm0.10}}    
\newcommand{\hatcurSMEizfehshorteccenxxxxxxC}{\ensuremath{0.10}}      
\newcommand{\hatcurSMEiloggeccenxxxxxxC}{\ensuremath{4.750\pm0.050}}  
\newcommand{\hatcurSMEivsineccenxxxxxxC}{\ensuremath{5.0\pm5.0}}      
\newcommand{\hatcurSMEivmaceccenxxxxxxC}{\ensuremath{nff\pmnff}}      
\newcommand{\hatcurSMEivmiceccenxxxxxxC}{\ensuremath{nff\pmnff}}      
\newcommand{\hatcurextraerrMJeccenxxxxxxC}{\ensuremath{0\pm0}}        
\newcommand{\hatcurextraerrMJtwosiglimeccenxxxxxxC}{\ensuremath{<0.0200}} 
\newcommand{\hatcurextraerrMHeccenxxxxxxC}{\ensuremath{0\pm0}}        
\newcommand{\hatcurextraerrMHtwosiglimeccenxxxxxxC}{\ensuremath{<0.0200}} 
\newcommand{\hatcurextraerrMKseccenxxxxxxC}{\ensuremath{0\pm0}}       
\newcommand{\hatcurextraerrMKstwosiglimeccenxxxxxxC}{\ensuremath{<0.0200}} 
\newcommand{\hatcurextraerrMGeccenxxxxxxC}{\ensuremath{0\pm0}}        
\newcommand{\hatcurextraerrMGtwosiglimeccenxxxxxxC}{\ensuremath{<0.0200}} 
\newcommand{\hatcurextraerrMRPeccenxxxxxxC}{\ensuremath{0\pm0}}       
\newcommand{\hatcurextraerrMRPtwosiglimeccenxxxxxxC}{\ensuremath{<0.0200}} 
\newcommand{\hatcurextraerrMgeccenxxxxxxC}{\ensuremath{0\pm0}}        
\newcommand{\hatcurextraerrMgtwosiglimeccenxxxxxxC}{\ensuremath{<0.0200}} 
\newcommand{\hatcurextraerrMreccenxxxxxxC}{\ensuremath{0\pm0}}        
\newcommand{\hatcurextraerrMrtwosiglimeccenxxxxxxC}{\ensuremath{<0.0200}} 
\newcommand{\hatcurextraerrMieccenxxxxxxC}{\ensuremath{0\pm0}}        
\newcommand{\hatcurextraerrMitwosiglimeccenxxxxxxC}{\ensuremath{<0.0200}} 
\newcommand{\hatcurextraerrMWoneeccenxxxxxxC}{\ensuremath{0\pm0}}     
\newcommand{\hatcurextraerrMWonetwosiglimeccenxxxxxxC}{\ensuremath{<0.0200}} 
\newcommand{\hatcurextraerrMWtwoeccenxxxxxxC}{\ensuremath{0\pm0}}     
\newcommand{\hatcurextraerrMWtwotwosiglimeccenxxxxxxC}{\ensuremath{<0.0200}} 
\newcommand{\hatcurLBiBeccenxxxxxxC}{\ensuremath{0.5256}}             
\newcommand{\hatcurLBiiBeccenxxxxxxC}{\ensuremath{0.3086}}            
\newcommand{\hatcurLBiVeccenxxxxxxC}{\ensuremath{0.5054}}             
\newcommand{\hatcurLBiiVeccenxxxxxxC}{\ensuremath{0.3250}}            
\newcommand{\hatcurLBiReccenxxxxxxC}{\ensuremath{0.4394}}             
\newcommand{\hatcurLBiiReccenxxxxxxC}{\ensuremath{0.3031}}            
\newcommand{\hatcurLBiIeccenxxxxxxC}{\ensuremath{0.39\pm0.10}}        
\newcommand{\hatcurLBiiIeccenxxxxxxC}{\ensuremath{0.15\pm0.15}}       
\newcommand{\hatcurLBiueccenxxxxxxC}{\ensuremath{0.4650}}             
\newcommand{\hatcurLBiiueccenxxxxxxC}{\ensuremath{0.3204}}            
\newcommand{\hatcurLBigeccenxxxxxxC}{\ensuremath{0.46\pm0.11}}        
\newcommand{\hatcurLBiigeccenxxxxxxC}{\ensuremath{0.37\pm0.15}}       
\newcommand{\hatcurLBireccenxxxxxxC}{\ensuremath{0.591\pm0.087}}      
\newcommand{\hatcurLBiireccenxxxxxxC}{\ensuremath{0.12\pm0.13}}       
\newcommand{\hatcurLBiieccenxxxxxxC}{\ensuremath{0.384\pm0.088}}      
\newcommand{\hatcurLBiiieccenxxxxxxC}{\ensuremath{0.30\pm0.14}}       
\newcommand{\hatcurLBizeccenxxxxxxC}{\ensuremath{0.274\pm0.085}}      
\newcommand{\hatcurLBiizeccenxxxxxxC}{\ensuremath{0.25\pm0.13}}       
\newcommand{\hatcurLBiJeccenxxxxxxC}{\ensuremath{0.1069}}             
\newcommand{\hatcurLBiiJeccenxxxxxxC}{\ensuremath{0.2591}}            
\newcommand{\hatcurLBiHeccenxxxxxxC}{\ensuremath{0.0865}}             
\newcommand{\hatcurLBiiHeccenxxxxxxC}{\ensuremath{0.2899}}            
\newcommand{\hatcurLBiKeccenxxxxxxC}{\ensuremath{0.0778}}             
\newcommand{\hatcurLBiiKeccenxxxxxxC}{\ensuremath{0.2615}}            
\newcommand{\hatcurLBiTeccenxxxxxxC}{\ensuremath{0.19\pm0.12}}        
\newcommand{\hatcurLBiiTeccenxxxxxxC}{\ensuremath{0.15\pm0.18}}       
\newcommand{\hatcurLBikepeccenxxxxxxC}{\ensuremath{0.3334}}           
\newcommand{\hatcurLBiikepeccenxxxxxxC}{\ensuremath{0.3865}}          
\newcommand{\hatcurLBiCeccenxxxxxxC}{\ensuremath{0.2805}}             
\newcommand{\hatcurLBiiCeccenxxxxxxC}{\ensuremath{0.3966}}            
\newcommand{\hatcurLBiMeccenxxxxxxC}{\ensuremath{0.4437}}             
\newcommand{\hatcurLBiiMeccenxxxxxxC}{\ensuremath{0.3598}}            
\newcommand{\hatcurLBiSoneeccenxxxxxxC}{\ensuremath{0.0686}}            
\newcommand{\hatcurLBiiSoneeccenxxxxxxC}{\ensuremath{0.1876}}           
\newcommand{\hatcurLBiStwoeccenxxxxxxC}{\ensuremath{0.0500}}            
\newcommand{\hatcurLBiiStwoeccenxxxxxxC}{\ensuremath{0.1526}}           
\newcommand{\hatcurLBiSthreeeccenxxxxxxC}{\ensuremath{0.0535}}            
\newcommand{\hatcurLBiiSthreeeccenxxxxxxC}{\ensuremath{0.1327}}           
\newcommand{\hatcurLBiSfoureccenxxxxxxC}{\ensuremath{0.0603}}            
\newcommand{\hatcurLBiiSfoureccenxxxxxxC}{\ensuremath{0.0964}}           
\newcommand{\hatcurISOmeccenxxxxxxC}{\ensuremath{0.521\pm0.027}}      
\newcommand{\hatcurISOmshorteccenxxxxxxC}{\ensuremath{0.52}}          
\newcommand{\hatcurISOmlongeccenxxxxxxC}{\ensuremath{0.521\pm0.027}}  
\newcommand{\hatcurISOreccenxxxxxxC}{\ensuremath{0.4945\pm0.0091}}    
\newcommand{\hatcurISOrshorteccenxxxxxxC}{\ensuremath{0.49}}          
\newcommand{\hatcurISOrlongeccenxxxxxxC}{\ensuremath{0.4945\pm0.0091}} 
\newcommand{\hatcurISOrhoeccenxxxxxxC}{\ensuremath{6.07\pm0.17}}      
\newcommand{\hatcurISOrholongeccenxxxxxxC}{\ensuremath{6.07\pm0.17}}  
\newcommand{\hatcurISOloggeccenxxxxxxC}{\ensuremath{4.767\pm0.012}}   
\newcommand{\hatcurISOlumeccenxxxxxxC}{\ensuremath{0.0391\pm0.0035}}  
\newcommand{\hatcurISOlumshorteccenxxxxxxC}{\ensuremath{0.04}}        
\newcommand{\hatcurISOteffeccenxxxxxxC}{\ensuremath{3653\pm76}}       
\newcommand{\hatcurISOzfeheccenxxxxxxC}{\ensuremath{0.387\pm0.081}}   
\newcommand{\hatcurISOageeccenxxxxxxC}{\ensuremath{10.3\pm5.6}}       
\newcommand{\hatcurISOspececcenxxxxxxC}{M}                            
\newcommand{\hatcurRVKeccenxxxxxxC}{\ensuremath{167.2\pm4.4}}         
\newcommand{\hatcurRVKtwosiglimeccenxxxxxxC}{\ensuremath{<174.3}}     
\newcommand{\hatcurRVrkeccenxxxxxxC}{\ensuremath{-0.079\pm0.077}}     
\newcommand{\hatcurRVrheccenxxxxxxC}{\ensuremath{-0.044\pm0.078}}     
\newcommand{\hatcurRVkeccenxxxxxxC}{\ensuremath{-0.0089_{-0.0176}^{+0.0095}}} 
\newcommand{\hatcurRVheccenxxxxxxC}{\ensuremath{-0.0046_{-0.0155}^{+0.0092}}} 
\newcommand{\hatcurRVtroneeccenxxxxxxC}{\ensuremath{0\pm0}}           
\newcommand{\hatcurRVtrtwoeccenxxxxxxC}{\ensuremath{0\pm0}}           
\newcommand{\hatcurRVgammaAeccenxxxxxxC}{\ensuremath{4.8\pm8.6}}      
\newcommand{\hatcurRVjitterAeccenxxxxxxC}{\ensuremath{18.0\pm8.9}}    
\newcommand{\hatcurRVjittertwosiglimAeccenxxxxxxC}{\ensuremath{<37.3}} 
\newcommand{\hatcurRVfitrmsAeccenxxxxxxC}{\ensuremath{0.0}}           
\newcommand{\hatcurRVgammaBeccenxxxxxxC}{\ensuremath{-0.9\pm7.6}}     
\newcommand{\hatcurRVjitterBeccenxxxxxxC}{\ensuremath{0.1\pm8.1}}     
\newcommand{\hatcurRVjittertwosiglimBeccenxxxxxxC}{\ensuremath{<21.7}} 
\newcommand{\hatcurRVfitrmsBeccenxxxxxxC}{\ensuremath{0.0}}           
\newcommand{\hatcurRVgammaCeccenxxxxxxC}{\ensuremath{-77.9\pm5.5}}    
\newcommand{\hatcurRVjitterCeccenxxxxxxC}{\ensuremath{0.0\pm4.5}}     
\newcommand{\hatcurRVjittertwosiglimCeccenxxxxxxC}{\ensuremath{<8.4}} 
\newcommand{\hatcurRVfitrmsCeccenxxxxxxC}{\ensuremath{0.0}}           
\newcommand{\hatcurRVecceneccenxxxxxxC}{\ensuremath{0.016\pm0.014}}   
\newcommand{\hatcurRVeccentwosiglimeccenxxxxxxC}{\ensuremath{<0.047}} 
\newcommand{\hatcurRVomegaeccenxxxxxxC}{\ensuremath{204\pm69}}        
\newcommand{\hatcurPPieccenxxxxxxC}{\ensuremath{87.820\pm0.095}}      
\newcommand{\hatcurPPgeccenxxxxxxC}{\ensuremath{17.35\pm0.60}}        
\newcommand{\hatcurPPloggeccenxxxxxxC}{\ensuremath{3.239\pm0.015}}    
\newcommand{\hatcurPPareccenxxxxxxC}{\ensuremath{11.43\pm0.10}}       
\newcommand{\hatcurPPareleccenxxxxxxC}{\ensuremath{0.02629\pm0.00046}} 
\newcommand{\hatcurPPrhoeccenxxxxxxC}{\ensuremath{0.873\pm0.039}}     
\newcommand{\hatcurPPmeccenxxxxxxC}{\ensuremath{0.691\pm0.030}}       
\newcommand{\hatcurPPmtwosiglimeccenxxxxxxC}{\ensuremath{<0.74}}      
\newcommand{\hatcurPPmshorteccenxxxxxxC}{\ensuremath{0.69}}           
\newcommand{\hatcurPPmlongeccenxxxxxxC}{\ensuremath{0.691\pm0.030}}   
\newcommand{\hatcurPPmeeccenxxxxxxC}{\ensuremath{219.7\pm9.6}}        
\newcommand{\hatcurPPmeshorteccenxxxxxxC}{\ensuremath{219.7}}         
\newcommand{\hatcurPPmelongeccenxxxxxxC}{\ensuremath{219.7\pm9.6}}    
\newcommand{\hatcurPPreccenxxxxxxC}{\ensuremath{0.994\pm0.019}}       
\newcommand{\hatcurPPrshorteccenxxxxxxC}{\ensuremath{0.99}}           
\newcommand{\hatcurPPrlongeccenxxxxxxC}{\ensuremath{0.994\pm0.019}}   
\newcommand{\hatcurPPreeccenxxxxxxC}{\ensuremath{11.14\pm0.22}}       
\newcommand{\hatcurPPreshorteccenxxxxxxC}{\ensuremath{11.1}}          
\newcommand{\hatcurPPrelongeccenxxxxxxC}{\ensuremath{11.14\pm0.22}}   
\newcommand{\hatcurPPmrcorreccenxxxxxxC}{\ensuremath{0.65}}           
\newcommand{\hatcurPPteffeccenxxxxxxC}{\ensuremath{763\pm16}}         
\newcommand{\hatcurPPthetaeccenxxxxxxC}{\ensuremath{0.0699\pm0.0023}} 
\newcommand{\hatcurPPfluxperieccenxxxxxxC}{\ensuremath{7.97\pm0.72}}  
\newcommand{\hatcurPPfluxperidimeccenxxxxxxC}{\ensuremath{7}}         
\newcommand{\hatcurPPfluxapeccenxxxxxxC}{\ensuremath{7.44\pm0.67}}    
\newcommand{\hatcurPPfluxapdimeccenxxxxxxC}{\ensuremath{7}}           
\newcommand{\hatcurPPfluxavgeccenxxxxxxC}{\ensuremath{7.70\pm0.66}}   
\newcommand{\hatcurPPfluxavgdimeccenxxxxxxC}{\ensuremath{7}}          
\newcommand{\hatcurPPfluxavglogeccenxxxxxxC}{\ensuremath{7.886\pm0.037}} 
\newcommand{\hatcurXsecphaseeccenxxxxxxC}{\ensuremath{0.4943\pm0.0095}} 
\newcommand{\hatcurXsecondaryeccenxxxxxxC}{\ensuremath{2459671.192\pm0.020}} 
\newcommand{\hatcurXsecdureccenxxxxxxC}{\ensuremath{0.0673\pm0.0013}} 
\newcommand{\hatcurXsecingdureccenxxxxxxC}{\ensuremath{0.01372\pm0.00035}} 
\newcommand{\hatcurPPphiconjeccenxxxxxxC}{\ensuremath{-0.28_{-0.12}^{+0.33}}} 
\newcommand{\hatcurPPperieccenxxxxxxC}{\ensuremath{2459670.73\pm0.55}} 
\newcommand{\hatcurPPaequiveccenxxxxxxC}{\ensuremath{0.1330\pm0.0057}} 
\newcommand{\hatcurPPtcirceccenxxxxxxC}{\ensuremath{39.2\pm2.5}}      
\newcommand{\hatcurPPtinfalleccenxxxxxxC}{\ensuremath{3020\pm180}}    
\newcommand{\hatcurXdisteccenxxxxxxC}{\ensuremath{113.10\pm0.28}}     
\newcommand{\hatcurXAveccenxxxxxxC}{\ensuremath{0.044\pm0.015}}       
\newcommand{\hatcurXdistredeccenxxxxxxC}{\ensuremath{113.13\pm0.28}}  
\newcommand{\hatcurXEBVeccenxxxxxxC}{\ensuremath{0.0140\pm0.0050}}    
\newcommand{\hatcurCCpmraeccenxxxxxxC}{\ensuremath{19.826\pm0.025}}   
\newcommand{\hatcurCCpmdececcenxxxxxxC}{\ensuremath{-70.762\pm0.020}} 
\newcommand{\hatcurCCpmeccenxxxxxxC}{\ensuremath{73.487\pm0.032}}     

%% file: TOI4201.eccen.tex
\newcommand{\hatcurhtreccenxxxxxxD}{TOI4201}                          
\newcommand{\hatcurfieldeccenxxxxxxD}{\ensuremath{string}}            
\newcommand{\hatcurCCraeccenxxxxxxD}{\ensuremath{06^{\mathrm h}01^{\mathrm m}53.9212{\mathrm s}}}                   
\newcommand{\hatcurCCdececcenxxxxxxD}{\ensuremath{-13{\arcdeg}27{\arcmin}41.0292{\arcsec}}}                 
\newcommand{\hatcurCCmageccenxxxxxxD}{15.297}                         
\newcommand{\hatcurCCtwomasseccenxxxxxxD}{2MASS~06015391-1327410}     
\newcommand{\hatcurCCgsceccenxxxxxxD}{GSC~NULL}                       
\newcommand{\hatcurCCgaiaeccenxxxxxxD}{GAIA~2997312059307263616}      
\newcommand{\hatcurCCgaiadrtwoeccenxxxxxxD}{GAIA~DR2~2997312063605005056} 
\newcommand{\hatcurCCtassmveccenxxxxxxD}{\ensuremath{15.297\pm0.024}} 
\newcommand{\hatcurCCtassmvshorteccenxxxxxxD}{\ensuremath{15.3}}      
\newcommand{\hatcurCCtassmBeccenxxxxxxD}{\ensuremath{16.70\pm0.14}}   
\newcommand{\hatcurCCtassmBshorteccenxxxxxxD}{\ensuremath{16.7}}      
\newcommand{\hatcurCCtassmIeccenxxxxxxD}{\ensuremath{nff\pmnff}}      
\newcommand{\hatcurCCtassmIshorteccenxxxxxxD}{\ensuremath{0.0}}       
\newcommand{\hatcurCCtassmgeccenxxxxxxD}{\ensuremath{16.001\pm0.055}} 
\newcommand{\hatcurCCtassmgshorteccenxxxxxxD}{\ensuremath{16.0}}      
\newcommand{\hatcurCCtassmreccenxxxxxxD}{\ensuremath{14.655\pm0.064}} 
\newcommand{\hatcurCCtassmrshorteccenxxxxxxD}{\ensuremath{14.7}}      
\newcommand{\hatcurCCtassmieccenxxxxxxD}{\ensuremath{13.911\pm0.084}} 
\newcommand{\hatcurCCtassmishorteccenxxxxxxD}{\ensuremath{13.9}}      
\newcommand{\hatcurCCparallaxeccenxxxxxxD}{\ensuremath{5.291\pm0.019}} 
\newcommand{\hatcurCCgaiamGeccenxxxxxxD}{\ensuremath{14.4805\pm0.0028}} 
\newcommand{\hatcurCCgaiamBPeccenxxxxxxD}{\ensuremath{15.4724\pm0.0034}} 
\newcommand{\hatcurCCgaiamRPeccenxxxxxxD}{\ensuremath{13.4948\pm0.0039}} 
\newcommand{\hatcurCCtwomassJmageccenxxxxxxD}{\ensuremath{12.258\pm0.021}} 
\newcommand{\hatcurCCtwomassHmageccenxxxxxxD}{\ensuremath{11.564\pm0.024}} 
\newcommand{\hatcurCCtwomassKmageccenxxxxxxD}{\ensuremath{11.368\pm0.025}} 
\newcommand{\hatcurCCcitJmageccenxxxxxxD}{\ensuremath{12.247\pm0.023}} 
\newcommand{\hatcurCCcitHmageccenxxxxxxD}{\ensuremath{11.556\pm0.025}} 
\newcommand{\hatcurCCcitKmageccenxxxxxxD}{\ensuremath{11.392\pm0.026}} 
\newcommand{\hatcurCCbbJmageccenxxxxxxD}{\ensuremath{12.339\pm0.025}} 
\newcommand{\hatcurCCbbHmageccenxxxxxxD}{\ensuremath{11.581\pm0.026}} 
\newcommand{\hatcurCCbbKmageccenxxxxxxD}{\ensuremath{11.412\pm0.026}} 
\newcommand{\hatcurCCesoJmageccenxxxxxxD}{\ensuremath{12.348\pm0.030}} 
\newcommand{\hatcurCCesoHmageccenxxxxxxD}{\ensuremath{11.579\pm0.038}} 
\newcommand{\hatcurCCesoKmageccenxxxxxxD}{\ensuremath{11.408\pm0.027}} 
\newcommand{\hatcurCCesoJHmageccenxxxxxxD}{\ensuremath{0.769\pm0.045}} 
\newcommand{\hatcurCCesoJKmageccenxxxxxxD}{\ensuremath{0.940\pm0.038}} 
\newcommand{\hatcurCCesoHKmageccenxxxxxxD}{\ensuremath{0.171\pm0.045}} 
\newcommand{\hatcurCCWonemageccenxxxxxxD}{\ensuremath{11.272\pm0.024}} 
\newcommand{\hatcurCCWtwomageccenxxxxxxD}{\ensuremath{11.301\pm0.021}} 
\newcommand{\hatcurCCWthreemageccenxxxxxxD}{\ensuremath{11.28\pm0.15}} 
\newcommand{\hatcurCCWfourmageccenxxxxxxD}{\ensuremath{nff\pmnff}}    
\newcommand{\hatcurLCdipeccenxxxxxxD}{\ensuremath{20.0}}              
\newcommand{\hatcurLCrprstareccenxxxxxxD}{\ensuremath{0.1894\pm0.0015}} 
\newcommand{\hatcurLCbsqeccenxxxxxxD}{\ensuremath{0.176_{-0.022}^{+0.023}}} 
\newcommand{\hatcurLCimpeccenxxxxxxD}{\ensuremath{0.419_{-0.028}^{+0.027}}} 
\newcommand{\hatcurLCzetaeccenxxxxxxD}{\ensuremath{26.34\pm0.18}}     
\newcommand{\hatcurLCdureccenxxxxxxD}{\ensuremath{0.09311\pm0.00045}} 
\newcommand{\hatcurLCdurshorteccenxxxxxxD}{\ensuremath{0.0931}}       
\newcommand{\hatcurLCdurhreccenxxxxxxD}{\ensuremath{2.235\pm0.011}}   
\newcommand{\hatcurLCdurhrshorteccenxxxxxxD}{\ensuremath{2.235}}      
\newcommand{\hatcurLCqeccenxxxxxxD}{\ensuremath{0.02600\pm0.00013}}   
\newcommand{\hatcurLCqshorteccenxxxxxxD}{\ensuremath{0.026}}          
\newcommand{\hatcurLCingdureccenxxxxxxD}{\ensuremath{0.01754\pm0.00055}} 
\newcommand{\hatcurLCPeccenxxxxxxD}{\ensuremath{3.5819110\pm0.0000017}} 
\newcommand{\hatcurLCPprececcenxxxxxxD}{\ensuremath{3.5819110}}       
\newcommand{\hatcurLCPshorteccenxxxxxxD}{\ensuremath{3.5819}}         
\newcommand{\hatcurLCTeccenxxxxxxD}{\ensuremath{2459867.91018\pm0.00012}} 
\newcommand{\hatcurLCTAeccenxxxxxxD}{\ensuremath{2458470.96488\pm0.00059}} 
\newcommand{\hatcurLCTBeccenxxxxxxD}{\ensuremath{2460007.60471\pm0.00017}} 
\newcommand{\hatcurLChatnetmeccenxxxxxxD}{\ensuremath{13.57618\pm0.00027}} 
\newcommand{\hatcurLCiblendeccenxxxxxxD}{\ensuremath{0.981\pm0.025}}  
\newcommand{\hatcurLCrhoeccenxxxxxxD}{\ensuremath{4.18_{-0.17}^{+0.11}}} 
\newcommand{\hatcurSMEiteffeccenxxxxxxD}{\ensuremath{3937\pm70}}      
\newcommand{\hatcurSMEizfeheccenxxxxxxD}{\ensuremath{0.390\pm0.090}}  
\newcommand{\hatcurSMEizfehshorteccenxxxxxxD}{\ensuremath{0.39}}      
\newcommand{\hatcurSMEiloggeccenxxxxxxD}{\ensuremath{4.60\pm0.50}}    
\newcommand{\hatcurSMEivsineccenxxxxxxD}{\ensuremath{5.4\pm1.0}}      
\newcommand{\hatcurSMEivmaceccenxxxxxxD}{\ensuremath{nff\pmnff}}      
\newcommand{\hatcurSMEivmiceccenxxxxxxD}{\ensuremath{nff\pmnff}}      
\newcommand{\hatcurextraerrMJeccenxxxxxxD}{\ensuremath{0\pm0}}        
\newcommand{\hatcurextraerrMJtwosiglimeccenxxxxxxD}{\ensuremath{<0.0200}} 
\newcommand{\hatcurextraerrMHeccenxxxxxxD}{\ensuremath{0\pm0}}        
\newcommand{\hatcurextraerrMHtwosiglimeccenxxxxxxD}{\ensuremath{<0.0200}} 
\newcommand{\hatcurextraerrMKseccenxxxxxxD}{\ensuremath{0\pm0}}       
\newcommand{\hatcurextraerrMKstwosiglimeccenxxxxxxD}{\ensuremath{<0.0200}} 
\newcommand{\hatcurextraerrMGeccenxxxxxxD}{\ensuremath{0\pm0}}        
\newcommand{\hatcurextraerrMGtwosiglimeccenxxxxxxD}{\ensuremath{<0.0200}} 
\newcommand{\hatcurextraerrMRPeccenxxxxxxD}{\ensuremath{0\pm0}}       
\newcommand{\hatcurextraerrMRPtwosiglimeccenxxxxxxD}{\ensuremath{<0.0200}} 
\newcommand{\hatcurextraerrMgeccenxxxxxxD}{\ensuremath{0\pm0}}        
\newcommand{\hatcurextraerrMgtwosiglimeccenxxxxxxD}{\ensuremath{<0.0200}} 
\newcommand{\hatcurextraerrMreccenxxxxxxD}{\ensuremath{0\pm0}}        
\newcommand{\hatcurextraerrMrtwosiglimeccenxxxxxxD}{\ensuremath{<0.0200}} 
\newcommand{\hatcurextraerrMieccenxxxxxxD}{\ensuremath{0\pm0}}        
\newcommand{\hatcurextraerrMitwosiglimeccenxxxxxxD}{\ensuremath{<0.0200}} 
\newcommand{\hatcurextraerrMWoneeccenxxxxxxD}{\ensuremath{0\pm0}}     
\newcommand{\hatcurextraerrMWonetwosiglimeccenxxxxxxD}{\ensuremath{<0.0200}} 
\newcommand{\hatcurextraerrMWtwoeccenxxxxxxD}{\ensuremath{0\pm0}}     
\newcommand{\hatcurextraerrMWtwotwosiglimeccenxxxxxxD}{\ensuremath{<0.0200}} 
\newcommand{\hatcurLBiBeccenxxxxxxD}{\ensuremath{0.5515}}             
\newcommand{\hatcurLBiiBeccenxxxxxxD}{\ensuremath{0.2609}}            
\newcommand{\hatcurLBiVeccenxxxxxxD}{\ensuremath{0.4683}}             
\newcommand{\hatcurLBiiVeccenxxxxxxD}{\ensuremath{0.3143}}            
\newcommand{\hatcurLBiReccenxxxxxxD}{\ensuremath{0.4169}}             
\newcommand{\hatcurLBiiReccenxxxxxxD}{\ensuremath{0.3015}}            
\newcommand{\hatcurLBiIeccenxxxxxxD}{\ensuremath{0.2483}}             
\newcommand{\hatcurLBiiIeccenxxxxxxD}{\ensuremath{0.3232}}            
\newcommand{\hatcurLBiueccenxxxxxxD}{\ensuremath{0.5807}}             
\newcommand{\hatcurLBiiueccenxxxxxxD}{\ensuremath{0.2281}}            
\newcommand{\hatcurLBigeccenxxxxxxD}{\ensuremath{0.42\pm0.11}}        
\newcommand{\hatcurLBiigeccenxxxxxxD}{\ensuremath{0.35\pm0.16}}       
\newcommand{\hatcurLBireccenxxxxxxD}{\ensuremath{0.524\pm0.095}}      
\newcommand{\hatcurLBiireccenxxxxxxD}{\ensuremath{0.20\pm0.14}}       
\newcommand{\hatcurLBiieccenxxxxxxD}{\ensuremath{0.37\pm0.13}}        
\newcommand{\hatcurLBiiieccenxxxxxxD}{\ensuremath{0.29\pm0.17}}       
\newcommand{\hatcurLBizeccenxxxxxxD}{\ensuremath{0.124\pm0.074}}      
\newcommand{\hatcurLBiizeccenxxxxxxD}{\ensuremath{0.42\pm0.13}}       
\newcommand{\hatcurLBiJeccenxxxxxxD}{\ensuremath{0.178\pm0.093}}      
\newcommand{\hatcurLBiiJeccenxxxxxxD}{\ensuremath{0.16\pm0.14}}       
\newcommand{\hatcurLBiHeccenxxxxxxD}{\ensuremath{0.1161}}             
\newcommand{\hatcurLBiiHeccenxxxxxxD}{\ensuremath{0.2593}}            
\newcommand{\hatcurLBiKeccenxxxxxxD}{\ensuremath{0.0883}}             
\newcommand{\hatcurLBiiKeccenxxxxxxD}{\ensuremath{0.2403}}            
\newcommand{\hatcurLBiTeccenxxxxxxD}{\ensuremath{0.38\pm0.15}}        
\newcommand{\hatcurLBiiTeccenxxxxxxD}{\ensuremath{0.33\pm0.17}}       
\newcommand{\hatcurLBikepeccenxxxxxxD}{\ensuremath{0.3544}}           
\newcommand{\hatcurLBiikepeccenxxxxxxD}{\ensuremath{0.3567}}          
\newcommand{\hatcurLBiCeccenxxxxxxD}{\ensuremath{0.3188}}             
\newcommand{\hatcurLBiiCeccenxxxxxxD}{\ensuremath{0.3589}}            
\newcommand{\hatcurLBiMeccenxxxxxxD}{\ensuremath{0.4340}}             
\newcommand{\hatcurLBiiMeccenxxxxxxD}{\ensuremath{0.3442}}            
\newcommand{\hatcurLBiSoneeccenxxxxxxD}{\ensuremath{0.0669}}            
\newcommand{\hatcurLBiiSoneeccenxxxxxxD}{\ensuremath{0.1859}}           
\newcommand{\hatcurLBiStwoeccenxxxxxxD}{\ensuremath{0.0568}}            
\newcommand{\hatcurLBiiStwoeccenxxxxxxD}{\ensuremath{0.1468}}           
\newcommand{\hatcurLBiSthreeeccenxxxxxxD}{\ensuremath{0.0550}}            
\newcommand{\hatcurLBiiSthreeeccenxxxxxxD}{\ensuremath{0.1281}}           
\newcommand{\hatcurLBiSfoureccenxxxxxxD}{\ensuremath{0.0630}}            
\newcommand{\hatcurLBiiSfoureccenxxxxxxD}{\ensuremath{0.1031}}           
\newcommand{\hatcurISOmeccenxxxxxxD}{\ensuremath{0.659\pm0.035}}      
\newcommand{\hatcurISOmshorteccenxxxxxxD}{\ensuremath{0.66}}          
\newcommand{\hatcurISOmlongeccenxxxxxxD}{\ensuremath{0.659\pm0.035}}  
\newcommand{\hatcurISOreccenxxxxxxD}{\ensuremath{0.607\pm0.012}}      
\newcommand{\hatcurISOrshorteccenxxxxxxD}{\ensuremath{0.61}}          
\newcommand{\hatcurISOrlongeccenxxxxxxD}{\ensuremath{0.607\pm0.012}}  
\newcommand{\hatcurISOrhoeccenxxxxxxD}{\ensuremath{4.18_{-0.17}^{+0.11}}} 
\newcommand{\hatcurISOrholongeccenxxxxxxD}{\ensuremath{4.18_{-0.17}^{+0.11}}} 
\newcommand{\hatcurISOloggeccenxxxxxxD}{\ensuremath{4.692\pm0.014}}   
\newcommand{\hatcurISOlumeccenxxxxxxD}{\ensuremath{0.0790\pm0.0059}}  
\newcommand{\hatcurISOlumshorteccenxxxxxxD}{\ensuremath{0.08}}        
\newcommand{\hatcurISOteffeccenxxxxxxD}{\ensuremath{3931\pm61}}       
\newcommand{\hatcurISOzfeheccenxxxxxxD}{\ensuremath{0.495\pm0.095}}   
\newcommand{\hatcurISOageeccenxxxxxxD}{\ensuremath{0.59_{-0.36}^{+2.65}}} 
\newcommand{\hatcurISOspececcenxxxxxxD}{M}                            
\newcommand{\hatcurRVKeccenxxxxxxD}{\ensuremath{473\pm18}}            
\newcommand{\hatcurRVKtwosiglimeccenxxxxxxD}{\ensuremath{<503.9}}     
\newcommand{\hatcurRVrkeccenxxxxxxD}{\ensuremath{0.067_{-0.074}^{+0.055}}} 
\newcommand{\hatcurRVrheccenxxxxxxD}{\ensuremath{-0.186_{-0.039}^{+0.060}}} 
\newcommand{\hatcurRVkeccenxxxxxxD}{\ensuremath{0.013\pm0.013}}       
\newcommand{\hatcurRVheccenxxxxxxD}{\ensuremath{-0.038\pm0.016}}      
\newcommand{\hatcurRVtroneeccenxxxxxxD}{\ensuremath{0\pm0}}           
\newcommand{\hatcurRVtrtwoeccenxxxxxxD}{\ensuremath{0\pm0}}           
\newcommand{\hatcurRVgammaeccenxxxxxxD}{\ensuremath{77\pm12}}         
\newcommand{\hatcurRVjittereccenxxxxxxD}{\ensuremath{32\pm13}}        
\newcommand{\hatcurRVjittertwosiglimeccenxxxxxxD}{\ensuremath{<58.8}} 
\newcommand{\hatcurRVfitrmseccenxxxxxxD}{\ensuremath{.1fym}}          %
\newcommand{\hatcurRVecceneccenxxxxxxD}{\ensuremath{0.043\pm0.014}}   
\newcommand{\hatcurRVeccentwosiglimeccenxxxxxxD}{\ensuremath{<0.062}} 
\newcommand{\hatcurRVomegaeccenxxxxxxD}{\ensuremath{289\pm34}}        
\newcommand{\hatcurPPieccenxxxxxxD}{\ensuremath{88.36\pm0.10}}        
\newcommand{\hatcurPPgeccenxxxxxxD}{\ensuremath{53.5\pm2.5}}          
\newcommand{\hatcurPPloggeccenxxxxxxD}{\ensuremath{3.729\pm0.021}}    
\newcommand{\hatcurPPareccenxxxxxxD}{\ensuremath{14.17_{-0.19}^{+0.12}}} 
\newcommand{\hatcurPPareleccenxxxxxxD}{\ensuremath{0.03991\pm0.00071}} 
\newcommand{\hatcurPPrhoeccenxxxxxxD}{\ensuremath{2.40\pm0.14}}       
\newcommand{\hatcurPPmeccenxxxxxxD}{\ensuremath{2.70\pm0.14}}         
\newcommand{\hatcurPPmtwosiglimeccenxxxxxxD}{\ensuremath{<2.94}}      
\newcommand{\hatcurPPmshorteccenxxxxxxD}{\ensuremath{2.70}}           
\newcommand{\hatcurPPmlongeccenxxxxxxD}{\ensuremath{2.70\pm0.14}}     
\newcommand{\hatcurPPmeeccenxxxxxxD}{\ensuremath{857\pm46}}           
\newcommand{\hatcurPPmeshorteccenxxxxxxD}{\ensuremath{857.1}}         
\newcommand{\hatcurPPmelongeccenxxxxxxD}{\ensuremath{857\pm46}}       
\newcommand{\hatcurPPreccenxxxxxxD}{\ensuremath{1.118\pm0.024}}       
\newcommand{\hatcurPPrshorteccenxxxxxxD}{\ensuremath{1.12}}           
\newcommand{\hatcurPPrlongeccenxxxxxxD}{\ensuremath{1.118\pm0.024}}   
\newcommand{\hatcurPPreeccenxxxxxxD}{\ensuremath{12.54\pm0.26}}       
\newcommand{\hatcurPPreshorteccenxxxxxxD}{\ensuremath{12.5}}          
\newcommand{\hatcurPPrelongeccenxxxxxxD}{\ensuremath{12.54\pm0.26}}   
\newcommand{\hatcurPPmrcorreccenxxxxxxD}{\ensuremath{0.52}}           
\newcommand{\hatcurPPteffeccenxxxxxxD}{\ensuremath{739\pm12}}         
\newcommand{\hatcurPPthetaeccenxxxxxxD}{\ensuremath{0.291\pm0.013}}   
\newcommand{\hatcurPPfluxperieccenxxxxxxD}{\ensuremath{7.37\pm0.47}}  
\newcommand{\hatcurPPfluxperidimeccenxxxxxxD}{\ensuremath{7}}         
\newcommand{\hatcurPPfluxapeccenxxxxxxD}{\ensuremath{6.21\pm0.50}}    
\newcommand{\hatcurPPfluxapdimeccenxxxxxxD}{\ensuremath{7}}           
\newcommand{\hatcurPPfluxavgeccenxxxxxxD}{\ensuremath{6.76\pm0.45}}   
\newcommand{\hatcurPPfluxavgdimeccenxxxxxxD}{\ensuremath{7}}          
\newcommand{\hatcurPPfluxavglogeccenxxxxxxD}{\ensuremath{7.830\pm0.029}} 
\newcommand{\hatcurXsecphaseeccenxxxxxxD}{\ensuremath{0.5084\pm0.0084}} 
\newcommand{\hatcurXsecondaryeccenxxxxxxD}{\ensuremath{2459869.731\pm0.030}} 
\newcommand{\hatcurXsecdureccenxxxxxxD}{\ensuremath{0.0871\pm0.0022}} 
\newcommand{\hatcurXsecingdureccenxxxxxxD}{\ensuremath{0.01597\pm0.00050}} 
\newcommand{\hatcurPPphiconjeccenxxxxxxD}{\ensuremath{0.413_{-0.782}^{+0.053}}} 
\newcommand{\hatcurPPperieccenxxxxxxD}{\ensuremath{2459866.4\pm1.3}}  
\newcommand{\hatcurPPaequiveccenxxxxxxD}{\ensuremath{0.1419\pm0.0047}} 
\newcommand{\hatcurPPtcirceccenxxxxxxD}{\ensuremath{892\pm68}}        
\newcommand{\hatcurPPtinfalleccenxxxxxxD}{\ensuremath{4730_{-350}^{+260}}} 
\newcommand{\hatcurXdisteccenxxxxxxD}{\ensuremath{189.10\pm0.66}}     
\newcommand{\hatcurXAveccenxxxxxxD}{\ensuremath{0.056\pm0.017}}       
\newcommand{\hatcurXdistredeccenxxxxxxD}{\ensuremath{189.11\pm0.66}}  
\newcommand{\hatcurXEBVeccenxxxxxxD}{\ensuremath{0.0180\pm0.0056}}    
\newcommand{\hatcurCCpmraeccenxxxxxxD}{\ensuremath{11.731\pm0.017}}   
\newcommand{\hatcurCCpmdececcenxxxxxxD}{\ensuremath{6.053\pm0.018}}   
\newcommand{\hatcurCCpmeccenxxxxxxD}{\ensuremath{13.201\pm0.025}}     

%% file: TOI5344.eccen.tex
\newcommand{\hatcurhtreccenxxxxxxE}{TOI5344}                          
\newcommand{\hatcurfieldeccenxxxxxxE}{\ensuremath{string}}            
\newcommand{\hatcurCCraeccenxxxxxxE}{\ensuremath{04^{\mathrm h}13^{\mathrm m}03.8474{\mathrm s}}}                   
\newcommand{\hatcurCCdececcenxxxxxxE}{\ensuremath{+20{\arcdeg}54{\arcmin}54.9086{\arcsec}}}                 
\newcommand{\hatcurCCmageccenxxxxxxE}{15.275}                         
\newcommand{\hatcurCCtwomasseccenxxxxxxE}{2MASS~04130384+2054550}     
\newcommand{\hatcurCCgsceccenxxxxxxE}{GSC~NULL}                       
\newcommand{\hatcurCCgaiaeccenxxxxxxE}{GAIA~52359533989321472}        
\newcommand{\hatcurCCgaiadrtwoeccenxxxxxxE}{GAIA~DR2~52359538285081728} 
\newcommand{\hatcurCCtassmveccenxxxxxxE}{\ensuremath{15.275\pm0.069}} 
\newcommand{\hatcurCCtassmvshorteccenxxxxxxE}{\ensuremath{15.3}}      
\newcommand{\hatcurCCtassmBeccenxxxxxxE}{\ensuremath{16.880\pm0.069}} 
\newcommand{\hatcurCCtassmBshorteccenxxxxxxE}{\ensuremath{16.9}}      
\newcommand{\hatcurCCtassmIeccenxxxxxxE}{\ensuremath{nff\pmnff}}      
\newcommand{\hatcurCCtassmIshorteccenxxxxxxE}{\ensuremath{0.0}}       
\newcommand{\hatcurCCtassmgeccenxxxxxxE}{\ensuremath{16.106\pm0.055}} 
\newcommand{\hatcurCCtassmgshorteccenxxxxxxE}{\ensuremath{16.1}}      
\newcommand{\hatcurCCtassmreccenxxxxxxE}{\ensuremath{14.622\pm0.043}} 
\newcommand{\hatcurCCtassmrshorteccenxxxxxxE}{\ensuremath{14.6}}      
\newcommand{\hatcurCCtassmieccenxxxxxxE}{\ensuremath{13.603\pm0.062}} 
\newcommand{\hatcurCCtassmishorteccenxxxxxxE}{\ensuremath{13.6}}      
\newcommand{\hatcurCCparallaxeccenxxxxxxE}{\ensuremath{7.305\pm0.023}} 
\newcommand{\hatcurCCgaiamGeccenxxxxxxE}{\ensuremath{14.3186\pm0.0028}} 
\newcommand{\hatcurCCgaiamBPeccenxxxxxxE}{\ensuremath{15.5037\pm0.0037}} 
\newcommand{\hatcurCCgaiamRPeccenxxxxxxE}{\ensuremath{13.2419\pm0.0039}} 
\newcommand{\hatcurCCtwomassJmageccenxxxxxxE}{\ensuremath{11.799\pm0.021}} 
\newcommand{\hatcurCCtwomassHmageccenxxxxxxE}{\ensuremath{11.087\pm0.022}} 
\newcommand{\hatcurCCtwomassKmageccenxxxxxxE}{\ensuremath{10.860\pm0.018}} 
\newcommand{\hatcurCCcitJmageccenxxxxxxE}{\ensuremath{11.786\pm0.023}} 
\newcommand{\hatcurCCcitHmageccenxxxxxxE}{\ensuremath{11.078\pm0.024}} 
\newcommand{\hatcurCCcitKmageccenxxxxxxE}{\ensuremath{10.884\pm0.019}} 
\newcommand{\hatcurCCbbJmageccenxxxxxxE}{\ensuremath{11.881\pm0.025}} 
\newcommand{\hatcurCCbbHmageccenxxxxxxE}{\ensuremath{11.104\pm0.025}} 
\newcommand{\hatcurCCbbKmageccenxxxxxxE}{\ensuremath{10.904\pm0.019}} 
\newcommand{\hatcurCCesoJmageccenxxxxxxE}{\ensuremath{11.891\pm0.031}} 
\newcommand{\hatcurCCesoHmageccenxxxxxxE}{\ensuremath{11.103\pm0.040}} 
\newcommand{\hatcurCCesoKmageccenxxxxxxE}{\ensuremath{10.900\pm0.021}} 
\newcommand{\hatcurCCesoJHmageccenxxxxxxE}{\ensuremath{0.788\pm0.046}} 
\newcommand{\hatcurCCesoJKmageccenxxxxxxE}{\ensuremath{0.991\pm0.034}} 
\newcommand{\hatcurCCesoHKmageccenxxxxxxE}{\ensuremath{0.203\pm0.043}} 
\newcommand{\hatcurCCWonemageccenxxxxxxE}{\ensuremath{10.739\pm0.024}} 
\newcommand{\hatcurCCWtwomageccenxxxxxxE}{\ensuremath{10.728\pm0.020}} 
\newcommand{\hatcurCCWthreemageccenxxxxxxE}{\ensuremath{10.55\pm0.11}} 
\newcommand{\hatcurCCWfourmageccenxxxxxxE}{\ensuremath{nff\pmnff}}    
\newcommand{\hatcurLCdipeccenxxxxxxE}{\ensuremath{40.0}}              
\newcommand{\hatcurLCrprstareccenxxxxxxE}{\ensuremath{0.1652\pm0.0015}} 
\newcommand{\hatcurLCbsqeccenxxxxxxE}{\ensuremath{0.538_{-0.020}^{+0.017}}} 
\newcommand{\hatcurLCimpeccenxxxxxxE}{\ensuremath{0.733_{-0.013}^{+0.012}}} 
\newcommand{\hatcurLCzetaeccenxxxxxxE}{\ensuremath{36.11\pm0.42}}     
\newcommand{\hatcurLCdureccenxxxxxxE}{\ensuremath{0.07376\pm0.00073}} 
\newcommand{\hatcurLCdurshorteccenxxxxxxE}{\ensuremath{0.0738}}       
\newcommand{\hatcurLCdurhreccenxxxxxxE}{\ensuremath{1.770\pm0.017}}   
\newcommand{\hatcurLCdurhrshorteccenxxxxxxE}{\ensuremath{1.770}}      
\newcommand{\hatcurLCqeccenxxxxxxE}{\ensuremath{0.01940\pm0.00020}}   
\newcommand{\hatcurLCqshorteccenxxxxxxE}{\ensuremath{0.019}}          
\newcommand{\hatcurLCingdureccenxxxxxxE}{\ensuremath{0.02064\pm0.00097}} 
\newcommand{\hatcurLCPeccenxxxxxxE}{\ensuremath{3.7926219\pm0.0000066}} 
\newcommand{\hatcurLCPprececcenxxxxxxE}{\ensuremath{3.7926219}}       
\newcommand{\hatcurLCPshorteccenxxxxxxE}{\ensuremath{3.7926}}         
\newcommand{\hatcurLCTeccenxxxxxxE}{\ensuremath{2459909.67226\pm0.00017}} 
\newcommand{\hatcurLCTAeccenxxxxxxE}{\ensuremath{2459477.31337\pm0.00076}} 
\newcommand{\hatcurLCTBeccenxxxxxxE}{\ensuremath{2459951.39110\pm0.00019}} 
\newcommand{\hatcurLChatnetmAeccenxxxxxxE}{\ensuremath{-0.00024\pm0.00014}} 
\newcommand{\hatcurLCiblendAeccenxxxxxxE}{\ensuremath{0.902\pm0.047}} 
\newcommand{\hatcurLChatnetmBeccenxxxxxxE}{\ensuremath{-0.00030\pm0.00012}} 
\newcommand{\hatcurLCiblendBeccenxxxxxxE}{\ensuremath{0.820\pm0.044}} 
\newcommand{\hatcurLCrhoeccenxxxxxxE}{\ensuremath{4.31\pm0.16}}       
\newcommand{\hatcurSMEiteffeccenxxxxxxE}{\ensuremath{3766\pm70}}      
\newcommand{\hatcurSMEizfeheccenxxxxxxE}{\ensuremath{0.390\pm0.090}}  
\newcommand{\hatcurSMEizfehshorteccenxxxxxxE}{\ensuremath{0.39}}      
\newcommand{\hatcurSMEiloggeccenxxxxxxE}{\ensuremath{4.67\pm0.50}}    
\newcommand{\hatcurSMEivsineccenxxxxxxE}{\ensuremath{5.1\pm1.0}}      
\newcommand{\hatcurSMEivmaceccenxxxxxxE}{\ensuremath{nff\pmnff}}      
\newcommand{\hatcurSMEivmiceccenxxxxxxE}{\ensuremath{nff\pmnff}}      
\newcommand{\hatcurextraerrMJeccenxxxxxxE}{\ensuremath{0\pm0}}        
\newcommand{\hatcurextraerrMJtwosiglimeccenxxxxxxE}{\ensuremath{<0.0200}} 
\newcommand{\hatcurextraerrMHeccenxxxxxxE}{\ensuremath{0\pm0}}        
\newcommand{\hatcurextraerrMHtwosiglimeccenxxxxxxE}{\ensuremath{<0.0200}} 
\newcommand{\hatcurextraerrMKseccenxxxxxxE}{\ensuremath{0\pm0}}       
\newcommand{\hatcurextraerrMKstwosiglimeccenxxxxxxE}{\ensuremath{<0.0200}} 
\newcommand{\hatcurextraerrMGeccenxxxxxxE}{\ensuremath{0\pm0}}        
\newcommand{\hatcurextraerrMGtwosiglimeccenxxxxxxE}{\ensuremath{<0.0200}} 
\newcommand{\hatcurextraerrMRPeccenxxxxxxE}{\ensuremath{0\pm0}}       
\newcommand{\hatcurextraerrMRPtwosiglimeccenxxxxxxE}{\ensuremath{<0.0200}} 
\newcommand{\hatcurextraerrMgeccenxxxxxxE}{\ensuremath{0\pm0}}        
\newcommand{\hatcurextraerrMgtwosiglimeccenxxxxxxE}{\ensuremath{<0.0200}} 
\newcommand{\hatcurextraerrMreccenxxxxxxE}{\ensuremath{0\pm0}}        
\newcommand{\hatcurextraerrMrtwosiglimeccenxxxxxxE}{\ensuremath{<0.0200}} 
\newcommand{\hatcurextraerrMieccenxxxxxxE}{\ensuremath{0\pm0}}        
\newcommand{\hatcurextraerrMitwosiglimeccenxxxxxxE}{\ensuremath{<0.0200}} 
\newcommand{\hatcurextraerrMWoneeccenxxxxxxE}{\ensuremath{0\pm0}}     
\newcommand{\hatcurextraerrMWonetwosiglimeccenxxxxxxE}{\ensuremath{<0.0200}} 
\newcommand{\hatcurextraerrMWtwoeccenxxxxxxE}{\ensuremath{0\pm0}}     
\newcommand{\hatcurextraerrMWtwotwosiglimeccenxxxxxxE}{\ensuremath{<0.0200}} 
\newcommand{\hatcurLBiBeccenxxxxxxE}{\ensuremath{0.4527}}             
\newcommand{\hatcurLBiiBeccenxxxxxxE}{\ensuremath{0.3417}}            
\newcommand{\hatcurLBiVeccenxxxxxxE}{\ensuremath{0.4083}}             
\newcommand{\hatcurLBiiVeccenxxxxxxE}{\ensuremath{0.3705}}            
\newcommand{\hatcurLBiReccenxxxxxxE}{\ensuremath{0.3642}}             
\newcommand{\hatcurLBiiReccenxxxxxxE}{\ensuremath{0.3397}}            
\newcommand{\hatcurLBiIeccenxxxxxxE}{\ensuremath{0.32\pm0.14}}        
\newcommand{\hatcurLBiiIeccenxxxxxxE}{\ensuremath{0.32\pm0.16}}       
\newcommand{\hatcurLBiueccenxxxxxxE}{\ensuremath{0.4343}}             
\newcommand{\hatcurLBiiueccenxxxxxxE}{\ensuremath{0.3245}}            
\newcommand{\hatcurLBigeccenxxxxxxE}{\ensuremath{0.23\pm0.13}}        
\newcommand{\hatcurLBiigeccenxxxxxxE}{\ensuremath{0.14\pm0.16}}       
\newcommand{\hatcurLBireccenxxxxxxE}{\ensuremath{0.26\pm0.14}}        
\newcommand{\hatcurLBiireccenxxxxxxE}{\ensuremath{0.29\pm0.16}}       
\newcommand{\hatcurLBiieccenxxxxxxE}{\ensuremath{0.23\pm0.13}}        
\newcommand{\hatcurLBiiieccenxxxxxxE}{\ensuremath{0.12\pm0.15}}       
\newcommand{\hatcurLBizeccenxxxxxxE}{\ensuremath{0.18\pm0.11}}        
\newcommand{\hatcurLBiizeccenxxxxxxE}{\ensuremath{0.21\pm0.15}}       
\newcommand{\hatcurLBiJeccenxxxxxxE}{\ensuremath{0.17\pm0.11}}        
\newcommand{\hatcurLBiiJeccenxxxxxxE}{\ensuremath{0.03\pm0.13}}       
\newcommand{\hatcurLBiHeccenxxxxxxE}{\ensuremath{0.0809}}             
\newcommand{\hatcurLBiiHeccenxxxxxxE}{\ensuremath{0.2652}}            
\newcommand{\hatcurLBiKeccenxxxxxxE}{\ensuremath{0.0716}}             
\newcommand{\hatcurLBiiKeccenxxxxxxE}{\ensuremath{0.2389}}            
\newcommand{\hatcurLBiTeccenxxxxxxE}{\ensuremath{0.20\pm0.13}}        
\newcommand{\hatcurLBiiTeccenxxxxxxE}{\ensuremath{0.17\pm0.16}}       
\newcommand{\hatcurLBikepeccenxxxxxxE}{\ensuremath{0.2833}}           
\newcommand{\hatcurLBiikepeccenxxxxxxE}{\ensuremath{0.4135}}          
\newcommand{\hatcurLBiCeccenxxxxxxE}{\ensuremath{0.2423}}             
\newcommand{\hatcurLBiiCeccenxxxxxxE}{\ensuremath{0.4143}}            
\newcommand{\hatcurLBiMeccenxxxxxxE}{\ensuremath{0.3683}}             
\newcommand{\hatcurLBiiMeccenxxxxxxE}{\ensuremath{0.4059}}            
\newcommand{\hatcurLBiSoneeccenxxxxxxE}{\ensuremath{0.0614}}            
\newcommand{\hatcurLBiiSoneeccenxxxxxxE}{\ensuremath{0.1756}}           
\newcommand{\hatcurLBiStwoeccenxxxxxxE}{\ensuremath{0.0463}}            
\newcommand{\hatcurLBiiStwoeccenxxxxxxE}{\ensuremath{0.1424}}           
\newcommand{\hatcurLBiSthreeeccenxxxxxxE}{\ensuremath{0.0506}}            
\newcommand{\hatcurLBiiSthreeeccenxxxxxxE}{\ensuremath{0.1245}}           
\newcommand{\hatcurLBiSfoureccenxxxxxxE}{\ensuremath{0.0577}}            
\newcommand{\hatcurLBiiSfoureccenxxxxxxE}{\ensuremath{0.0933}}           
\newcommand{\hatcurISOmeccenxxxxxxE}{\ensuremath{0.619\pm0.034}}      
\newcommand{\hatcurISOmshorteccenxxxxxxE}{\ensuremath{0.62}}          
\newcommand{\hatcurISOmlongeccenxxxxxxE}{\ensuremath{0.619\pm0.034}}  
\newcommand{\hatcurISOreccenxxxxxxE}{\ensuremath{0.587\pm0.011}}      
\newcommand{\hatcurISOrshorteccenxxxxxxE}{\ensuremath{0.59}}          
\newcommand{\hatcurISOrlongeccenxxxxxxE}{\ensuremath{0.587\pm0.011}}  
\newcommand{\hatcurISOrhoeccenxxxxxxE}{\ensuremath{4.31\pm0.16}}      
\newcommand{\hatcurISOrholongeccenxxxxxxE}{\ensuremath{4.31\pm0.16}}  
\newcommand{\hatcurISOloggeccenxxxxxxE}{\ensuremath{4.693\pm0.015}}   
\newcommand{\hatcurISOlumeccenxxxxxxE}{\ensuremath{0.0612\pm0.0046}}  
\newcommand{\hatcurISOlumshorteccenxxxxxxE}{\ensuremath{0.06}}        
\newcommand{\hatcurISOteffeccenxxxxxxE}{\ensuremath{3749\pm63}}       
\newcommand{\hatcurISOzfeheccenxxxxxxE}{\ensuremath{0.431\pm0.086}}   
\newcommand{\hatcurISOageeccenxxxxxxE}{\ensuremath{6.7_{-6.0}^{+7.9}}} 
\newcommand{\hatcurISOspececcenxxxxxxE}{M}                            
\newcommand{\hatcurRVKeccenxxxxxxE}{\ensuremath{74.4\pm8.2}}          
\newcommand{\hatcurRVKtwosiglimeccenxxxxxxE}{\ensuremath{<86.6}}      
\newcommand{\hatcurRVrkeccenxxxxxxE}{\ensuremath{0.04\pm0.10}}        
\newcommand{\hatcurRVrheccenxxxxxxE}{\ensuremath{-0.018\pm0.096}}     
\newcommand{\hatcurRVkeccenxxxxxxE}{\ensuremath{0.003_{-0.012}^{+0.022}}} 
\newcommand{\hatcurRVheccenxxxxxxE}{\ensuremath{-0.001\pm0.018}}      
\newcommand{\hatcurRVtroneeccenxxxxxxE}{\ensuremath{0\pm0}}           
\newcommand{\hatcurRVtrtwoeccenxxxxxxE}{\ensuremath{0\pm0}}           
\newcommand{\hatcurRVgammaeccenxxxxxxE}{\ensuremath{-9.6\pm6.2}}      
\newcommand{\hatcurRVjittereccenxxxxxxE}{\ensuremath{15\pm36}}        
\newcommand{\hatcurRVjittertwosiglimeccenxxxxxxE}{\ensuremath{<23.9}} 
\newcommand{\hatcurRVfitrmseccenxxxxxxE}{\ensuremath{.1fym}}          %
\newcommand{\hatcurRVecceneccenxxxxxxE}{\ensuremath{0.018\pm0.018}}   
\newcommand{\hatcurRVeccentwosiglimeccenxxxxxxE}{\ensuremath{<0.054}} 
\newcommand{\hatcurRVomegaeccenxxxxxxE}{\ensuremath{210\pm120}}       
\newcommand{\hatcurPPieccenxxxxxxE}{\ensuremath{87.180\pm0.076}}      
\newcommand{\hatcurPPgeccenxxxxxxE}{\ensuremath{11.6\pm1.3}}          
\newcommand{\hatcurPPloggeccenxxxxxxE}{\ensuremath{3.063\pm0.047}}    
\newcommand{\hatcurPPareccenxxxxxxE}{\ensuremath{14.86\pm0.18}}       
\newcommand{\hatcurPPareleccenxxxxxxE}{\ensuremath{0.04057\pm0.00074}} 
\newcommand{\hatcurPPrhoeccenxxxxxxE}{\ensuremath{0.614\pm0.074}}     
\newcommand{\hatcurPPmeccenxxxxxxE}{\ensuremath{0.415\pm0.048}}       
\newcommand{\hatcurPPmtwosiglimeccenxxxxxxE}{\ensuremath{<0.49}}      
\newcommand{\hatcurPPmshorteccenxxxxxxE}{\ensuremath{0.41}}           
\newcommand{\hatcurPPmlongeccenxxxxxxE}{\ensuremath{0.415\pm0.048}}   
\newcommand{\hatcurPPmeeccenxxxxxxE}{\ensuremath{132\pm15}}           
\newcommand{\hatcurPPmeshorteccenxxxxxxE}{\ensuremath{131.8}}         
\newcommand{\hatcurPPmelongeccenxxxxxxE}{\ensuremath{132\pm15}}       
\newcommand{\hatcurPPreccenxxxxxxE}{\ensuremath{0.943\pm0.021}}       
\newcommand{\hatcurPPrshorteccenxxxxxxE}{\ensuremath{0.94}}           
\newcommand{\hatcurPPrlongeccenxxxxxxE}{\ensuremath{0.943\pm0.021}}   
\newcommand{\hatcurPPreeccenxxxxxxE}{\ensuremath{10.57\pm0.24}}       
\newcommand{\hatcurPPreshorteccenxxxxxxE}{\ensuremath{10.6}}          
\newcommand{\hatcurPPrelongeccenxxxxxxE}{\ensuremath{10.57\pm0.24}}   
\newcommand{\hatcurPPmrcorreccenxxxxxxE}{\ensuremath{0.21}}           
\newcommand{\hatcurPPteffeccenxxxxxxE}{\ensuremath{688\pm12}}         
\newcommand{\hatcurPPthetaeccenxxxxxxE}{\ensuremath{0.0575\pm0.0065}} 
\newcommand{\hatcurPPfluxperieccenxxxxxxE}{\ensuremath{5.28\pm0.40}}  
\newcommand{\hatcurPPfluxperidimeccenxxxxxxE}{\ensuremath{7}}         
\newcommand{\hatcurPPfluxapeccenxxxxxxE}{\ensuremath{4.86\pm0.39}}    
\newcommand{\hatcurPPfluxapdimeccenxxxxxxE}{\ensuremath{7}}           
\newcommand{\hatcurPPfluxavgeccenxxxxxxE}{\ensuremath{5.07\pm0.35}}   
\newcommand{\hatcurPPfluxavgdimeccenxxxxxxE}{\ensuremath{7}}          
\newcommand{\hatcurPPfluxavglogeccenxxxxxxE}{\ensuremath{7.705\pm0.030}} 
\newcommand{\hatcurXsecphaseeccenxxxxxxE}{\ensuremath{0.502\pm0.013}} 
\newcommand{\hatcurXsecondaryeccenxxxxxxE}{\ensuremath{2459911.576\pm0.049}} 
\newcommand{\hatcurXsecdureccenxxxxxxE}{\ensuremath{0.07361\pm0.00082}} 
\newcommand{\hatcurXsecingdureccenxxxxxxE}{\ensuremath{0.0204\pm0.0013}} 
\newcommand{\hatcurPPphiconjeccenxxxxxxE}{\ensuremath{0.12_{-0.48}^{+0.24}}} 
\newcommand{\hatcurPPperieccenxxxxxxE}{\ensuremath{2459909.2\pm1.2}}  
\newcommand{\hatcurPPaequiveccenxxxxxxE}{\ensuremath{0.1639\pm0.0057}} 
\newcommand{\hatcurPPtcirceccenxxxxxxE}{\ensuremath{395\pm54}}        
\newcommand{\hatcurPPtinfalleccenxxxxxxE}{\ensuremath{39100\pm4800}}  
\newcommand{\hatcurXdisteccenxxxxxxE}{\ensuremath{136.90\pm0.44}}     
\newcommand{\hatcurXAveccenxxxxxxE}{\ensuremath{0.542\pm0.070}}       
\newcommand{\hatcurXdistredeccenxxxxxxE}{\ensuremath{136.87\pm0.44}}  
\newcommand{\hatcurXEBVeccenxxxxxxE}{\ensuremath{0.175\pm0.023}}      
\newcommand{\hatcurCCpmraeccenxxxxxxE}{\ensuremath{40.325\pm0.028}}   
\newcommand{\hatcurCCpmdececcenxxxxxxE}{\ensuremath{-22.194\pm0.020}} 
\newcommand{\hatcurCCpmeccenxxxxxxE}{\ensuremath{46.029\pm0.034}}     

%% file: newcommand_spec_multi_TOI519.tex
\newcommand{\hatcurxxxxxxA}{TOI~519}
\newcommand{\hatcurbxxxxxxA}{TOI~519\,b}
\newcommand{\hatcurcxxxxxxA}{TOI~519\,c}

\newcommand{\hatcurplanetnumxxxxxxA}{519}

\newcommand{\hatcurCCtwomassshortxxxxxxA}{08182567-1939465}
\newcommand{\hatcurCCgaiadrtwoshortxxxxxxA}{5707485527450614656}
\newcommand{\hatcurCCtoixxxxxxA}{519}
\newcommand{\hatcurCCticxxxxxxA}{218795833}

\newcommand{\hatcurRVgammaabsxxxxxxA}{\hatcurRVgamma{\hatcurplanetnumxxxxxxA}}                           

\newcommand{\hatcurRVgammarelxxxxxxA}{\hatcurRVgamma{\hatcurplanetnumxxxxxxA}}                           

\newcommand{\hatcurCCtassvixxxxxxA}{\ensuremath{NULL\pm NULL}}                  

\newcommand{\hatcurSMEversionxxxxxxA}{i}                                       

\newcommand{\hatcurisoshortxxxxxxA}{YY}
\newcommand{\hatcurisofullxxxxxxA}{Yonsei-Yale (YY)}
\newcommand{\hatcurisocitexxxxxxA}{yi:2001}

\newcommand{\hatcurlumindxxxxxxA}{\arstar}

\newcommand{\hatcurjhkfilsetxxxxxxA}{ESO}

\newcommand{\hatcurSMEteffxxxxxxA}{\ifthenelse{\equal{\hatcurSMEversionxxxxxxA}{i}}{\hatcurSMEiteff{\hatcurplanetnumxxxxxxA}}{\hatcurSMEiiteff{\hatcurplanetnumxxxxxxA}}}
\newcommand{\hatcurSMEzfehxxxxxxA}{\ifthenelse{\equal{\hatcurSMEversionxxxxxxA}{i}}{\hatcurSMEizfeh{\hatcurplanetnumxxxxxxA}}{\hatcurSMEiizfeh{\hatcurplanetnumxxxxxxA}}}
\newcommand{\hatcurSMEzfehshortxxxxxxA}{\ifthenelse{\equal{\hatcurSMEversionxxxxxxA}{i}}{\hatcurSMEizfehshort{\hatcurplanetnumxxxxxxA}}{\hatcurSMEiizfehshort{\hatcurplanetnumxxxxxxA}}}
\newcommand{\hatcurSMEloggxxxxxxA}{\ifthenelse{\equal{\hatcurSMEversionxxxxxxA}{i}}{\hatcurSMEilogg{\hatcurplanetnumxxxxxxA}}{\hatcurSMEiilogg{\hatcurplanetnumxxxxxxA}}}
\newcommand{\hatcurSMEvsinxxxxxxA}{\ifthenelse{\equal{\hatcurSMEversionxxxxxxA}{i}}{\hatcurSMEivsin{\hatcurplanetnumxxxxxxA}}{\hatcurSMEiivsin{\hatcurplanetnumxxxxxxA}}}
\newcommand{\hatcurSMEvmacxxxxxxA}{\ifthenelse{\equal{\hatcurSMEversionxxxxxxA}{i}}{\hatcurSMEivmac{\hatcurplanetnumxxxxxxA}}{\hatcurSMEiivmac{\hatcurplanetnumxxxxxxA}}}
\newcommand{\hatcurSMEvmicxxxxxxA}{\ifthenelse{\equal{\hatcurSMEversionxxxxxxA}{i}}{\hatcurSMEivmic{\hatcurplanetnumxxxxxxA}}{\hatcurSMEiivmic{\hatcurplanetnumxxxxxxA}}}

%% file: newcommand_spec_multi_TOI3629.tex
\newcommand{\hatcurxxxxxxB}{TOI~3629}
\newcommand{\hatcurbxxxxxxB}{TOI~3629\,b}
\newcommand{\hatcurcxxxxxxB}{TOI~3629\,c}

\newcommand{\hatcurplanetnumxxxxxxB}{3629}

\newcommand{\hatcurCCtwomassshortxxxxxxB}{23591015+3918514}
\newcommand{\hatcurCCgaiadrtwoshortxxxxxxB}{2881820324294985856}
\newcommand{\hatcurCCtoixxxxxxB}{3629}
\newcommand{\hatcurCCticxxxxxxB}{455784423}

\newcommand{\hatcurRVgammaabsxxxxxxB}{\hatcurRVgammaB{\hatcurplanetnumxxxxxxB}}                           

\newcommand{\hatcurRVgammarelxxxxxxB}{\hatcurRVgammaB{\hatcurplanetnumxxxxxxB}}                           

\newcommand{\hatcurCCtassvixxxxxxB}{\ensuremath{NULL\pm NULL}}                  

\newcommand{\hatcurSMEversionxxxxxxB}{i}                                       

\newcommand{\hatcurisoshortxxxxxxB}{YY}
\newcommand{\hatcurisofullxxxxxxB}{Yonsei-Yale (YY)}
\newcommand{\hatcurisocitexxxxxxB}{yi:2001}

\newcommand{\hatcurlumindxxxxxxB}{\arstar}

\newcommand{\hatcurjhkfilsetxxxxxxB}{ESO}

\newcommand{\hatcurSMEteffxxxxxxB}{\ifthenelse{\equal{\hatcurSMEversionxxxxxxB}{i}}{\hatcurSMEiteff{\hatcurplanetnumxxxxxxB}}{\hatcurSMEiiteff{\hatcurplanetnumxxxxxxB}}}
\newcommand{\hatcurSMEzfehxxxxxxB}{\ifthenelse{\equal{\hatcurSMEversionxxxxxxB}{i}}{\hatcurSMEizfeh{\hatcurplanetnumxxxxxxB}}{\hatcurSMEiizfeh{\hatcurplanetnumxxxxxxB}}}
\newcommand{\hatcurSMEzfehshortxxxxxxB}{\ifthenelse{\equal{\hatcurSMEversionxxxxxxB}{i}}{\hatcurSMEizfehshort{\hatcurplanetnumxxxxxxB}}{\hatcurSMEiizfehshort{\hatcurplanetnumxxxxxxB}}}
\newcommand{\hatcurSMEloggxxxxxxB}{\ifthenelse{\equal{\hatcurSMEversionxxxxxxB}{i}}{\hatcurSMEilogg{\hatcurplanetnumxxxxxxB}}{\hatcurSMEiilogg{\hatcurplanetnumxxxxxxB}}}
\newcommand{\hatcurSMEvsinxxxxxxB}{\ifthenelse{\equal{\hatcurSMEversionxxxxxxB}{i}}{\hatcurSMEivsin{\hatcurplanetnumxxxxxxB}}{\hatcurSMEiivsin{\hatcurplanetnumxxxxxxB}}}
\newcommand{\hatcurSMEvmacxxxxxxB}{\ifthenelse{\equal{\hatcurSMEversionxxxxxxB}{i}}{\hatcurSMEivmac{\hatcurplanetnumxxxxxxB}}{\hatcurSMEiivmac{\hatcurplanetnumxxxxxxB}}}
\newcommand{\hatcurSMEvmicxxxxxxB}{\ifthenelse{\equal{\hatcurSMEversionxxxxxxB}{i}}{\hatcurSMEivmic{\hatcurplanetnumxxxxxxB}}{\hatcurSMEiivmic{\hatcurplanetnumxxxxxxB}}}

%% file: newcommand_spec_multi_TOI3714.tex
\newcommand{\hatcurxxxxxxC}{TOI~3714}
\newcommand{\hatcurbxxxxxxC}{TOI~3714\,b}
\newcommand{\hatcurcxxxxxxC}{TOI~3714\,c}

\newcommand{\hatcurplanetnumxxxxxxC}{3714}

\newcommand{\hatcurCCtwomassshortxxxxxxC}{04381253+3927299}
\newcommand{\hatcurCCgaiadrtwoshortxxxxxxC}{178924390478792320}
\newcommand{\hatcurCCtoixxxxxxC}{3714}
\newcommand{\hatcurCCticxxxxxxC}{155867025}

\newcommand{\hatcurRVgammaabsxxxxxxC}{\hatcurRVgammaB{\hatcurplanetnumxxxxxxC}}                           

\newcommand{\hatcurRVgammarelxxxxxxC}{\hatcurRVgammaB{\hatcurplanetnumxxxxxxC}}                           

\newcommand{\hatcurCCtassvixxxxxxC}{\ensuremath{NULL\pm NULL}}                  

\newcommand{\hatcurSMEversionxxxxxxC}{i}                                       

\newcommand{\hatcurisoshortxxxxxxC}{YY}
\newcommand{\hatcurisofullxxxxxxC}{Yonsei-Yale (YY)}
\newcommand{\hatcurisocitexxxxxxC}{yi:2001}

\newcommand{\hatcurlumindxxxxxxC}{\arstar}

\newcommand{\hatcurjhkfilsetxxxxxxC}{ESO}

\newcommand{\hatcurSMEteffxxxxxxC}{\ifthenelse{\equal{\hatcurSMEversionxxxxxxC}{i}}{\hatcurSMEiteff{\hatcurplanetnumxxxxxxC}}{\hatcurSMEiiteff{\hatcurplanetnumxxxxxxC}}}
\newcommand{\hatcurSMEzfehxxxxxxC}{\ifthenelse{\equal{\hatcurSMEversionxxxxxxC}{i}}{\hatcurSMEizfeh{\hatcurplanetnumxxxxxxC}}{\hatcurSMEiizfeh{\hatcurplanetnumxxxxxxC}}}
\newcommand{\hatcurSMEzfehshortxxxxxxC}{\ifthenelse{\equal{\hatcurSMEversionxxxxxxC}{i}}{\hatcurSMEizfehshort{\hatcurplanetnumxxxxxxC}}{\hatcurSMEiizfehshort{\hatcurplanetnumxxxxxxC}}}
\newcommand{\hatcurSMEloggxxxxxxC}{\ifthenelse{\equal{\hatcurSMEversionxxxxxxC}{i}}{\hatcurSMEilogg{\hatcurplanetnumxxxxxxC}}{\hatcurSMEiilogg{\hatcurplanetnumxxxxxxC}}}
\newcommand{\hatcurSMEvsinxxxxxxC}{\ifthenelse{\equal{\hatcurSMEversionxxxxxxC}{i}}{\hatcurSMEivsin{\hatcurplanetnumxxxxxxC}}{\hatcurSMEiivsin{\hatcurplanetnumxxxxxxC}}}
\newcommand{\hatcurSMEvmacxxxxxxC}{\ifthenelse{\equal{\hatcurSMEversionxxxxxxC}{i}}{\hatcurSMEivmac{\hatcurplanetnumxxxxxxC}}{\hatcurSMEiivmac{\hatcurplanetnumxxxxxxC}}}
\newcommand{\hatcurSMEvmicxxxxxxC}{\ifthenelse{\equal{\hatcurSMEversionxxxxxxC}{i}}{\hatcurSMEivmic{\hatcurplanetnumxxxxxxC}}{\hatcurSMEiivmic{\hatcurplanetnumxxxxxxC}}}

%% file: newcommand_spec_multi_TOI4201.tex
\newcommand{\hatcurxxxxxxD}{TOI~4201}
\newcommand{\hatcurbxxxxxxD}{TOI~4201\,b}
\newcommand{\hatcurcxxxxxxD}{TOI~4201\,c}

\newcommand{\hatcurplanetnumxxxxxxD}{4201}

\newcommand{\hatcurCCtwomassshortxxxxxxD}{06015391-1327410}
\newcommand{\hatcurCCgaiadrtwoshortxxxxxxD}{2997312063605005056}
\newcommand{\hatcurCCtoixxxxxxD}{4201}
\newcommand{\hatcurCCticxxxxxxD}{95057860}

\newcommand{\hatcurRVgammaabsxxxxxxD}{\hatcurRVgamma{\hatcurplanetnumxxxxxxD}}                           

\newcommand{\hatcurRVgammarelxxxxxxD}{\hatcurRVgamma{\hatcurplanetnumxxxxxxD}}                           

\newcommand{\hatcurCCtassvixxxxxxD}{\ensuremath{NULL\pm NULL}}                  

\newcommand{\hatcurSMEversionxxxxxxD}{i}                                       

\newcommand{\hatcurisoshortxxxxxxD}{YY}
\newcommand{\hatcurisofullxxxxxxD}{Yonsei-Yale (YY)}
\newcommand{\hatcurisocitexxxxxxD}{yi:2001}

\newcommand{\hatcurlumindxxxxxxD}{\arstar}

\newcommand{\hatcurjhkfilsetxxxxxxD}{ESO}

\newcommand{\hatcurSMEteffxxxxxxD}{\ifthenelse{\equal{\hatcurSMEversionxxxxxxD}{i}}{\hatcurSMEiteff{\hatcurplanetnumxxxxxxD}}{\hatcurSMEiiteff{\hatcurplanetnumxxxxxxD}}}
\newcommand{\hatcurSMEzfehxxxxxxD}{\ifthenelse{\equal{\hatcurSMEversionxxxxxxD}{i}}{\hatcurSMEizfeh{\hatcurplanetnumxxxxxxD}}{\hatcurSMEiizfeh{\hatcurplanetnumxxxxxxD}}}
\newcommand{\hatcurSMEzfehshortxxxxxxD}{\ifthenelse{\equal{\hatcurSMEversionxxxxxxD}{i}}{\hatcurSMEizfehshort{\hatcurplanetnumxxxxxxD}}{\hatcurSMEiizfehshort{\hatcurplanetnumxxxxxxD}}}
\newcommand{\hatcurSMEloggxxxxxxD}{\ifthenelse{\equal{\hatcurSMEversionxxxxxxD}{i}}{\hatcurSMEilogg{\hatcurplanetnumxxxxxxD}}{\hatcurSMEiilogg{\hatcurplanetnumxxxxxxD}}}
\newcommand{\hatcurSMEvsinxxxxxxD}{\ifthenelse{\equal{\hatcurSMEversionxxxxxxD}{i}}{\hatcurSMEivsin{\hatcurplanetnumxxxxxxD}}{\hatcurSMEiivsin{\hatcurplanetnumxxxxxxD}}}
\newcommand{\hatcurSMEvmacxxxxxxD}{\ifthenelse{\equal{\hatcurSMEversionxxxxxxD}{i}}{\hatcurSMEivmac{\hatcurplanetnumxxxxxxD}}{\hatcurSMEiivmac{\hatcurplanetnumxxxxxxD}}}
\newcommand{\hatcurSMEvmicxxxxxxD}{\ifthenelse{\equal{\hatcurSMEversionxxxxxxD}{i}}{\hatcurSMEivmic{\hatcurplanetnumxxxxxxD}}{\hatcurSMEiivmic{\hatcurplanetnumxxxxxxD}}}

%% file: newcommand_spec_multi_TOI5344.tex
\newcommand{\hatcurxxxxxxE}{TOI~5344}
\newcommand{\hatcurbxxxxxxE}{TOI~5344\,b}
\newcommand{\hatcurcxxxxxxE}{TOI~5344\,c}

\newcommand{\hatcurplanetnumxxxxxxE}{519}

\newcommand{\hatcurCCtwomassshortxxxxxxE}{04130384+2054550}
\newcommand{\hatcurCCgaiadrtwoshortxxxxxxE}{52359538285081728}
\newcommand{\hatcurCCtoixxxxxxE}{5344}
\newcommand{\hatcurCCticxxxxxxE}{16005254}

\newcommand{\hatcurRVgammaabsxxxxxxE}{\hatcurRVgamma{\hatcurplanetnumxxxxxxE}}                           

\newcommand{\hatcurRVgammarelxxxxxxE}{\hatcurRVgamma{\hatcurplanetnumxxxxxxE}}                           

\newcommand{\hatcurCCtassvixxxxxxE}{\ensuremath{NULL\pm NULL}}                  

\newcommand{\hatcurSMEversionxxxxxxE}{i}                                       

\newcommand{\hatcurisoshortxxxxxxE}{YY}
\newcommand{\hatcurisofullxxxxxxE}{Yonsei-Yale (YY)}
\newcommand{\hatcurisocitexxxxxxE}{yi:2001}

\newcommand{\hatcurlumindxxxxxxE}{\arstar}

\newcommand{\hatcurjhkfilsetxxxxxxE}{ESO}

\newcommand{\hatcurSMEteffxxxxxxE}{\ifthenelse{\equal{\hatcurSMEversionxxxxxxE}{i}}{\hatcurSMEiteff{\hatcurplanetnumxxxxxxE}}{\hatcurSMEiiteff{\hatcurplanetnumxxxxxxE}}}
\newcommand{\hatcurSMEzfehxxxxxxE}{\ifthenelse{\equal{\hatcurSMEversionxxxxxxE}{i}}{\hatcurSMEizfeh{\hatcurplanetnumxxxxxxE}}{\hatcurSMEiizfeh{\hatcurplanetnumxxxxxxE}}}
\newcommand{\hatcurSMEzfehshortxxxxxxE}{\ifthenelse{\equal{\hatcurSMEversionxxxxxxE}{i}}{\hatcurSMEizfehshort{\hatcurplanetnumxxxxxxE}}{\hatcurSMEiizfehshort{\hatcurplanetnumxxxxxxE}}}
\newcommand{\hatcurSMEloggxxxxxxE}{\ifthenelse{\equal{\hatcurSMEversionxxxxxxE}{i}}{\hatcurSMEilogg{\hatcurplanetnumxxxxxxE}}{\hatcurSMEiilogg{\hatcurplanetnumxxxxxxE}}}
\newcommand{\hatcurSMEvsinxxxxxxE}{\ifthenelse{\equal{\hatcurSMEversionxxxxxxE}{i}}{\hatcurSMEivsin{\hatcurplanetnumxxxxxxE}}{\hatcurSMEiivsin{\hatcurplanetnumxxxxxxE}}}
\newcommand{\hatcurSMEvmacxxxxxxE}{\ifthenelse{\equal{\hatcurSMEversionxxxxxxE}{i}}{\hatcurSMEivmac{\hatcurplanetnumxxxxxxE}}{\hatcurSMEiivmac{\hatcurplanetnumxxxxxxE}}}
\newcommand{\hatcurSMEvmicxxxxxxE}{\ifthenelse{\equal{\hatcurSMEversionxxxxxxE}{i}}{\hatcurSMEivmic{\hatcurplanetnumxxxxxxE}}{\hatcurSMEiivmic{\hatcurplanetnumxxxxxxE}}}